\documentclass{elsart}

\usepackage{amssymb} 
\usepackage{amsmath}    
\usepackage{epsfig}
\usepackage{graphicx}

\newcommand{\aap}{A\&A}
\newcommand{\aaps}{A\&A Supl.}

\newcommand{\aj}{AJ}
\newcommand{\apj}{ApJ}
\newcommand{\apjl}{ApJL}
\newcommand{\apjs}{ApJSS}

\newcommand{\mnras}{MNRAS}

\newcommand{\pasj}{PASJ}
\newcommand{\hst}{{\it HST}}

\newcommand{\flex}{Flexion1}
\newcommand{\sflex}{Flexion2}
\newcommand{\fNL}{f_{\rm NL}}

\begin{document}

\begin{frontmatter}
\title{Cluster Lenses}
\author[ad1]{Jean-Paul Kneib\thanksref{corr}} \&
\author[ad2,ad3]{Priyamvada Natarajan\thanksref{corr}}
\address[ad1]{Laboratoire d'Astrophysique de Marseille, CNRS-Universit\'e Aix-Marseille, 38 rue F. Joliot-Curie \cty 13388 Marseille Cedex 13\cny, France}
\address[ad2]{Department of Astronomy, Yale University, \cty New Haven CT 06511, 
\cny USA}
\address[ad3]{Department of Physics, Yale University, \cty New Haven CT 06520, \cny USA}
\thanks[corr]{Author emails: jean-paul.kneib@oamp.fr; priyamvada.natarajan@yale.edu\\
Published in Open Access at: http://www.springerlink.com/content/j183018170485723/}

\begin{abstract}
Clusters of galaxies are the most recently assembled, massive, bound
structures in the Universe. As predicted by General Relativity, given their 
masses, clusters strongly deform space-time in their vicinity.  Clusters act as some of 
the most powerful gravitational lenses in the Universe. Light rays traversing 
through clusters from distant sources are hence deflected, and the resulting images of these 
distant objects therefore appear distorted and magnified. 
Lensing by clusters occurs in two regimes, each with unique observational signatures. 
The  {\bf strong lensing} regime is characterized by effects readily seen by eye, namely, the 
production of giant arcs, multiple-images, and arclets. The {\bf weak lensing} regime is 
characterized by small deformations in the shapes of background galaxies only detectable 
statistically. Cluster lenses have been exploited successfully to address several important current 
questions in cosmology: {\it (i)} {\em the study of the lens(es)} - understanding cluster mass 
distributions and issues pertaining to cluster formation and evolution, as well as
constraining the nature of dark matter; {\it (ii)} {\em the study of
the lensed objects} - probing the properties of the background lensed
galaxy population -- which is statistically at higher redshifts and of 
lower intrinsic luminosity thus enabling the probing of galaxy formation 
at the earliest times right up to the Dark Ages; and {\it (iii)} {\em the study of 
the geometry of the Universe} - as the strength of lensing depends on the 
ratios of angular diameter distances between the lens, source and observer, lens deflections 
are sensitive to the value of cosmological parameters and offer a powerful geometric tool to 
probe Dark Energy. In this review, we present the basics of cluster lensing and provide a 
current status report of the field. 
\end{abstract}

\end{frontmatter}

\section{Introduction and historical perspective}

In the early days of modern cosmology, soon after it was realized that
the Universe was expanding (Hubble 1929; Lema\^{\i}tre 1931; Hubble 1931); Zwicky (1933) suggested 
that some unseen matter was the likely dominant mass component in clusters of 
galaxies. With remarkable prescience,  Zwicky (1937) further noted that gravitational 
lensing by clusters would be an invaluable tool to: {\it (i)} trace and measure the amount of this
unseen mass, now referred to as dark matter and currently thought to pervade the
 cosmos; and {\it (ii)} study magnified distant objects lying behind clusters. Zwicky's bold 
 predictions were based on a profound and intuitive understanding of
the properties of gravitational lensing. However at that time,
inadequate imaging technology coupled with the lack of theoretical
understanding of structure formation in the Universe hampered further
observational progress and discoveries of gravitational lensing effects.

Although the existence of clusters of galaxies has been recognized for nearly 
two centuries - they were first recognized by Messier and
Herschel as "remarkable concentrations of nebulae on the sky" (see the
review of Biviano 2000 and references therein) the study of clusters
began in earnest only really in the 1950s. In particular, the
publication of the first comprehensive cluster catalog of the nearby
Universe by Abell in 1958, can be considered as a milestone that
spurred the study of clusters of galaxies transforming it into an active 
observational research area.

In comparison, gravitational lensing theory developed much later in
the 1960s with early theoretical studies demonstrating the usefulness
of lensing for astronomy. In particular, Sjur Refsdal derived the
basic equations of gravitational lens theory (Refsdal 1964a) and
subsequently showed how the gravitational lensing phenomenon can be used to
determine Hubble's constant by measuring the time delay between two
lensed images (Refsdal 1964b). Following the discovery of quasars,
Barnothy (1965) proposed gravitational lensing as a tool for the study
of quasars. With the discovery of the first double quasar Q0957+561 by
Walsh, Carswell \& Weymann (1979) gravitational lensing really emerged
in astronomy as an active observational field of study.

The study of clusters of galaxies as astronomical objects on the other hand, 
came of age in the 1970s and early 1980s specially with the discovery of the
X-ray emitting intra-cluster medium (Lea et al. 1973; Gull \&
Northover 1976; Bahcall \& Sarazin 1977; Serlemistos et al. 1977;
Cavaliere \& Fusco-Femiano 1978) and the numerous studies of the
stellar populations of galaxies in clusters (Bautz \& Morgan 1970;
Sandage 1976; Leir \& van den Bergh 1977; Hoessl, Gunn \& Thuan 1980;
Dressler 1980). However, there was no discussion of their lensing properties in
theoretical papers till the 1980s. The paper by
Narayan, Blandford \& Nityananda (1984) is one of the
earliest theoretical papers that explored in detail the possibility that clusters 
can act as powerful lenses. As an example, this paper explained large separation 
multiple quasars as likely "cluster-assisted" lensing systems. Although such a 
possibility had been already proposed by Young et al. (1980), who discovered a 
cluster of galaxies near the first double quasar Q0957+561, it was not so obvious 
for most other systems.

The likely explanation for the lack of interest in cluster
lensing research was probably the belief that clusters were rather
diffuse/extended systems and therefore not dense enough to act as
powerful light deflectors. Only with the establishment of the role of cold
dark matter in structure formation, did it become clear that clusters
are indeed repositories of vast amounts of dark matter that enable
them to act as efficient lenses in the Universe. The theory of
structure formation in the context of a cold dark matter dominated
Universe was developed in a seminal paper by Blumenthal et
al. (1984). An attractive feature of this cold dark matter hypothesis
was its considerable predictive power: the post-recombination
fluctuation spectrum was calculable, and it in turn governs the
formation of galaxies and clusters. At that time, good agreement with
the data was obtained for a Zel'dovich spectrum of primordial
fluctuations. Several decades later, a version of this paradigm the 
$\Lambda$ Cold Dark Matter ($\Lambda$CDM hereafter) model
which postulates the existence of a non-zero cosmological constant 
$\Omega_{\Lambda}$ is currently well 
established and is in remarkable agreement with a wide range of current 
observations on cluster and galaxy-scales.

\begin{figure}
\includegraphics[width=\textwidth]{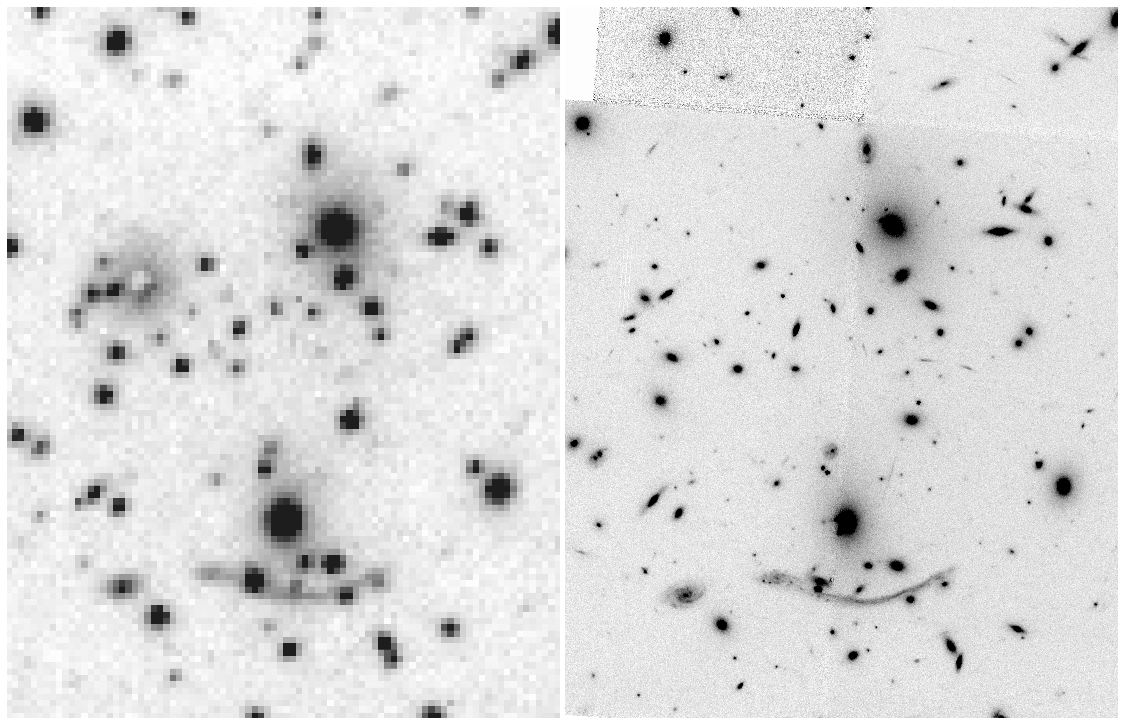} 
\caption{The galaxy cluster Abell 370 as observed by CFHT in 1985 (left) with
one of the first CCD cameras (R-band), in which the first gravitationally
lensed arc was later identified (Lynds \& Petrosian 1986; Soucail et
al. 1987a, 1987b).  For comparison, the image on the right shows the 
Hubble Space Telescope image of the same cluster Abell 370 taken with 
the WFPC-2 camera with the F675W filter in December 1995 (Soucail et al 1999). Most of the bright galaxies seen 
are cluster members at $z = 0.375$, whereas the arc, i.e. the highly elongated 
feature, is the image of a background galaxy at redshift $z = 0.724$ (Soucail et al. 1988). The
image is oriented such that North is on top, East to the left and the
field of view is roughly 40 x 60 arcsec$^2$.\label{figa370} }
\end{figure}

Nevertheless, it still came as quite a surprise when in 1986, Lynds \&
Petrosian (1986) and Soucail et al. (1987) independently discovered
the first "giant arcs": the strongly elongated images of distant background galaxies in the core of
massive clusters (see Figure \ref{figa370}). This new phenomenon was then immediately identified
by Paczyn\'ski (1987) as the consequence of gravitational lensing by the
dense centers of clusters, and was soon confirmed by the measurement of
the redshift of the arc in Abell~370 (Soucail et al.~1988). The
discovery of giant arcs revealed the existence of the strong lensing
regime, however as we know now, it only represents the tip of the iceberg!  

Coupled with the
growing theoretical understanding of the structure and assembly history of clusters, this observational 
discovery of cluster lensing opened up an entire new vista to probe the detailed distribution of dark matter 
in these systems. In 1990, Antony Tyson while obtaining deep CCD imaging of clusters,
identified a "systematic alignment" of faint galaxies around cluster
cores (Tyson, Wenk \& Valdes 1990). He then suggested that this weak alignment produced 
by the distortion due to lensing by clusters could be used to map dark matter at larger radii in
clusters than strong lensing afforded. These two key discoveries of strong 
and weak lensing respectively opened up a rich, new field in astronomy, the 
study of "cluster lenses", which we discuss further in this review.

These observational discoveries stoked the theoretical community to
produce a number of key papers in the first half of the 1990s that
developed the theoretical framework for strong and weak lensing
techniques. Several of the seminal papers date from this period, and
theorists delved into quantifying this new territory of gravitational
lensing.  Some of the significant early papers are: Schneider (1984);
Blandford \& Narayan (1986); Blandford, Kochanek, Kovner \& Narayan
(1989); Kochanek (1990); Miralda-Escude (1991); Kaiser (1992); Kaiser
\& Squires (1993). It is important to underline that significant advances in
technology spurred the field dramatically during these years. The
discovery of the lensing phenomenon in clusters was made possible thanks
to the successful development of CCD imaging that allowed deeper and
sharper optical images of the sky, as well as deep spectroscopy -
essential to measure the spectrum and the redshifts of the faint lensed background 
galaxies. Another technological revolution was in preparation at that time, a telescope
above the atmosphere: the {\it Hubble Space Telescope} (\hst). \hst\
has dramatically impacted cluster lensing studies, and, in particular,
that of the strong lensing regime. Although launched in 1991, \hst\
did not make a strong impact at first, as its unforeseen "blurred
vision" made the faint images of distant galaxies inadequate for
lensing work. Nevertheless, even with the first \hst-WFPC1 (Wide Field Planetary
Camera) images of Abell 370 and AC114 one could already see the potential power of {\it
Hubble} for lensing studies.

In December 1993, with the first successful servicing mission and the installation 
of the odd shaped WFPC2 camera, Hubble recovered its image sharpness, and it is
not surprising that one of the first image releases following the
installation of WFPC2 was the astonishing view of the cluster lens
Abell 2218 (Kneib et al. 1996), which is iconic and has been included in most 
recent introductory astronomy textbooks.

Image sharpness is one of the key pre-requisites for studying lensing by
clusters (e.g. Smail \& Dickinson 1995), and unsurprisingly another requirement is a large image field
of view. The strong lensing regime in clusters corresponds to the
inner one arc-minute region around the cluster center. Typically, 
cluster virial radii are of the order of a few Mpc, which corresponds to $\sim
15$\,arcminutes for a cluster at $z\sim 0.2$. Therefore, to go beyond the inner regions 
and to measure the weak lensing signal from cluster outskirts, cameras with a
sufficiently large field of view are required to ideally cover the full size of a
cluster in one shot (e.g. Kaiser et al 1998, Joffre et al 2000).

From the second half of the 1990s we have seen the rapid development of wide field
imaging cameras such as: the UH8k followed by CFHT12k at CFHT (Canada France Hawaii
Telescope); Suprime at the Subaru Telescope; WFI at the 2.2m telescope at ESO (European Southern
Observatory); the Megacam camera at CFHT; 
the Gigacam of Pan-STARRS (PS-1); the OmegaCam of the VST
and soon the Dark Energy Camera at CTIO (Cerro Tololo Inter-American Observatory).  
These cameras are composed of a mosaic of
many large format CCDs (4k$\times$2k or larger) allowing coverage of a large field (ranging 
from a quarter of a square degree up to a few square degrees).  The making of these
instruments was strongly motivated by the detection of the weak lensing
distortion of faint galaxies produced by foreground clusters and intervening large scale
structure, the latter effect is commonly referred to as "cosmic shear".

In parallel, techniques to accurately measure the gravitational shear
were also developed. The most well documented is the ``KSB'' technique
(Kaiser, Squires \& Broadhurst 1995) which is implemented in the
commonly used {\sc imcat} software package,\footnote {IMCAT software is available at 
http://www.ifa.hawaii.edu/~kaiser/imcat/} which has been since improved by 
several groups. The accuracy of shape measurements for distorted background images
 is key to exploiting lensing effects. The difficulty in the shear measurement arises as galaxy
  ellipticities need to be measured extremely accurately given that there are other 
  confounding sources that generate distortions. Spurious distortions are induced by 
  the spatially and temporally  variable PSF (Point Spread Function) 
   as well as by intrinsic shape correlations that are 
unrelated to lensing (Crittenden et al. 2001; 2002). Corrections that carefully take into account these 
additional and variable sources of image distortion have been incorporated into shape
measurement algorithms like {\sc lensfit}\footnote{LENSFIT software is available at - http://www.physics.ox.ac.uk/lensfit/ } (Miller et al. 2007; Kitching et al. 2008). Although the ``KSB'' technique has been quite popular due to its speed and
efficiency, many new implementations for extracting the shear signal with the rapid increase in 
the speed and processing power of computers are currently available.

The first weak lensing measurements of clusters were reported with 
relatively small field of view cameras (Fahlman et al. 1994; Bonnet et
al. 1994) but were soon extended to the larger field of view
mosaic cameras (e.g. Dahle et al. 2002; Clowe \& Schneider 2001, 2002;
Bardeau et al. 2005, 2007). Two-dimensional dark matter mapping gets
rapidly noisy as one extends over more than $\sim$2 arcminutes from the
cluster center due to a rapidly diminishing lensing signal. However, radial 
averaging of the shear field provides an effective way to probe
the mass profile of clusters out to their virial radius and even beyond. This
technique of inverting the measured shear profile to constrain the
mass distribution of clusters is currently widely used. Combining constraints
from the strong and weak lensing regime has enabled us to derive the dark matter 
density profile over a wide range of physical scales. As a consequence, gravitational lensing 
has become a powerful method to address fundamental questions pertinent to cluster growth and assembly.

Theoretically, as it is known that clusters are dominated by dark matter, enormous progress has been
made in tracking their formation and evolution using large cosmological N-body simulations since 
the 1980s. Gravitational lensing is sensitive to the total mass of clusters, thereby enabling detailed
comparison of the mass distribution and properties inferred observationally with simulated clusters. 
Lensing observations have therefore allowed important tests of the standard structure formation paradigm.   

At the turn of the second millennium the new role of lensing clusters
is its growing use as natural telescopes to study very high-redshift galaxies
 that formed during the infancy of the Universe
({\it e.g.} Franx et al. 1997; Pell{\' o} et al. 1999; Ellis et al. 2001).
This became possible with deep spectroscopy on 4~m and then 8-10~m
class telescopes that enable probing the high-redshift Universe, primarily 
by exploiting the lensing amplification and magnification\footnote{the magnification refers to the spatial stretching of the images by the gravitational lensing effect, however the magnification cannot be recognized when the lensed object is not resolved by the observations (if the object is compact or if the PSF is broad) leading to an apparent amplification of the flux of the lensed object. In some cases, a lensed object can be tangentially magnified but radially amplified, the use of the terms magnification and amplification are thus sometimes mixed.}
 produced by these natural telescopes (Pello et al 2001). Capitalizing
on the achromatic nature of cluster lensing, various observatories functioning at 
different wavelengths of the electromagnetic spectrum have been deployed for these 
studies. In particular, the discovery and study of the population of sub-millimeter galaxies 
using SCUBA at the James Clerk Maxwell Telescope (JCMT hereafter; see the reviews by Blain et al. 2002; Smail et al. 2002;
Kneib et al. 2004; Knudsen et al. 2005; Borys et al. 2005, Knudsen et al. 2008), the Caltech
interferometer at Owens Valley ({\it e.g.} Frayer et al. 1998; Sheth et al. 2004), the IRAM
interferometer ({\it e.g.} Neri et al. 2003; Kneib et al. 2005),  the Very
Large Array (VLA) ({\it e.g.} Smail et al. 2002; Ivison et al. 2002; Chapman et
al. 2002) and Sub-Millimeter Array (SMA) ({\it e.g.} Knudsen et al. 2010) greatly benefited from the boost 
provided by the magnification effect of gravitational lensing in
cluster fields.  Similarly, observation of lensed galaxies in the
mid-infrared with the ISOCAM mid-infrared camera on the Infra-red Space Observatory (ISO) satellite
(Altieri et al. 1999; Metcalfe et al. 2003), followed with the {\it
Spitzer} observatory (Egami et al. 2005) and now with the {\it Herschel} 
space observatory (Egami et al. 2010; Altieri et al. 2010) have pushed the
limits of our knowledge of distant galaxies further. Gravitational lensing is
now recognized as a powerful technique to count the faintest galaxies
in their different classes: Extremely Red Objects (Smith et al. 2001);
Lyman-$\alpha$ emitters at $z\sim 4-6$ (Hu et al. 2002; Santos et al.
2004, Stark et al. 2007); Lyman-break galaxies at $z\sim 6-10$ (Richard et al. 2008) as
well as to study in detail the rare, extremely magnified individual sources
(Pettini et al. 2000; Kneib et al. 2004; Egami et al. 2005, Smail et al. 2007, Swinbank et al. 2007, 2010) 
in the distant Universe.

\begin{figure}
\includegraphics[width=\textwidth]{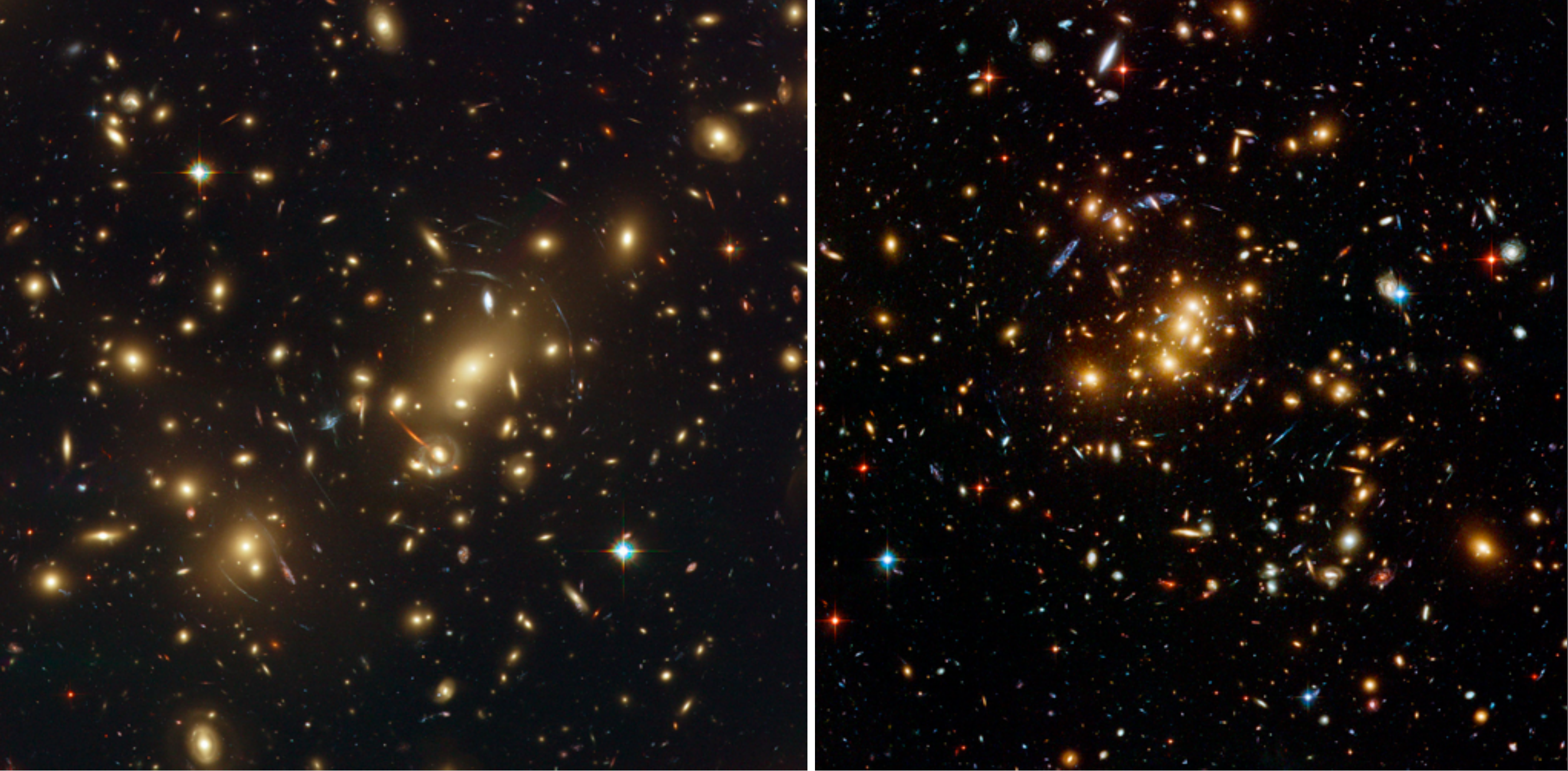}
\caption{Color image of two cluster lenses observed by \hst-ACS: Left panel - Abell 2218 at $z=0.175$ and Right panel - 
Cl0024+1654 at $z=0.395$.\label{fig:color2218_0024} }
\end{figure}

Since March 2002, the installation of the new ACS camera onboard \hst\ has provided
further observational advances in the study and unprecedented use of cluster
lenses (see Figure~\ref{fig:color2218_0024}). These are exemplified in the very deep and spectacular 
ACS images of Abell 1689 (Broadhurst et al. 2005; Halkola, Seitz \& Pannella  2006). This color 
image reveals more than 40 multiple-image systems in the core of this cluster
(Limousin et al. 2007) and well over a hundred lensed images in total. The
dramatic increase in the number of strong lensing constraints that these observations provide
 in the cluster core has spurred important and significant new developments in mass 
 reconstruction techniques ({\it e.g.} Diego et al. 2005a,b; Jullo et al. 2007, 2009; Coe et al. 2008). With this
 amount of high quality data the construction of extremely high-resolution mass
models of the cluster core are now possible. Mass models with high precision have enabled 
the use of this cluster to constrain the cosmological parameters $\Omega_{\rm m}$ and $\Omega_{\Lambda}$ 
(Link and Pierce 1998; Golse et al. 2004; Gilmore \& Natarajan 2009; Jullo et al. 2010; D'Aloisio \& Natarajan 2011). First 
observational constraints were attempted by Soucail et al. (2004), and more recent work by Jullo et al. (2010) has demonstrated the 
feasibility of this technique involving detailed modeling of deep ACS images coupled with comprehensive 
redshift determinations for the numerous multiple-image systems. Combining these cosmological constraints 
from the cluster lens Abell 1689 with those obtained from independent X-ray measurements and a flat 
Universe prior from WMAP,  Jullo et al. (2010) find results that are competitive with the other more 
established methods like SuperNovae (Riess et al. 1998; Perlmutter et al. 1999) and Baryonic Acoustic 
Oscillations (Eisenstein et al. 2005).  Therefore, in the very near future cluster strong lensing is likely to 
provide us with a viable complementary technique to constrain the geometry of the Universe and probe the equation of 
state of Dark Energy, which is a key unsolved problem in cosmology today.

This brief and non-exhaustive historical account of cluster lensing
research summarizes some of the important scientific results gathered
up to now and demonstrates the growing importance of cluster lensing
in modern cosmology. This review is organized as follows: we
first describe the key features of gravitational lensing in clusters
of galaxies, starting with strong lensing, and then summarize the
various weak lensing techniques as well as some recent developments in
the intermediate lensing regime. We also dedicate a section to the
lensing effect and measurements of galaxy halos in clusters which has
provided new insights into the granularity of the dark matter distribution. 
The potency here arises from the ability to directly compare lensing inferred properties for 
substructure directly with results from high-resolution cosmological N-body simulations.
We then present the different uses of cluster lenses in modern cosmology.  We
start with the study of the lens: its mass distribution, and the
relation of the lensing mass to other mass estimates for clusters. We
then discuss the use of cluster lenses as natural telescopes to study
faint and distant background galaxy populations. And lastly, we
discuss the potential use of clusters to constrain cosmological
parameters. Finally, we recap the important developments that are
keenly awaited in the field, and describe some of the exciting science
that will become possible in the next decade, focusing on future
facilities and instruments. Cluster lensing is today a rapidly evolving and 
observationally driven field.

When necessary, we adopt a flat world model with a Hubble constant
$H_0 = 70$\,km\,s$^{-1}$\,Mpc$^{-1}$, a density parameter in matter
$\Omega_{\rm m} = 0.3$ and a cosmological constant $\Omega_\lambda =
0.7$. Magnitudes are expressed in the AB system.


\section{Lensing theory as applied to clusters of galaxies}

\subsection{General description}

\begin{figure}
\epsfig{file=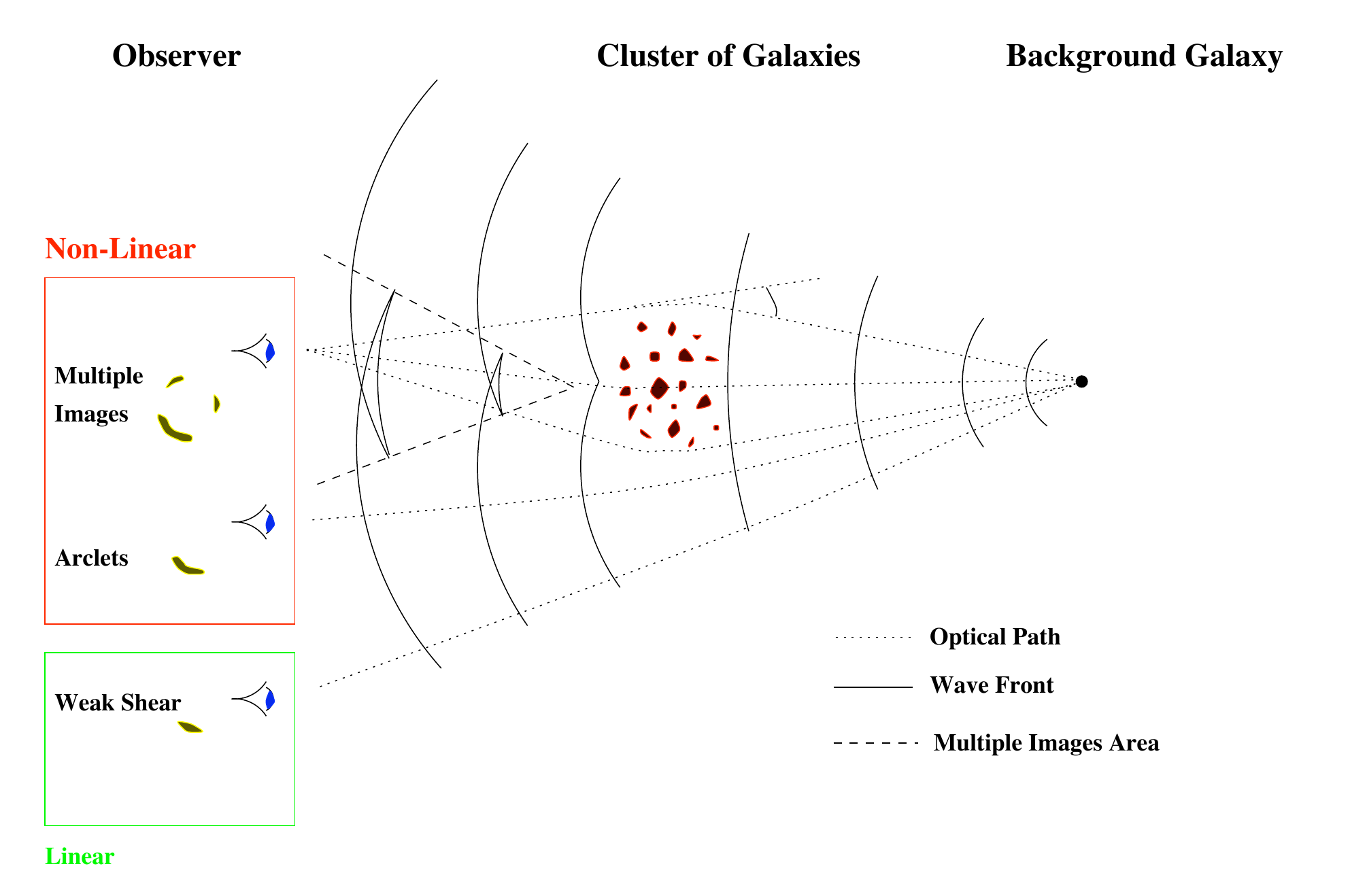,width=\textwidth}
\caption{Gravitational lensing in clusters: A simple schematic of
how lensed images are produced delineating the various regimes: strong, 
intermediate and weak lensing (see text for a detailed description).
\label{figjpk1}
}
\end{figure}

Clusters of galaxies are the largest and most massive bound structures
in the Universe.  Due to their large mass, galaxy clusters (as do
galaxies) locally deform space-time (see Figure~\ref{figjpk1}).
Therefore, the wave front of light emitted by a distant source
traversing a foreground galaxy cluster will be distorted. This
distortion occurs regardless of the wavelength of light as the effect
is purely geometric. Moreover, for the most massive clusters the mass
density in the inner regions is high enough to break the wave front coming from
a distant source into several pieces, thereby occasionally producing multiple-images of the
same single background source. Background galaxies multiply imaged in
this fashion tend to form the observed extraordinary gravitational
giant arcs that characterize the so-called {\it strong lensing}
domain. Strongly lensed distant galaxies will thus appear distorted and highly magnified.
They are often referred to as arclets due to their noticeably elongated
shape and preferential tangential alignment around the cluster center. 
Note however that their observed distorted shape is a combination of 
their intrinsic shape and the distortion induced by the lensing effect of the cluster.

When the alignment between the observer, a cluster and distant background galaxies is
less perfect, then the distortion induced by the cluster will be less
important and cannot be recognized clearly. Statistical methods
are required to detect this change in shape of background galaxies seen in the {\it weak regime}.  
In the weak lensing regime, the observed shapes of background galaxies in the field of the 
cluster are typically dominated by their intrinsic ellipticities or even worse 
by the distortion of the imaging camera optics and the imaging 
point spread function (PSF) which is a function of position on the detector and may
 also vary with time. Thus, only a careful statistical 
analysis correcting the observed images for the various non-lensing induced distortion effects can 
reveal the true weak lensing signal. The shape changes induced in the outskirts
of clusters in the weak regime are at the few percent level, while the strong
lensing distortions are often larger, and are typically at the 10\% - 20\% level. 

\subsection{Gravitational Lens Equation}

Before proceeding to the elegant mathematics of lensing, we first
recap the assumptions needed to derive the basic lens equation.  First, it
is assumed that the "Cosmological Principle" (i.e. the Universe is
homogeneous and isotropic) holds on large scales. The scales
under consideration here are the ones relevant to the long-range
gravitational force:
\begin{equation}
L \sim { c\over \sqrt{G\bar\rho}} \sim 2 {\rm \,Gpc},
\end{equation}
where $c$ is the speed of light, $G$ is the gravitational constant and
$\bar\rho$ is the mean density of the Universe. The large scale
distribution of galaxies as determined by surveys like the 2 degree Field 
survey (2dF), the Sloan Digital Sky Survey (SDSS) and the Cosmic 
Microwave Background (CMB) as revealed by the Cosmic Background Explorer (COBE), and the
Wilkinson Anisotropy Probe (WMAP) satellites are in good agreement with the cosmological
principle. The assumption of homogeneity and isotropy imposes strong 
symmetries on the metric that describes the Universe and allows solutions that
correspond to both expansion and contraction. Symmetries restrict the metric that 
describes space-time to the following form:
\begin{equation}
ds^2=c^2dt^2-a^2(t)\Biggl({dr^2 \over 1-kr^2} +r^2d\theta^2
+r^2\sin^2\theta d\varphi^2 \Biggr),
\end{equation}
where $a(t)$ is the scale factor, and $k$ defines the curvature of the
Universe.

This metric will be locally perturbed by the presence of any dense 
mass concentration, such as individual stars, black holes, galaxies or 
clusters of galaxies. The Schwarzschild solution (e.g. Weinberg 1992) gives
 the form of the metric near a point mass, and is easy to generalize for a continuous 
 mass distribution in the stationary weak field limit corresponding to $\Phi<<c^2$:
\begin{equation}
ds^2 =(1+{2\Phi\over c^2})c^2dt^2- (1-{2\Phi\over c^2})dr^2,
\label{eq_metric}
\end{equation}
where $\Phi$ is the 3D gravitational potential of the mass
distribution under consideration.

\begin{figure}
\epsfig{file=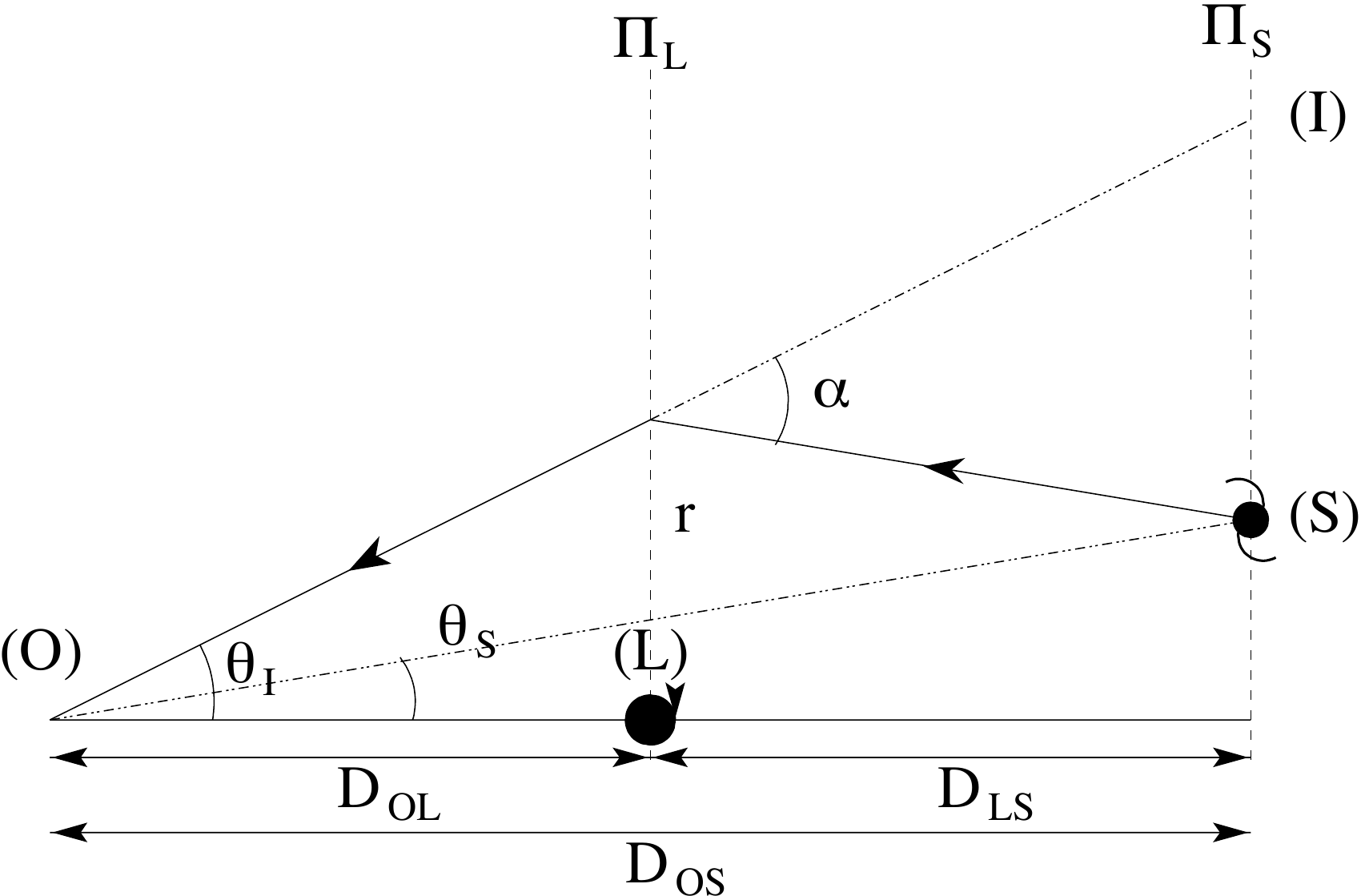,width=\textwidth}
\caption{A single deflector lensing configuration showing the relevant angles
and distances that appear in the lens equation.
\label{fig_geometry}}
\end{figure}

If we consider a simple configuration of a single thin deflecting lens 
(Figure~\ref{fig_geometry}), the observer (O) will see the image (I)
of the source (S) deflected by the lens (L). The geometric equation
relating the position of the source $\vec\theta_S$ to the position of
the image $\vec\theta_I$ depends on the deflection angle $\vec\alpha$
and the relevant intervening angular diameter distances $D_{ij}$ in this case between 
the lens and source (denoted by $D_{LS}$) and the observer and source (denoted by 
$D_{OS})$:
\begin{equation}
\vec\theta_I = \vec\theta_S + {D_{LS}\over D_{OS}}  \vec\alpha(\vec\theta_{I}).
\label{eq_geometry}
\end{equation}

The value of $\vec\alpha$ depends on the local perturbation of the
mass on space-time measured at the location of $\vec\theta_{I}$. 
The photon path follows a null geodesic that is
defined by $ds^2=0$. Hence from Equation \ref{eq_metric}, one can
determine the travel time $t_T$ for a given path length which in turn, is a
function of the angle $\vec\alpha$. By applying Fermat's principle,
which states that light follows the path with a stationary travel
time, i.e. $ {dt_T / d\vec{\theta}_I}=\vec 0$, we can derive the value of the deflection 
$\vec\alpha$ as a function of the local Newtonian gravitational
potential:
\begin{equation}
\vec \alpha(\vec{\theta}_I)=
{2\over c^2}{D_{LS}\over D_{OS}}\
\vec\nabla_{\vec{\theta}_I}\phi^{2D}_N(\vec{\theta}_I),
\label{eq_defalpha}
\end{equation}
where  $\phi^{2D}_N$ is the Newtonian gravitational potential projected in the lens plane.

Combining Equations \ref{eq_geometry} and \ref{eq_defalpha} we 
derive the {\sl lens equation} under the thin lens approximation, which holds 
for a wide range of deflector masses, from stars to galaxies to clusters of galaxies 
(see Schneider, Ehlers \& Falco 1992 for a more detailed derivation):

\begin{equation}
{\vec \theta_S} = {\vec \theta_I} - 
   {2 {\cal E} \over c^2}   {\vec \nabla}\phi^{2D}_N({\vec \theta_I})
= {\vec \theta_I} - {\vec \nabla}\varphi({\vec \theta_I}).
\label{eq_lensdeflection}
\end{equation}

\begin{figure}
\epsfig{file=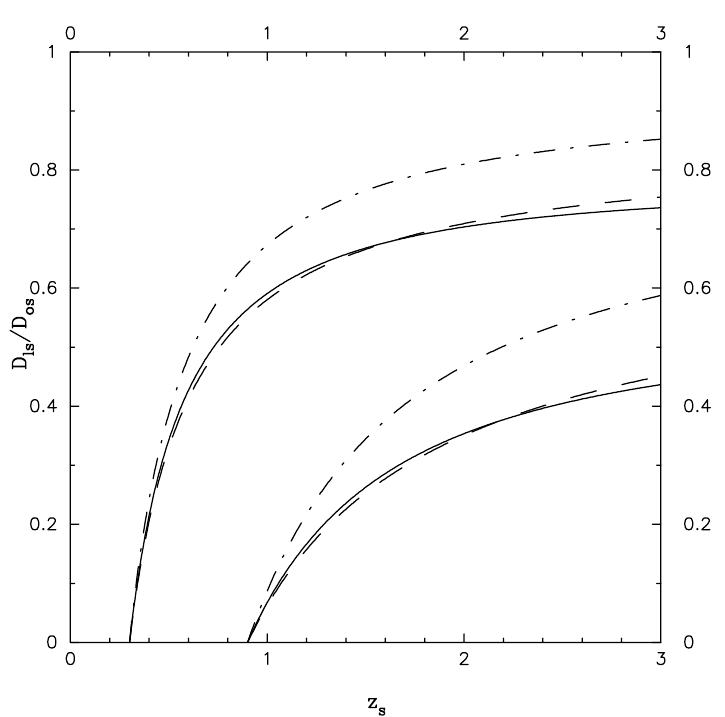,width=\textwidth}
\caption{Lensing efficiency ${\cal E} = D_{LS}/D_{OS}$ for a given
lens as a function of source redshift $z_S$ for different cosmologies. The two sets
of curves correspond to two different lens redshifts $z_L=0.3$ and $z_L=0.9$ and the  
solid lines correspond to $\Omega_{\rm m}=0.1$, $\Omega_{\Lambda}=0$;
the dashed line to $\Omega_{\rm m}=1$, $\Omega_{\Lambda}=0$; 
and the dashed-dotted line to $\Omega_{\rm m}=0.1$, $\Omega_{\Lambda}=0.9$.
\label{fig_efficiency}
}
\end{figure}

The thin lens approximation is valid when the distances from the 
observer to the lens and source are significantly larger than the 
physical extent of the lens, an assumption that is strictly true for
all galaxies and clusters. Above we define $\varphi$ as the lensing 
potential - a lensing normalized version of the Newtonian projected 
potential, and the distance ratio ${\cal E} = D_{LS}/D_{OS}$ which depends on the
redshift of the cluster $z_L$ and the background source $z_S$, as well
as - but only weakly - on the cosmological parameters $\Omega_{\rm m}$ and
$\Omega_\lambda$.  The distance ratio ${\cal E}$ measures the
efficiency of a given lens at redshift $z_L$. The factor ${\cal E}$ is
an increasing function of the source redshift $z_S$ (Figure
\ref{fig_efficiency}); therefore the larger the background source redshift, the stronger
the deflection and distortion. This relation can be slightly more
complex for sources located in the strong lensing regions. Note also that ${\cal E}$ is
independent of the Hubble constant, therefore lensing
deflection angles and deformations are independent of the value of $H_0$.

It has also been shown that going beyond the thin lens approximation,
the above lensing equation can be derived in the more general case (with 
Equation \ref{eq_geometry} being the limiting case for Einstein de-Sitter space-time) 
by simply calculating the null geodesics intersecting an observer's
world-line without partitioning light paths into near and far lens
regions (see Pyne \& Birkinshaw 1996 for a detailed derivation).  The particularly interesting 
case is when more than one lensing deflector is responsible
for producing the observed magnification and shear.  Observations
suggest that the lensing effect of most clusters are likely further amplified due to 
the existence of multiple additional mass concentrations aligned along
the line of sight. Therefore, multiple lens planes will ultimately need to be 
taken into account for accurate mass modeling of cluster lenses. The precise 
coupling between the lensing effects of two adjacent masses depends on their transverse separation. Examining 
the two-screen gravitational lens, Kochanek \& Apostolakis
(1988) find, albeit for galaxy-scale lenses, that their effects
interact significantly for transverse separations less than $4 \times
r_0$ where $r_0$ is the radius of the outer critical line of the
singular potential.\footnote{Critical lines and caustics are defined in the next
  subsection.} Independent lenses that are close
in redshift almost always interact and these interactions can lead to
either an increase or a decrease in the total cross section relative to
the cross section of two isolated lenses depending on the system's
geometry. The resultant image geometries in such cases are dominated by the effects 
of fold caustics. The deflection can be calculated for the two-screen lens configuration
numerically and most current lens equation solvers are adapted to do
so.

\subsection{Gravitational Lens Mapping}

\begin{figure}
\centerline{
\epsfig{file=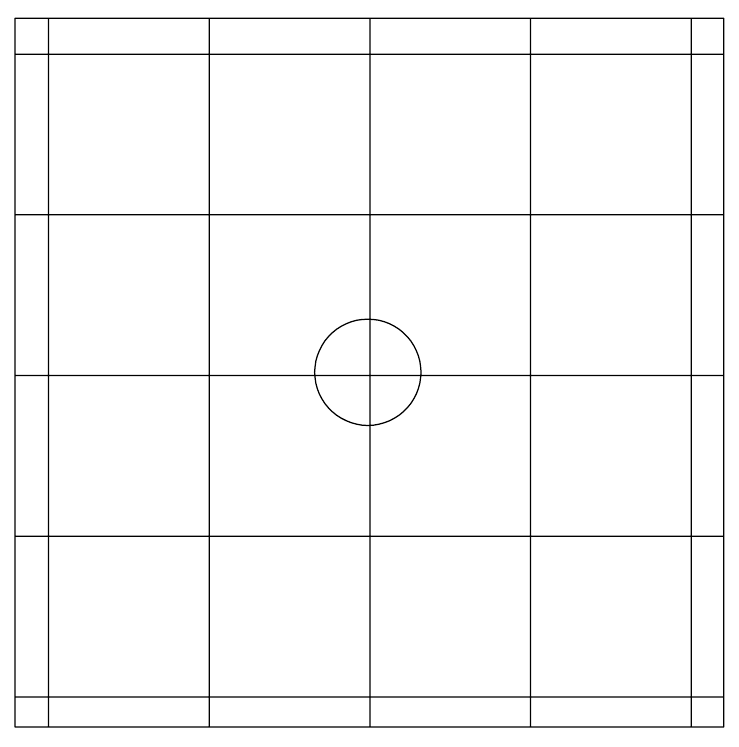,width=0.5\textwidth}
\epsfig{file=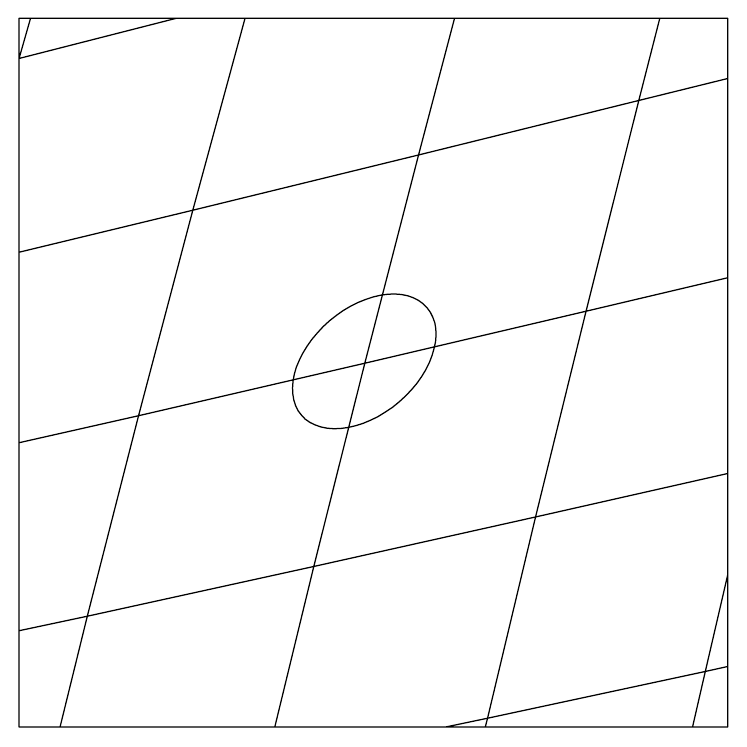,width=0.5\textwidth}}
\caption{Illustration of the effect of lensing: local deformation of a regular grid and a circle (left: source map) by a lens with
constant value of the convergence $\kappa$ and the shear $\gamma$ over the region (right: image map).
\label{lensmapping}}
\end{figure}

The effect of gravitational lensing can be modeled as a mathematical transformation
of source shapes into observed image shapes. The lensing transformation is thus a 
mapping from the source plane (S) to the image plane (I) [See Figure \ref{lensmapping}]. In the case of a single lens plane,
the Hessian of this transformation (also called the magnification matrix) relates to first order 
a source element of the image ($d\vec \theta_I$) to the source plane ($d\vec
\theta_S$) in the following way, in Cartesian and polar coordinates, respectively:
\begin{equation}
{{d\vec \theta_S} \over {d\vec \theta_I}}= {\cal A}^{-1} =
\begin{pmatrix}
1 - \partial_{xx} \varphi & - \partial_{xy} \varphi\cr
- \partial_{xy} \varphi & 1 - \partial_{yy} \varphi \cr
\end{pmatrix}
=
\begin{pmatrix}
1-\partial_{rr}\varphi &
        -\partial_{r}\left({1\over r}\partial_{\theta}\varphi\right)\cr
-\partial_{r}\left({1\over r}\partial_{\theta}\varphi\right)&
 1-{1\over r}\partial_{r}\varphi -{1\over r^2}\partial_{\theta\theta}\varphi\cr .
\end{pmatrix}
\end{equation}
This matrix is referred to as the magnification/amplification matrix and it is
conventionally written as:
\begin{equation}
{\cal A}^{-1} =
\begin{pmatrix}
1 - \kappa - \gamma_1 & -\gamma_2 \cr
-\gamma_2 & 1 - \kappa + \gamma_1 \cr,
\end{pmatrix}
\end{equation}
where the {\em convergence} is defined as
$\kappa=\Delta\varphi/2=\Sigma/\Sigma_{crit}$ and the {\em shear} vector
(also often denoted as a complex number) $\vec\gamma =
(\gamma_1,\gamma_2)$ as:
\begin{equation}
\gamma_1 = (\partial_{xx}\varphi-\partial_{yy}\varphi)/2\ \ \ \ \
\gamma_2 = \partial_{xy}\varphi,
\end{equation}
and the norm is given by:
\begin{equation}
2\gamma =
 \sqrt{(\partial_{xx}\varphi-\partial_{yy}\varphi)^2+(2\partial_{xy}\varphi)^2}.
\end{equation}
The term $\Sigma_{crit}$ is the critical lensing surface density defined as:
\begin{equation}
\Sigma_{crit}= {c^{2}\over 4\pi G}{D_{OS}\over D_{LS}D_{OL}} = {c H_{0}\over 4\pi G}{D_{OS}\over D_{LS}}
{c/H_{0}\over D_{OL}}.
\end{equation}
It can be clearly seen that the critical surface mass density scales as:
\begin{equation}
\Sigma_{crit} \simeq 0.162\,
\left({H_0\over 70 {\rm kms^{-1}Mpc^{-1}}}\right)
\left({D_{OS}\over D_{LS}}\right)
\left({c/H_{0}\over D_{OL}}\right)\,\, {\rm g\,cm^{-2}}.
\end{equation}

For instance, given a cluster lens at $z_{L}=0.3$ and a source at redshift $z_{S}=1.0$, ${D_{OS}\over D_{LS}}=1.567$ and
${c/H_{0}\over D_{OL}}=4.661$, yielding:
\begin{equation}
\Sigma_{crit} \simeq 1.18\,
\left({H_0\over 70\, {\rm kms^{-1}Mpc^{-1}}}\right)\,\, {\rm g\,cm^{-2}}
\,.
\end{equation}

Thus, for a cluster with a depth of $\sim$300 kpc, the 3D mass density needed 
to reach the lensing critical surface mass density 
is about 10$^{-24}$g/cm$^{3}$, which corresponds to a density that is $\sim$10,000 times the critical 
density of the Universe $\rho_{crit}$. Background galaxies viewed via a cluster region where the 
surface mass density is critical or higher are likely to be multiply imaged.

We can readily see that the magnification matrix is real and
symmetric, therefore, it can be diagonalized, and can be written in its
principal axes as follows:
\begin{equation}
{\cal A}^{-1}= \left(
 \begin{array}{cc}
1-\kappa+\gamma & 0\\
0 & 1-\kappa-\gamma
\end{array}
\right)
= (1-\kappa)\ \left[ \left(
 \begin{array}{cc}
1 & 0\\
0 & 1
\end{array}
\right)
+{\gamma\over 1-\kappa}
\left(
 \begin{array}{cc}
1 & 0\\
0 & -1
\end{array}
\right)
\right].
\end{equation}

From this equation, we see that $1-\kappa$ describes the isotropic
deformation, and the shear $\gamma$
describes the anisotropic deformation. 
Note that the quantity that is most directly measured from 
faint galaxy shapes is the reduced shear $g$ defined as: $g = \gamma/(1 - \kappa)$.

The direction of the
deformation (or equivalently of the shear) can be written as:
\begin{equation}
\tan 2\theta_{\rm shear} 
= {2\partial_{xy}\varphi \over \partial_{yy}\varphi-\partial_{xx}\varphi }.
\end{equation}

As the direction of the shear is a ratio of the components of the
lensing potential, the shear direction $\theta_{\rm shear}$ will be
independent (modulo 90 degrees) of the distance ratio ${\cal
E}=D_{LS}/D_{OS}$ and thus will be independent of the source redshift
$z_S$. Only the intensity or magnitude of the shear will change with
the source redshift $z_S$. 

\subsection{Critical and caustic lines}

The magnification $\mu$ is defined as the determinant of the
magnification matrix and can be expressed as a function of $\kappa$
and $\gamma$ as:
\begin{equation}
\mu^{-1}=\det({\cal A}^{-1})=(1-\kappa)^2-\gamma^2 = (1-\kappa)^2(1-g^{2}).
\label{magdef}
\end{equation}

\begin{figure}
\centerline{
\epsfig{file=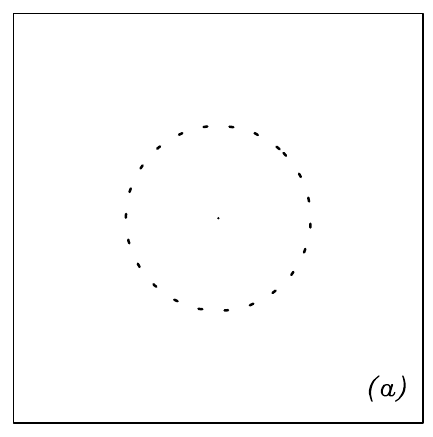}
\epsfig{file=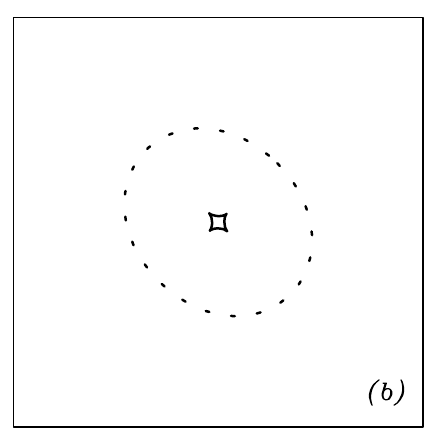}
\epsfig{file=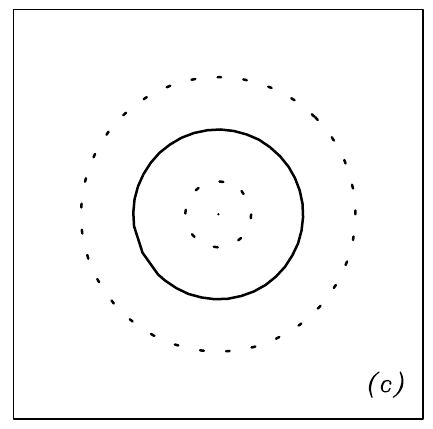}}
\centerline{
\epsfig{file=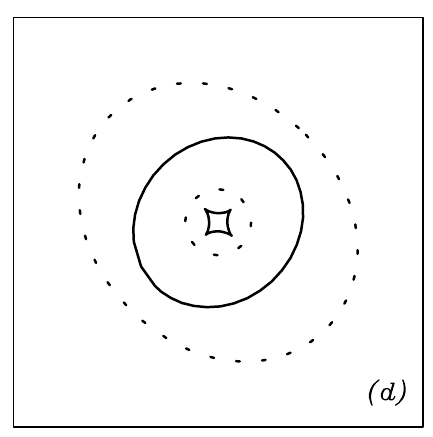}
\epsfig{file=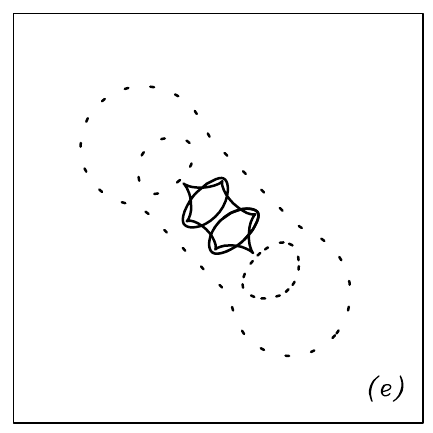}
\epsfig{file=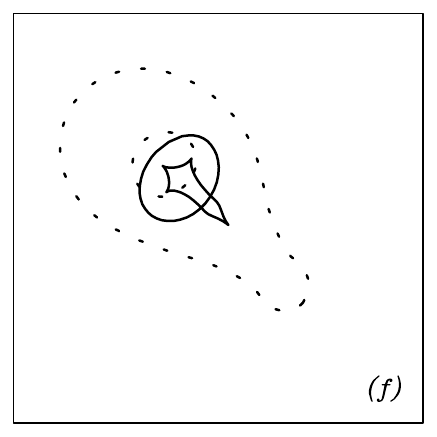}}
\caption{Critical lines (dashed) and caustics (solid) for different
classes of mass models: (a) for a singular isothermal circular mass
distribution, the radial critical line is the central point, and the
corresponding caustic line is at infinity, (b) a singular isothermal
elliptical mass distribution, the tangential caustic line is an
astroid, (c) a circular mass distribution with an inner slope shallower
than isothermal mass distribution, in this case a radial critical
curve appears, and both caustics are circles.  (d) same as (c) but for
an elliptical mass distribution, the relative size of both caustic
lines will depend on the mass profile and the ellipticity of the mass
distribution, (e) a bimodal mass distribution with two clumps of equal mass, similar to
(d), and (f) for a bimodal distribution with unequal masses.
\label{cclines}}
\end{figure}

The magnification is infinite if one of the principal values of the
magnification matrix is equal to zero, which implies that the reduced
shear $g$ is equal to $1$ or $-1$. Thus, the locus in the image plane
of infinite magnification defines two closed lines that do not
intersect (as $g$ cannot be equal to $1$ and $-1$ at the same location)
and these are called the "critical lines". The corresponding lines in the
source plane are called "caustic lines", they are also closed lines but
contrary to the critical lines, they can intersect each other. In
general, for a simple mass distribution, we can easily distinguish the two
critical lines: the external critical line where the deformations
are tangential, and the internal critical line where the
deformations are radial. Note that these simple geometries for the critical and 
caustic lines do not hold strictly for more complex mass distributions (Figure \ref{cclines} for 
examples of critical and caustic lines for different simple mass dsitributions).

For a circularly symmetric mass distribution, the equations for the critical lines
are simple. The magnification matrix in polar coordinates simplifies
to:
\begin{equation} 
A^{-1}=
\begin{pmatrix}
1-\partial_{rr}\varphi & 0 \cr
0 & 1-{1\over r}\partial_{r}\varphi \cr
\end{pmatrix}
\,.
\end{equation}

Thus both the critical and caustic lines (if they exist) are
circles. In fact, substituting the equation of the tangential critical
line: $r=\partial_{r}\varphi$ into the lensing equation to compute the
caustic line, we find that the tangential caustic line is always
restricted and reduces to a single point in the case of a circular mass
distribution.  It is also relatively easy to demonstrate that for a
well behaved mass distribution the radial critical line is always located
within the tangential critical line (Kneib 1993).

It is important to notice that for a circularly symmetric mass distribution, the
projected mass enclosed within the radius $r$ can be written as:
\begin{equation}
M(r)={c^2\over 4G}{D_{OS}D_{OL}\over D_{LS}} r\partial_r \varphi(r)
 = \pi \Sigma_{crit}\,r\,\partial_r \varphi(r).
\end{equation}
At the tangential critical radius we have:
$r_{ct}=\partial_{r}\varphi(r_{ct})$, thus the mass within the 
tangential critical radius (also referred to as the  Einstein radius $r_E$) is:
\begin{equation}
M(r_E)=\pi \Sigma_{crit} r_E^2.
\end{equation}
The critical surface mass density $\Sigma_{crit}$ corresponds to the
mean surface density enclosed within the Einstein radius. Thus the higher
the mass concentration, the larger the Einstein radius. For a given
surface mass density profile, the size of the Einstein radius will
depend on the redshift of the lens and the source as well as the
underlying cosmology. The variation of $\Sigma_{crit}$ for a given source 
redshift as a function of the lens redshift shows that for a given lens mass 
distribution the most effective lens is placed at roughly less than half the source
redshift.

Furthermore, the radial critical curve is defined as:
\begin{equation}
\partial_{rr}\varphi(r) = \partial_{r} \left({M(r)\over \pi\Sigma_{crit}  r}
\right)=1,
\end{equation}
thus, the position of the radial critical line depends on the gradient
of the mass profile.

The above equations suggest that: {\sl i)} from the tangential critical curve location, the
total mass enclosed within a circular aperture can be measured precisely, and {\sl ii)}
from the radial critical curve, the slope of the mass profile near the cluster center 
can be strongly constrained. However, for an accurate estimate of the mass enclosed the redshifts of the cluster
and the arc need to be known precisely. Furthermore, note that only the mass normalization 
scales directly with the value of $H_{0}$, 
but not the derived mass profile slope.

\begin{figure}
\epsfig{file=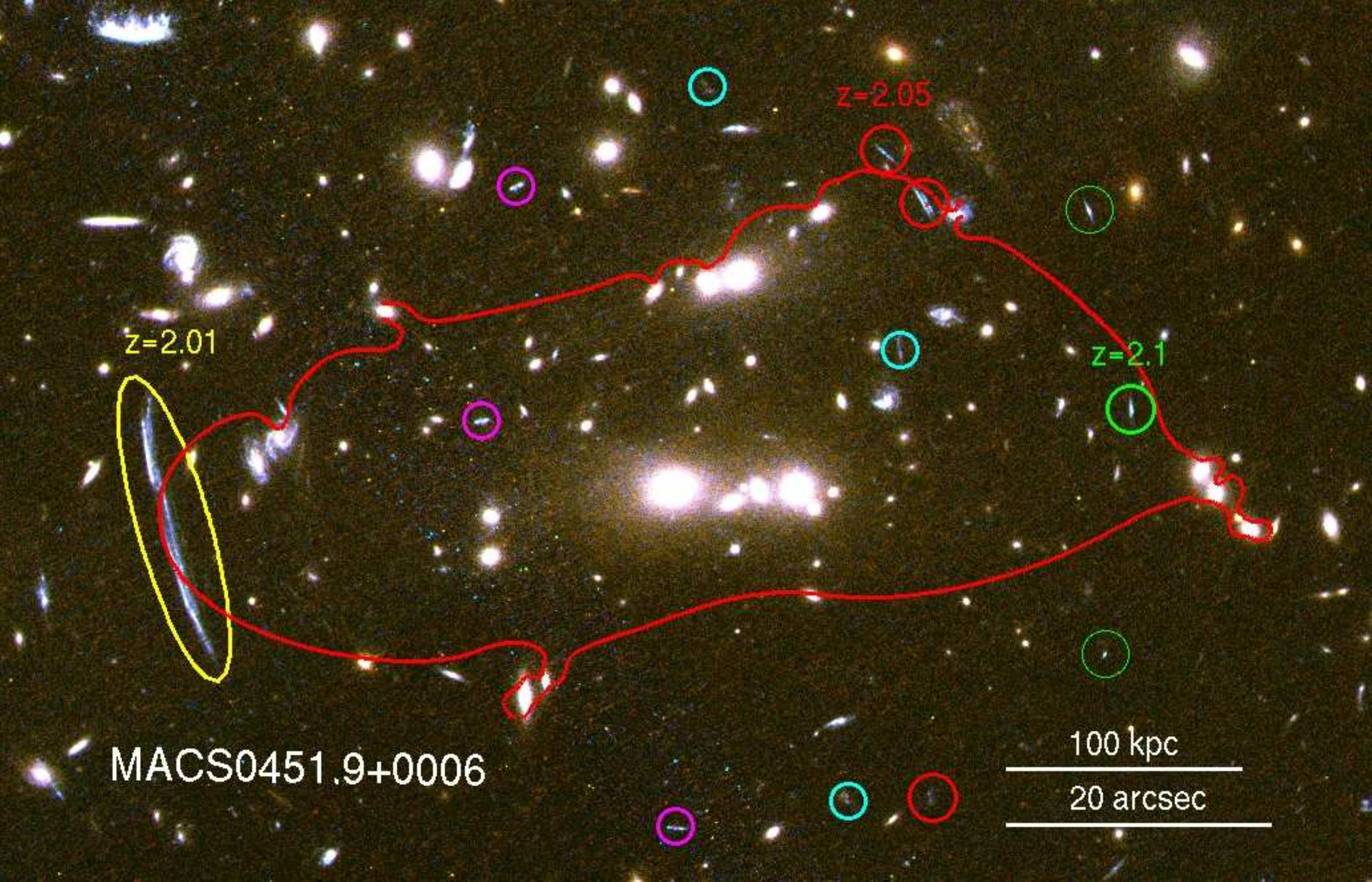,width=\textwidth}
\caption{\hst/ACS color image of MACSJ0451+00. The red curve shows the
location of the critical line for a source at $z=2$. A giant arc at $z=2.01$, as well 
as different sets of multiple  images are identified  (each system of images is marked 
with a circle of the same color - the cyan and magenta identified multiple-image have no spectroscopic redshift measurement yet) [Richard et al. private communication].
}
\label{macs0451-02}
\end{figure}

\begin{figure}
\centerline{ \includegraphics[width=0.8\textwidth]{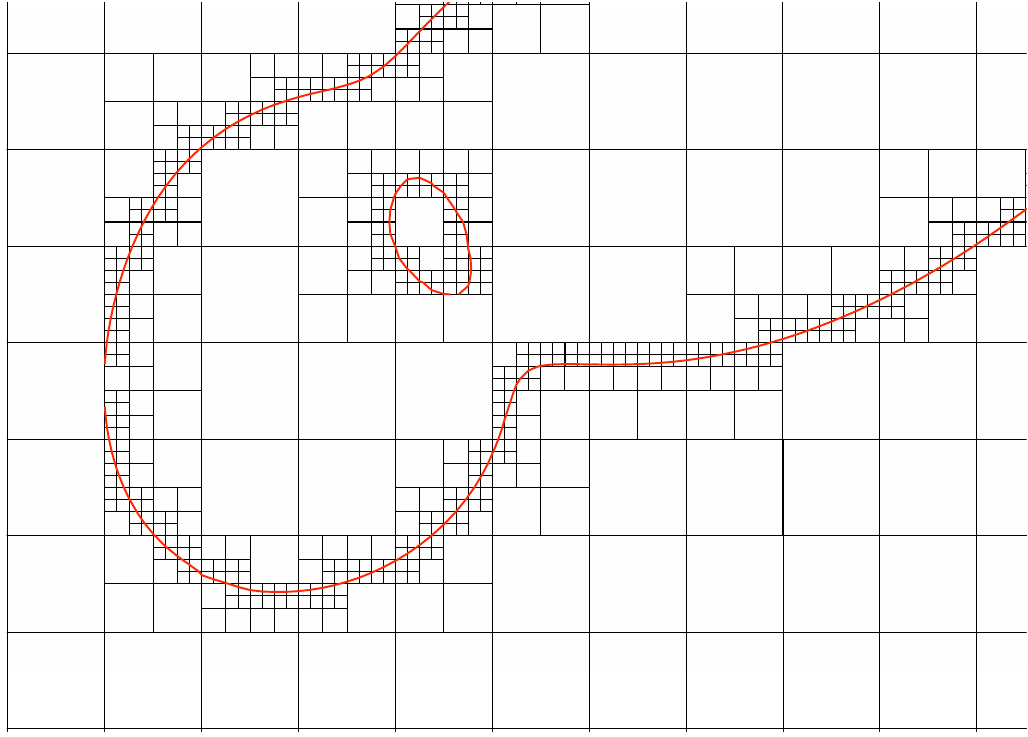} }
\caption{\label{critic-ms}
Multi-scale marching square field splitting to map critical lines: the boxes represent the
splitting squares and the red lines chart the critical curve contour. The
imposed upper and lower limits for the box sizes are 10'' (corresponding to the largest box shown) and 1''
respectively. The 1'' boxes are not plotted here for clarity. Figure adapted 
from  Jullo et al. (2007) where more details may be found.}
\end{figure}

For the general non-circular case, the determination of the critical
lines cannot be addressed analytically except for certain simple
elliptical mass profiles (e.g. Kneib 1993). The complexity of the shapes of critical lines 
can be seen for the lens model of MACS0451-02 (Figure~\ref{macs0451-02}). Indeed, 
to solve for the critical line in complex lens mass models, one has to resort to 
numerical methods. Iterative methods are more economical in terms of CPU time. 
For example, Jullo et al. (2007) have implemented the ``Marching Square'' technique 
for computing critical lines (see illustration in Figure~\ref{critic-ms}).

The above property linking the total mass within the
critical line to the area within the critical line does not hold
exactly for the more general cases but it is still a good approximation
if the mass distribution is not too different from the circularly symmetric case 
(Kassiola \& Kovner 1993). Hence identifying the characteristic sizes of the critical
lines both radial and tangential in an observed cluster is the first important step 
toward measuring the mass and its degree of concentration in the inner regions.

\subsection{Multiple-images}

\subsubsection{Definition}

Critical lines are virtual lines, and thus cannot be directly mapped.
However, multiple-images that straddle critical lines can easily be identified in high
resolution images.  For instance, tangentially distorted images are found near tangential 
critical lines and radially distorted ones near the radial critical lines. One often
refers to tangential pairs or radial pairs, which are simple configurations that are 
easily recognizable (e.g. Miralda-EscudŽ \& Fort 1993). For example,  one can have triplets,
quadruplets, quintuplets or even larger multiplicities of images of the same source
depending on the complexity of the mass distribution.

The number of multiple-images produced is simply the number of solutions of 
Equation \ref{eq_lensdeflection}. It can be estimated easily using
catastrophe theory (Thom 1989, Zeeman 1977, Erdl \& Schneider 1993), 
according to which each time one crosses a caustic
line in the source plane two additional lensed images are produced. For a non-singular
mass distribution we expect to always have an odd number
of multiple-images (Burke 1981). However, some images are likely to be less
magnified, or in fact, demagnified so that they are not observable, thereby complicating
at first the task of counting the total number of multiple-images produced. Often, the presence of 
a bright central galaxy in clusters scuppers the detection of the central demagnified image.

\subsubsection{Multiple-image symmetry}

Multiple-images have different symmetries which can be summarized by the signs of the eigenvalues of the 
magnification matrix, we can thus in principle have three possibilities for the parities, which correspond to the symmetry of 
the source, denoted as: (+,+), (+,-) and (-,-). For example we often talk about ``mirror'' symmetry, when we recognize 
a counter image as the flipped image of galaxy with a remarkably similar morphology. The image symmetry property 
is generally used to identify multiple-images in what turns out to be a secure way, as we see in the pair configuration of Figure~\ref{ac114pair}.

\begin{figure}
\epsfig{file=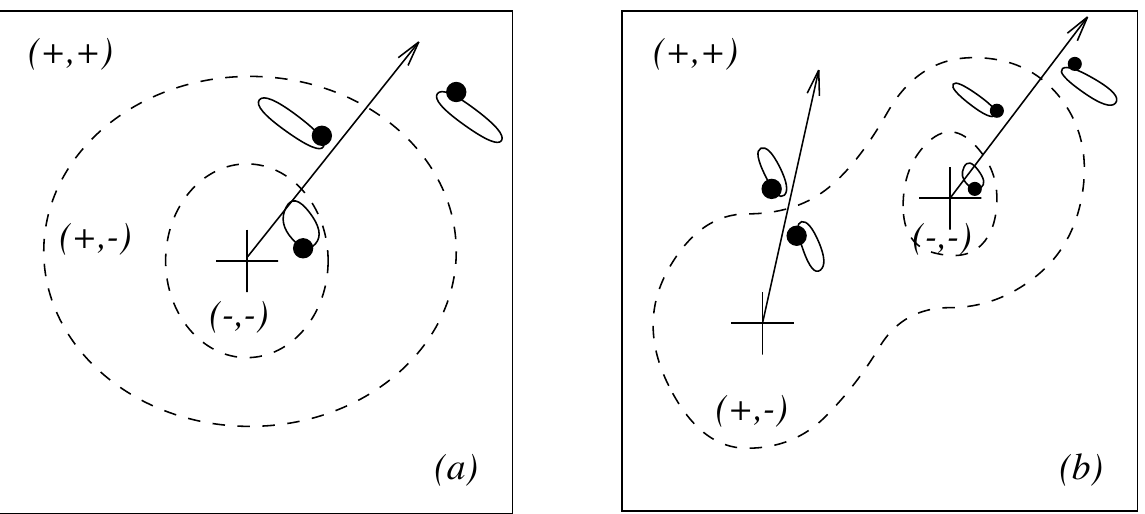,width=0.95\textwidth}
\caption{Area in the image plane showing different image parities (indicated by the signs in parentheses): (a) in the case of a simple elliptical 
mass distribution, (b) in the case of a bimodal mass distribution. The dashed lines correspond to the critical lines. The arrow is just an indication of the radial direction of the closest mass clump. We note that while the deformations shown in this figure are completely arbitrary, the orientation of the images is portrayed
accurately.
\label{parity}
}
\end{figure}

Indeed, each time, one crosses a critical line (this corresponds to a change in sign of one of the eigenvalues of the 
magnification matrix), the parity of the image changes (Blandford \& Narayan 1986;  Schneider, Ehlers \& Falco 1992). For simple 
mass distributions, only three parities described above by the notation (+,+), (-,-) and (+,-) can be observed as shown in Figure~\ref{parity}. 
Since for a simple mass distribution the radial critical line is always inside the tangential critical line, the parity (-,+) is not physically allowed.

\subsubsection{Examples of multiple-image systems}

Massive clusters frequently produce multiple-images, and this happens when
the surface mass density of the cluster core is close to or larger
than the {\it critical surface mass density}: 
$$\Sigma_{crit}= {c^2\over 4\pi G}{D_{OS}
D_{OL} \over D_{LS} },$$ for given lens and source redshifts. The detailed configuration of multiple-images can 
be used to unravel the structure of the mass distribution. 

\begin{figure}
\epsfig{file=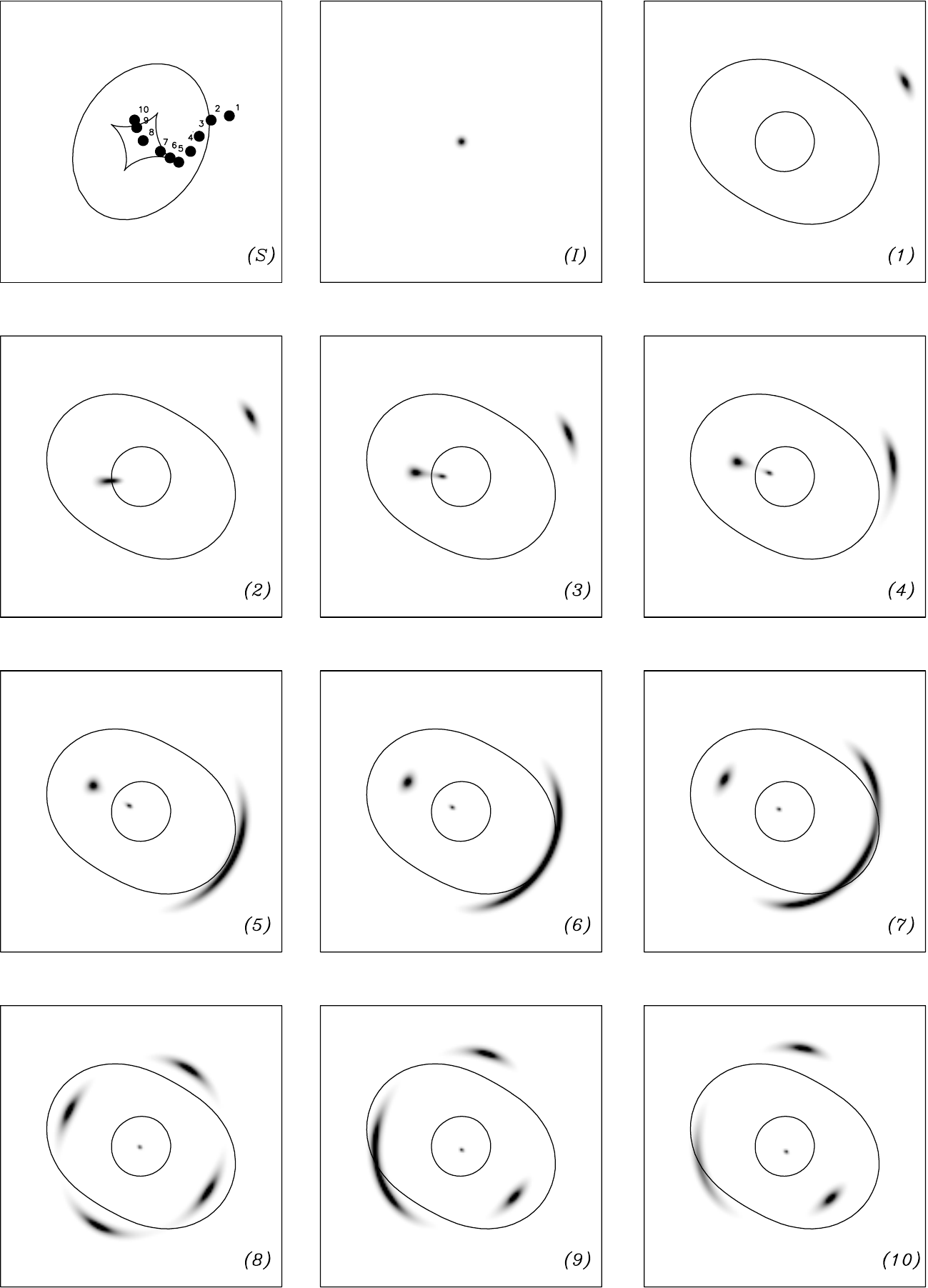,width=\textwidth}
\caption{Multiple-image configurations produced by a simple elliptical mass
distribution. The panel (S) shows the caustic lines in the source
plane and the positions numbered 1 to 10 denote the source position 
relative to the caustic lines.  The panel (I) shows the image of the source
without lensing.  The panels (1) to (10) show the resulting lensed
images for the various source positions. Certain configurations are
very typical and are named as follows: (3)
radial arc, (6) cusp arc, (8) Einstein cross, (10) fold arc.}
\label{figjpk2}
\end{figure}

\begin{figure}
\epsfig{file=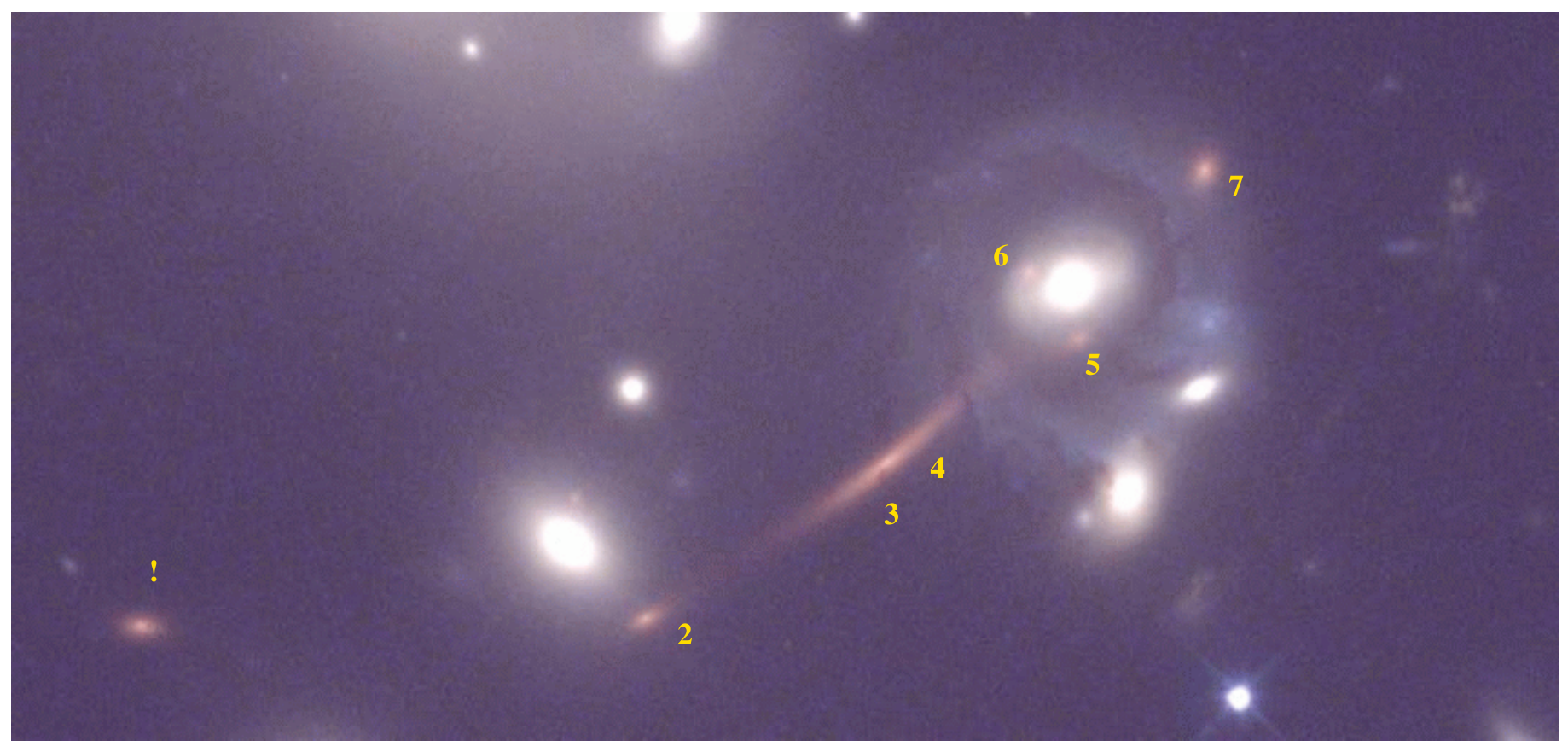,width=\textwidth}
\caption{A spectacular set of multiple-images seen in the cluster Abell
  2218 in the composite B, R, and I-band \hst\  image. A distant 
  E/S0 galaxy at z=0.702 is lensed into a 7-image configuration. } \label{figjpk3}
\end{figure}

A cluster with one dominant clump of mass will produce (for the range of multiple-image 
configurations see Figure~\ref{figjpk2}) {\it
fold}, {\it cusp} or {\it radial} arcs ({\it e.g} MS2137.3-2353: Fort
et al.\ 1992, Mellier et al.\ 1993; AC114: Natarajan et al.\ 1998;
A383: Smith et al.\ 2001, 2003); a bimodal cluster can produce
straight arcs ({\it e.g} A2390: Pello et al 1991, Cl2236-04: Kneib et al.\ 1994), triplets
(A370: Kneib et al. 1993, Bezecourt et al.\ 1999); 
a very complex structure with lots of massive halos in the core
can produce multiple-image systems with 7 or more images of the same
source ({\it e.g} A2218, see Figure~\ref{figjpk3}).  The presence
of every nearby perturbing mass can typically add two extra images to a simple
configuration if that mass is well positioned relative to the central core.
Very elongated/elliptical clusters with appropriate inner density profile slopes 
can produce hyperbolic-umbilic catastrophes producing quintuple arc configurations 
such as the one seen in  Abell 1703 (Limousin et al. 2008). A thorough description of 
exotic configurations has been discussed quite extensively in a paper by 
Orban de Xivry \& Marshall (2009).

\subsubsection{Multiple-image identification}

\begin{figure}
\centerline{\epsfig{file=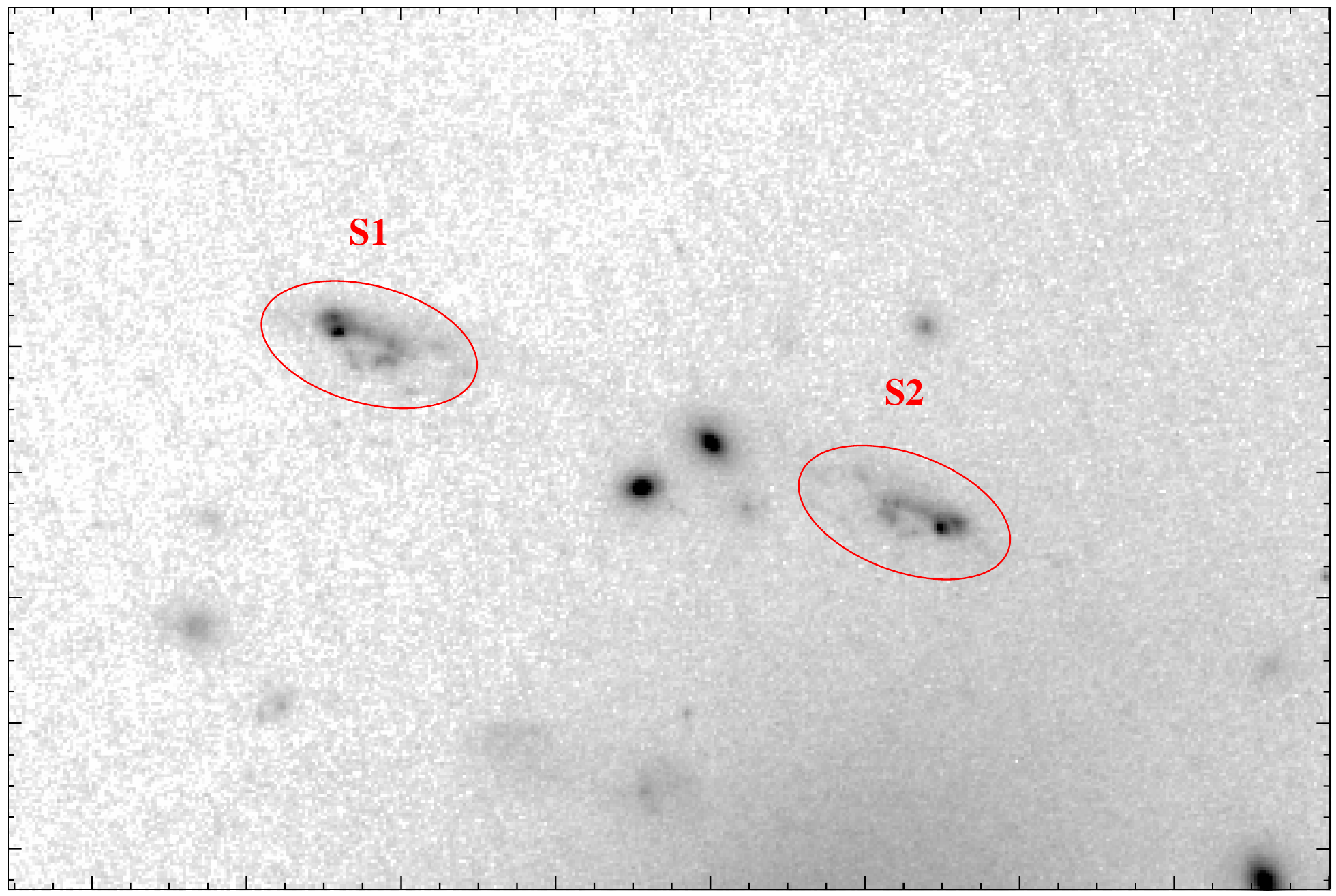,width=0.7\textwidth}}
\caption{ The lensed pair S1-S2 in AC114. This galaxy at $z = 1.867$
displays the surprising morphology of a hook, with an obvious change in 
parity (Smail et al 1995a, Campusano et al 2001).
\label{ac114pair}
}
\end{figure}

Multiple-images can be identified by their distinct
properties. Traditionally, multiple-images have been recognized as the
images forming the giant arcs (3 images in the case of Abell 370, but
only two images in the case of MS2137-17 or Cl2244-04). However, not
every giant arc is composed of multiple-images, for example it is most
likely that the northern giant arc in Abell 963 is only a single image, and 
that the southern arc in Abell 963 is composed of two or
three arclets (single images) from sources at different redshifts as revealed
by their different colors. Multiple-images can be recognized in terms of their (mirror) symmetry,
which is of course best visible with high-resolution \hst\ data.
One of the classic examples is the ``hook-pair'' in AC114 (Figure~\ref{ac114pair}) 
where the image symmetry is readily identified. Furthermore, as lensing is achromatic, multiple-images can be recognized by the 
similarity of their colors, or by their extreme brightness at a specific wavelength like
 in the  sub-mm or in mid-infra red.

Finally, the secure way to identify and confirm the existence of a
multiple-image system, is through the detailed modeling of the cluster
lens itself. This allows one in principle to test if a set of images
having similar morphology and colors can actually be multiple
images of the same source.  Calibrated lens models can predict
the location of counter images and also predict the redshift of the
multiply lensed source (Kneib et al. 1993, 1996).

Ultimately, for studying a large sample of massive clusters, one would 
likely need to develop automatic techniques to identify multiple-image 
systems based on their morphology, color and more sophisticated 
lens modeling  software. Although such robotic processes are being developed (Sharon 
private communication), further developments
are needed to make them completely user friendly.

\subsubsection{Multiple-image regions}

\begin{figure}
\centerline{
\epsfig{file=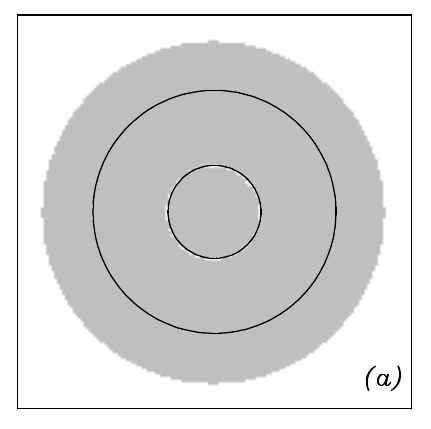,width=0.3\textwidth} 
\epsfig{file=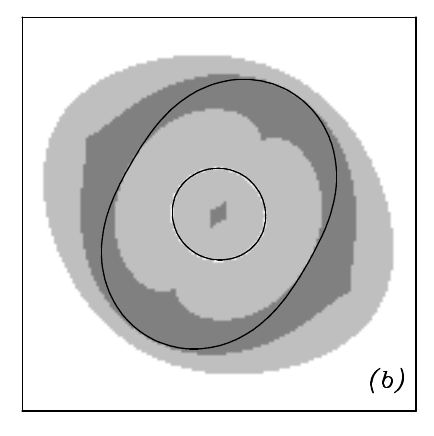,width=0.3\textwidth} 
\epsfig{file=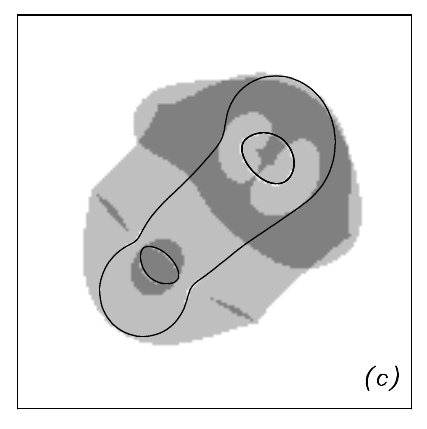,width=0.3\textwidth} 
}
\caption{Diagram showing the regions wherein multiple-images are produced with the location 
of critical lines marked, for the case of different mass models (a) circular mass 
distribution; (b) an elliptical mass distribution and (c) a bimodal mass distribution. The 
grey scale areas indicate regions behind which we can expect a single background galaxy to 
be imaged into three (light grey) or five (dark grey)  multiple-images.} 
\label{zone_mult}
\end{figure}

\begin{figure}
\centerline{ \epsfig{file=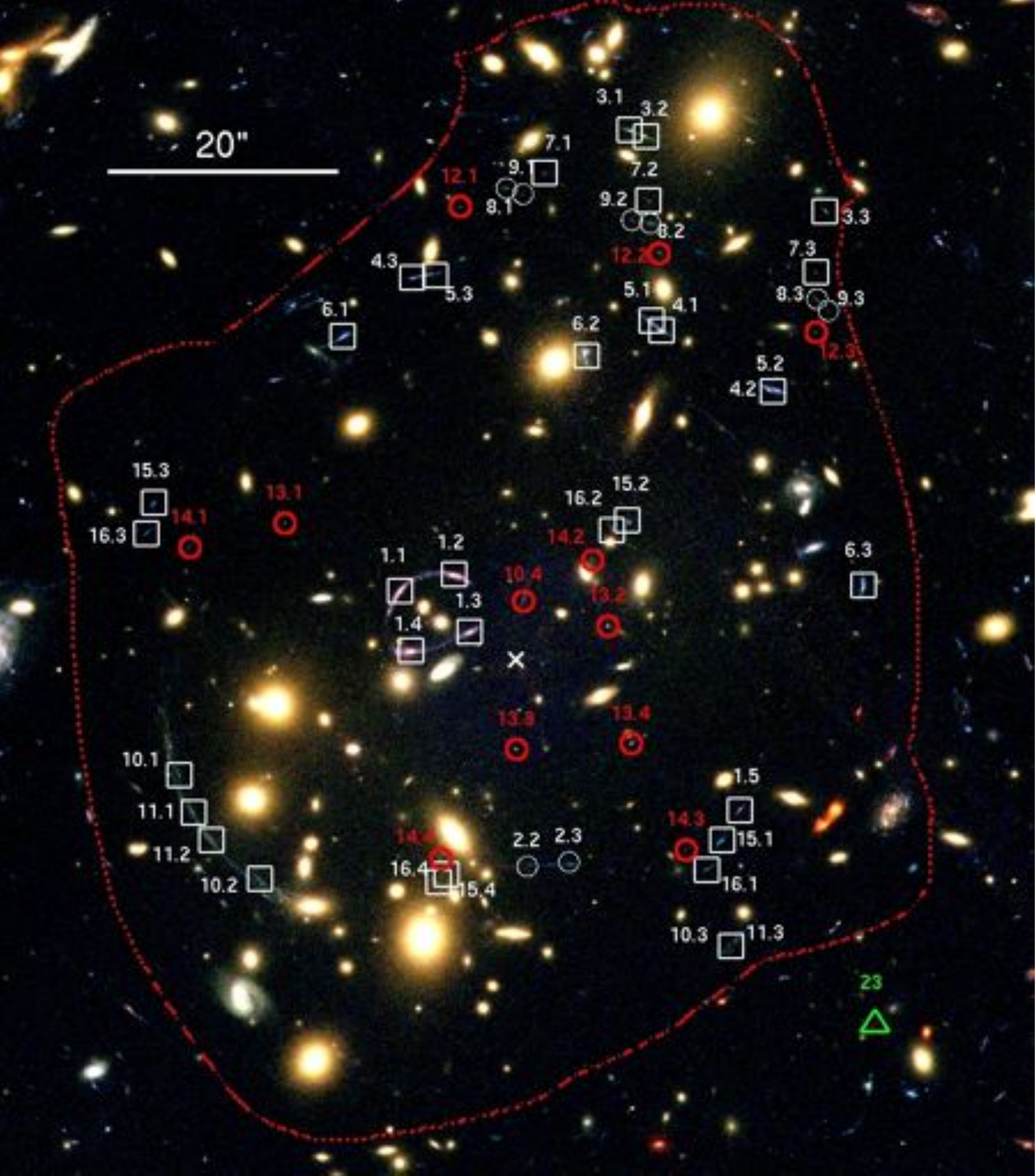,width=0.75\textwidth} }
\caption{Hubble ACS color image of Abell 1703 (image shown is from the combination of the F450W, F606W and F850LP filters), 
showing the location of all the multiply imaged systems.
The white cross at the center of the image marks the location of the brightest cluster galaxy, which has been 
subtracted from this image for clarity.
The red dashed line outlines the limit of the region where we expect multiple-images 
from sources out to $z=6$ (Figure from Richard et al. 2009).} 
\label{richard_a1703}
\end{figure}

Multiple-images are located in the central regions of clusters where the surface mass density
is close to or higher than the lensing critical surface density. For a given source redshift one can compute
the region conducive to multiple-imaging and the expected multiplicity. This is easily computed, as for any given image position 
$\theta_{I}$, we can determine the source position $\theta_{S}$ using the lensing equation (Equation 4).
Given a source position it is straightforward to determine whether $\theta_{S}$ lies within a caustic curve or not. The expected number of
images is given by $(1+2 N_{c})$ where $N_{c}$ is the minimum number of caustic curves that need to be crossed 
to reach the position $\theta_{S}$ (Figure \ref{zone_mult}).  The calculation of multiple-imaging regions can be useful 
to help observationally identify multiple-image counterparts (Figure \ref{richard_a1703}) particularly in the case of complicated mass distributions.

\subsection{First order shape deformations - Shear}

Distant sources are only multiply imaged in the central regions of 
cluster where the surface mass density is sufficiently high. However,  
every observed galaxy image in the field of the cluster is deformed by 
lensing, typically in the weak regime. To first order, one can approximate 
the light distribution of a galaxy as an object with elliptical isophotes. In this 
event, the shape and size of galaxies can be defined in terms of the axis ratio 
and the area enclosed by a defined boundary isophote.

\begin{figure}
\centerline{
\epsfig{file=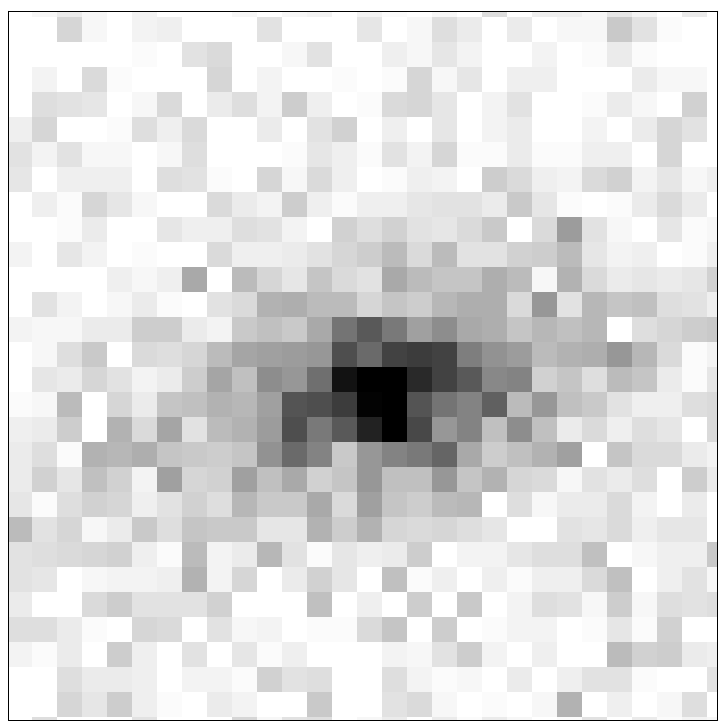,width=0.5\textwidth}
\
\epsfig{file=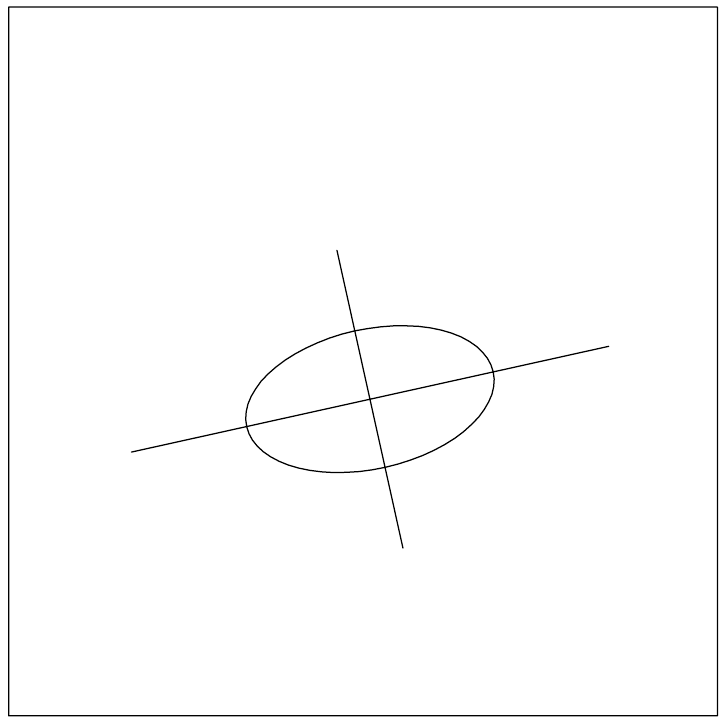,width=0.5\textwidth}
}
\caption{A typical faint galaxy observed on a CCD image (left), and
the equivalent ellipse defined from the second order moments (right).
\label{faintgalellipse}
}
\end{figure}

However, the real shapes of faint galaxies can be quite irregular and not
well approximated by ellipses. We thus need to express the shape of a
galaxy in terms of its pixelized surface brightness as measured on a
digital detector. For this purpose, we use the moments of the
light distribution to define shape parameters.  If $I(\vec\theta)$ is the
surface brightness distribution of the galaxy under consideration, we can
define the center of the image $\vec{\theta_c}$=($\theta^C_i,\theta^C_j$) using the
first moment of the $I(\vec\theta)$ distribution:
\begin{equation}
\vec{\theta_c} = {\int W(I(\vec\theta))\,\vec\theta\,d\vec\theta
		\over \int W(I(\vec\theta)) d\vec\theta }.
\end{equation}
Note that W(I) is a weight/window function, that allows the integrals above to 
be finite in the case of noisy data. The simplest choice for the
function W(I) is the heaviside step function $H(I-I_{iso})$ which
is equal to 1 for $I(\vec\theta)>I_{iso}$ where $I_{iso}$ is the
isophote limiting the detection of the object, and 0 otherwise. The image
center found is then taken as the center of the detection isophote. Another 
popular weight function that is frequently adopted is: $W(I)=I\times H(I-I_{iso})$,
where the window function is now weighted by the light distribution within the
isophote.

The second order moment matrix of the light distribution centered on
$\vec{\theta_c}$:
\begin{equation}
M_{ij} = {\int\int W(I(\vec\theta)) (\theta_i-\theta_i^C)(\theta_j-\theta_j^C)\,
			 d\theta_{i}\,d\theta_{j}
                \over \int\int W(I(\vec\theta)) d\theta_{i}\theta_{j} },
\end{equation}
allows us to define the size, the axis ratio and the orientation of the
corresponding approximated ellipse. Indeed the moment matrix $M$ is positive definite 
and can be written in its principal axes as:

\begin{equation}
M_{ij} = R_\theta 
\begin{pmatrix}
a^2 & 0 \cr
0 & b^2 \cr
\end{pmatrix}
R_{-\theta},
\end{equation}
where $a$ and $b$ are the semi-major and semi-minor axes respectively,  and 
$\theta$ the position angle of the equivalent ellipse, and $R_{\theta}$ is the rotation matrix of an angle $\theta$. Thus the moment matrix $M$ contains
three parameters: the size of the galaxy, its ellipticity and its orientation (see illustration in Figure \ref{faintgalellipse})

It is useful to define a complex ellipticity which encodes both the shape
parameter and the orientation of an observed galaxy. 

There are however a
number of ways to define the norm of the complex ellipticity, and the lensing community has
experimented several notations:
\begin{equation}
|\varepsilon| ={a^2-b^2 \over a^2+b^2}\ \ \
|\delta| ={a^2+b^2 \over 2ab}\ \ \
|\tau| ={a^2-b^2 \over 2ab}\ \ \
|\epsilon| ={a-b \over a+b}.
\end{equation}
With the complex ellipticity defined for example as:
\begin{equation}
{\bf \varepsilon} =|\varepsilon| e^{2i\theta}.
\end{equation}

The notation $\varepsilon$ was the first to be introduced, as it emerges naturally from the
moment calculation, then $\tau$ and $\delta$ were introduced in the context of
cluster lensing by Kneib (1993) and Natarajan \& Kneib (1997). The
advantage of this form is that the lens mapping can be written as a
simple linear transformation from the image plane to the source
plane which is mathematically convenient. Subsequently $\epsilon$ was adopted, and it has now become the
standard definition, essentially because it is a direct estimator (modulo the
PSF correction) of the measured quantity, which is the reduced shear $g$ as we show below.  All
ellipticity parameters are of course linked to each other, and in
particular we have $\varepsilon = 2\epsilon/(1+|\epsilon|^2)$.

With the various definitions in hand for the relevant parameters, we
can now explicitly express the transformation produced by
gravitational lensing on the shape of a background galaxy.  First, 
it can be shown that the image
of the center of the source corresponds to the center of the image in
the case where the magnification matrix does not change significantly
across the size of the image (Kochanek 1990; Miralda-Escud\'e 1991). 
This is generally adopted as the definition 
of the weak lensing regime as such a simplification does not hold in the strong
lensing regime. To demonstrate this explicitly, one has to use the
fact that the surface brightness is conserved by gravitational lensing
as was first demonstrated by Etherington (1933), namely,
$I(\vec\theta_I)=I(\vec\theta_S)$.

The lens mapping will transform the shape of the galaxy,
by magnifying it and stretching it along the shear direction. This
transformation can be written in terms of the moment matrix $M^{S}$ (for the galaxy in the source plane)
and $M^{I}$ (for the galaxy in the image plane that is, as observed) as follows:
\begin{equation}
M^S=\ A^{-1}\ M^I\ {}^tA^{-1},
\end{equation}
or if the matrix $A^{-1}$ is not singular:
\begin{equation}
M^I=\ A\ M^S\ {}^tA.
\end{equation}
Note that ${}^tA$ is the transpose of matrix $A$.

These equations describe how the ellipse defining the source shape is
mapped onto the equivalent ellipse of the image or vice versa.  If we
consider the size $\sigma =\pi a\times b$ of the equivalent ellipse, we
can write:
\begin{equation}
\sigma_S^2=\det M^S = \det M^I . (\det A^{-1})^2 = \sigma_I^2.\mu^{-2}.
\end{equation}
Thus the overall size $\sigma_S$ of the source is enlarged by the
magnification factor $\mu$. Similarly, it is possible to write the lensing transformation for the
complex ellipticity, which of course will depend on the ellipticity
estimator chosen. 
For the ellipticity $\epsilon$, and using the 
complex notation, we have:
\begin{equation}
\epsilon_S = {\epsilon_I - g \over 1 - g \epsilon_I},\,\,\,\,\,\,{\rm for}\,|g|<1,
\end{equation}
which corresponds to the region external to the critical lines, and
\begin{equation}
\epsilon_S = {1 - g^*\epsilon_I^* \over  \epsilon_I^* -g^*},  {\rm for}\,\,\,\,\,\, |g|>1,
\end{equation}
which corresponds to the region inside the critical lines (the
notation $^*$ denotes the transpose of a complex number).

In the weak regime, where the distortions are small ($|g|<<1$) the
lensing equation simplifies to:
\begin{equation}
\epsilon_I = \epsilon_S + g.
\end{equation}
Thus the ellipticity of the image is just a linear sum of the intrinsic source
ellipticity and the lensing distortion in this limit.  Thus averaging
the above equation over a number of sources yields the convenient 
fact that image ellipticities are a direct measure of the reduced shear $g$. 
Note, however that these simplified equations mask observational
limitations such as the PSF/seeing convolution and pixelization, all effects
that contribute to and contaminate observed image shapes.

\subsection{Mass-sheet degeneracy}

The "mass sheet degeneracy" problem was recognized as soon as mass distributions began to
be mapped using lensing observations  ({\it e.g.} Falco et al. 1985)
and the issue has been
discussed in detail in Schneider, Ehlers \& Falco (1992) and
Schneider \& Bartelmann  (1997) and Bradac et al (2004)  in the context of weak lensing mass measurement. 
This degeneracy  arises due to the lack of
information needed to calibrate the total mass of clusters in the
absence of a normalization scheme due to the simple fact that the
addition of a constant surface mass density sheet leaves the measured
shear unaltered. 

Expressed mathematically, the magnification and shear
are invariant under the following transformation:
\begin{equation}
\kappa^\prime = (1-\lambda) \kappa +\lambda
\end{equation}
and
\begin{equation}
\gamma^\prime = (1-\lambda) \gamma,
\end{equation}
where $\lambda$ is the mass-sheet (denoting a sheet of constant surface mass
density) added to the lens plane. Expressing the reduced shear with
the above two equations we can show that:
\begin{equation}
g^\prime = g,
\end{equation}
thus the reduced shear is conserved under this transformation. This means that for a given
observed reduced shear field, one can only extract the surface mass
density distribution $\kappa$ up to a constant factor given by the 
unknown value of $\lambda$.

There are several ways to break the mass sheet degeneracy, the
obvious way is to use lensed sources from different source redshift
planes.  Indeed with, $\kappa(z_1) = {\cal E}(z_1)/{\cal E}(z_2)
\kappa(z_2)$, such a transformation is incompatible with the above
invariance. Other methods to break the mass-sheet degeneracy, 
such as the inclusion of constraints from the strong lensing regime are discussed
further in Section 3.4.3 on lens modeling.

\subsection{Higher order shape deformations - Flexion}

The equations in the previous section assume that $\kappa$ and
$\gamma$ and as a consequence the reduced shear $g$ are all constant across an
image. This assumption fails when an image is physically large and/or
when it is close to critical regions where the lensing distortion is
changing rapidly. There are basically two effects that lensing
produces on a background elliptical source: a shift in the peak flux
at the center of the image compared to that of the fainter isophotes 
(while still preserving the surface brightness from the source to the image),
and the distortion of the elliptical shape into an extended "banana"
shape. Therefore, there is additional, valuable information that can be gleaned from 
higher order lensing effects.

\begin{figure}
\epsfig{figure=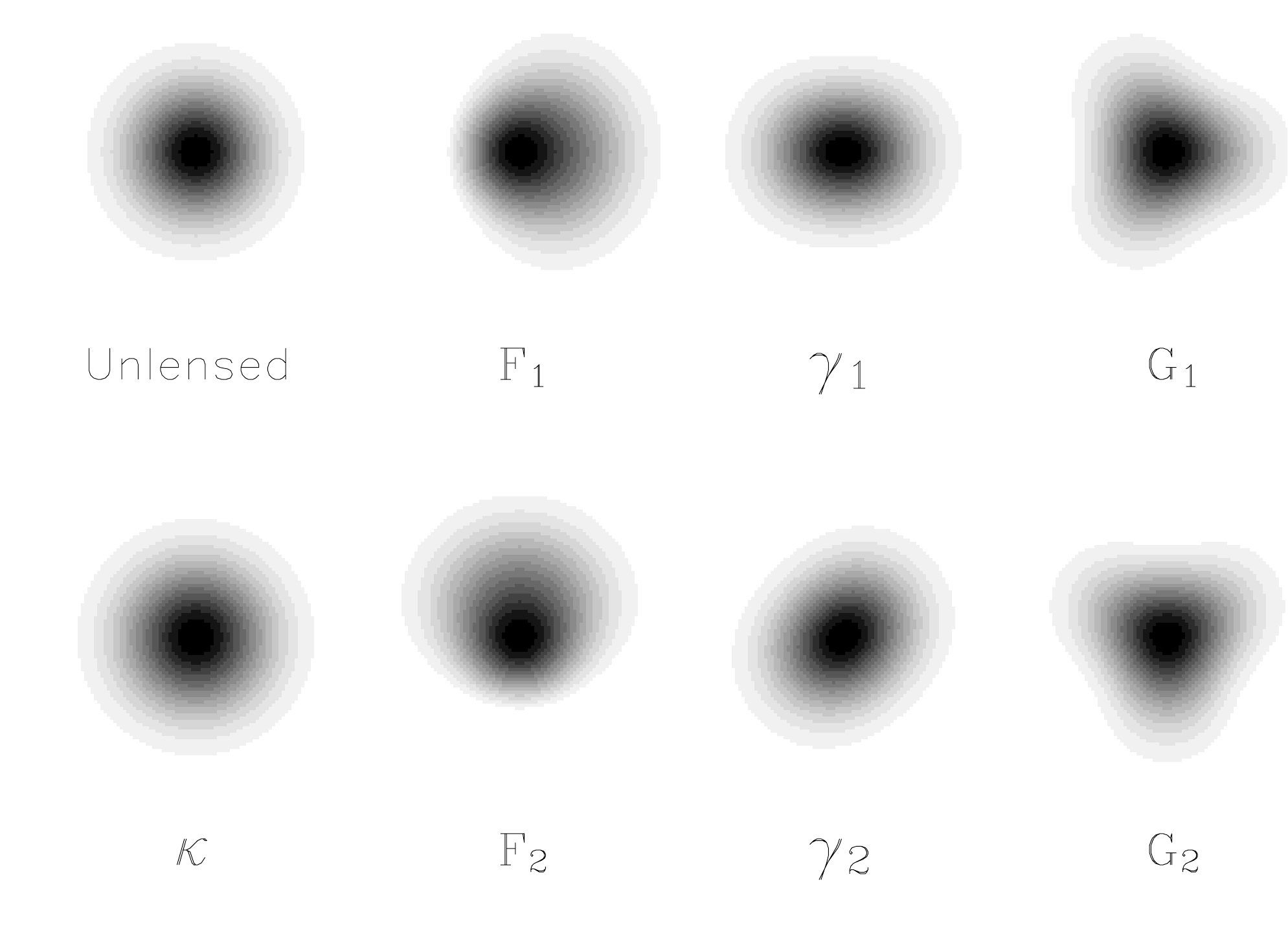, width=\textwidth}
\caption{Decomposition of weak lensing distortions, illustrated for an
 unlensed Gaussian galaxy with a radius of 1 arcsec. The source has been distorted
 with 10\% convergence/shear, and 0.28 arcsec$^{-1}$ flexion. The
 convergence $\kappa$, and two components of the first flexion ($F_1$
 and $F_2$); shear ($\gamma_1$, $\gamma_2$); and second flexion $G_1$
 and $G_2$ are shown. As mentioned, flexion causes the arci-ness or
 elongation of weakly lensed arcs once one combines $F$, $G$ and $\gamma$ (Figure from Bacon et al. 2005).
 \label{figurebacon2005}}
\end{figure}

To determine these higher order effects numerically, one needs to use
higher orders of the lensing transformation using the Taylor expansion
of the image shape.  This was first investigated by Goldberg \&
Natarajan (2002), and followed up by Goldberg \& Bacon (2005). A
recent summary of the formalism and applications is reviewed in Bacon
et al. (2005). Flexion is the significant third-order weak gravitational lensing
effect responsible for the skewed and arc-like appearance of lensed
galaxies. Flexion has two components: the first flexion, which is
essentially the derivative of the shear field which contains local
information about the gradient of the matter density (Goldberg \&
Natarajan 2002) and the second flexion which contains non-local
information (Bacon et al. 2005 and Figure \ref{figurebacon2005}). Flexion measurements can be used to
measure density profiles and these reconstructions can be combined
with those derived from the shear alone. One key advantage of using
the flexion estimator is that it is not plagued by the mass sheet
degeneracy as it is a higher order term, while its dispersion measure
is comparable to that of the shear. Recent successful applications of 
flexion to map mass distributions can be found in Okura et al. (2008); 
Leonard, King \& Goldberg (2011) and Er, Li \& Schneider (2011).

The shear and convergence vary within a galaxy image, flexion is the higher order
derivative and in order to derive it, expansion to the second order is required:

\begin{equation}
\theta'_i\simeq A_{ij}\theta_j +\frac{1}{2}D_{ijk} \theta_j \theta_k,
\end{equation}
with 
\begin{equation}
D_{ijk}=\partial_k A_{ij}.
\end{equation}
Using results from Kaiser (1995), we find that
\begin{eqnarray}
D_{ij1}&=&\left(
\begin{array}{cc}
-2\gamma_{1,1}-\gamma_{2,2} & \ \ -\gamma_{2,1} \\
-\gamma_{2,1} & \ \ -\gamma_{2,2} 
\end{array}
\right), \\ \nonumber
D_{ij2}&=& \left(
\begin{array}{cc}
-\gamma_{2,1} & \ \ -\gamma_{2,2}\\
-\gamma_{2,2} & \ \ 2\gamma_{1,2}-\gamma_{2,1}
\end{array}
\right).
\label{eq:ddef}
\end{eqnarray}
Using the equations above, the surface brightness of the imagecan be expanded in a Taylor series. In the weak lensing regime
we can approximate the brightness to second order as follows:\begin{equation}
f({\bf \theta})\simeq\left\{1+
	\left[ (A-I)_{ij}\theta_j+\frac{1}{2}D_{ijk}\theta_j\theta_k\right]
	\partial_i \right\}	f'({\bf \theta}) \ .
\label{eq:2ndorder}
\end{equation}
Therefore the expression for flexion can be written in terms
of derivatives of the shear field. Using the notation of Kaiser (1995)
we can write flexion in terms of the gradient of the convergence:
 
\begin{eqnarray}
{\bf {\cal F}}&\equiv& (\gamma_{1,1}+\gamma_{2,2}){\bf i}+
(\gamma_{2,1}-\gamma_{1,2}){\bf j} \\
&=&\nabla \kappa\\
&=&|{\cal F}| e^{i\phi}.
\label{eq:f1def}
\end{eqnarray}

We need to be able to measure the derivatives of the shear field $\gamma_{i,j}$ with a
high degree of accuracy from images in order to measure flexion. This is becoming
$\gamma_{i,j}$ with sufficient accuracy. This is becoming increasingly feasible with the availability of
high quality imaging data. The first flexion probes the local density via the
gradient of the shear field and quantifies the variation of the center of the different isophotal
contours. The second flexion probes the nonlocal part of the gradient of the shear
field and quantifies the shape variation and departure from elliptical symmetry.

Flexion has been incorporated as an additional constraint in the
cluster mass reconstructions only recently as extremely high quality
data is required to extract the flexion field and this is very
challenging (see Leonard, King \& Goldberg 2011 for the case of Abell 1689).
This higher order shape estimator however offers a powerful probe provided it can be 
measured accurately from observations (Leonard \& King 2010; Er, Li \& Schneider 2011).
As an illustration, we present the calculation of the flexion for the SIS model in the Appendix (A.4).


\section{Constraining cluster mass distributions}

In most cases, intermediate redshift $z\sim 0.2-0.5$ massive clusters are the most 
significant mass distribution along the line of sight, thus they can be represented 
by a single lens plane in concordance with the thin lens approximation. In the $\Lambda$CDM model, the 
probability of finding two massive clusters extremely well aligned along the line of sight (albeit separated in redshift) is 
extremely unlikely as clusters are very rare objects.  Lensing deflections and distortions probe the two dimensional projected cluster mass 
along the line of sight. This allows us to constrain the two dimensional Newtonian 
potential, $\phi(x,y)$, resulting from the three-dimensional density distribution
$\rho(x,y,z)$ projected onto the lens plane. The related projected surface
mass density $\Sigma(x,y)$ is then given by:
\begin{equation}
\label{equ1}
4\pi\mathrm{G}\Sigma(x,y)=\nabla^2\phi(x,y).
\end{equation}

Often we are interested in the two-dimensional projected mass inside an
aperture radius $R$ (particularly when comparing different mass estimators), 
which is defined explicitly as follows:
\begin{equation}
\label{equ2}
M_{aper}(R)=2\pi\int_0^{R}\Sigma(x)xdx,
\end{equation}
and the mean surface density inside the radius $R$ is given by:
\begin{equation}
\overline{\Sigma}(R)=\displaystyle{\frac{1}{\pi R^2}\int\limits_{0}^{R} 2\pi x\Sigma(x)dx}.
\label{sigma_mean}
\end{equation}

The important quantities for lensing in clusters are primarily 
the deflection angle $\vec{\alpha}$ between the image and 
the source, the convergence $\kappa$, and the shear $\gamma$, which 
can all be conveniently expressed in terms of the projected potential:
\begin{equation}
\left\lbrace
\begin{array}{l}
\vec{\alpha}(\vec{\theta})=\vec{\nabla}_{\vec{\theta}}\varphi(\vec{\theta})\\
\kappa(\vec{\theta})=\displaystyle{\frac{1}{2}\left(\frac{\partial^2\varphi}
{\partial\theta_1^2}
+\frac{\partial^2\varphi}{\partial\theta_2^2}\right)}\\
\gamma^2(\vec{\theta})=\|\vec{\gamma}(\vec{\theta})\|^2=\displaystyle{
\frac{1}{4}\left(\frac{\partial^2\varphi}{\partial\theta_1^2}
-\frac{\partial^2\varphi}{\partial\theta_2^2}\right)^2+
\left(\frac{\partial^2\varphi}{\partial\theta_1\partial\theta_2}\right)^2}.
\end{array}
\right.
\label{deflection_amplification}
\end{equation}

For a radially symmetric mass distribution, these expressions can be written as:
\begin{equation}
\left\lbrace
\begin{array}{lcl}
\kappa(x) & = & 
\displaystyle{\frac{\Sigma(x)}{\Sigma_\mathrm{crit}}}
\\
\gamma(x) & = &
\displaystyle{\frac{\overline{\Sigma}(x)-\Sigma(x)}{\Sigma_\mathrm{crit}}}
\\
\vec{\alpha}(x)& = & \theta \, 
\displaystyle{\frac{\overline{\Sigma}(x)}{\Sigma_\mathrm{crit}}}
= \theta \, (\kappa(x) + \gamma(x))
\end{array}
\right.
\label{eqcirc}
\end{equation}
where $x=D_{OL}\theta$ is the radial physical distance. From this equation, we note that one can derive $\gamma$ directly from 
$\alpha$ and $\kappa$. This formulation is particularly useful when trying to compute an analytic
expression of the lensing produced by a given mass profile.

\subsection{Strong lensing modeling}

\subsubsection{Modeling approaches}

Traditionally, modeling of the cluster mass distribution in the 
strong lensing regime is done using ``parametric models'' ({\it e.g.} Kneib et al. 1996; Natarajan \& Kneib 1997). 
In these schemes the mass distribution is described by a finite number of mass 
clumps; some small scale (galaxy components) and some large scale (to represent the dark matter, X-ray gas
in the Intra-Cluster Medium), each of which are described in turn by a finite 
number of parameters contingent upon the choice of mass profile deployed. The simplest mass distribution 
that is commonly employed is the circular Singular Isothermal Sphere (SIS), which is described by 
three parameters. The parameters are the position of its center $(x,y)$ and the value of the velocity 
dispersion $\sigma$, which in this case is a constant. Other mass distributions such as the PIEMD (Pseudo Isothermal Elliptical 
Mass Distribution) or NFW (Navarro-Frenk-White) profiles are often used in
lensing analysis and their relevant parameters are described in the Appendix (see A.1 - A.3).

\begin{figure}
\centerline{\epsfig{file=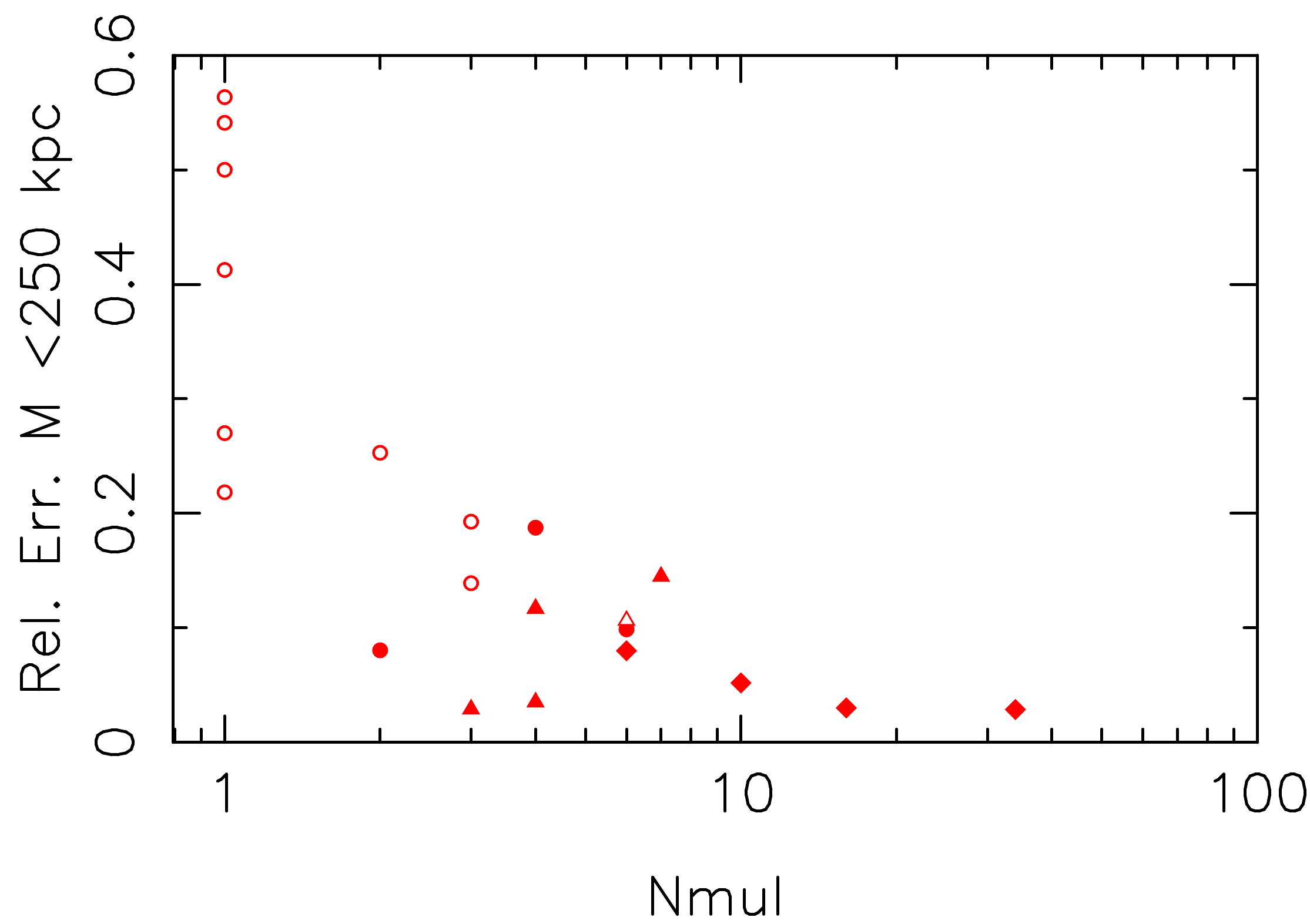,width=0.75\textwidth}}
\caption{
Relative aperture mass error as a function of the number of multiple-images as measured in Richard et al. (2010). Open symbols: clusters observed with WFPC2. Filled symbols: clusters observed with ACS. Symbols reflect the number of filters used to image the cluster (circle: 1 filter, triangle: 2 filters, diamond: 3 or more filters). One can see that with multi-band ACS data we can uncover more than 10 multiple-image systems for the most massive clusters, and thus achieve mass accuracy within 10\% or so in cluster cores.
}
\label{mul_err}
\end{figure}

Using a simple mass model makes sense when there are not many available 
observational constraints indeed one needs to balance the number of model 
parameters to the number of observational constraints available in order to compute a sensible best-fit model. However,  
with recent deep images of cluster cores from \hst\  a very large number of multiple-images 
can readily be identified. For instance, more than 40 multiple-image systems have been 
identified in the massive cluster Abell 1689 by Limousin et al. (2007). The discovery and identification
of such a large number of multiple-images has dramatically increased the number of constraints 
available for mass modeling of massive clusters in the last decade. With the availability of a larger number 
of multiply imaged systems with redshift measurements, more accurate mass models (Figure \ref{mul_err}) 
are now possible.

Therefore, the number of allowable parameters required to describe 
the mass distribution of a cluster has also increased, leading obviously to a more accurate description 
of the mass profile in the cluster core. This is for example evident upon comparing the model of Kneib et al. (1993) with that of 
Richard et al. (2010) for the cluster Abell~370. In the case of ``parametric models'', the increase in the number of constraints translates 
to the fact  that cluster mass distributions can now
be  described by a larger number of mass clumps and each of these clumps can 
be more complex ({\it e.g.} having elliptical mass distributions rather than circular, and a radial profile 
described by more parameters). 

The increase in the number of available constraints has 
also lead to the development  of new ``non-parametric'' methods, where no (or few) external 
priors are required to describe the mass distribution of clusters (Diego et al. 2005a; Saha \& Williams 1997; 
Coe et al. 2010).  Generically, in most of these 
Ònon-parametricÓ methods, the mass distribution is typically tessellated 
into a regular grid of smaller mass elements. Further details of such methods 
are discussed later on in this section.

\subsubsection{From simple to more complex mass determinations}

A particularly useful and popular mass estimate in the strong lensing
regime is the mass enclosed within the Einstein radius $\theta_E$, given by: 
\begin{equation}
M(<\theta_{E}) = \pi\Sigma_{crit} D_{OL}^{2}\theta_E^2,
\end{equation} 
where $\theta_E$ is the location of the tangential critical
line for a circular mass distribution, usually approximated by the tangential arc
radius $\theta_{arc}$. It is a very handy expression that is independent of the
mass profile for circularly symmetric cases. However, caution needs to be
exercised when using this expression as often the arc used to derive the mass has
an unknown redshift (thus $\Sigma_{crit}$ is not well defined),
or the arc is a single image and therefore does not
trace the Einstein radius or the mass distribution is very complex with
a lot of substructure. Note that, however, for a singular isothermal sphere model, a
single image cannot be closer than twice the Einstein radius since it will then
have a counter image. In conclusion, this estimator tends to typically 
overestimate the mass in instances where the tangential arc is not multiply imaged or its
redshift is unknown. 

The radial critical line can be constrained when a radial arc is
observed in the cluster core, this has now been done for a number
of cluster lenses ({\it e.g.} Fort et al.\ 1992; Smith et al.\ 2001; Sand et al.\
2002; Gavazzi et al.\ 2003).  These features are important as they lie
very close to the cluster core, and thus provide a {\it unique} probe of the surface mass 
density in the very center. Baryons are highly
 concentrated in the inner regions of clusters and they are expected to play
 an important role in possibly modifying the dark matter distribution on the
 smallest scales. The scales on which these effects are expected are accessible effectively with lensing data. 
 Radial arcs have  been used to probe the dark matter slope 
 in the inner most regions of clusters (Sand et al. 2005, 2008;
Newman et al. 2011), the results of these studies will be discussed in detail 
later on in this review.

The proper way to accurately constrain the mass in cluster cores is thus to
use multiple-images with preferably measured spectroscopic redshifts to absolutely
calibrate the mass. To do this, one generally defines a likelihood $L$ for the observed data 
$D$ and parameters ${\mathbf p}$ of the model: 
\begin{equation}
\mathcal{L} = \mathrm{Pr}(D|{\mathbf p})
            = \prod_{i=1}^N \frac{1}{\prod_{j=1}^{n_{i}} \sigma_{ij} \sqrt{2\pi}}\exp^{-\frac{\chi^2_i}{2}}\;,
\end{equation}

 \noindent where $N$ is the number of systems, and $n_i$ the number
 of multiple-images for the system $i$. The contribution from the multiple-image system $i$ to the overall
 $\chi^2$ can be simply given by:

\begin{equation}
\chi^2_i = \sum_{j=1}^{n_i}
        \frac{[\theta^j_\mathrm{obs} - \theta^j({\mathbf p})]^2}{\sigma_{ij}^2}\;,
\end{equation}

\noindent where $\theta^j({\mathbf p})$ is the position of image $j$ predicted
by the current model, whose parameters are ${\mathbf p}$ and where $\sigma_{ij}$
is the error on the position of image $j$.

The accurate determination of $\sigma_{ij}$ depends on the signal-to-noise of the image S/N
ratio. For extended images, a pixellated approach is the only
accurate way to take the S/N ratio of each pixel into account
(Dye \& Warren 2005;  Suyu et al. 2006) but this is not optimal for cluster lenses
with a large number of multiple-image systems. However, to a first approximation, the positional 
error of images can be determined by fitting a 2D Gaussian profile to the image
surface brightness, which assumes that the
background galaxy is compact and its surface brightness profile is
smooth enough so that the brightest point in the source plane can be reliably matched 
to the brightest point in the image plane.

A major issue in the $\chi^2$ computation is how to match the
predicted and observed images one by one. In models producing
different configurations of multiple-images (e.g. a radial system
instead of a tangential system), the $\chi^2$ computation will fail and the 
corresponding model will then be rejected. This usually happens when the model is not
yet well determined, and this can slow down the convergence of the modeling
significantly. To get around this complexity, one often computes the $\chi^2$ in the
source plane (by computing the difference in the source position for a
given parameter sample ${\mathbf p}$) instead of  doing so in the image plane. The
source plane $\chi^2$ is written as:

\begin{equation}
\chi^2_{S_i} = \sum_{j=1}^{n_i}
        \frac{[\theta^j_\mathrm{S}({\mathbf p})-<\theta^j_\mathrm{S}({\mathbf p})>]^2}
               {\mu_j^{-2}\sigma_{ij}^2}\;,
\end{equation}

\noindent where $\theta^j_\mathrm{S}({\mathbf p})$ is the corresponding source position of
the observed image $j$, $<\theta^j_\mathrm{S}({\mathbf p})>$ is the barycenter
position of all the $n_i$ source positions, and $\mu_j$ is the
magnification for image $j$. Written in this way, there is no need to
solve the lensing equation repeatedly and so the calculation of the $\chi^2$ is very fast.
However, in the case where only a small number of multiple-image systems are used, 
source plane optimization may lead to a biased solution, typically favoring mass
models with large ellipticity.

It is important to use physically well motivated representations of the mass distribution and
adjust these in order to best reproduce the different families of observed multiple 
images ({\it e.g.} Kneib et al.\ 1996; Smith et al.\ 2001) iteratively. Indeed, once a set of 
multiple-images is securely identified, other multiple-image systems can in turn be discovered using
morphological/color/redshift-photometric criteria, or on the basis of the lens model predictions.
Better data, or data at different wavelengths may also bring new information enabling new
multiple-images to be identified increasing the number of constraints for modeling and hence 
the accuracy of mass models.

\subsubsection{Modeling the various cluster mass components}

In a cluster, the positions of multiple-images are known to great accuracy and they are usually
scattered at different locations within the cluster inner regions. A simple mass model
with one clump cannot usually successfully reproduce observed image configurations.

We know that galaxies in general are more massive than represented by 
their stellar content alone. In fact, the visible stellar-mass represents only a small 
portion (likely 10 - 20\%) of their total mass. The existence of an extended dark matter halo around
 individual galaxies has been established for disk galaxies with the measurement of their 
 flat and spatially extended rotation curves ({\it e.g.}  van Albada et al.\ 1985).  The existence of a 
 dark matter halo has been accepted for ellipticals only relatively recently ({\it e.g.} Kochanek et al.\ 
1995; Rix et al.\ 1997). These studies found that while the stellar content
dominates the central parts of galaxies, at distances larger
than the effective radius the dark matter halo dominates the total mass inventory.
What is less obvious in clusters of galaxies, given their dense environments, is how far 
the dark halos of individual early-type galaxies extend. One expects  tidal``stripping'' of extended 
dark matter halos to occur as cluster galaxies fall in and traverse through cluster cores during the assembly process. 
This is borne out qualitatively  by the strong morphological evolution observed in cluster galaxies ({\it e.g.} Lewis, Smail 
\& Couch 2000; Kodama et al.\ 2002; Treu et al.\ 2003). In fact, lensing
offers a unique probe of the mass distribution on these smaller scales within cluster environments.

The lensing effects of individual galaxies in clusters was first noted by
Kassiola et al. (1992) who detected that the lengths of the triple arc in
Cl0024+1654 can only be explained if the galaxies near the `B' image
were massive enough. Detailed treatment of the individual galaxy contribution to
the overall cluster mass distribution became critical with the refurbishment of 
the \hst\  as first shown by Kneib et al.\ (1996). It was found that
cluster member galaxies and their associated individual dark matter halos need 
to be taken into account to accurately model the observed strong lensing features 
in the core of Abell 2218.

The theory of what is now referred to as  
galaxy-galaxy lensing in clusters was first formulated and discussed in detail by
Natarajan \& Kneib (1997), and its application to data followed shortly
(Natarajan et al.\ 1998; and Geiger \& Schneider 1999).  From their detailed analysis
of the cluster AC 114, Natarajan et al.\ (1998) concluded that dark matter 
distributed on galaxy-scales in the form of halos of cluster members contributes about
10\% of the total cluster mass. Analysis
of this effect in several  cluster lenses at various redshifts
seems to indicate that tidal stripping does in fact severely truncate the dark 
matter halos of infalling cluster galaxies. The dark matter halos of early-type galaxies 
in clusters is found to be truncated
compared to that of equivalent luminosity field galaxies (Natarajan et al.
2002, 2004). More recent work finds that tidal stripping is on average more 
efficient for late-type galaxies compared to early-type galaxies (Natarajan
et al. 2009) in the cluster environment.

Lens models need to include the contribution of small scale 
potentials in clusters like those associated with individual cluster galaxies to reproduce the
observed image configurations and positions. As there are only a finite number of
multiple-images, the number of constraints is limited. It is therefore
important to limit the number of free parameters of the model and keep
it physically motivated -- as in the end -- we are interested in 
deriving physical properties that characterize the cluster fully. 

Generally, in these parametric approaches, the cluster gravitational potential is decomposed in the following manner:
\begin{equation}
\label{eq:decomposition}
\phi_{tot} = \sum_i \phi_{c_i} + \sum_j \phi_{p_j}\;,
\end{equation}
\noindent where we distinguish smooth, large-scale
potentials $\phi_{c_i}$, and the sub-halo potentials $\phi_{p_j}$ that are associated with the halos 
of individual cluster galaxies as providing small perturbations (Natarajan \& Kneib 1997). The smooth 
cluster-scale halos usually represent both the dark matter and the
intra-cluster gas. However, combining with X-ray observations, each of these two
components could in principle be modeled separately. For complex systems, more than
one cluster-scale halo is often needed to fit the data. In fact,  this is the case for many clusters:
Abell 370, 1689, 2218 to name a few.

The galaxy-scale halos included in the model represent all the massive cluster member 
galaxies that are roughly within two times the Einstein radius of the cluster. This is generally 
achieved by selecting galaxies within the cluster red sequence and picking the brighter ones 
such that their lensing deflection is comparable to the spatial resolution of the lensing 
observation. Studies of galaxy-galaxy lensing in the field have shown that a strong correlation 
exists between the light and the mass profiles of elliptical galaxies (Mandelbaum et al. 2006). 
Consequently, to a first approximation, in  mass modeling the location, ellipticity  and 
orientation of the smaller galaxy halos are matched to their luminous counterparts.

Except for a few galaxy-scale sub-halos that do perturb the locations of  multiple-images in their vicinity or alter 
the multiplicity of lensed images in rare cases, the
vast majority of galaxy-scale sub-halos act to merely increase the total mass enclosed
within the Einstein radius. In order to reduce the number of parameters required to describe 
galaxy-scale halos, well motivated scaling relations with luminosity are often adopted. Following the work 
of Brainerd et al.(1996) for galaxy-galaxy lensing in the field,  galaxy-scale sub-halos within clusters are 
usually modeled with individual PIEMD potentials. The mass profile parameters for this model are the core
radius ($r_{core}$), cut-off radius ($r_{cut}$), and central velocity
dispersion ($\sigma_0$), which are scaled to the galaxy
luminosity $L$ in the following way:

\begin{equation}
\label{eqscaling}
\left\{ \begin{array}{l}
\sigma_0  =   \sigma_0^\star (\frac{L}{L^\star} )^{1/4}\;,  \\
r_{core}  =  r_{core}^\star (\frac{L}{L^\star} )^{1/2}\;, \\
r_{cut}  =   r_{cut}^\star (\frac{L}{L^\star} )^\alpha\;.  \\
\end{array}
\right.
\end{equation}

The total mass of a galaxy-scale sub-halo then scales as (Natarajan \& Kneib 1997):

\begin{equation}
 M = (\pi/G)(\sigma_0^\star)^2 r_{cut}^\star (L/L^\star)^{1/2+\alpha}\;,
\end{equation}

\noindent where $L^\star$ is the typical luminosity of a galaxy at
the cluster redshift. When $r_{core}^\star$
vanishes, the potential becomes a singular isothermal potential
truncated at the cut-off radius.

In the above scaling relations (Equation \ref{eqscaling}), the velocity dispersion scales with the
total luminosity in agreement with the empirically derived  Faber-Jackson 
relations for  elliptical galaxies (for spiral galaxies the Tully-Fisher should be used instead, 
but those are not numerous in cluster cores).
When $\alpha = 0.5$, the mass-to-light ratio is constant
and is independent of the galaxy luminosity, however,  if $\alpha = 0.8$, the mass-to-light ratio
scales with $L^{0.3}$ similar to the scaling seen in the fundamental plane
(Jorgensen et al. 1996; Halkola et al. 2006). Other scalings are of course permissible, and
a particularly interesting one that has been recently explored in field galaxy-galaxy lensing 
studies, is to scale the sub-halo mass distribution directly with the stellar-mass (see Leauthaud et al. 2011).  

\subsubsection{Bayesian modeling}

State of the art parametric modeling is done in the  context of a fully 
Bayesian framework (Jullo et al. 2007), where the prior is well defined and the 
marginalization is done over all the relevant model parameters that represent the cluster mass distribution. 
Indeed, the Bayesian approach is better suited than regression
techniques in situations where the data by themselves do not
sufficiently constrain the model. In this case, prior knowledge about
the Probability Density Function (PDF) of parameters helps to reduce
degeneracies in the model. The Bayesian approach is well-suited to strong
lens modeling given the few constraints generally available to
optimize a model. The Bayesian approach provides two levels of inference rather efficiently: 
parameter space exploration, and model comparison. Bayes Theorem can be written as:

\begin{equation}
\label{bayesTheorem}
\mathrm{Pr}({\mathbf p}|D,M) = \frac{\mathrm{Pr}(D|{\mathbf p},M) \mathrm{Pr}({\mathbf p}|M)}{\mathrm{Pr}(D|M)}\;,
\end{equation}

\noindent where $\mathrm{Pr}({\mathbf p}|D,M)$ is the posterior PDF,
$\mathrm{Pr}(D|{\mathbf p},M)$ is the likelihood of getting the observed data $D$
given the parameters ${\mathbf p}$ of the model $M$, $\mathrm{Pr}({\mathbf p}|M)$ is
the prior PDF for the parameters, and $\mathrm{Pr}(D|M)$ is the evidence.

The value of the posterior PDF will be the highest for the set of parameters
${\mathbf p}$ which gives the best-fit and that is consistent with the prior
PDF, regardless of the complexity of the model $M$.  Meanwhile, the
evidence $\mathrm{Pr}(D|M)$  is the probability of getting the data $D$ given
the assumed model $M$. It measures the complexity of model $M$, and,
when used in model selection, it acts as Occam's razor.\footnote{ ``All
things being equal, the simplest solution tends to be the best one.''}

Jullo et al. (2007) have implemented in \textsc{Lenstool}\footnote{LENSTOOL is available at - http://lamwws.oamp.fr/lenstool/} 
a model optimization based on a Bayesian Markov Chain Monte Carlo (MCMC) approach, which is currently widely 
used. Since this approach involves marginalizing over all relevant parameters,
it offers a clearer picture of all the model degeneracies.

\subsection{Probing the radial profile of the mass in cluster cores}

\begin{figure*}
\begin{center}
\mbox{
\mbox{\epsfig{file=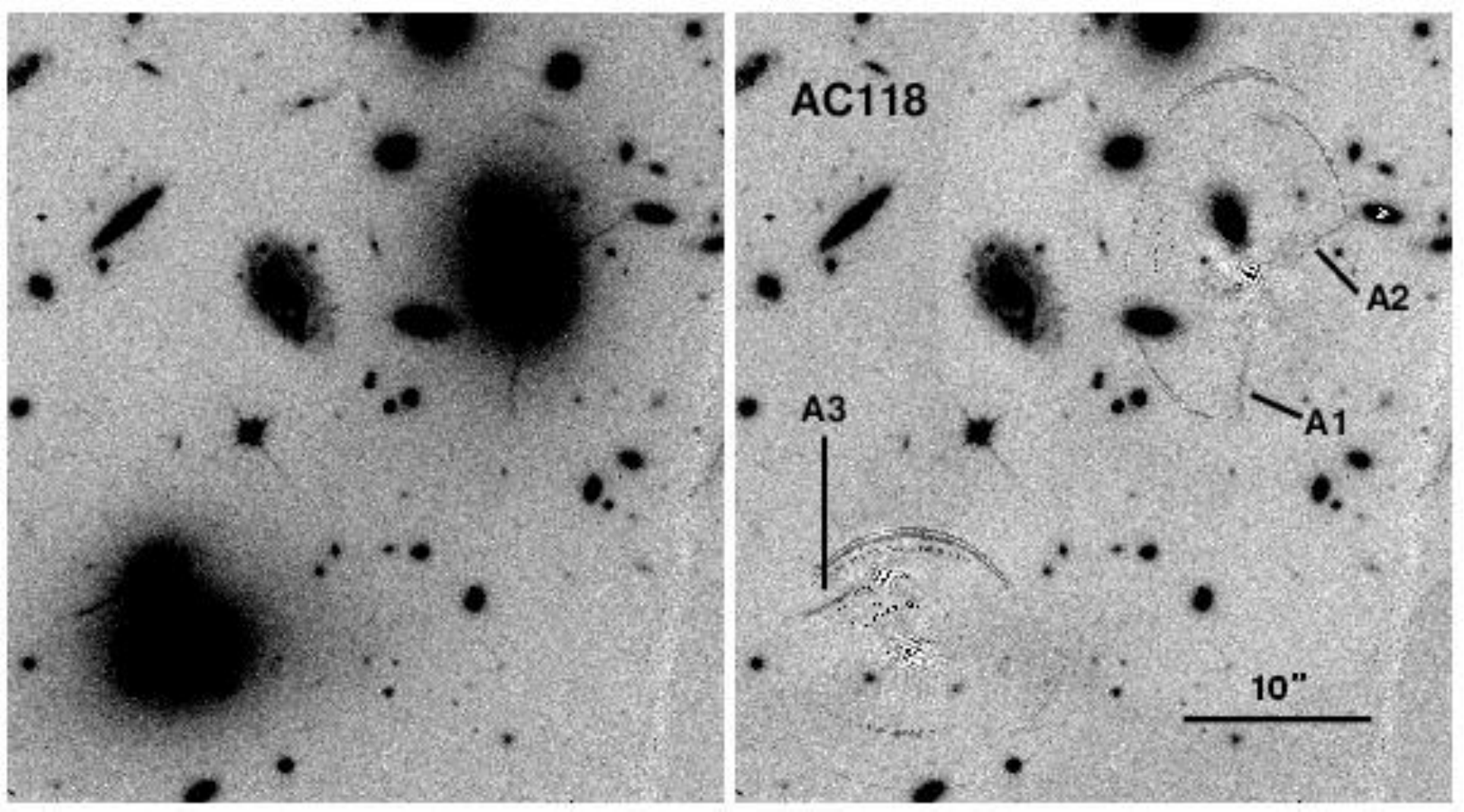,width=0.49\textwidth}}
\mbox{\epsfig{file=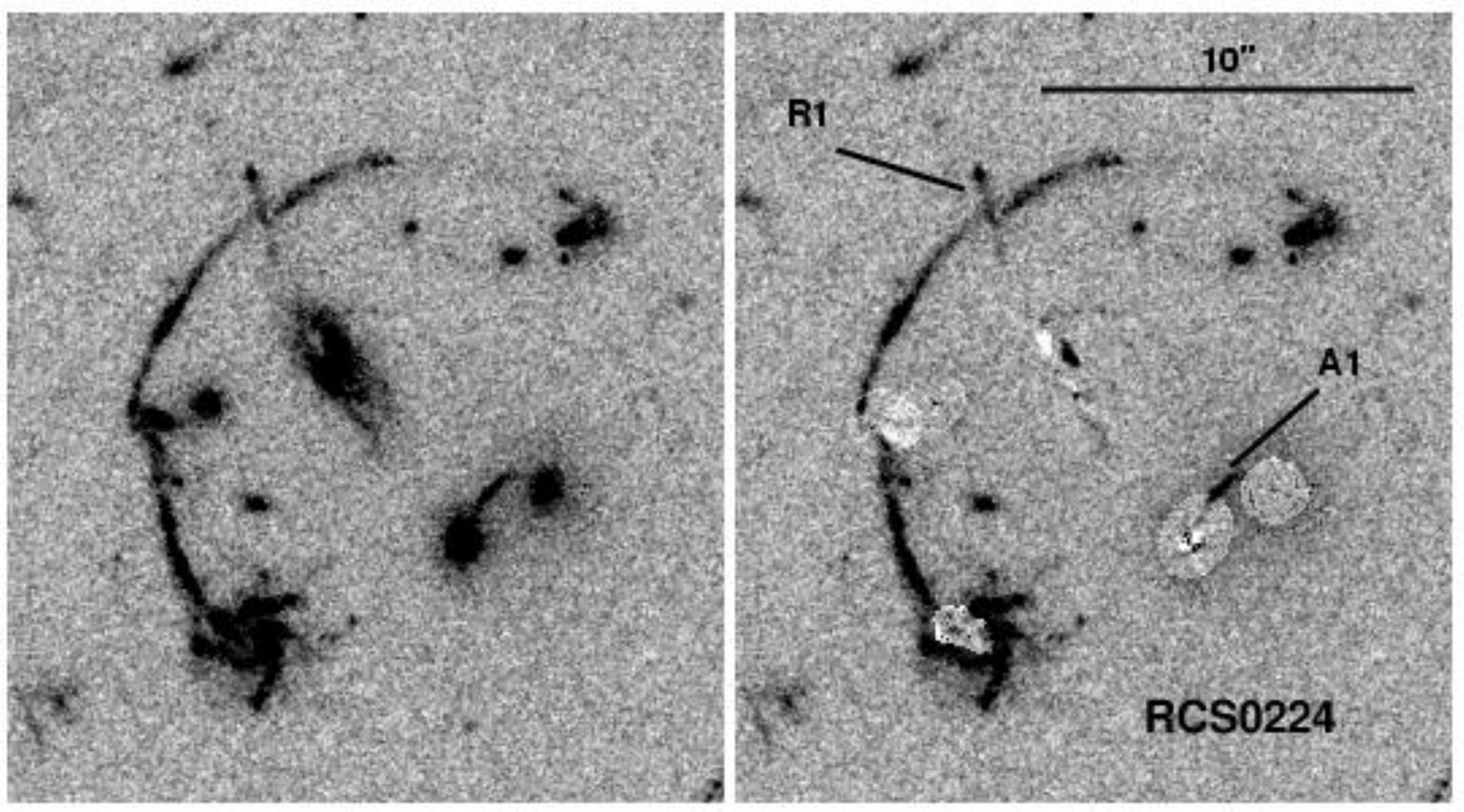,width=0.49\textwidth}}
}
\mbox{
\mbox{\epsfig{file=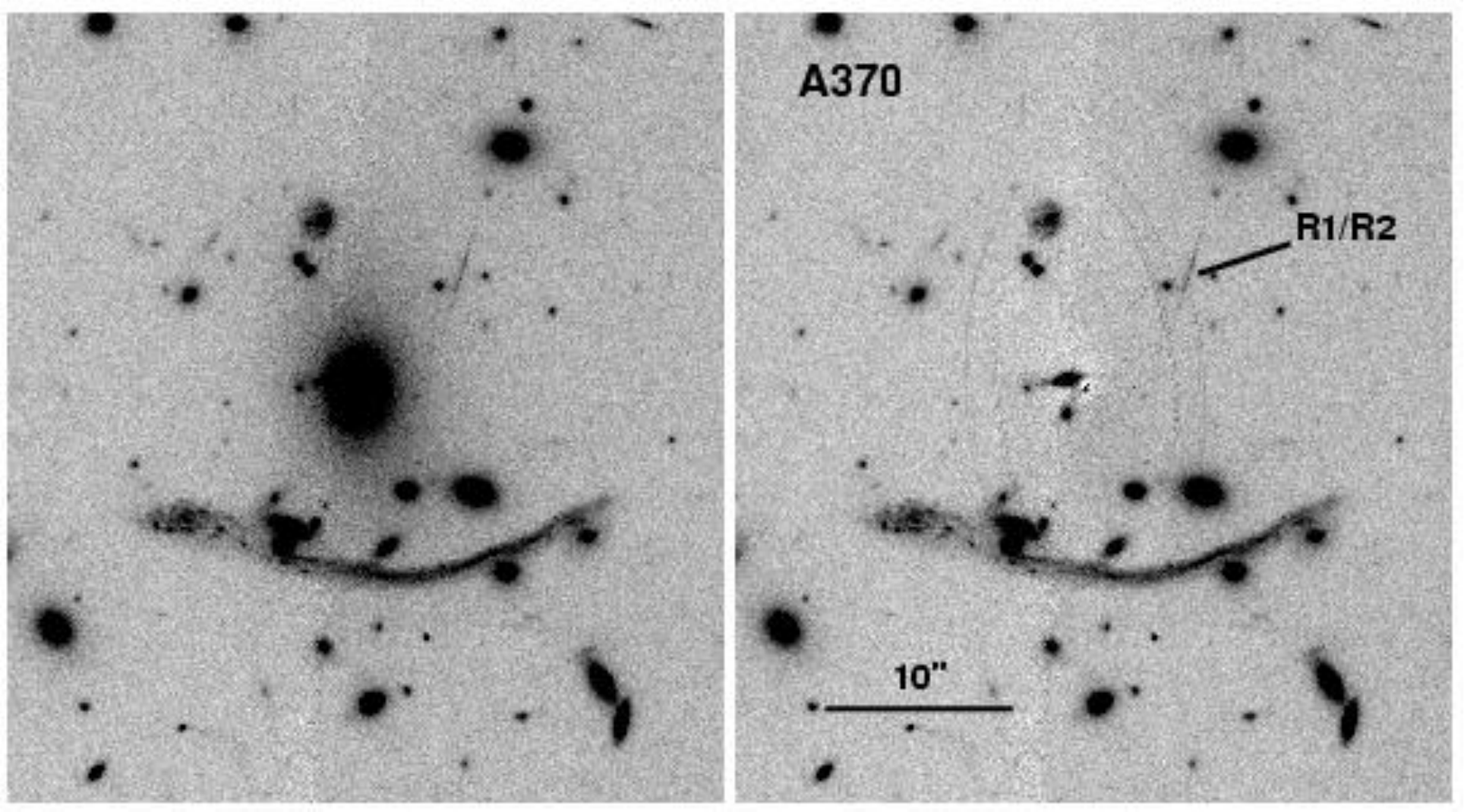,width=0.49\textwidth}}
\mbox{\epsfig{file=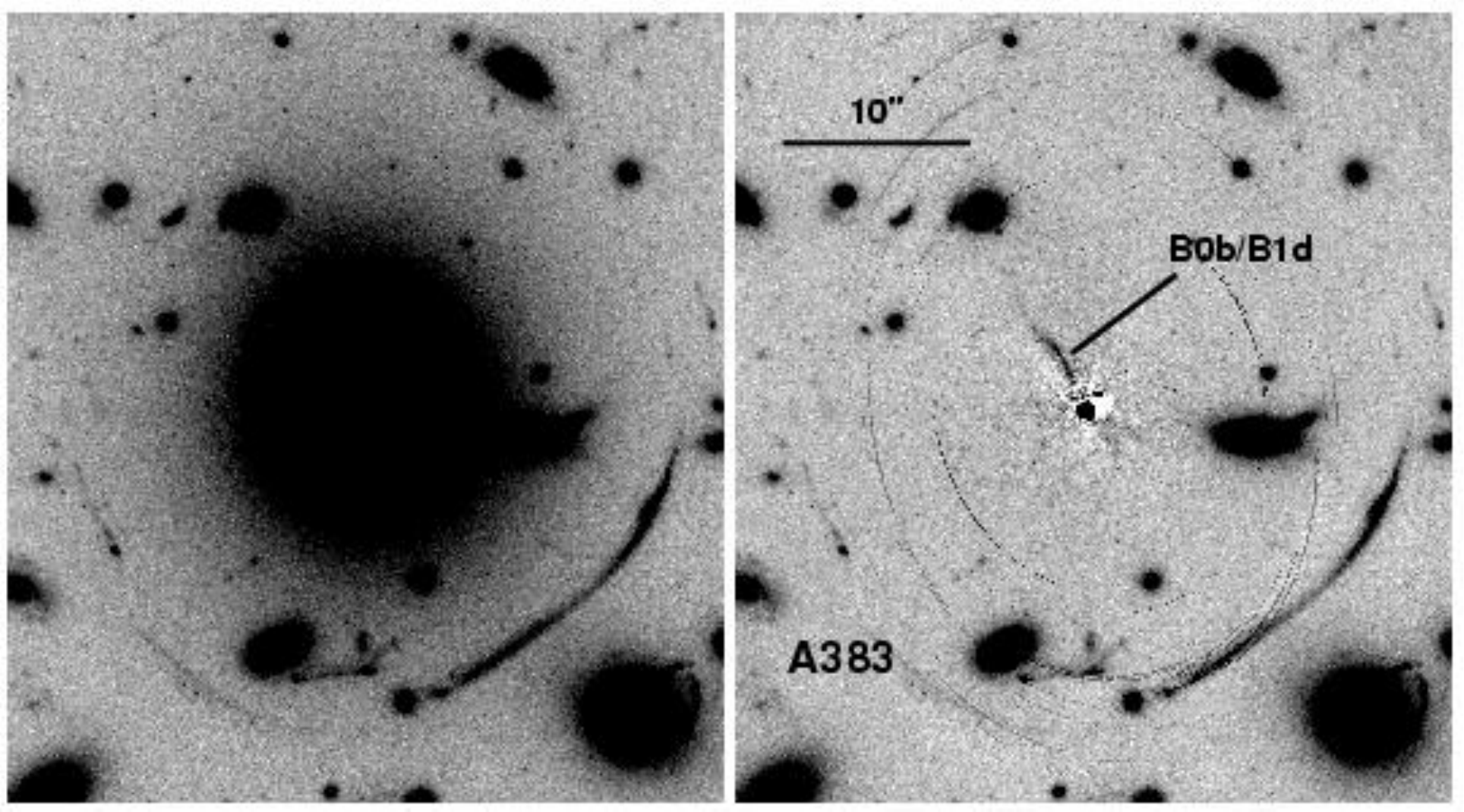,width=0.49\textwidth}}
}
\caption{Example of radial arcs found in the 4 cluster AC118, RCS0224, A370 and A383 (from Sand et al. 2005). The right side of each panel shows the BCG subtracted images. \label{fig:radarcs}}
\end{center}
\end{figure*}

One important prediction from {\it dark matter only} numerical
simulations of structure formation and evolution in a $\Lambda$CDM Universe, is 
the value of the slope $\beta$ of the density profile 
$\rho_{dark matter}\propto r^{-\beta}$ in the central part of relaxed gravitational systems.
Although there has been
ongoing debate for the past decade on the exact value of the inner 
slope (Navarro, Frenk \& White 1997 ($\beta=1$); Moore et al.\ 1998 ($\beta=1.5$)), 
the real limitation of such predictions is the lack of baryonic matter in these
simulations. Baryons dominate the mass budget and the gravitational potential in the inner most regions
of clusters and need to be taken into account while trying to constrain the inner slope of the dark 
matter density profile. This is expected to change in the near future with better numerical simulations.
Even in dark matter only simulations,
it has been found that non-singular three-parameter models, {\it e.g.} the Einasto profile has a better performance than the singular two-parameter NFW model in the fitting of a wide range of dark matter halo structures (Navarro et al. 2010).
Nevertheless, the radial slope of the total mass profile
is a quantity that lensing observations can uniquely constrain (e.g. Miralda-EscudŽ 1995). This was first 
attempted for Abell 2218 by Natarajan \& Kneib (1996) and subsequently by Smith et al.\ (2001) for the cluster Abell 383, by 
modeling the cluster center as the sum of a cD halo and a large-scale cluster component. Abell 383 is an interesting and unique system wherein both a tangential and a 
radial arc with similar redshift are observed. Such a configuration provides a particularly 
good handle on the inner slope (here considered as the sum of the stellar and Dark Matter component), which in the case of Abell 383 was found to be steeper 
than the NFW prediction.

\begin{figure}
\centerline{ \epsfig{file=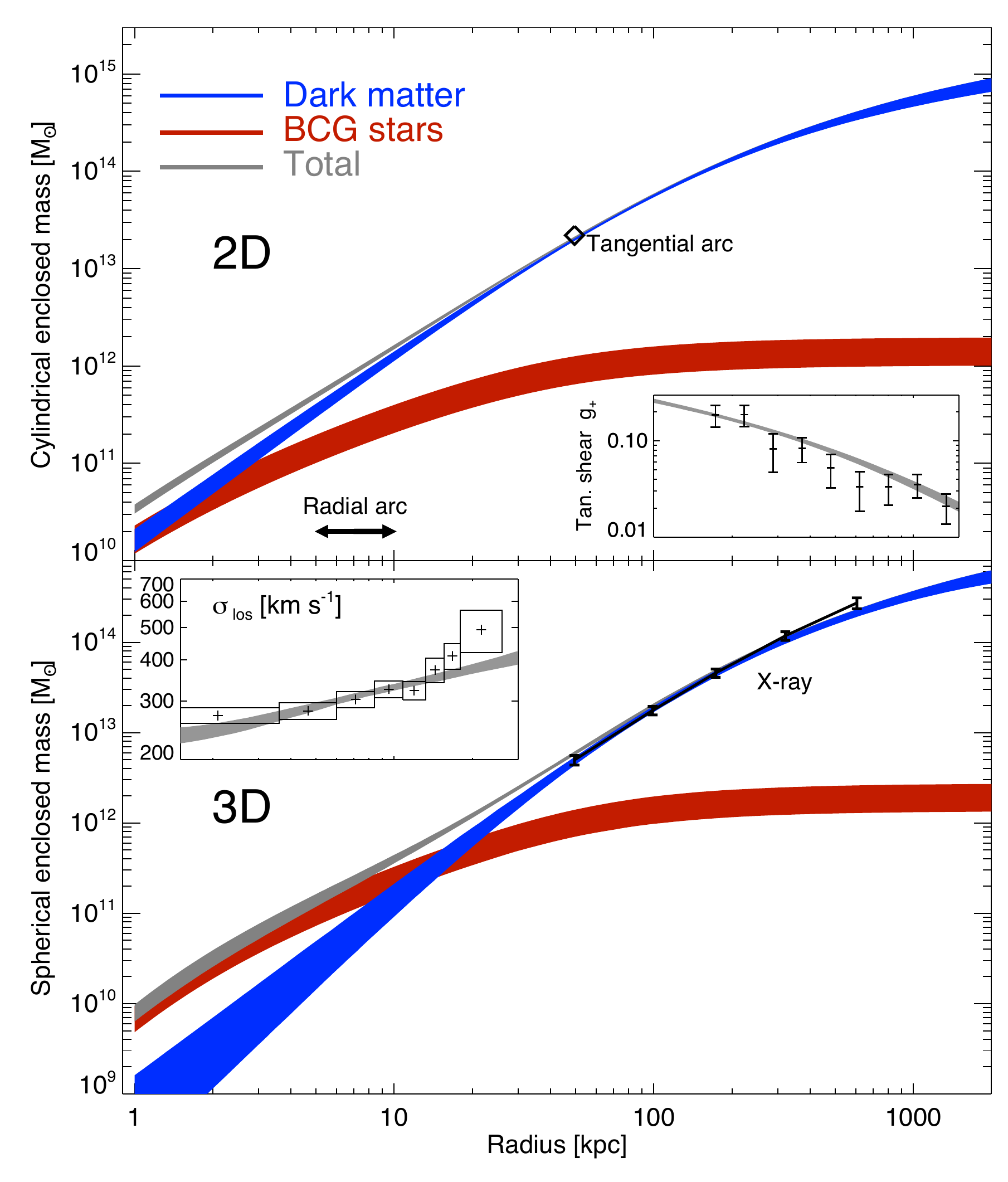,width=0.75\textwidth} }
\caption{
\emph{Top:} Projected mass for the dark matter and stellar components, as
well as the total mass distribution, with tangential reduced shear ($g$) data inset at the
same radial scale for the cluster Abell 383. \emph{Bottom:} 3D Mass, with velocity dispersion data
inset and X-ray constraints overlaid. All bands show 68\% confidence regions. The models 
acceptably fit all constraints ranging from the smallest spatial scales $\simeq 2$~kpc to $\simeq1.5$~Mpc. 
This figure is taken from Newman et al. (2011).}
\label{newman_a383}
\end{figure}

Once tangential and radial arcs have been identified from  \hst\  images (see 
Figure \ref{fig:radarcs}),  the main observational limitation is the 
measurement of the redshift of the multiply imaged arc to firmly constrain the
radial mass profile.  Large telescopes (Keck/VLT/Gemini/Subaru)
have been playing a {\it key role} in cluster lensing by
measuring the redshifts of multiple-images. 
Furthermore, working at high
spectral resolution allows one to also probe the dynamics of cD galaxies
in the core of clusters (Natarajan \& Kneib 1996; Sand et al.\ 2002).  Thus combining constraints from 
stellar dynamics, in particular, measurements of the velocity dispersion of stars
and coupling that with lensing enables the determination of the mass distribution in cluster cores. This combination is very powerful as it weighs 
the different mass components individually: stellar-mass, X-ray gas and dark matter 
in the cores of clusters. Sand et al.\ (2005) applied this technique to the clusters 
MS2137-23 and Abell 383 and found that the dark matter component is best described
by a generalized NFW model with an inner slope that is {\it shallower} than the theoretically 
predicted canonical NFW profile. A similar analysis was also conducted by Gavazzi et al.\ (2003) 
with the same result.  It must be stressed that the comparison between numerical simulations and 
observations is not direct as the stars in the cD galaxies dominate the total mass budget in the very center 
and the additional contributions of these baryons are not accounted for in the dark matter only simulations.
It is widely believed that the significant presence of baryons in cluster cores likely
modifies the inner density profile slope of dark matter, although there is disagreement at present 
on how significant this adiabatic compression is likely to be (Blumenthal et al. 1986; Gnedin et al. 2004; 
Zappacosta et al. 2006). Radial arcs offer a unique and possible only handle to probe the 
inner slopes of density profiles. Several other clusters with radial arcs have been 
discovered from \hst\  archives recently, and are currently being followed up 
spectroscopically (Sand et al. 2005, 2008).

In a recent paper, Newman et al. (2011) have obtained high accuracy velocity 
dispersion measurements for the cD galaxy in Abell 383 out to a radius 
of $\sim$26~kpc for the first time in a lensing cluster.
Adopting a triaxial dark matter distribution, an axisymmetric dynamical model and using the
constraints from both strong and weak lensing, they demonstrate that
the logarithmic slope of the dark matter density at small radii is $\beta < 1.0$ (95\% confidence),
shallower than the NFW prediction (see Figure \ref{newman_a383}).
Similar analysis of other relaxed clusters, including constraints from small to large scales
will help improve our understanding of the mass distribution in cluster cores 
and help test the assumptions used in numerical simulations where both
dark matter and baryonic matter (stars and X-ray-gas) are explicitly included.

\subsection{Non-Parametric Strong Lensing modeling}

In addition to the use of the parametric analytic mass models described above, 
there has been considerable progress in developing non-parametric mass reconstruction
techniques in the past decade ({\it e.g.} Abdelsalam et al. 1998; Saha \& Williams 1997; 
Diego et al. 2005a,b; Jullo \& Kneib 2009; Coe et al. 2010; Zitrin et al. 2010). 
Non-parametric cluster mass reconstruction methods have become more popular with the increase in 
available observational constraints from the numerous multiple-images that are now more routinely 
found in deep \hst\  data ({\it e.g.} Broadhurst et al. 2005). Non-parametric models have increased 
flexibility which allows a more comprehensive exploration of allowed mass distributions. These 
schemes are particularly useful to model extremely complex mass distributions such as the ÒBullet ClusterÓ 
(Brada{\v c} et al. 2005).
 
Contrary to  the analytic profile driven``parametric'' methods,  in Ònon-parametricÓ schemes, the mass distribution is 
generally tessellated into a regular grid of small mass elements, referred often to as pixels  
(Saha \& Williams 1997; Diego et al. 2005a).  Alternatively, instead of starting with mass elements, 
Brada{\v c} et al. (2005) prefer tessellating the gravitational potential because its derivatives directly 
yield the surface density and other important lensing quantities that can be related more straightforwardly
to measurements. Pixels can also be replaced by radial basis functions (RBFs) that are real-valued functions with radial 
symmetry. Several RBFs for density profiles have 
been tested so far. Liesenborgs et al. (2007) use Plummer profiles, and Diego et al. (2007) use 
RBFs with Gaussian profiles. The properties of power law profiles, isothermal profiles as well 
as Legendre and Hermite polynomials have been explored as RBFs. These studies find that the use of 
compact profiles such as the Gaussian or Power law profiles are generally preferred as they are more
accurate in reproducing the surface mass density.

In more recent work, instead of using a regular grid, Coe et al.
(2008, 2010) and Deb et al. (2008) use the actual distribution of images
as an irregular grid. Then, they either place RBFs on this grid or
directly estimate the derivatives of the potential at the location of the images.
Whatever their implementation, the reproduction of multiple-images is 
generally greatly improved with respect to traditional ÒparametricÓ 
modeling with these techniques. However, the robustness of these models is
still a matter of debate given the current observational constraints available from data.
Indeed, due to the large amount of freedom that inevitably goes with the
large degree of flexibility afforded by a ``non-parametric'' approach, many 
models can fit perfectly the data and discriminating between models is challenging. 
To identify the best physically motivated model and eventually learn more about the  dark
matter distribution in galaxy clusters, additional external criteria (e.g. mass
positivity) or regularization terms (e.g. to avoid unwanted high
spatial frequencies) are necessary. 
Furthermore, galaxy mass scales are usually not taken into account in these non-parametric schemes, despite the fact 
that successful parametric modeling has clearly demonstrated that these smaller scale mass clumps 
do significantly affect the positions of observed multiple-images. This is a key limitation 
of most of the Ònon-parametricÓ models. 

\begin{figure} 
        \center 
        \includegraphics[width=0.75\linewidth]{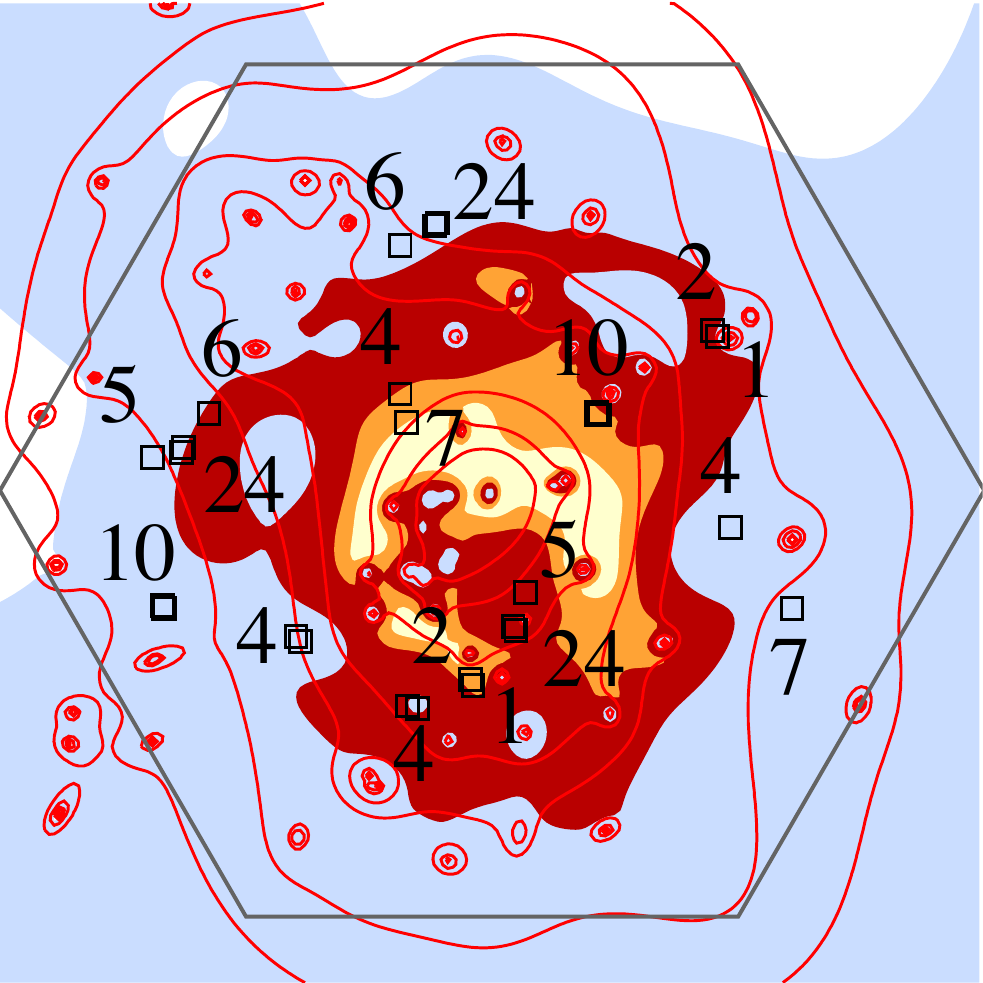}
        \caption{ \label{jullo2009sn} Map of S/N ratio for the A1689
        mass reconstruction using the JK09 ``non-parametric'' mass reconstruction.
        Colored contours bound regions with S/N greater than 300, 200, 100 and
        10. The highest S/N region is at the center where there are the most constraints. 
        Red contours are mean iso-mass contours. Black boxes mark the positions of the multiple-images
        used to constrain the mass distribution, and the numbers indicate the 
        different multiple-image systems (from JK09). }
\end{figure}

Jullo \& Kneib (2009) [JK09 hereafter] have proposed a novel modeling scheme that includes both a multi-scale grid of RBFs 
and a sample of analytically defined galaxy-scale dark matter halos, thus allowing combined modeling
of both complex large-scale mass components and galaxy-scale halos. In this hybrid scheme, similar to the
one adopted by Diego et al. (2005a), JK09 define a coarse multi-scale grid from a pixelated input mass map and 
recursively refine it in the densest regions. However, in contrast to previous work, they start from
a hexagonal grid (composed of triangles), on the grounds that it better fits the generally roundish shapes of galaxy 
clusters. For their RBF, they use truncated circular isothermal mass models, and truncated
PIEMD models to explicitly include galaxy-scale halos. Thus both components are modeled with similar analytical
functions which permits a simple combination for ready incorporation into the Bayesian MCMC optimization
scheme built into \textsc{LENSTOOL}. Figure \ref{jullo2009sn} shows 
the derived S/N convergence map for the model of the cluster A1689. As the S/N of the mass reconstruction is found 
to be larger than 10 everywhere inside the hexagon, the error in the convergence derived mass is less than 10\%, demonstrating
the power of this hybrid scheme.


\subsection{Cluster Weak lensing modeling}

As soon as  we look a little bit further out radially from the cluster core, the
lensing distortion gets smaller (distortions in shape get to be of the order
of a few percent at most), and very quickly the shape of faint
galaxies gets dominated by their intrinsic ellipticities (the dispersion of the intrinsic ellipticity 
distribution of observed galaxies $\sigma_{\epsilon}\sim 0.25$).
Thus the lensing distortion is no longer visible in individual images and can only
be probed in a statistical fashion, characteristic of the weak lensing regime (e.g. Bartelmann 1995). 
The nature of constraints provided by observations are fundamentally different in the weak
lensing regime compared to the strong lensing regime. In the strong regime, as we shown above 
every set of observed multiple-images provides strong constraints 
on the mass distribution. In the weak regime, however, what is measured are the 
{\it mean} ellipticities and/or the {\it mean} number density of faint galaxies in the frame.  In order to
relate these to the mean surface mass density $\kappa$ of the cluster,  these data need to be 
used statistically. There are two key sets of challenges in doing so:

 $\bullet$ {\it Observational} : How to best determine the `true' ellipticity of an observed faint
galaxy image which is smeared by a PSF of comparable size, and is not circular (as a result of camera 
distortions, variable focus across the image, tracking and guiding errors) and not stable in time?  
How best to estimate and isolate the variation in the number density of faint galaxies due to 
lensing, while taking into account the crowding effect due to the presence of
cluster members and the intrinsic spatial fluctuations in the distribution of
galaxies; and the unknown redshift distribution of background sources?\\
$\bullet$ {\it Theoretical} : what is the optimal 
method to reconstruct the surface mass density distribution $\kappa$ (as a mass map or
a radial mass profile) using either the `reduced shear field' $\vec g$ and/or
the amplification?\\

Various approaches have been proposed to solve these sets of problems, and
two distinct families of methods can be distinguished: {\bf direct} and {\bf
inverse} methods. We describe them in detail in what follows.

\begin{figure}
\centerline{\epsfig{file=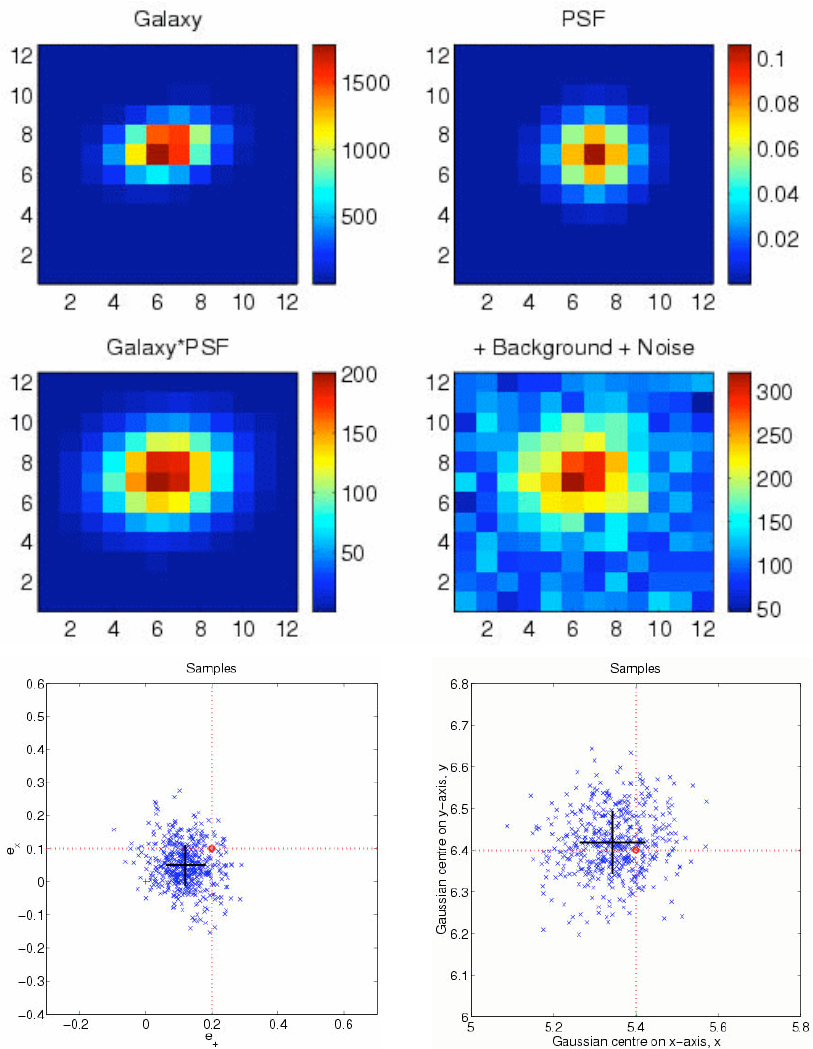,width=\textwidth}}
\caption{ Simulated images demonstrating the various sources of noise in weak lensing data: 
the galaxy model, the PSF model, the convolution of the two, and the final image when noise
is added. Simple simulations allow the production of model images that include
observational sources of noise and can therefore be compared directly to the 
shapes of observed faint galaxies. The best galaxy model is then found by taking 
into account the convolution effect of the measured PSF that best-fits the data.  
{\it (Bottom panels):} the MCMC samples fitting the final image in terms of its ellipticity vector - with coordinates [$e_{+}=e_{1}$] and [$e_{\times}=e_{2}$] - (on the left) and 
in terms of the position (x,y) of the image center (right), 
from which the best-fit fiducial models with errors can be
extracted (black crosses). One of the first implementations of this $inverse$ technique was
done in the \textsc{Im2shape} software (Bridle et al. 2002).
\label{figjpk6}
}
\end{figure}

\subsubsection{Weak lensing observations}

For observers, before any data handling, the first step is to
choose the telescope (and instrument) that will minimize the sources of noise in the
determination of the ellipticity of faint galaxies. Although \hst\  has the best characteristics in terms of the PSF size, it has a very limited field of view $\sim10$ square arcminutes for the ACS camera. 
Furthermore, $Hubble$ is ``breathing'' as it is 
orbiting around the Earth, which affects the focus of its instruments, and the most recent ACS 
imaging data appear to be suffering from Charge Transfer Inefficiency (CTI)
which needs special post-processing for removal of these instrumental effects (Massey 2010).
Although, as we have seen \hst\ is ideal for picking out strong lensing features (e.g. Gioia et al 1998), it  is not the most appropriate instrument to probe the large scale distribution of a 
cluster extending out to and beyond the virial radius. Note however, this limitation becomes less of a 
problem when observing high-redshift clusters (e.g. Hoekstra et al 2002, Jee et al 2005, Lombardi et al 2005).

On the ground a number of wide-field imaging cameras have been used to conduct 
weak lensing measurements in cluster fields. The most productive ones in the last decades have been:
the CFHT12k camera ({\it e.g.} Bardeau et al. 2007, Hoekstra 2007) and the more recent Megacam camera ({\it e.g.} Gavazzi \& Soucail 2007, Shan et al. 2010) at CFHT, and the Suprimecam on Subaru ({\it e.g.} Okabe et al. 2010). 
However a number of studies have also been done with other instruments, in particular, 
the VLT/FORS ({\it e.g.} Cypriano et al. 2001, Clowe et al. 2004), 2.2m/WFI ({\it e.g.} Clowe \& Schneider 2002), and more recently with 
the LBT camera (Romano et al. 2010).

\subsubsection{Galaxy shape measurement}

Once the data have been carefully taken either from the ground or space with utmost care to minimize 
contaminating distortions and hopefully under the best seeing conditions, the next step is to convert the 
image of the cluster into a catalog where the PSF corrected shapes of galaxies are computed.

Before measuring the galaxy shape a number of steps are usually undertaken: 1) masking of the data and 
identifying the regions in the image that suffer from observational defects: bleeding stars,
 satellite tracks, hot pixels, spurious reflections; 2) source identification  and catalog production 
 which is usually done using the \textsc{Sextractor} software (Bertin \& Arnouts 1996); 3) identification of the  stellar objects  in order to accurately compute the PSF as a function of position in the image.
Once this is done, the shapes of the stellar objects and galaxies can be computed 
to derive the PSF corrected shapes of galaxies which will then be used for weak lensing measurements.
For accomplishing this crucial next step, a popular and direct approach is often used, to convert galaxy shapes to 
shear measurements using the \textsc{Imcat} software package. This implementation is based on the Kaiser, 
Squires and Broadhurst (1995) methodology, but has been subsequently improved by various other 
groups ({\it e.g.} Luppino \& Kaiser 1997;  Rhodes et al.\ 2000; Hoekstra et al. 2000), providing variants of the original KSB technique.
To correct the galaxy shape from the PSF anisotropy and circularization, the KSB technique uses the 
weighted moment of the object's surface brightness to find its center,  and shape to measure
higher order components that can be used to improve the PSF correction. As a result the correction is fast
and processing a large amount of data can be done swiftly and efficiently. 

Alternatively, one can use the inverse approach using maximum likelihood methods or Bayesian techniques to find the optimal 
galaxy shape that when convolved with the local PSF best reproduces the observed galaxy shape ({\it e.g.}  Kuijken, 1999; 
{\sc Im2Shape}: Bridle et al.\ 2002). A recent implementation of this inverse method is available in the \textsc{Lensfit}
software package (Miller et al. 2007; Kitching et al. 2008) and one of its key advantages is that it works directly on the individual exposures of a given field. 
\textsc{Lensfit} has been developed in the context of the CFHT-LS survey, but is flexible and can be easily adapted for use with other
observations. 

These inverse approaches have the advantage that they provide a direct estimate of the 
uncertainty in the parameter recovery as illustrated in Figure~\ref{figjpk6}. Further extension of these inverse techniques,
has led to the use of {\sc Shapelets} (Refregier 2003; Refregier \& Bacon 2003) that offer a more sophisticated basis set to 
characterize the two-dimensional shapes of  the PSF and faint galaxies. The versatility of shapelets has made this technique 
quite popular for lensing measurements. Nevertheless, it has been realized that these different shape measurement recipes need to be tuned, compared  
and calibrated amongst each other in order to obtain accurate, unbiased and robust shear measurements. This calibration work has been done in the context 
of various numerical challenges, wherein different research groups measure the shapes of the same set of simulated images as part of STEP (Heymans et al. 2006; Massey et al. 2007);  the GREAT08 and GREAT10 challenges (Bridle et al. 2010; Kitching et al. 2011). These challenges have proven to be very useful exercises for the
community as they have enabled calibration of the several independent techniques employed to derive shear from observed shapes.

\subsubsection{From galaxy shapes to mass maps}

From the catalog of shape measurements of faint galaxies, a mass map
can be derived. And here again direct and inverse methods have been explored.
The {\bf direct} approaches are: ({\it i)} the Kaiser \& Squires (1993)
method - this is an integral method, that expresses $\kappa$ as the convolution
of $\vec\gamma$ with a kernel and subsequent refinements thereof ({\it e.g.}
Seitz et al.\ 1995, 1996, Wilson et al 1996a); and ({\it ii)} a local inversion method
(Kaiser 1995; Schneider 1995; Lombardi \& Bertin 1998) that involves the 
integration of  the gradient of $\vec \gamma$ within the boundary of the observed
field to derive $\kappa$. This technique is particularly relevant for datasets that 
are limited to a small field of view.

The {\bf inverse} approach works for both the $\kappa$ field and the lensing
potential $\varphi$ and uses a maximum likelihood ({\it e.g.} Bartelmann et al.\
1996; Schneider et al.\ 2000; King \& Schneider 2001), maximum entropy
method ({\it e.g.} Bridle et al.\ 1998; Marshall et al.\ 2002) or atomic
inference approaches coupled with MCMC optimization 
techniques (Marshall 2006) to determine the most likely mass distribution
(as a 2D mass map or a 1D mass profile) that reproduces the reduced
shear field $\vec g$ and/or the variation in the faint galaxy number
densities. These inverse methods are of great interest as they enable
quantifying the errors in the resultant mass maps or mass estimates
({\it e.g.} Kneib et al. 2003), and in principle, can cope with the addition of 
further external constraints from strong lensing or X-ray data simultaneously.
Wavelet approaches that use the multi-scale entropy concept have also been 
extremely powerful in producing multi-scale mass maps (Starck, Pires \& Refregier 2006, Pires et al 2009).

\begin{figure}

\centerline{\epsfig{file=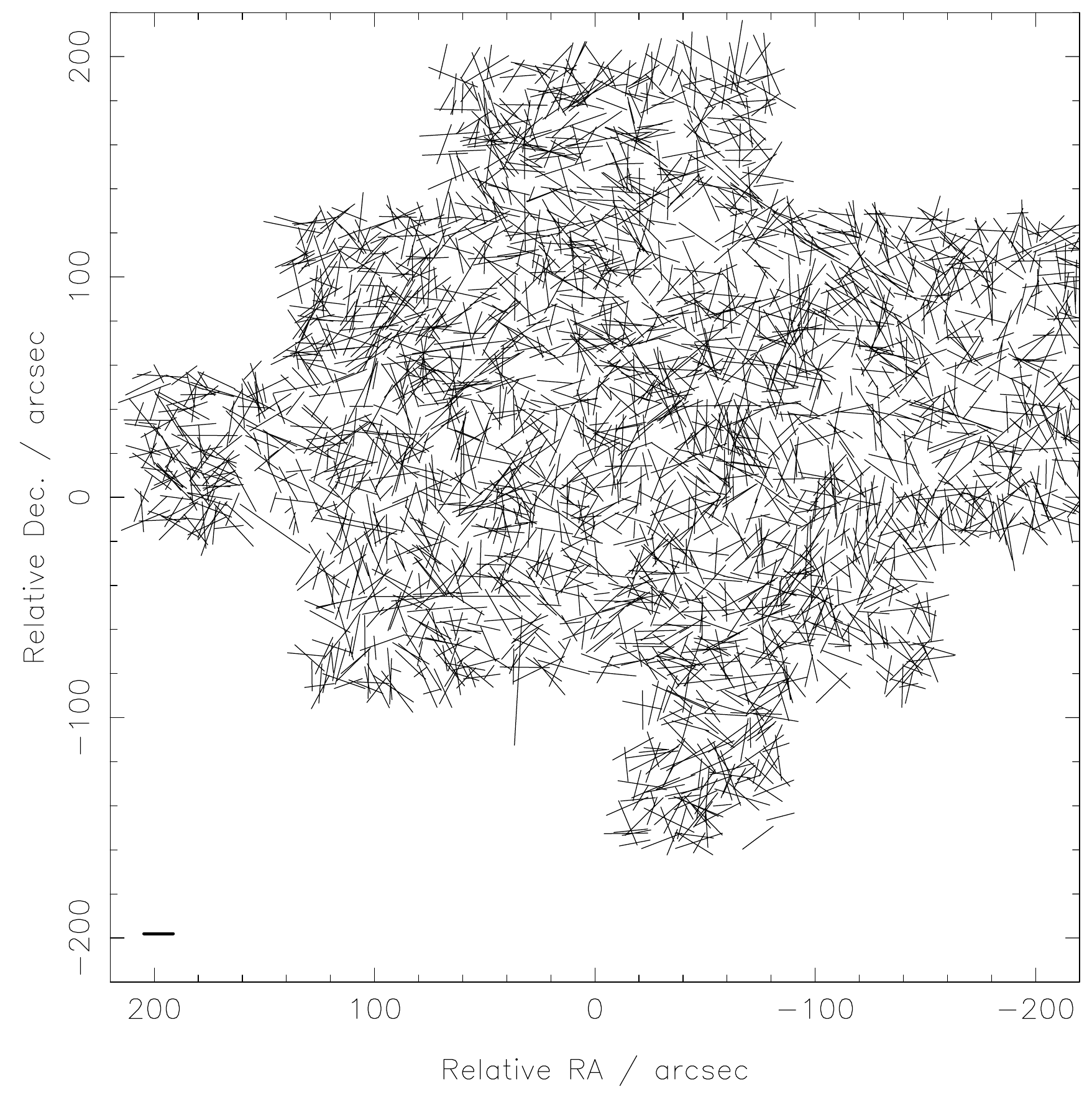,width=0.45\textwidth}\ \ \ \ \epsfig{file=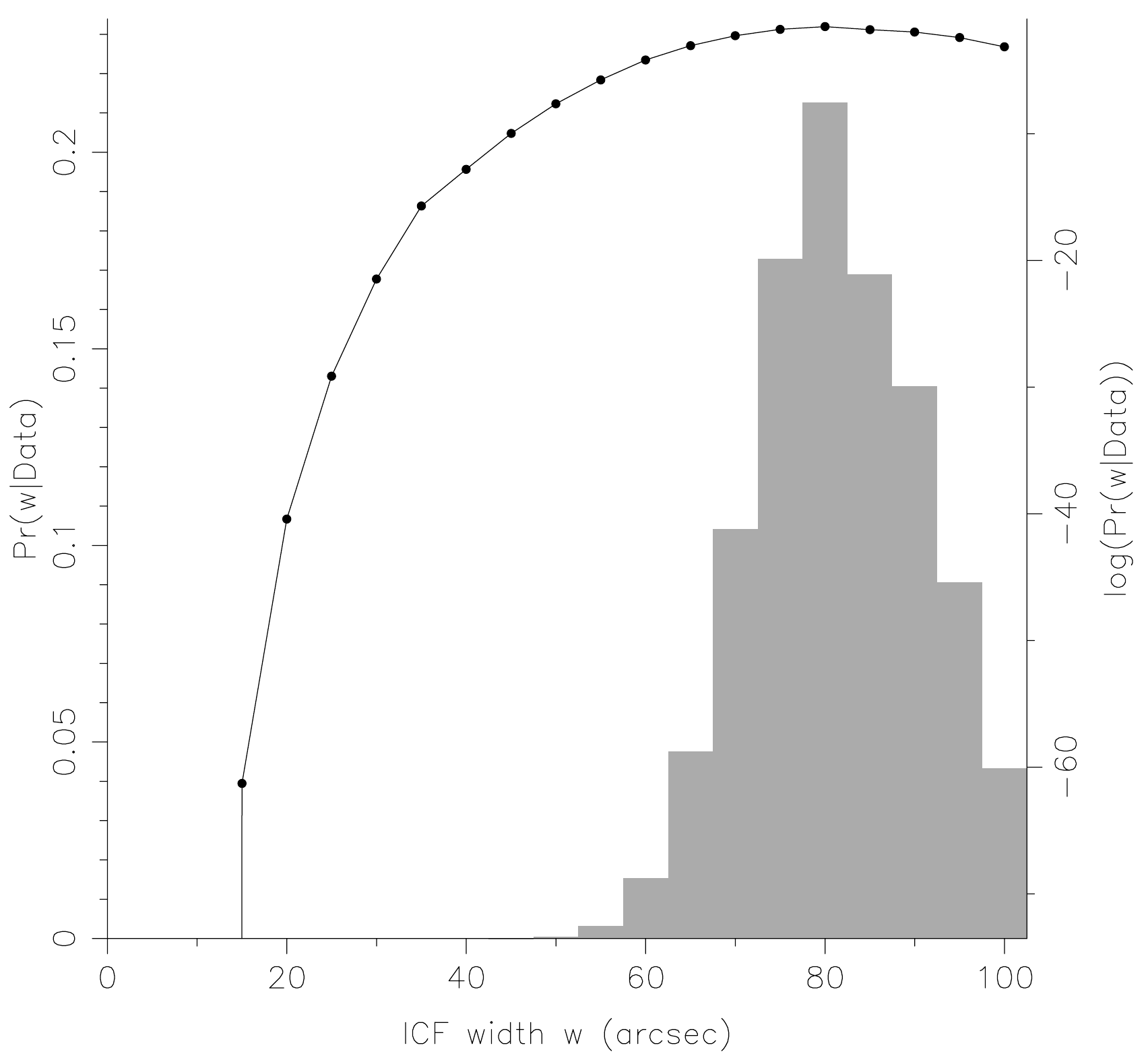,width=0.45\textwidth}}
\ 
\centerline{\epsfig{file=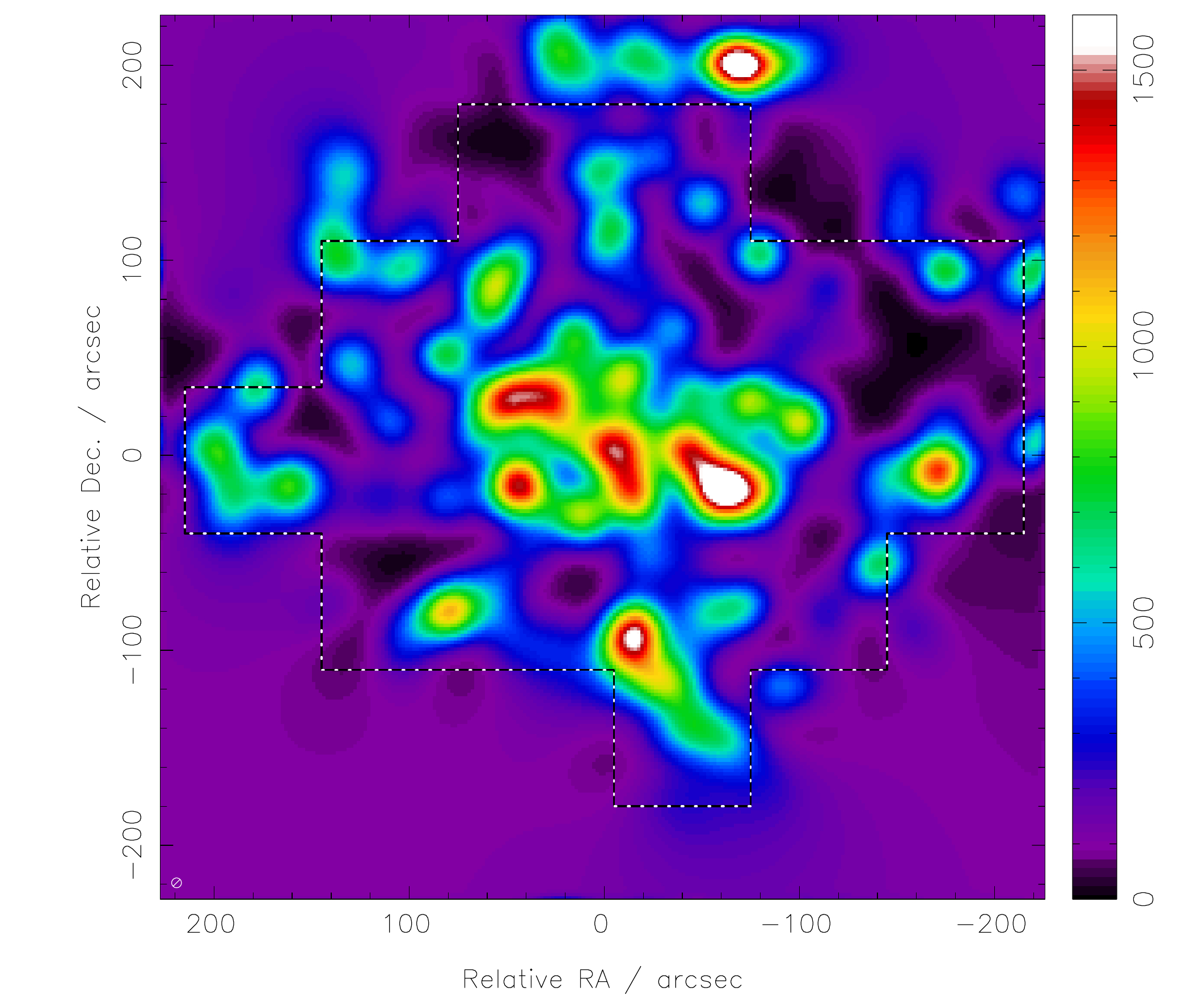,width=0.45\textwidth}\ \ \ \ \epsfig{file=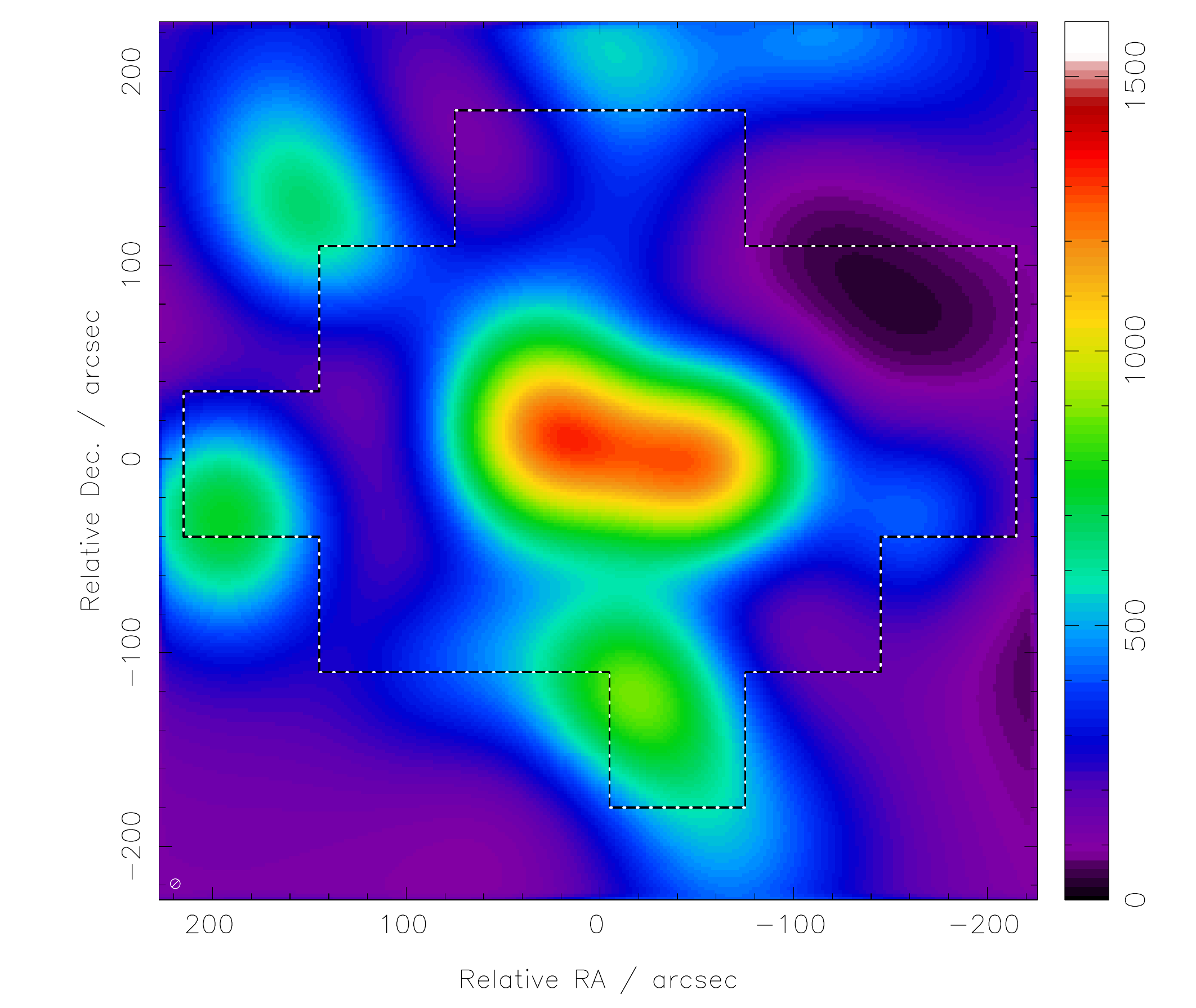,width=0.45\textwidth}}
\caption{
Maximum entropy mass reconstruction (Marshall et al. 2002, Marshall 2003)
of the X-ray luminous cluster MS1054 at $z=0.83$ using Hoekstra's \hst\  dataset
(Hoekstra et al.\ 2000). (Top Left) Distribution of the positions
of galaxies used in the mass reconstruction. (Top Right) Evidence values
for different sizes of the Intrinsic Correlation Function (ICF). (Bottom) Two mass 
reconstructions illustrate the case of 2 different values for the ICF: (left) 
small ICF with a low evidence value, (right) large ICF with the largest evidence.}
\label{figjpk7}
\end{figure}

An important issue for producing mass maps is the resolution 
at which the 2D lensing mass map can be reconstructed. 
Generally, mass maps are reconstructed on a fixed size
grid, which then automatically defines the minimum mass resolution that can be
obtained.  By comparing the likelihood of different resolution
mass maps, we can calculate the Bayesian evidence of each to determine the optimal
resolution (Figure~\ref{figjpk7}). However, it is most likely that the optimal
scale to which a mass map can be reconstructed is adaptive, and 
is determined by the strength of the lensing signal. As we are limited by the width of the 
intrinsic ellipticity distribution, it is only by averaging over a large
number of galaxies that we can reach lower shear levels. Thus low
shear levels can only be probed on relatively large scales by
averaging over a large number of galaxies. Furthermore, as the projected 
surface mass density of clusters on large scales falls off relatively quickly scaling as $1/R$ to $1/R^2$, 
respectively, for an SIS or a NFW profile, mass maps may quickly lose spatial resolution.

\begin{figure}
\centerline{\epsfig{file=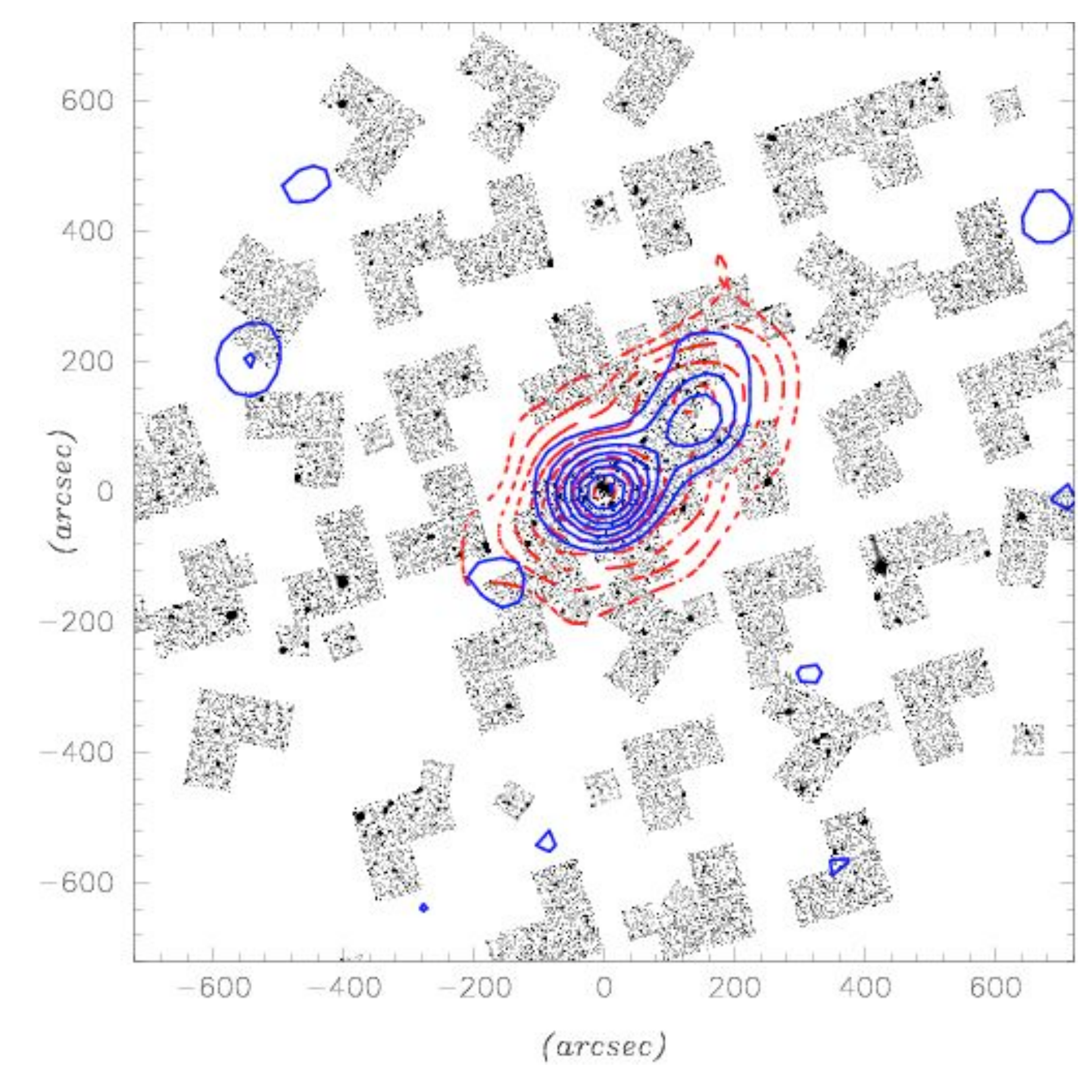,width=\textwidth}}
\caption{The 39 WFPC2/F814W, and the 38 STIS/50CCD pointings
sparsely covering the Cl0024+1654 cluster.  The (red) dashed contours
represent the number density of cluster members as derived by Czoske
et al. (2001). The blue solid contour is the mass map built from the
joint WFPC2/STIS analysis derived using the LensEnt software (Bridle
et al. 1998; Marshall et al. 2002). }
\label{hst0024}
\end{figure}

\begin{figure}
\centerline{\epsfig{file=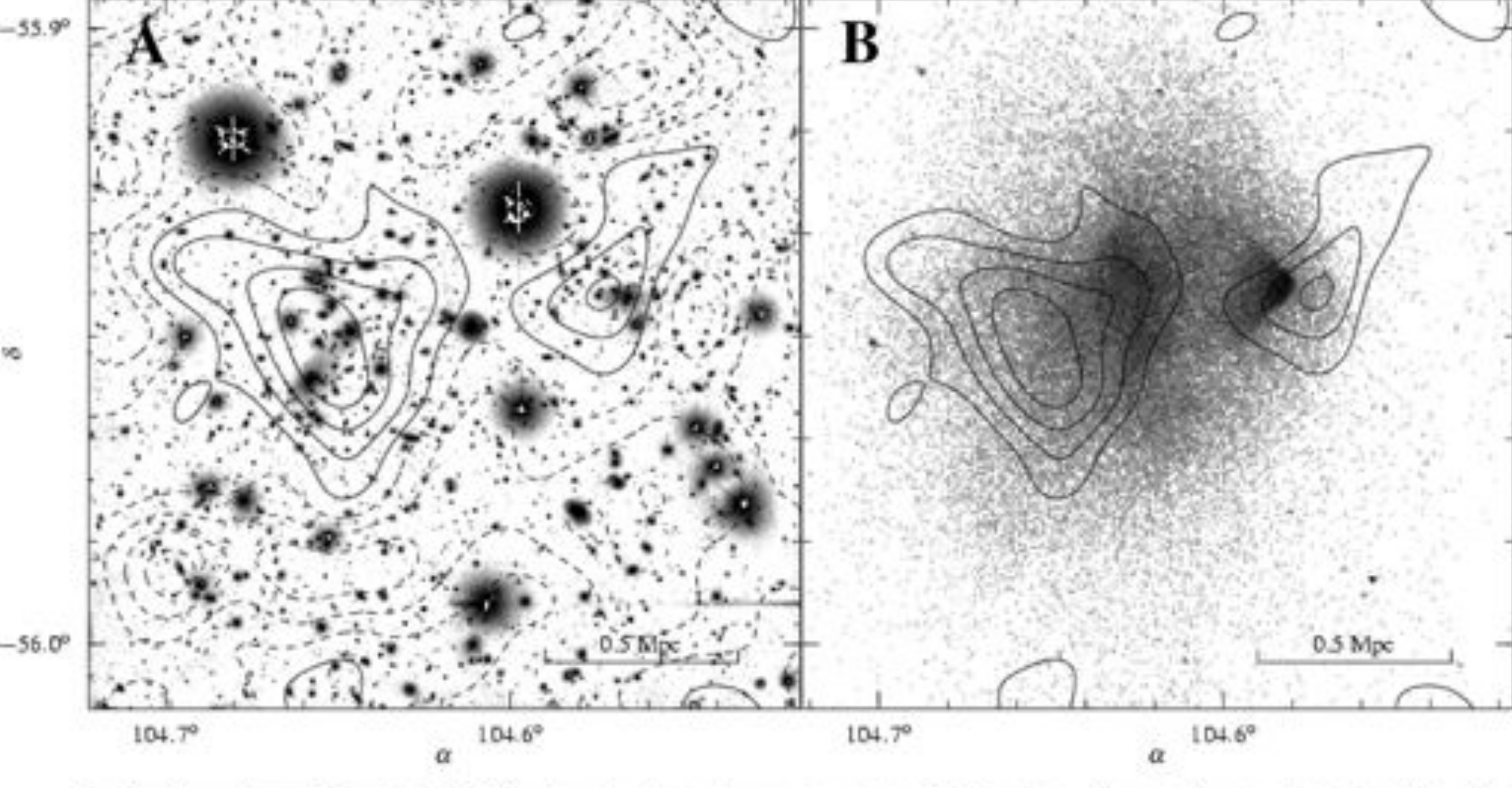,width=0.98\textwidth}}
\caption{Right Panel: Shown in greyscale is the $I$-band VLT/FORS image of the Bullet cluster used to measure
  galaxy shapes.  Over-plotted in black contours is the weak lensing mass reconstruction 
  with solid contours for positive
  mass, dashed contours for negative mass, and the dash-dotted contour for the
  zero-mass level. Left Panel:  Shown in greyscale is the
  Chandra X-ray image from Markevitch et al. (2002) with the same
  weak lensing contours as in the Right Hand panel. (Figure from Clowe et al. 2004).}
\label{clowe_bullet}
\end{figure}

Although lensing mass maps may quickly loose information content outside the cluster core, 
they can be very useful in identifying possible (unexpected) substructures on scales larger than the typical weak 
lensing smoothing scale ($\sim 1$ arcminute). This has been the case for several cluster lenses such as 
the cluster Cl0024+1654 ({\it e.g.} Kneib et al. 2003 and Figure \ref{hst0024}; Okabe et al. 2009, 2010) and the``Bullet Cluster'' (see 
Figure \ref{clowe_bullet}), and more recently in the so-called ``baby bullet'' cluster (Brada{\v c} 2009).

\subsubsection{Measuring total mass and mass profiles}

Weak lensing mass maps are useful to determine mass peaks, but they are of limited use for the
extraction of physical information. If a cluster has a relatively simple geometry (i.e. has a single mass clump in its
 2D mass map) one can easily extract the radial mass profile and compute the total mass enclosed within a given radial aperture. 
 Different approaches are currently available to compute the mass profile and the total mass.
The direct method is just to sum up the tangential weak shear as expressed in the aperture
mass densitometry first introduced by Fahlman et al. (1994) and then revised by Clowe et al. (1998). This
statistic is quite popular and has been used in several recent cluster modeling papers, including Hetterscheidt et al 2005,  Hoekstra (2007) 
and Okabe et al. (2010). Aperture mass densitometry measures the mass interior to a given radius following 
the $\zeta$ statistic, defined as:
\begin{equation}
\zeta(\theta_{1}) = \overline{\kappa}(\theta<\theta_{1}) - \overline{\kappa}(\theta_{1}<\theta<\theta_{max})
= 2(1 -{\theta_{1}^{2}\over \theta_{max}^{2}})\int_{\theta_{1}}^{\theta_{max}}<\gamma_{T}>d\ln \theta,
\end{equation}
which provides a lower bound on the mean convergence $\overline{\kappa}$ interior to radius $\theta_{1}$.
The mass within $\theta$ is then just given by:
\begin{equation}
M(<\theta)=\pi D_{OL}^{2}\theta^{2}\Sigma_{crit}\zeta(\theta).
\end{equation}
This statistic assumes however that all background galaxies are at a similar redshift, which can be a strong and severely limiting assumption, particularly for high 
redshift clusters.

Another semi-direct approach is to build the projected surface density contrast $\Delta\Sigma$ estimator as introduced by Mandelbaum et al. (2005):
\begin{equation}
\Delta\Sigma(r)= \overline{\Sigma}(<r) - \Sigma(r) = \gamma_{T}(r)\Sigma_{crit}(z_{S}).
\end{equation}
In practice, $\Delta\Sigma$ is measured by averaging over the galaxies at radius $r$ from the cluster center, and requires some information about the 
redshift distribution of background galaxies $z_{S}$. It can then be directly compared to the $ \Delta\Sigma(r)$ computed for a given parametrized mass model.
Mandelbaum et al. (2010) discuss and compare this cluster mass estimator with other proposed ones. In a recent paper, Gruen et al. (2011) compare the
use of various aperture mass estimators to calibrate mass-observable relations from weak lensing data. 

The alternative method is to directly fit the observables using simple parametric models similar to what is done in the strong 
lensing approach, for example using radially binned data ({\it e.g.} Fischer \& Tyson 1997; Clowe \& Schneider 2002; King et al 2002;  Kneib et al. 2003). The advantage of 
adopting such a method lies in its flexibility, {\it i.e.} allowing combination of strong and  weak lensing constraints. This direct 
approach also allows inclusion of external constraints such as those from X-ray data and the 
redshift distribution of background sources that can be estimated using photometric-redshift techniques. Of course, such parametric techniques 
require allowing sufficient freedom in the radial profile and the inclusion of substructures
substructures (e.g. Metzler 1999, 2001; King et al 2001) to closely match observed lensing distortions. 


\subsection{Cluster Triaxiality}

As lensing is sensitive to the integrated mass along
the line of sight, it is natural to expect mass overestimates due to fortuitous alignment of 
mass concentrations not physically related to the
cluster or alternatively departures of the cluster dark matter halo from spherical symmetry ({\it e.g.} Gavazzi 2005). Till 
recently, most studies of the dark matter distribution and
the intra-cluster medium (ICM) in galaxy clusters using X-ray data have been
limited due to the assumption of spherical symmetry. However, the Chandra and XMM-Newton X-ray telescopes have 
resolved the core of the clusters, and have detected departures from isothermality and spherical symmetry. Evidence
for a flattened triaxial dark matter halo around five Abell clusters had been reported early on
by Buote \& Canizares (1996). Furthermore, numerical simulations of cluster formation and evolution in a cold dark matter
dominated Universe do predict that dark matter halos have highly elongated axis ratios (Wang \& White 2009), disproving the assumption of spherical geometry.  In fact the departures from sphericity of a cluster may
help explain the discrepancy observed between cluster masses determined from 
X-ray and strong lensing observations (Gavazzi 2005). This suggests that clusters with observed prominent 
strong lensing features are likely to be typically preferentially elongated along the line of sight which might account for their enhanced 
lensing cross sections. This is definitely the case for the extreme strong lenses with large Einstein radii and therefore anomalously high concentrations. The galaxy cluster A1689 is a well-studied example with such a mass discrepancy (Andersson \& Madejski 2004; Lemze et al.
2008; Riemer-Sorensen et al. 2009; Peng et al. 2009). In the same vein, large values of the NFW model concentration parameters 
have also been reported for clusters with prominent strong lensing features (Comerford \& Natarajan 2007; Oguri et al. 2009). This can
again be explained by strong lensing cluster halos having their major axis preferentially oriented toward the line of 
sight (Corless et al. 2009).

Combining strong lensing constraints with high-resolution images of cluster cores 
in X-rays obtained with {\it Chandra} is an excellent way to probe the triaxiality of the 
mass distribution in cluster cores. Mahdavi et al. (2007)  provided a new framework for the 
joint analysis of cluster observations (JACO) using simultaneous fits to X-ray, Sunyaev-Zel'dovich (SZ), 
and weak lensing data. Their method fits the mass models simultaneously to all data, provides explicit 
separation of the gaseous, dark, and stellar components, and allows joint constraints on all measurable physical 
parameters. The JACO prescription includes additional improvements to previous X-ray techniques, such as the 
treatment of the cluster termination shock and explicit inclusion of the BCG's stellar-mass profile. Upon application of JACO 
to the rich galaxy cluster Abell 478 they report excellent agreement among the X-ray, lensing, and SZ data. 

Morandi et al. (2010) have 
used a triaxial halo model for the galaxy cluster MACS\,J1423.8+2404 to extract reliable information on 
the 3D shape and physical parameters, by combining X-ray and lensing
measurements. They found that this cluster is triaxial with dark matter halo
axial ratios 1.53$\pm$0.15 and 1.44$\pm$0.07 on the plane of the sky and along the line of sight, respectively. 
They report that such a geometry produces excellent agreement between the X-ray and lensing mass. 

These first results are very encouraging and pave the way for a better understanding of the 3D
distribution of the various mass constituents in clusters. Theoretically, according to the current dark matter 
dominated cosmological model for structure formation cluster halo shapes ought to be triaxial and a firm prediction 
is proffered for the distribution of axis ratios for clusters. More observational work needs to be done to test these predictions,
 and ultimately techniques that combine lensing, X-ray and Sunyaev-Zel'dovich decrement data 
 might be able to provide a complete 3-dimensional view of clusters.


\section{Mass distribution of cluster samples}

Although the careful modeling of individual cluster cores and extended regions offers a unique way
to characterize the mass distribution and understand cluster physics in detail,
analysis of cluster samples provides important insights into cluster assembly and evolution.
There have been several statistical studies focused on measuring cluster 
masses derived from lensing and comparing these with mass estimates from other 
measurements such as: richness,  X-ray luminosity, X-ray temperature, velocity dispersions of cluster galaxies, and 
the Sunyaev-Zel'dovich decrement. These multi-wavelength comparisons enable a deeper understanding of empirically
derived scaling relations between key physical properties of of clusters (e.g. Luppino \& Gioia 1992, Loeb \& Mao 1994, Miralda-Escud\'e \& Babul 1995, Allen 1998, Ota et al 1998, Ono et al 1999, Irgens et al 2002, Huterer \& White 2002). These studies also help
uncover how mass is partitioned between the different baryonic and non-baryonic
components on cluster scales. Studying cluster samples allows the probing of several fundamental questions with regard to the dynamical state of  clusters, namely, are clusters relaxed? 
How much substructure is present in clusters? How triaxial are clusters?  
How recently has a cluster had a major merger with another sub-cluster and what are the signatures of such an event? How important are projection effects in mass estimates? 
Are clusters in hydrostatic equilibrium? When did 
clusters start to assemble? And how have they evolved? Observationally derived answers to these questions from cluster samples can then be directly compared to numerical simulations, thus providing insights and tests  of the structure formation paradigm. 

 \subsection{Early Work}

Comprehensive multi-wavelength datasets that ideally span a range of spatial
scales in clusters are needed for such statistical studies. Collecting such datasets is a 
{\bf big} challenge as it requires coordination between researchers working with a range of 
observational techniques deploying many different resources. Some of the first studies of cluster 
samples did produce interesting cosmological results, as discussed in Luppino et al (1999), Allen et al (2001, 2002), Dahle et al. (2002)  and 
Smith et al. (2003).

\begin{figure}
\centerline{\epsfig{file=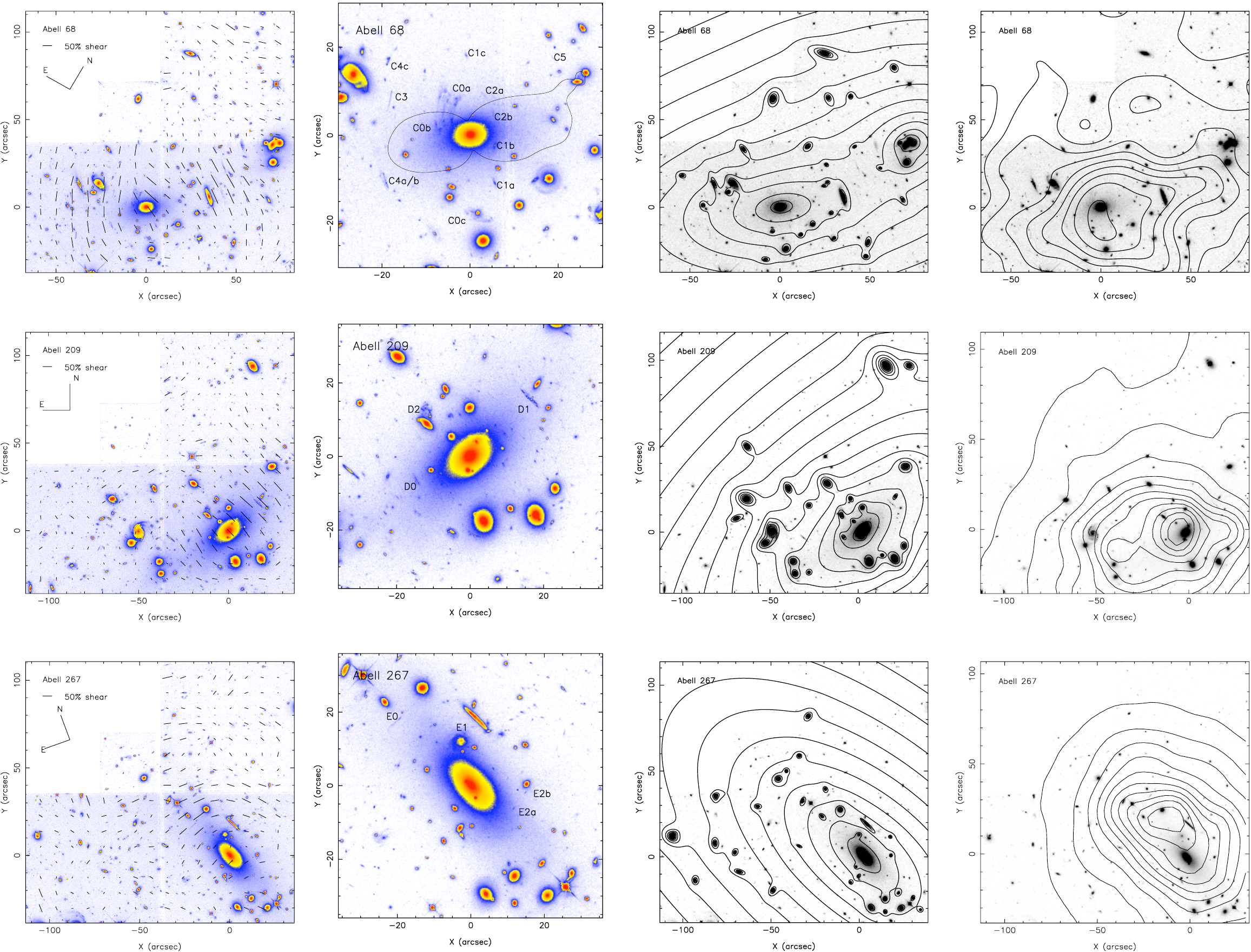,width=\textwidth}}
\caption{Cluster samples: 3 of the 12  $z\sim 0.2$ X-ray luminous clusters of galaxies selected
  from the XBACS catalog (Ebeling et al.\ 1996) 
  observed with the \hst/WFPC2 camera.  
  Top row is Abell 68, second row is Abell 209 and last row Abell 267.
  First column is the weak shear field as measured from the \hst\  data. The second column is a zoom of the cluster
  cores, and shows for Abell 68 the predicted critical lines (black lines). The third column is the strong lensing mass reconstruction and last column
  is the overlay of the {\it Chandra} X-ray map (Smith et al.\ 2003).}
\label{smith2003}
\end{figure}

\begin{figure}
\centerline{
\epsfig{file=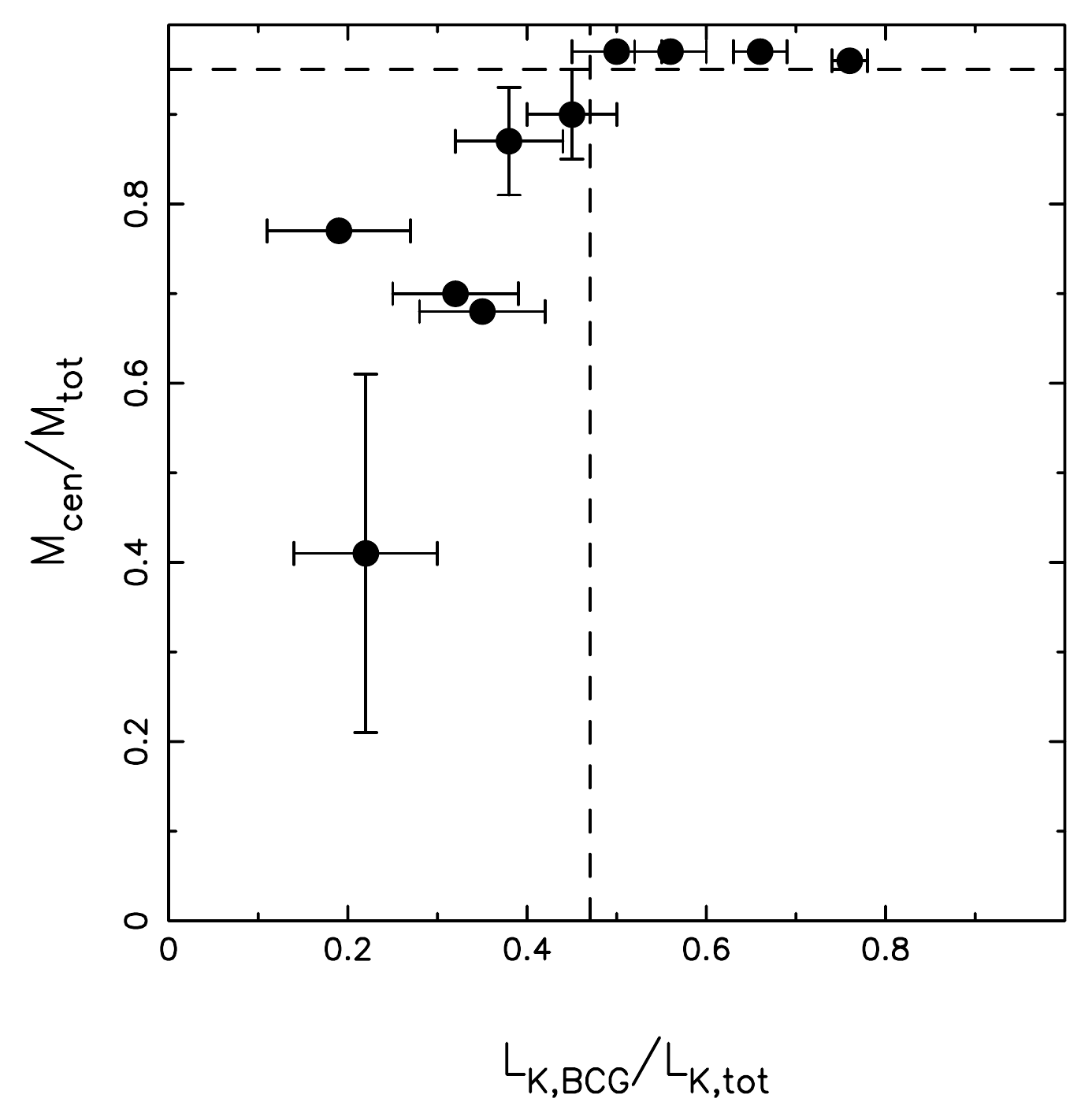,width=0.3\textwidth}
\epsfig{file=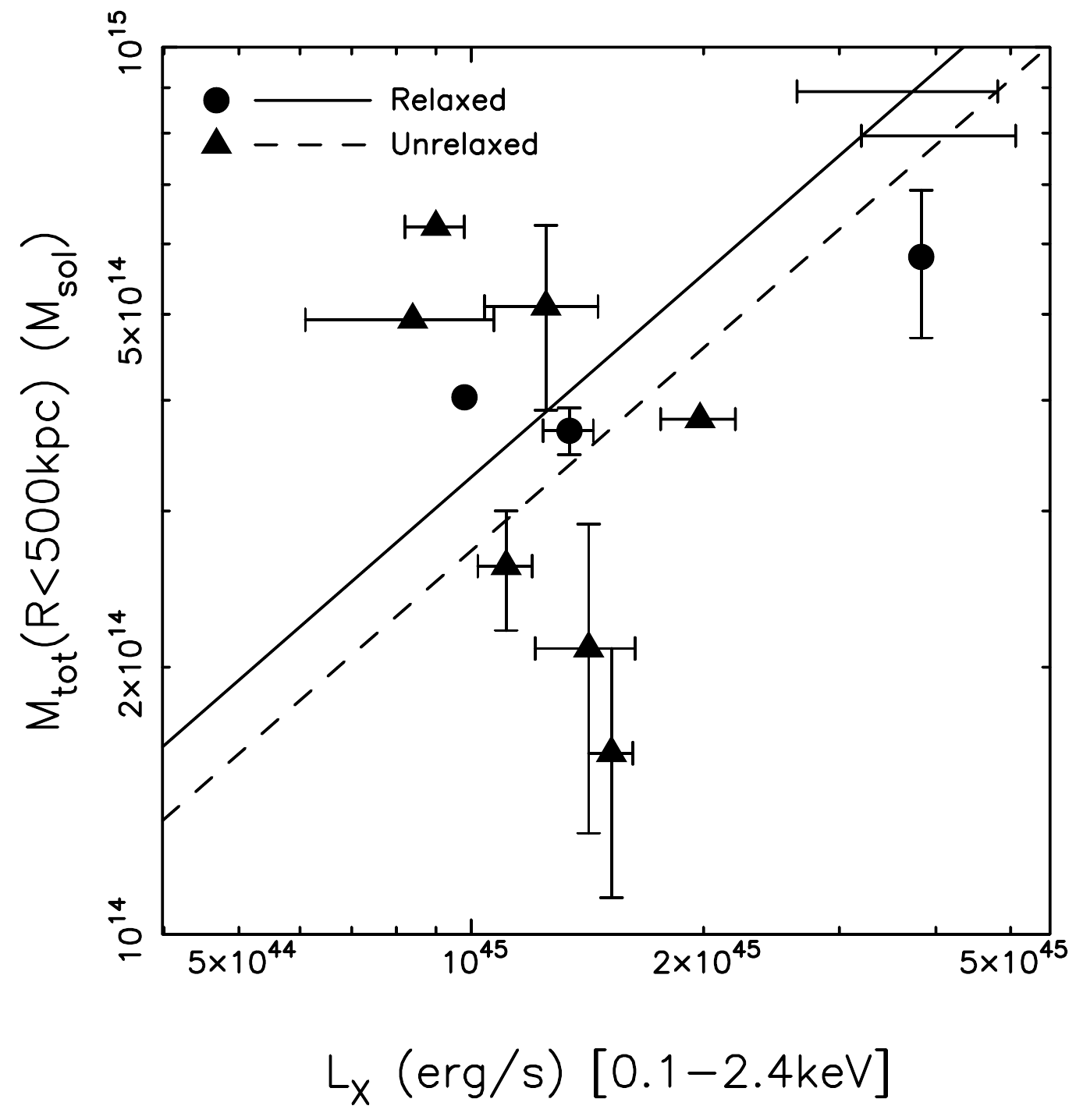,width=0.3\textwidth}
\epsfig{file=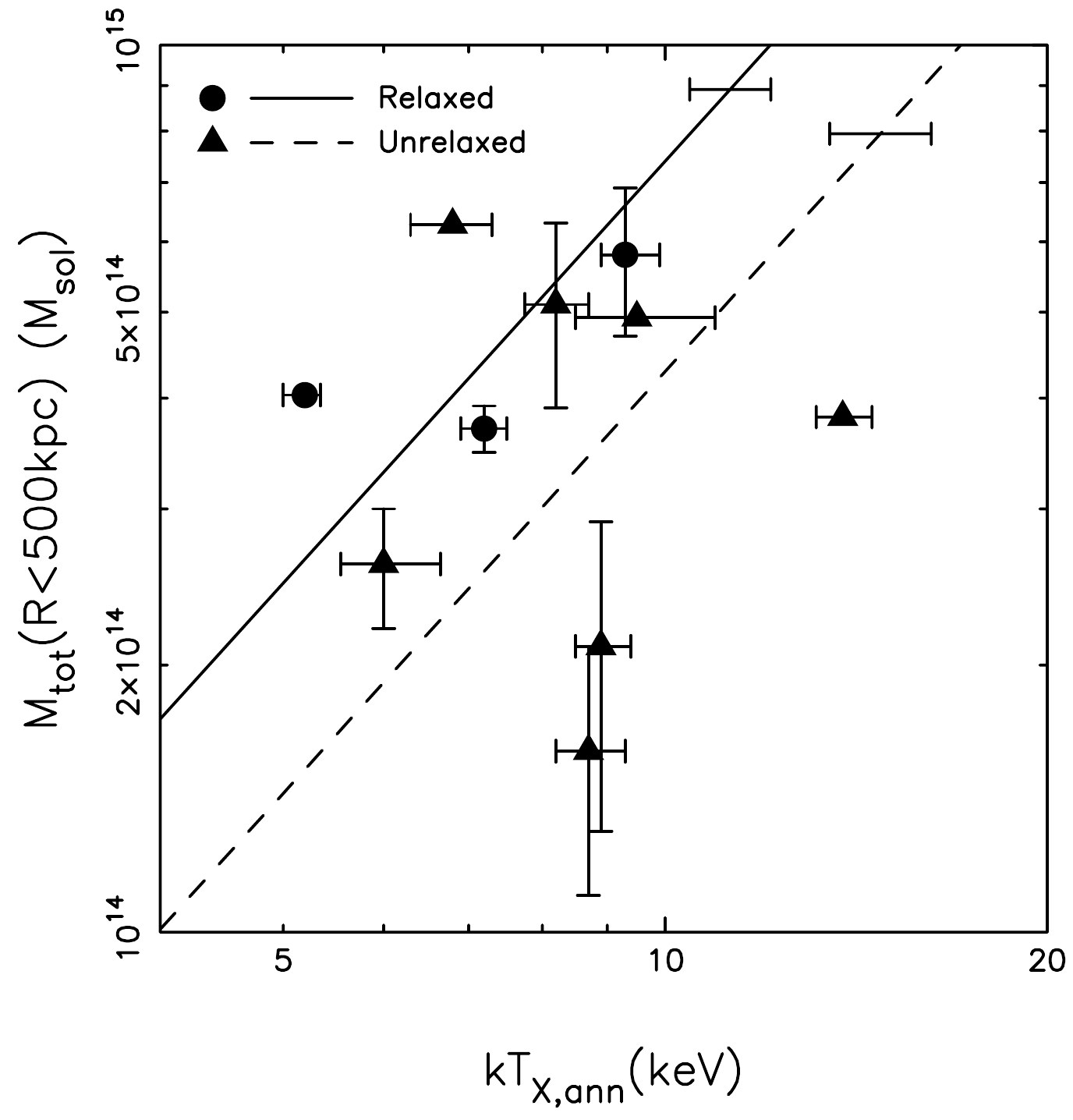,width=0.3\textwidth}
}
\caption{Left Panel: Central mass fraction (a measure of the dominance of the central dark matter halo), $M_{cen}/M_{tot}$ versus central K-band luminosity fraction (measures the dominance of the central galaxy), $L_{K,BCG}/L_{K,tot}$. There is a remarkably clean separation between a homogeneous population of centrally 
concentrated clusters ($M_{\rm cen}/M_{\rm tot}>0.95$, $L_{K,BCG}/L_{K,tot}> \sim 0.55$) and a much more diverse population of less concentrated clusters. 
Center \& Right Panels: Mass-$L_X$ and Mass-$T_X$ relations. The solid and dashed lines show the best-fit relations normalized by the relaxed and unrelaxed clusters respectively. The error bars on each line show the uncertainty on the normalizations. The scatter in the Mass-$L_X$ relation appears to be symmetric; in the mass--$T_X$ relation the normalization of the unrelaxed clusters appears to be 40\% hotter than the relaxed clusters at $2\sigma$ significance. Figures from Smith et al.  (2005).
}
\label{smith2005scaling}
\end{figure}

\begin{figure}
\centerline{
\epsfig{file=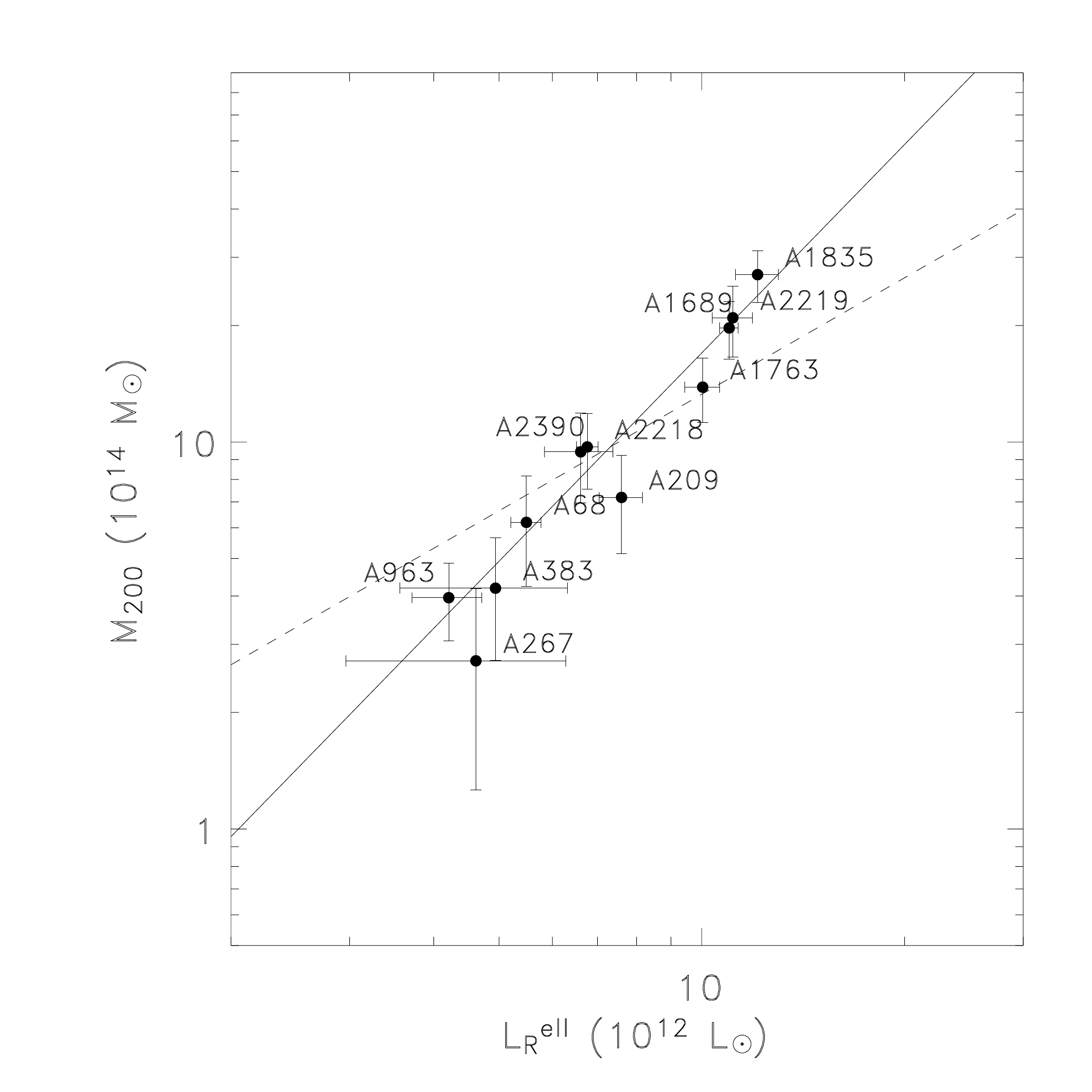,width=0.34\textwidth}
\epsfig{file=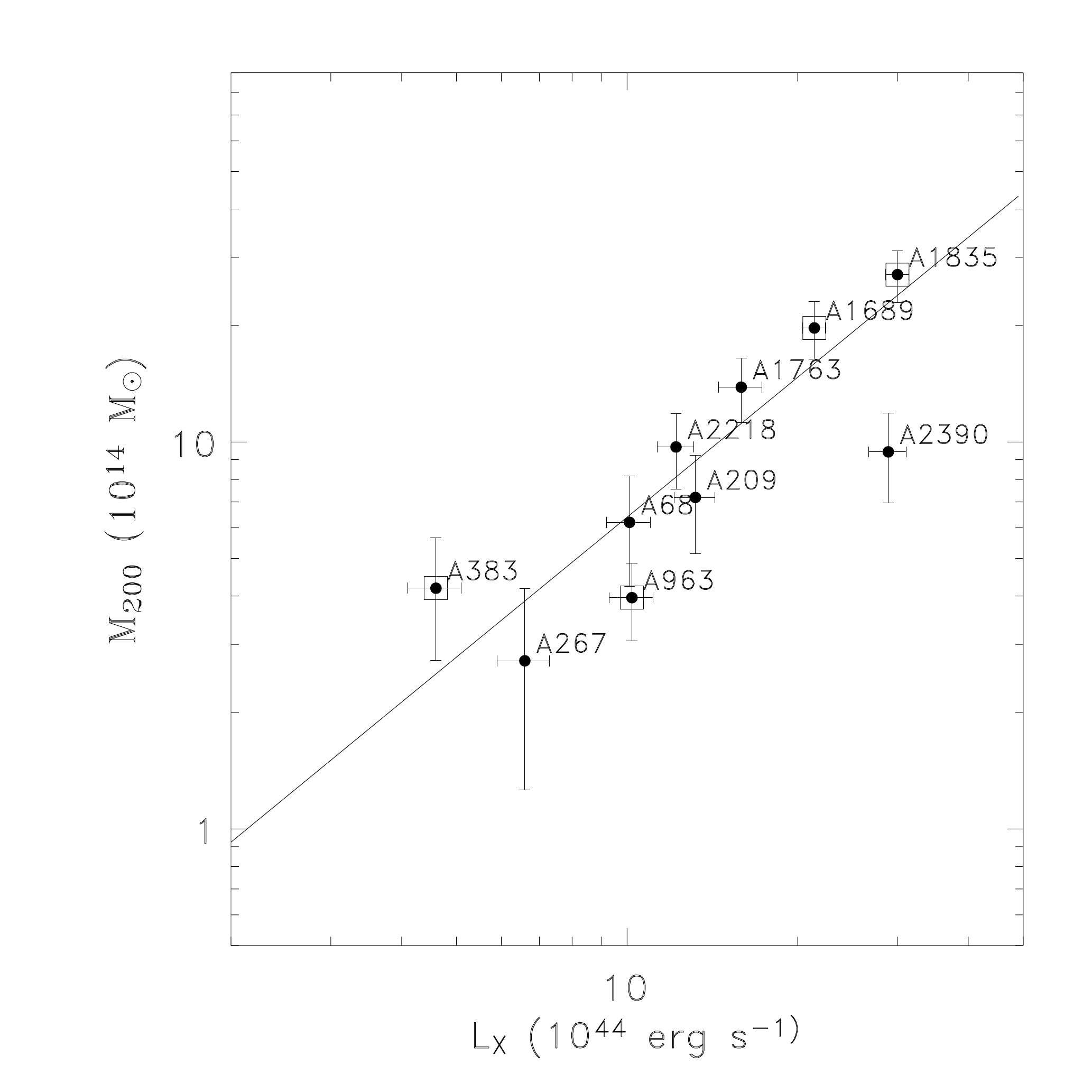,width=0.34\textwidth}
\epsfig{file=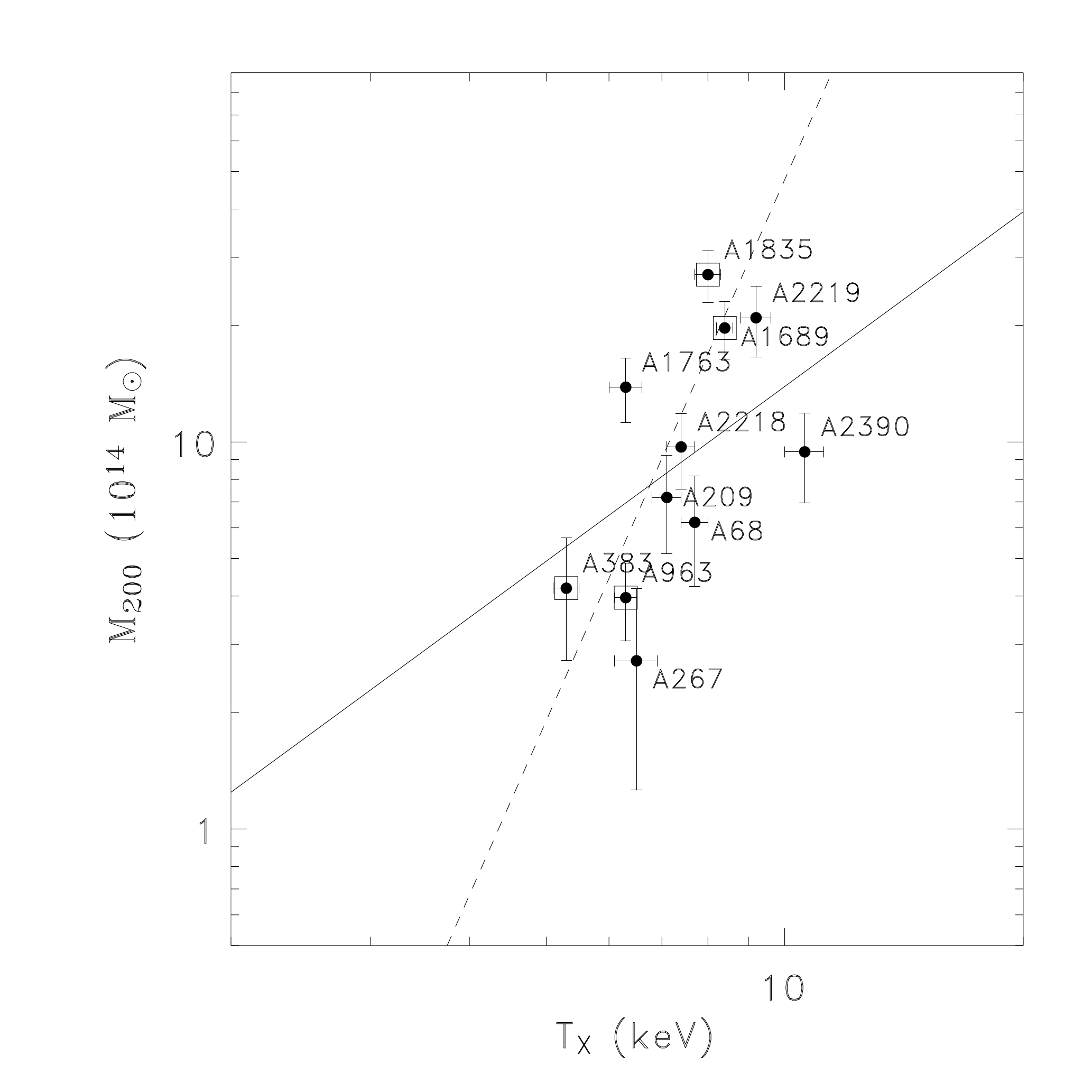,width=0.34\textwidth}
}
\caption{Left Panel: Lensing 2D mass versus optical luminosity for the clusters in the Bardeau et al.  (2007) sample (12 X-ray bright clusters selected to be 
at $z\sim 0.2$).
The lensing mass is computed at the virial radius $r_{200}$
    derived from the best weak lensing fits. The luminosity is computed in the
    $R$~band for the cluster red sequence galaxies. Dashed
    line represents a constant $M/L$ ratio of 133 in solar units.
    The solid line is the best-fit power law $M \propto L^{1.8}$.
    Center Panel: 
Weak lensing 3D virial mass $M_{200}$ versus
    X-ray luminosity. The best-fit line has a slope $\alpha = 1.20 \pm
    0.16$. 
    Right Panel: Weak lensing
    3D virial mass $M_{200}$ versus X-ray temperature.  The straight
    line corresponds to a $M_{200} \propto T^{3/2}$ relation while the
    dashed line corresponds to the best-fit power law relation $M
    \propto T^{4.6 \pm 0.7}$. Temperatures are derived from XMM
    data (Zhang et al.  2007), including A\,2219 from ASCA data (Ota et al.  2004).
    The 4 clusters with cooling core or
    relaxed properties are marked with empty boxes.
(From Bardeau et al.\ 2007).}
\label{bardeau2007}
\end{figure}

One of the key challenges for these statistical studies lies in the very definition of a sample with 
robust criteria, that will be complete and volume limited and be representative to avoid systematic biases. Starting from 
simple selection criteria is therefore very important. For instance, dramatic lensing clusters imaged by 
\hst\  are likely a biased sample of the most massive clusters at any redshift with enhanced strong 
lensing cross sections due to an excess of mass along the line of sight from either the cluster itself 
or the presence of other intervening structures. Most frequently cluster samples are therefore selected 
on the basis of their X-ray luminosities,  which should minimize projection effects that typically plague 
optically selected clusters. Since X-ray luminosity is proportional to the square of the electron density of 
the Intra-Cluster-Medium (ICM), this selection should pick genuinely virialized clusters, irrespective of the 
line of sight distribution of cluster member galaxies or additional background structures. One of the first systematic studies that combined
X-ray and lensing data was a sample of 12 $z\sim 0.2$ X-ray luminous clusters of
galaxies selected from the XBACS catalog (see Figures~\ref{smith2003} and \ref{smith2005scaling})
with $L_{X}>8\times 10^{44}$erg/s in the 0.1-2.4 keV band. These clusters have been imaged with the WFPC2 camera 
(Smith et al.\ 2001, 2005). It is found that the fraction of strong lensing clusters in this sample is 70\%. All of the cluster 
cores also have a significant weak lensing signal, providing independent lensing constraints on cluster masses.

Smith et al. (2005) defined a number of criteria to characterize whether clusters are relaxed and also quantified the amount of substructure in them. 
Out of 10 clusters, they found that three clusters form a homogeneous sub-sample that have mature, undisturbed gravitational potentials which satisfy the following criteria: a dominant 
central dark matter halo ($M_{\rm cen}/M_{\rm tot}>0.95$); a dominant central cluster galaxy K-band luminosity fraction ($L_{K,BCG}/L_{K,tot}> \sim0.5$); close alignment between the center of the 
mass distribution and the peak of the X-ray flux ($\Delta r_{peak}<3$kpc); a single cluster-scale dark matter halo best-fit for the lens model; and circular or mildly 
elliptical X-ray flux contours. The remaining seven clusters did not satisfy one or more of these criteria and were classified as disturbed. The disturbed clusters are 
much more diverse than the undisturbed clusters and typically have a bi- or tri-modal dark matter distribution, irregular X-ray morphology and an offset between 
X-ray and mass peaks. Comparison of these results with theoretical predictions indicates that the multi-modal dark matter distribution in disturbed clusters is 
due to recent infall of galaxy groups into the parent cluster since about $z=0.4$. The exact scaling relation between lensing mass and X-ray properties appears to be 
strongly dependent on the dynamical state of the cluster. Relaxed and unrelaxed clusters appear to follow slightly different scaling relations. 
Furthermore, this sample was also observed with the wide field CFHT12k camera 
in three bands (B,R,I) in order to probe the wide field mass distribution using the measured weak lensing shear signal out 
to the virial radius. However, the comparison of the weak lensing determined mass to the cluster
luminosity and X-ray mass estimates reported in Bardeau et al. (2007) [see
Figure~\ref{bardeau2007}] does not reveal an obvious difference between relaxed or unrelaxed clusters.
There are some strong limitations though with this dataset as there were scant constraints on the redshift
distribution of background sources, and some lingering inconsistencies between strong and weak lensing results. 
These first results with only 10 clusters set the stage for the need for larger cluster samples to understand the 
physical origin of such differences.

In a parallel paper, Hoekstra (2007) investigated the lensing versus X-ray mass relations for a sample of 
20 clusters including those of Bardeau et al. (2007), although their cluster selection was primarily driven by X-ray 
emission. This investigation has lead to a more ambitious project known as the Canadian Cluster Comparison Project (CCCP) 
that will add 30 more X-ray selected clusters observed with the CFHT12k or Megacam camera to the initial set of 20 clusters. 
Lensing results are however still pending at the time of writing this review.

\subsection{On-going and future cluster lensing surveys}

Clusters of galaxies are complicated systems that are rapidly assembling
and evolving, nevertheless they are considered to be very good tracers of the underlying cosmology (and in particular 
could probe Dark Energy) as well as a way to measure the growth of structure, thus potentially sensitive to gravity and to the nature 
of Dark Matter. A better understanding of clusters will be possible only with larger cluster samples, as earlier work and conclusions
therefrom were limited by statistics. The number of  massive clusters with published lensing data is steadily growing,
as is the number of cosmological surveys in which clusters can be studied with
strong and weak lensing techniques, either directly from the survey data or by further follow-ups.

Four techniques are avidly pursued to search for clusters:\\
\begin{itemize}
\item{Photometric searches that use wide field imaging surveys such as the Sloan Digital Sky 
Survey (SDSS), the Red-sequence Cluster Survey (RCS), and the CFHT Legacy Survey (CFHTLS). Furthermore, new photometric surveys have just started or will start in the next year, namely the VST KIDS survey, the Dark Energy Survey (DES), and the Subaru Hyper-Suprime Camera (HSC) survey.}\\
\item{X-ray selected cluster searches: i) based on the follow-up of the ROSAT All Sky Survey: such as the MAssive Cluster 
Survey (MACS) (Ebeling et al.\ 2001) and the REFLEX survey (Boehringer et al. 2004)  ii) based on dedicated (or serendipitous)  
X-ray ROSAT or XMM imaging surveys such as: WARPS (Scharf et al. 1997), SHARC (Collins et al. 1997), the ROSAT Deep Cluster Survey (Rosati et al. 2001), XDCS (the XMM Deep Cluster Survey - Fassbender et al. 2008), XCS (Romer 2008);   and XMM-LSS (Pierre et al. 2007).}\\
\item{SZ searches: {\it e.g.} Atacama Cosmology Telescope Cluster Survey (ACT) (see e.g. Hincks et al. 2010, Marriage et al 2011, Hand et al 2011), 
the South Pole Telescope Cluster Survey (SPT) (e.g. Chang et al. 2009, 
Vanderlinde et al 2010, Plagge et al 2010) and 
the {\it Planck} mission (Ade, P. A. R., et al. 2011).}\\
\item{Weak and Strong lensing searches based on photometric surveys, or following up X-ray or SZ selected clusters.}\\
\end{itemize}
We focus on the latter techniques in the following sub-sections.

\subsection{Targeted cluster surveys}

\subsubsection{The  Local Cluster Substructure Survey (LoCuSS)}

LoCuSS extends Smith et al.'s (2005) pilot study of 10 X-ray luminous clusters at $z=0.2$ to an order of magnitude larger sample at $0.15<z<0.3$, drawn from the ROSAT All-sky Survey Catalogues (Ebeling et al., 1998, 2000; Ebeling et al., 2004).  The main lensing-related goals of LoCuSS are to measure the mass, internal structure, and thermodynamics of a complete volume-limited sample of 80 clusters observable from Mauna Kea, and thus to obtain definitive results on the mass-observable scaling relations at low redshift.  The normalization, shape, scatter (and any structural segregation detected) of these scaling relations will calibrate the properties of low redshift clusters as an input to cluster-based cosmology experiments, and to help interpret high-redshift cluster samples.

To date LoCuSS has published weak lensing analysis of 30 clusters observed with Suprime-CAM on the Subaru 8.2-m telescope (Okabe et al. 2010; see also Oguri et al. 2010).  The main results from this statistical study are that (i) a simple color-magnitude selection of background galaxies yields samples that are statistically consistent with negligible residual contamination by faint cluster members, albeit with large uncertainties, (ii) cluster density profiles are curved (in log-log space), and statistically compatible with the Navarro Frenk \& White (1997) profile, and (iii) based on the NFW profile model fits, the normalization of the mass-concentration relation of X-ray selected clusters is consistent with theoretical $\Lambda$CDM-based predictions, although the slope of the observed relation may be steeper than predicted.  The last of these results is particularly interesting in the context of detailed studies of individual clusters selected to have a large Einstein radius.  As noted in Section 7, such objects are often found to have concentrations that exceed the CDM prediction by factors of ~2-3 (Comerford \& Natarajan 2007; Oguri et al. 2009). Okabe et al.'s results from 30 X-ray-selected clusters indicate that the large Einstein radius selection in earlier work introduces a strong bias.

Comparison of Okabe et al.'s weak lensing mass measurements with X-ray and Sunyaev-Zel'dovich (SZ) effect probes has so far been limited by the presence of outlier clusters in the small samples for which the relevant data are available.  For example, the well-known merging cluster A1914 strongly influences the results in the X-ray/lensing comparison of 12 clusters for which Subaru and XMM-Newton data are available (Okabe et al., 2010b; Zhang et al., 2010).  More recently, Marrone et al. (2011) presented the first weak lensing-based mass-SZ scaling relation based on Subaru and Sunyaev-Zel'dovich Array (SZA) observations of 18 clusters.  Encouragingly, this relation is consistent with self-similar predictions, although it presents 20\% scatter in natural log of mass at fixed integrated Y-parameter - a factor of 2 more scatter than found in studies that use X-ray data and assume hydrostatic equilibrium to infer cluster mass.  Indeed, the normalization of the $M_{\rm WL} - Y$ relation at $\Delta=500$ (roughly 1Mpc) for undisturbed clusters is 40\% higher in mass than that for disturbed clusters.  Marrone et al. identified several of the undisturbed clusters as likely prolate spheroids whose major axis is closely aligned with the line of sight as being largely responsible for this segregation.  These results highlight the feasibility and growing maturity of lensing-based studies of large cluster samples, and also emphasize that much important work remains to be done to fully understand the optimal methods for cluster mass measurement.

\subsubsection{The MAssive Cluster Survey}

\begin{figure}
\centerline{
\epsfig{file=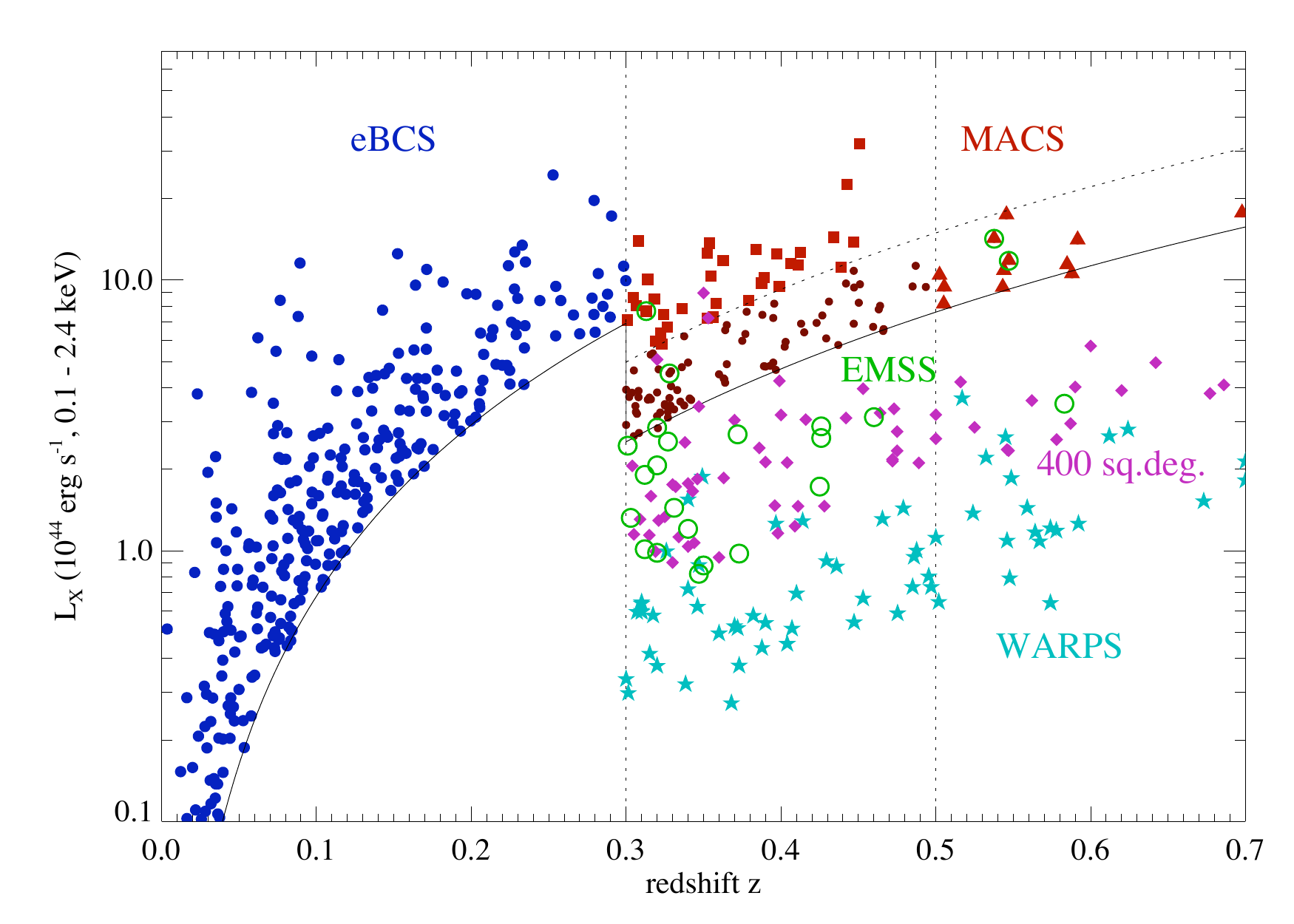,width=0.7\textwidth}}
\caption{The Luminosity versus redshift plot comparing the MACS surveys to a number of other X-ray surveys:
EMSS, eBCS, WARPS, the 400 square degree survey. It is evident from this figure that MACS is very efficient 
in selecting the most massive X-ray clusters at $z>0.3$.}
\label{macs_lx_z}
\end{figure}

The MAssive Cluster Survey (MACS) is an ongoing project aimed at the compilation and characterization of a statistically complete sample of very X-ray luminous (and thus, by inference, massive), distant clusters of galaxies. The primary goal of MACS was to increase the number of known massive clusters at $z > 0.3$ from a handful to a hundred. To achieve these goals, Ebeling et al. (2001) applied an X-ray flux and X-ray hardness-ratio cut to select distant cluster candidates from the ROSAT Bright Source catalog. Starting from a list of more than 5,000 X-ray sources within the survey area of 22,735 square degrees they use positional cross-correlations with public catalogs of Galactic and extragalactic objects, with reference to APM colors, visual inspection of Digitized Sky Survey images, extensive CCD imaging, and finally spectroscopic observations with the University of Hawaii's 2.2m and the Keck 10m telescopes to compile the final cluster sample. The MACS cluster sample comprises 124 spectroscopically confirmed clusters at $0.3 < z < 0.7$ (Figure 
\ref{macs_lx_z}). Comprehensive follow-up observations of MACS clusters include: weak lensing mass measurements using wide-field SUBARU imaging data, virial mass estimates based on cluster galaxy velocity dispersions measured with the CFHT and Keck, SZ observations with the BIMA mm-wave radio interferometer, measurements of the cluster gas and temperature distribution with $Chandra$, and both deep, multi-passband and snapshot images with \hst. A large number of MACS clusters are strong lenses and some of them have been studied in detail: MACS\,J1206-0847 (Ebeling et al.  2009);  MACS\,J1149.5+2223 (Smith et al.  2009); MACS\,J1423.8+2404 (Limousin et al. 2010; Morandi et al. 2010). MACS\,J0025.4-1222 (Brada{\v c} et al.  2008) was identified as a merging cluster with some similarity to the ``Bullet Cluster''. Zitrin et al.  (2011a) presented 
the results of a strong lensing analysis of the complete sample of the 12 MACS clusters at $z > 0.5$ using \hst\  images. The distribution of Einstein radii has a median value of $\sim$28 arcseconds (for a source redshift of $z_{S}\sim 2$), twice as large as other lower-z samples, making the MACS sample a truly massive cluster sample confirmed by the numerous strong lensing discoveries. One of the most extreme clusters known presently is likely MACS\,J0717.5+3745 (Ebeling et al. 2004) which
was recognized as a complex merger of 4 individual substructures, with a long tailed filamentary structure. The 4 substructures have all been identified in a recent lensing mass
reconstruction by Limousin et al.  (2011) and the filamentary structure was directly measured by weak lensing measurements with a 18-pointing \hst\ 
mosaic (Jauzac et al.  2012). Horesh et al.  (2010), investigated the statistics of strong lensed arcs in the X-ray selected MACS clusters versus the optically-selected RCS clusters (see below). They measured the lensed-arc statistics of 97 clusters imaged with \hst, identifying lensed arcs using two automated arc-detection algorithms.  They compile a catalog of 42 arcs in MACS and 7 arcs in the RCS. At $0.3<z<0.7$, MACS clusters have a significantly higher mean frequency of arcs, $1.2\pm$0.2 per cluster, versus 0.2$\pm$0.1 in RCS, which can easily be explained by the nature of the selection of these two different cluster samples.

\subsubsection{ESO distant cluster survey}

Nevertheless, optical selection is still common specially for high-redshift clusters ($z>0.6$) where X-ray selection is limited. A particular focused and productive survey is the ESO distant cluster survey (EDiSC, White et al. 2005). EDiSC is a survey of 20 fields containing distant galaxy clusters ($0.4<z<1.0$)
chosen amongst the brightest objects identified in the Las Campanas Distant Cluster Survey. They were confirmed by identifying red sequences in moderately deep two color data from VLT/FORS2, and further
investigations with VLT in spectroscopy, the ESO Wide Field Imager, and \hst/ACS mosaic images for 10 of the most distant clusters. Using the deep VLT/FORS2 data, Clowe et al.  (2006) measured the masses for the EDiSC clusters. In particular, they compared the mass measurements of  13 of the EDiSC clusters with luminosity measurements from cluster galaxies selected using photometric redshifts and find evidence of a dependence of the cluster mass-to-light ratio with redshift.

\subsubsection{Red-sequence cluster surveys}

Another important optically selected cluster survey is the 100\ deg$^{2}$ Red-Sequence Cluster Survey (RCS, Gladders  2002, Gladders \& Yee 2005) and its 1000\ deg$^{2}$ RCS-2 extension (Gilbank et al.  2011), that are based on shallow multi-color imaging with the CFHT12k and Megacam cameras. 
RCS-2 covers $\sim 1000$ deg$^{2}$ and includes the first RCS area, it reaches $5\sigma$ point-source limiting magnitudes in [g,r,i,z] = [24.4, 24.3, 23.7, 22.8], approximately 1-2 magnitudes deeper than the SDSS. RCS-2 is designed to detect clusters over the redshift range $0.1<z<1$, building a statistically complete, large ($\sim10^{4}$) sample of clusters, covering a sufficiently long redshift baseline to be able to place constraints on cosmological parameters probed via the evolution of the cluster mass function. 
Furthermore, a large sample of strongly lensed arcs associated with these clusters has been derived ({\it e.g.} Gladders et al.  2002, 2003), and weak lensing measurements from the most massive clusters detected in RCS-2 is likely possible.

\subsubsection{The Multi-Cluster Treasury: CLASH survey}

The recently approved MCT (Multi-Cluster Treasury) program on \hst\
will achieve multi-band imaging of a sample of 25 X-ray selected clusters (Postman et al 2011), thus providing detailed photometric redshift 
estimates for multiple-images.  This sample with appropriate ground based follow-up is likely to provide important insights into many of the
current unsolved problems in cluster assembly and evolution.
Dedicated lensing studies will enable detailed investigation of their mass distributions 
(Zitrin et al.  2011b, 2011c) and will help find some efficient lenses that can be exploited to study
the distant Universe by using them as gravitational telescopes (Richard et al. 2011) - a topic that will be discussed further 
in the next section.

\subsection{Cluster lenses in wide cosmological surveys}

The previous sub-section focused on targeted cluster surveys. However cluster lenses can also be found in wide cosmological surveys (e.g. Wittman et al 2001, 2003; Hamana et al 2004; Maturi et al 2005). 
We briefly outline some of the most representative surveys of this decade starting from the widest to the deepest.

\subsubsection{The Sloan Digital Sky Survey}

\begin{figure}
\centerline{
\epsfig{file=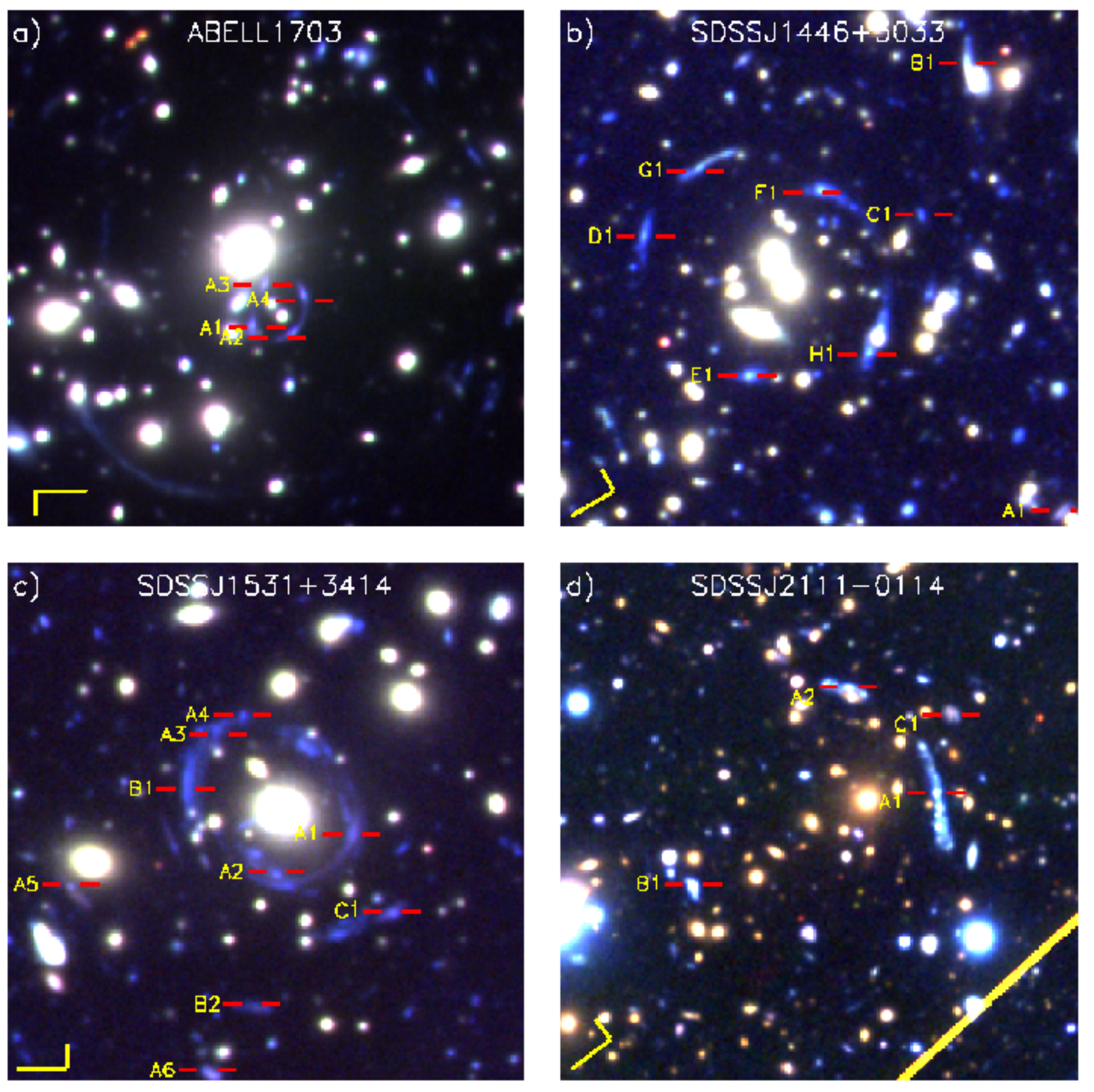,width=0.7\textwidth}
}
\caption{
SDSS discovered strong lensing clusters -- $a)$ Abell 1703, $b)$ SDSS J1446+3033,
$c)$ SDSS J1531+3414, and $d)$ SDSS J2111-0114. Color composite
images are made from $g,r,i$ imaging obtained with Subaru/SuprimeCam. All images are $75'' \times 75''$.
Background sources are bracketed by red lines and labeled. Source labels
with the same letter but different numbers (e.g. A1, A2, etc.) have
the same redshifts to within the measurement errors, and
are presumed to be the same source, multiply imaged (Figure from Bayliss et al.  2011).}
\label{baylis2011}
\end{figure}

The Sloan Digital Sky Survey (SDSS) is an imaging and spectroscopic survey covering 10,000 deg$^{2}$
(Aihara et al. 2011). Although, this survey was not designed or optimized to 
measure cluster lensing,  interesting results have been produced from detected strong and weak lensing measurements of clusters.
Estrada et al.  (2007) investigated the Sloan images of 825 SDSS galaxy clusters searching for giant arcs. Both a visual inspection of the images and an automated search were performed, and no arcs were found. They nevertheless report a serendipitous discovery of a bright arc in the Sloan images of an as  yet unknown cluster. Hennawi et al. (2008) presented the first results of a strong lensing imaging survey (using the WIYN and UH 2m telescope) targeting the richest clusters  (with $0.1<z<0.6$) selected from SDSS. From a total of 240 clusters followed-up, they uncovered 16 new lensing clusters with definite giant arcs, 12 systems for which the lensing interpretation is very likely, and 9 possible lenses which contain shorter arclets or candidate arcs which require further observations to confirm their lensing origin. The new lenses discovered in this survey will enable future systematic studies of the statistics of strong lensing and their implications for cosmology and the current structure formation paradigm. Kubo et al. (2009) and then Diehl et al. (2009) identified 10 strongly lensed galaxies as part of the ``Sloan Bright Arcs Survey''. Follow-up imaging identified  the lensing systems as group-scale lenses, an intermediate regime between isolated galaxies and galaxy clusters (see Cabanac et al. 2007). Baylis et al.  (2011) presented the results from a spectroscopic program targeting 26 strong lensing clusters ($0.2<z<0.65$) visually identified in SDSS or RCS-2 revealing 69 unique background sources with redshifts as high as $z=5.2$, which will enable robust strong lensing mass models to be constructed for these clusters (some of the most remarkable clusters discovered are presented in Figure \ref{baylis2011}).

On the weak lensing side, the first measurement was conducted by Sheldon et al (2001). 
Later on Rykoff et al. (2008) measured the scaling relation between X-ray luminosity and the total mass for 17,000 galaxy clusters in the SDSS maxBCG cluster sample. To achieve this, they stacked subsamples of clusters within fixed ranges of optical richness, and they measured the mean  X-ray luminosity $L_{X}$, and the weak lensing mean mass, $<M_{200}>$. For rich clusters, they found a power law
correlation between $L_{X}$ and $M_{200}$ with a slope compatible with previous estimates based on X-ray selected catalogs.
Furthermore, Rozo et al.  (2010) used the abundance and weak lensing mass measurements of the SDSS maxBCG cluster catalog to simultaneously constrain cosmology and the cluster richness-mass relation. Assuming a flat $\Lambda$CDM cosmology, they found that  $\sigma_{8}(\Omega_{\rm m}/0.25)^{0.41} = 0.832\pm 0.033$. These constraints are fully consistent with those derived from WMAP five-year data. With this remarkable consistency they claim that optically selected cluster samples may produce precision constraints on cosmological parameters in future wide-field imaging cosmological surveys.

\subsubsection{The CFHT-Legacy Survey}

Soon after the first light of the Megacam camera at CFHT, a legacy survey (LS) was started. It comprises
a deep $ugriz$ ($i\sim27.5$) survey of 4 square degrees in four independent fields spread across the sky, 
and a wide synoptic $u,g,r,i,z$ ($i\sim24.5$) survey of 170 square degrees in four patches of 25 to 72 square degrees. Due to the excellent 
seeing delivered by CFHT, the Legacy Survey has lead to intensive strong and weak lensing studies.

In particular, Cabanac et al. (2007) have searched for strong lensing arcs and Einstein rings around galaxies
in both the deep and wide part of the CFHT-LS. Most of the systems uncovered have deflection angles ranging between 2 and 15 arcseconds. Such samples have thus uncovered a large population of strong lenses from galaxy groups with typical halo masses of about $10^{13} h^{-1}$M$_{\odot}$. The 13 most massive systems have been studied in detail by Limousin et al.  (2009), and detailed analysis of the mass distribution on small and large scales has been investigated by Suyu \& Halkola (2010) and Limousin et al. (2010), respectively. A weak lensing search for galaxy clusters in the 4 
square degrees of the 4 CFHT-LS deep fields was performed and results are presented in Gavazzi \& Soucail (2007). Using deep $i$-band images they performed weak lensing mass reconstructions and identified high convergence 
peaks. They used galaxy photometric-redshifts to improve the weak lensing analysis. Among the 14 peaks found above 3.5$\sigma$, nine were considered as secure detections
upon cross-correlation studies with optical and X-ray catalogs.  Berge et al. (2008) conducted a joint weak lensing and  X-ray analysis of (only) 4 square degrees from the CFHTLS and XMM-LSS surveys. They identified 6 weak lensing-detected clusters of galaxies, and showed that their counts can be used to constrain the power-spectrum normalization $\sigma_{8} = 0.92^{+0.26}_{-0.30}$ for $\Omega_{\rm m} = 0.24$. They showed that deep surveys should be dedicated to the study of the physics of clusters and groups of galaxies, and wide surveys are preferred for the measurement of cosmological parameters. 
A first catalogue of lensing selected cluster has been recently published by Shan et al.(2011)
on the CFHT-LS W1 field. They perform a weak lensing mass map
reconstruction and identify high signal-to-noise ratio convergence peaks, that were then correlated with 
the optically selected cluster catalogue of Thanjavur et al.(2011). They then used tomographic techniques to 
validate their most significant detections and estimate a tomographic redshift.
More weak lensing cluster analyses are expected to be published from CFHT-LS in the near future.

\begin{figure}
\centerline{
\epsfig{file=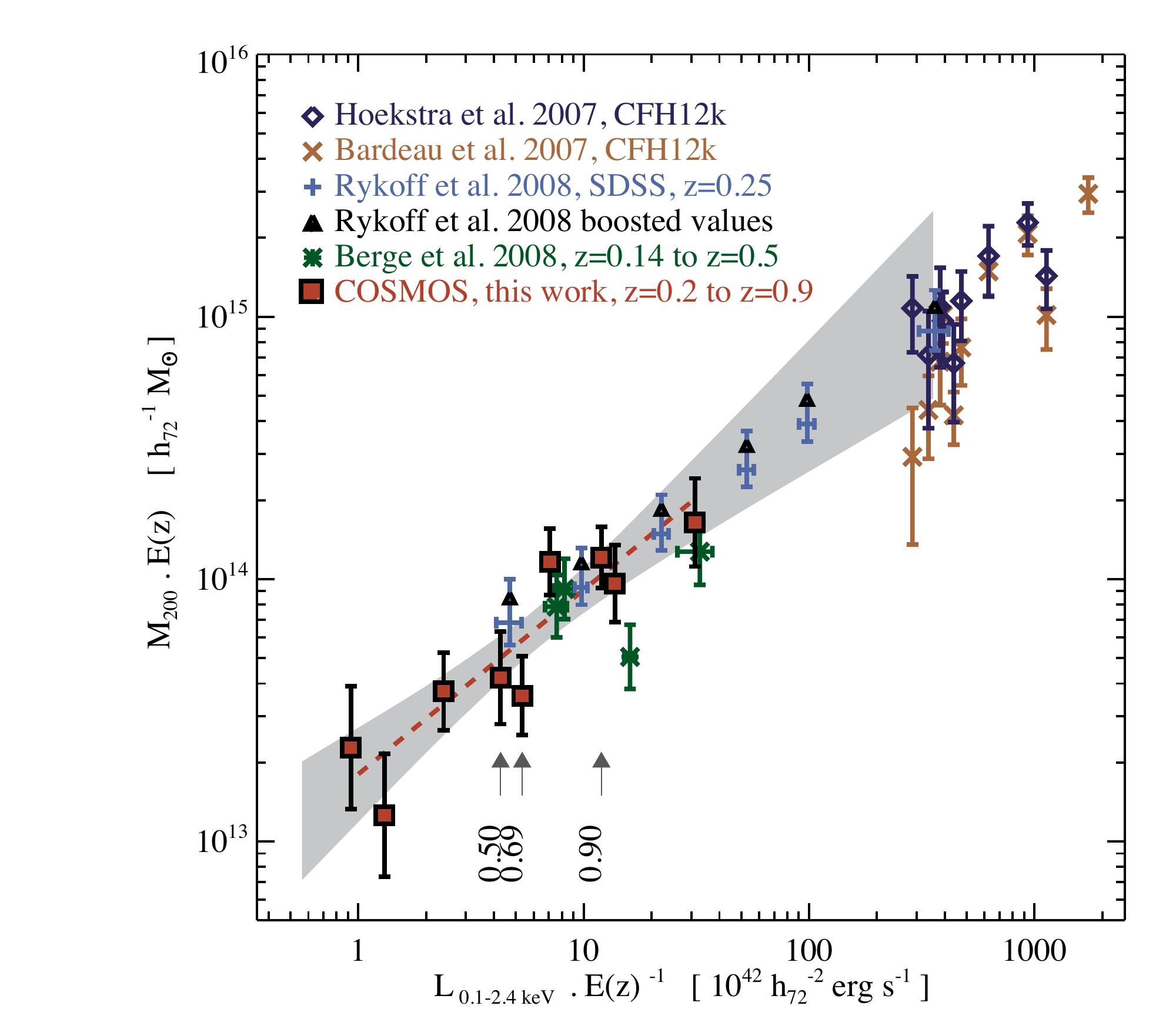,width=0.7\textwidth}
}
\caption{
The COSMOS $M-L_{X}$ relation from Leauthaud et al.  (2010). Dark blue diamonds show
  individually detected clusters from Hoekstra et al.  (2007) with updated masses from
  Madhavi et al. (2008). Sienna cross symbols show data points from
  Bardeau et al.  (2007). Light blue plus symbols represent the Rykoff et al.  (2008) results from a
  stacked analysis in the SDSS and black diamonds take into account a
  recent correction to these masses due to a new calibration of the
  source distribution. The upper error bars have been adjusted to
  account for the redshift uncertainty. Green asterisks show four data
  points at intermediate masses from Berge et al.  (2008). Finally, the red squares
  depict our COSMOS results which extend previous results to lower
  masses and to higher redshifts. Three arrows highlight the highest
  redshift COSMOS data points. The grey
  shaded region shows the upper and lower envelope of the ensemble of
  lines with a slope and intercept that lie within the 68 percent
  confidence region.
}
\label{leauthaud2010}
\end{figure}

\subsubsection{The COSMOS Survey}

With only 2 square degrees the COSMOS Survey focused on the relatively high-redshift Universe. Due to the 
relatively small volume probed, COSMOS is unlikely to find the most massive structures in the Universe, but it has delivered  
interesting constraints on the redshift evolution of clusters and the scaling relations between observables.
Thanks to the deep X-ray observation of COSMOS fields, clusters can be efficiently selected in principle out to $z\sim 2$. 
Taking advantage of the X-ray selected catalog, Leauthaud et al. (2010) have investigated the
scaling relation between X-ray luminosity ($L_{X}$) and the weak lensing halo mass ($M_{200}$) for
about  200 X-ray-selected galaxy groups. Weak lensing profiles and halo masses were derived for 9 sub-samples, narrowly 
binned in luminosity and redshift. The COSMOS data alone are well fit by a power law, $M_{200}\propto L_{X}^{\alpha}$, with 
a slope $\alpha= 0.66\pm0.14$. These observations significantly extend the dynamic range for which the halo masses of X-ray-selected 
structures have been measured with weak gravitational lensing as shown in Figure (\ref{leauthaud2010}). Combining with other measurements
 demonstrates that the $M-L_{X}$ relation is well described by a single power law with $\alpha=0.64\pm0.03$, over two decades in 
 mass: $M_{200}\sim 10^{13.5}-10^{15.5} h^{-1}_{72}$M$_{\odot}$. These results confirm
that clusters do not follow the self-similar evolution model with $\alpha=0.75$ proposed by Kaiser (1986).


\section{Cluster Lenses as Nature's Telescopes}

\subsection{Magnification due to Gravitational Lensing}

Cluster lenses magnify and distort the shapes of distant galaxies that lie behind them.
For strong lensing clusters, the amplification factor can in principle be infinite if the
source is compact enough and is located exactly behind the caustic, of course such an event is
infinitely rare! Nevertheless, in several strong lensing clusters, 
amplification factors larger than $40\times$ ($\sim$4 magnitudes) have been 
measured (Seitz et al. 1998), and in fact, amplification factors larger than $4\times$ ($\sim$1.5 magnitudes) 
are quite common (Richard et al.  2011).
It is thus not surprising that ``Cluster Lenses'' are often referred to as ``Nature's Telescopes'' or
``Cosmic Telescopes'' and have been rather effectively used to discover and study the most distant galaxies that lie behind them.

\begin{figure}
\epsfig{file=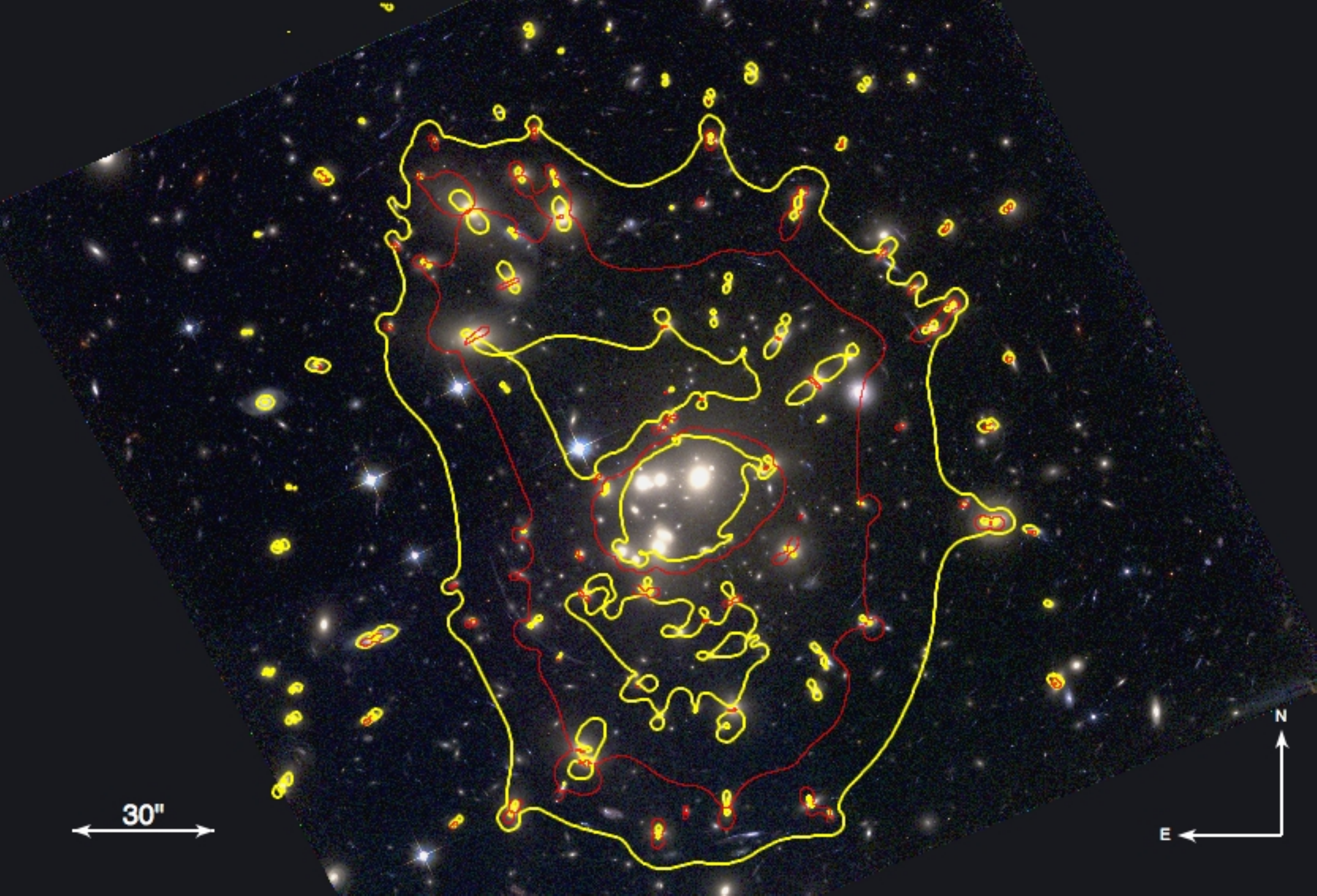,width=\textwidth}
\caption{ The high magnification critical region in the core of the 
massive cluster Abell 1689. Red lines are the critical line for $z=7$ (both radial and tangential lines are shown) and the yellow contours delimit the regions of the sky having a magnification larger than a factor of ten.
\label{figjpk19}
}
\end{figure}

The regions with the largest magnification are the regions closest to the critical lines in the image plane 
(typically less than 1 square arcminute), and  closest to the caustic line in the source plane (typically a few 
tens to hundreds square arcsecond). The cross section for high amplification will vary from cluster to cluster and depends
on the detailed mass distribution. To first order, the cross section scales with the square of the Einstein radius as well
as with the ellipticity or anisotropy of the projected mass distribution on the sky.

As the magnification is wavelength independent, the benefit of using cluster lenses as cosmic
telescopes has been exploited at various wavelengths,  from X-ray to the radio domain.
Lensing clusters were first used as cosmic telescopes in the optical/NIR domain, where
a large population of the most distant galaxies (at their time of discovery) 
were found behind these cluster magnified regions ({\it e.g} Yee et al. 1996;
Franx et al. 1997; Ellis et al.  2001; Hu et al. 2002; Kneib et al. 2004; Richard et al.  2011). 
Lensing clusters were also used at longer
wavelengths in sub-millimeter using $SCUBA$ at JCMT ({\it e.g} Smail et al. 1998) and in the mid-infrared domain 
using the ESA $ISO$ space telescope (Altieri et al.\ 1999; Metcalfe et al. 2003, Coia et al 2005a, 2005b) and now in the
far-infrared using the $Herschel$ space observatory (Egami et al. 2010; Altieri et al.  2010). 

There are two important and unprecedented advantages that cluster lenses offer as cosmic telescopes as they provide the largest field of view with high magnification: 
\begin{itemize}
\item the potential discovery of the most distant objects and low-luminosity objects that would otherwise remain undetected with 
similar blank field imaging, 
\item the possibility to study the morphology of distant galaxies which otherwise
would not be resolved and explore their physical properties that would otherwise
be impossible to characterize.
\end{itemize}

Furthermore, we note that as cluster lenses magnify they also distort the shapes of distant galaxies.
In general, the further the sources, the stronger the distortions. Hence to first 
order, the shape of a lensed galaxy (assuming it can be resolved),
and whether it is multiply imaged or not, can be used as a good distance indicator.

\subsection{Cosmic Telescope Surveys}

\begin{figure}
\epsfig{file=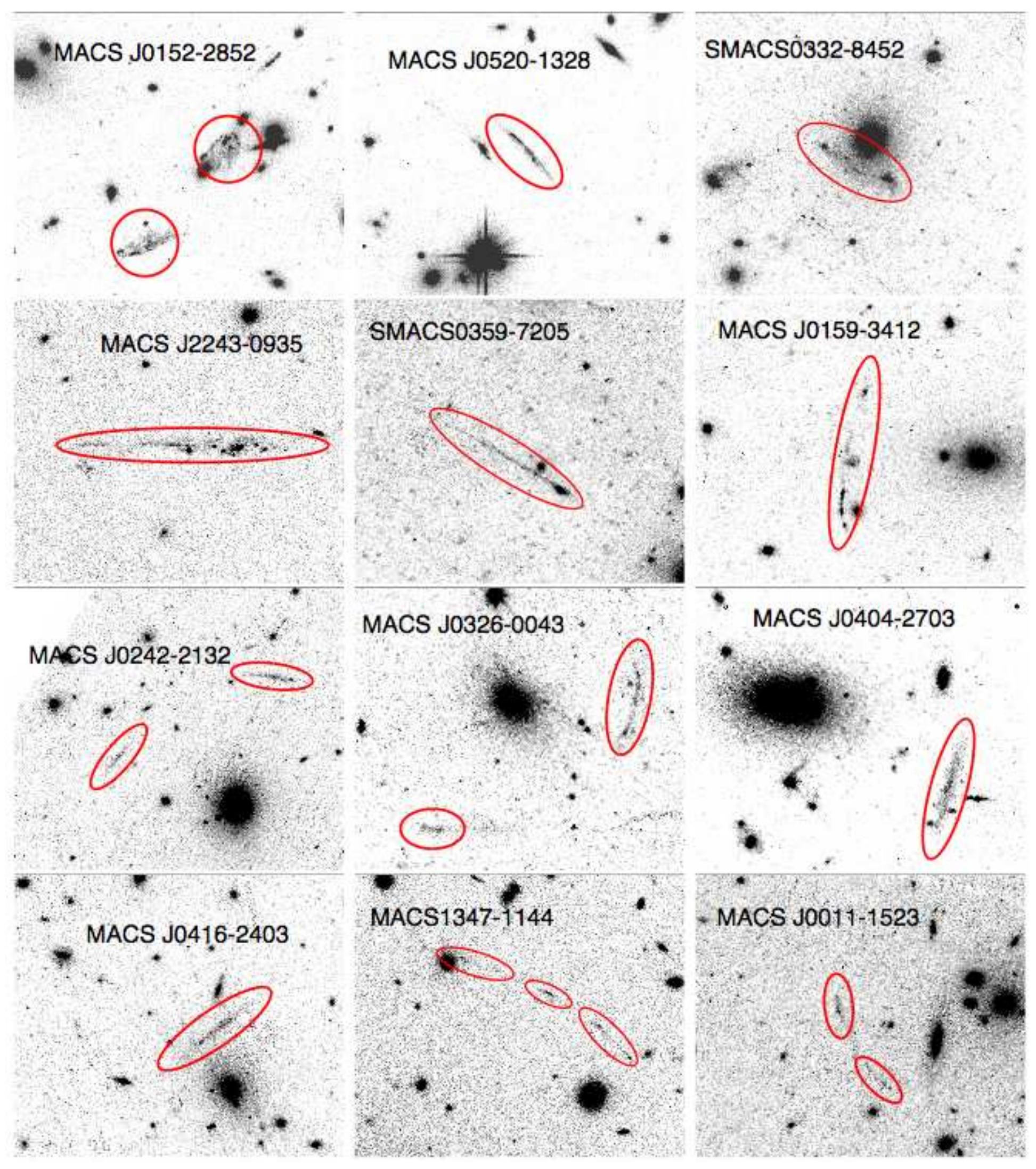,width=\textwidth}
\caption{ Newly discovered giant arcs or multiple-images as part at the SNAPSHOT \hst\ observations (PI. H. Ebeling) of the MACS cluster sample. These images show the diversity of morphology for these serendipitously discovered strongly lensed galaxies (Figure is from Richard et al. private communication). 
\label{mosaic}
}
\end{figure}

Rather similar to other galaxy surveys two distinct observational strategies that trade-off depth with area have been explored thus far:

1) deep mapping (in imaging or spectroscopy) of a few well modeled lensing clusters to search for distant lensed 
sources - this allows us to probe down the luminosity function of the targeted distant source population, 

2) shallow mapping on a large cluster sample to search for rare highly magnified background sources ({\it e.g.} Figure \ref{mosaic}), with the idea to thereafter 
conduct detailed follow-up observations of these sources benefiting from the high amplification/magnification to constrain important physical and morphological 
properties of high-redshift sources (e.g. Lemoine-Buserolle et al 2003).

As the strong magnification region covers typically at most about a few square arcminutes, surveys through cluster lenses are particularly 
adapted to those instruments/telescopes that have an instantaneous field of view comparable to the strong lensing region. The \hst\  
cameras are particularly well matched to cluster strong lensing regions and are thus very well adapted to use for the search and study of distant sources. 
Incidentally, this was also the case for the SCUBA instrument on JCMT, as well as the ISOCAM on the ISO space mission. In the near future ALMA, 
MUSE (the one square arcminute integral field unit on the VLT), and the forthcoming JWST instruments are all facilities that will most effectively exploit lensing 
magnification. 

Of course when conducting a detailed follow-up study of highly magnified sources, the most effective instruments are high-resolution imagers and spectrographs. 
In particular, because of the extended nature of the most amplified sources, integral field spectrographs are more adapted compared to long-slit instruments, and it is 
thus natural to conduct follow-up studies with instruments such as SINFONI on VLT or OSIRIS on Keck for the extremely magnified, rare, lensed sources.

Finally, an other particular observation strategy of cosmic telescope is that of 
 ``critical line mapping''.  In this case, one specifically targets regions 
near the critical lines using dedicated instruments such as a long slit spectrograph ({\it e.g.} Santos et al.  2004; Stark et al. 2007), an integral field spectrograph (Clement et al. 2011 in preparation) or a
millimeter wave interferometer to blindly probe the distant Universe. In this case the effective field of
view of the instrument is small compared to the critical region, thus requiring a mapping strategy to
cover the region with the highest amplification.

\subsection{``Lens redshift'' measurement}

As lensing distortion and magnification are a function of the redshift of the background sources, once a cluster mass distribution is well known, the lens model 
can be used to  predict the redshifts for the newly identified multiple systems ({\it e.g.} Kneib et al.\ 1993; Natarajan et al.\ 1998; Ellis et al.\ 2001) as well as for
 the arclets (Kneib et al.\ 1994b; 1996). 

For multiple-image systems, the relative positions of the different images is a strong function of the redshift of the background source. Although the redshift sensitivity decreases 
with the redshift of the background source (because the $D_{LS}/D_{S}$ variation is smaller with increasing redshift) it can nevertheless be used to easily distinguish between low and high-redshift solution. In this respect, one can easily discriminate $z\sim $1--2 obscured galaxies from $z>4$ high-redshift lensed galaxies, and this property has been 
used many times very effectively (Ellis et al. 2001; Kneib et al. 2004; Richard et al. 2008; Richard et al. 2011).
For the arclets, the redshift prediction is based on the fact that on average a distant galaxy is randomly orientated, and its ellipticity follows a relatively peaked
ellipticity distribution ($\sigma_{\epsilon}\sim 0.25$). Hence, by conducting high-resolution imaging ({\it e.g.} with Hubble), and by measuring the ellipticity and 
orientation of the background lensed sources in the core of massive cluster lenses, one can statistically derive the redshift distribution of the background lensed 
population. Such measurements were first introduced by Kneib et al. (1993), and developed further in Kneib et al.(1996). These predictions were tested and verified 
by Ebbels et al. (1998) using spectroscopy in the case of the lensing cluster Abell 2218. Despite the successful demonstration of the technique, it never became 
popular due to the following limitations. First, the derived probability distribution $p(z)$ distribution is relatively broad, particularly, at high-redshift.  Therefore the method is 
not really competitive with photometric redshift determinations, except may be for disentangling catastrophic photometric redshifts. Second, the cluster galaxy contamination 
is high in the optical/near-infrared domain and statistical estimates always have limited utility. Third, galaxy sizes decrease rapidly with redshift, and accurate measurement 
of the galaxy shape can only be done efficiently with deep Hubble imaging yet again limiting the use of this technique. Finally, as there are many other good science drivers to obtain multi-band information on these massive clusters, and as photometric redshift determination methods are rapidly improving, the statistical lens redshift measurements 
never became attractive and/or popular.

\subsection{Lensing Surveys in the Sub-millimeter}

\begin{figure}
\centerline{
\epsfig{file=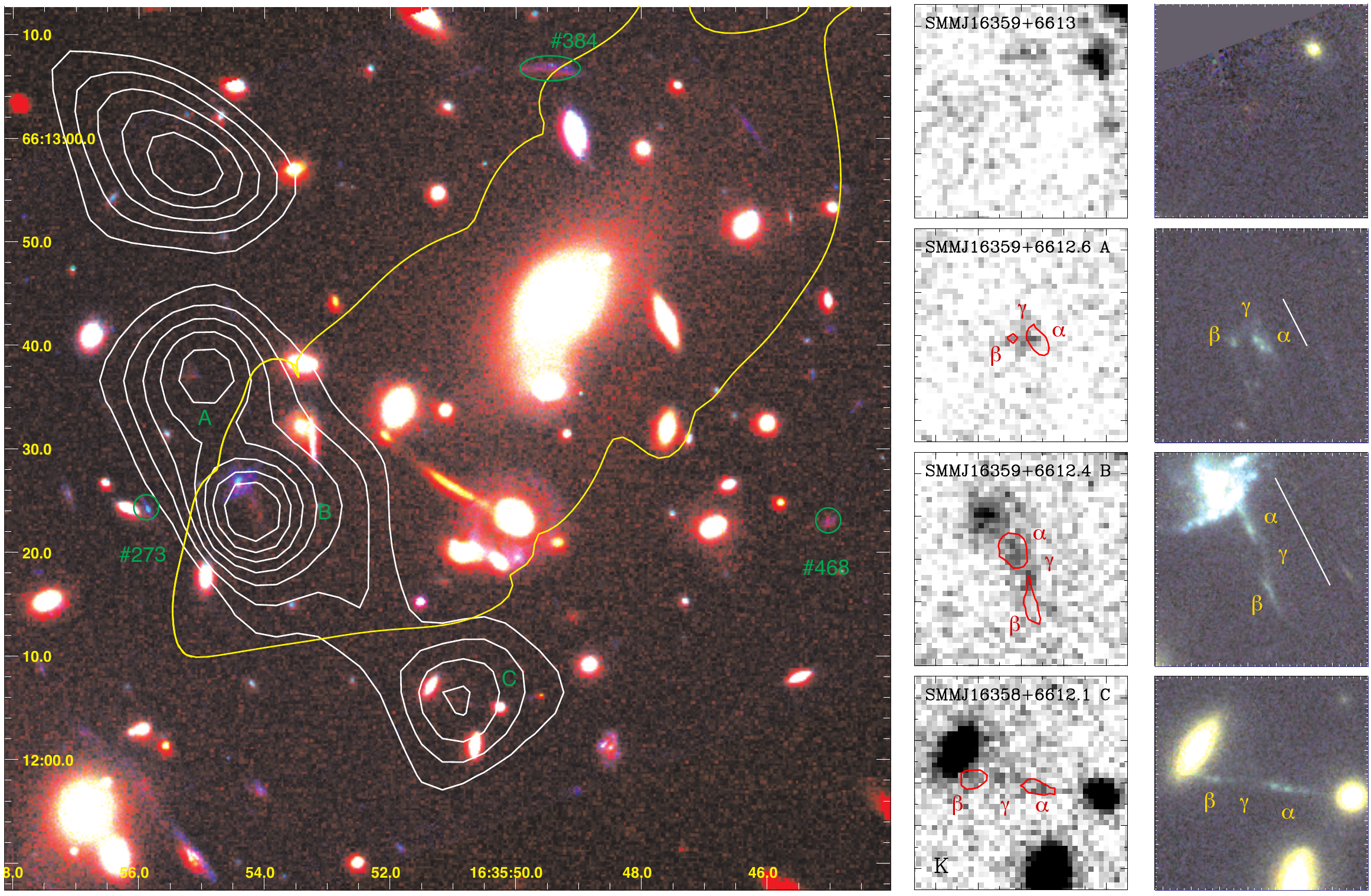,width=\textwidth}
}
\caption{(Left) A true-color image of the core of A\,2218 (blue:\hst\  F450W, green:\hst\  F814W  and red: WHT/INGRID $K_s$).  The 
850\,$\mu$m sub-mm image from SCUBA is overlayed as white contours. The three images of the multiply imaged
sub-mm galaxy are annotated as A, B and C.  The sub-mm contour at the top left corresponds to a $z=4.04$ sub-mm galaxy 
discussed in Knudsen et al.(2010). Two other galaxies at $z\sim 2.5$, are identified: the single-image \#273 and the
fold-image \#384 and its counter-image \#468. The yellow line shows the critical line at $z=2.515$. (Right) Panel of
10''$\times$10'' images showing the INGRID $K_s$-band (left column) and \hst\  true color image from F450W/F606W/F814W (right
column) of the four submm sources in the core of A\,2218.  Note how each of the sub-mm sources, SMM\,J16359+6612.6, SMM\,J16359+6612.4 and
SMM\,J16358+6612.1, comprises a NIR source ($\gamma$) which is bracketed by two features in the F814W image ($\alpha$ and $\beta$).
[From Kneib et al. 2004].}
\label{a2218smm}
\end{figure}

The SCUBA (the JCMT Submillimeter Common-User Bolometer Array) Lensing survey was likely one of the first systematic surveys to exploit distant 
lensed galaxies using massive clusters. This survey first started with the mapping of two massive clusters: Abell\,370 at $z=0.37$ and Cl\,2244-02 at $z=0.33$ (Smail et al. 
1997) and continued to map the region behind five similarly massive clusters covering a total area of 0.01 square degree (Smail et al. 1998). Each SCUBA continuum map from this cluster lens 
survey covered a total area of about 5 square arcminutes to 1$\sigma$ noise levels less than 14 mJy/beam and 2 mJy/beam at 
450 and 850 micron wavelengths respectively.

Since SCUBA was a new instrument that achieved a sensitivity 2--3 orders of magnitude deeper than was previously possible and thanks to the cluster magnification, 
Smail et al. (1997) were the first to find the distant sub-millimeter (sub-mm) selected galaxy population. In total 17 sources brighter than the 50\% completeness limits (10 brighter than the 80\% limit) were discovered (Smail et al. 1998). The sub-millimeter spectral properties of these first sources indicated that the majority lie at high-redshift ($1<z<5.5$), which was confirmed later with redshift measurements. Measured redshifts for a  large number of these submm-selected galaxies placed the bulk of this population at $z\sim 2.5$ (Ivison et al. 1998; Barger et al. 2002; Chapman et al. 2005).

The use of cluster lenses in the case of the sub-millimeter high-redshift searches was strongly motivated by the fact that cluster galaxy members are not sub-millimeter sources and are therefore transparent at this wavelength, making clusters perfect telescopes to preferentially probe the distant galaxy population (Blain 1997). Importantly, the use of cluster lenses increases the sensitivity of sub-mm maps and reduces the effects of source confusion (which plagues bolometer surveys in sub-mm and mm wavelengths) due to the dilution produced by lensing clusters. With accurate lens models Blain et al. (1998) first corrected the observed sub-mm source counts for lensing amplification using the SCUBA lensing survey data on the first seven clusters, thus pushing the 
850 micron counts down below the SCUBA confusion limit; for example at 1 mJy, 7900$\pm$3000 galaxies per square degree were found. Down to the 0.5 mJy limit, the resolved 850 micron background radiation intensity was measured to be (5$\pm$2)$\times$10$^{-10}$ W\,m$^{-2}$\,sr$^{-1}$, comparable to the current COBE estimate of the background, indicating for the first time that the bulk of the 850 micron background radiation is effectively produced by distant ultra-luminous galaxies. These first sub-mm galaxy counts were confirmed later with a larger sample of clusters mapped by SCUBA (Cowie et al. 2002; Knudsen et al. 2008) reaching a lens-corrected flux limit of 0.1 mJy. The first sub-mm multiple-images were found in Abell 2218 
(Kneib et al. 2004) and MS0451-03 (Borys et al. 2004) identified at $z=2.516$ and $z\sim 2.9$ respectively.
In particular, the source SMM J16359+6612 is gravitationally lensed by Abell 2218 into three discrete images with a total amplification factor of $\sim$45, implying that this galaxy has an unlensed 850-micron flux density of only 0.8 mJy. Furthermore, SMM J16359+6612 shows a complex morphology with three sub-components arguing for either a strong dust (lane) absorption or a merger. Interestingly, these sub-mm sources are surrounded by two other highly amplified galaxies at almost identical redshifts within a $\sim$100-kpc region suggesting this sub-mm galaxy is located in a dense high-redshift group (see Figure~\ref{a2218smm}). Further mapping at the IRAM Plateau de Bure interferometer indicated that this source is a compact merger of two typical Lyman-break galaxies with a maximal separation between the two nuclei of about 3 kpc, thus it bears a close similarity to comparable luminosity, dusty starbursts that result from lower-mass mergers in the local Universe.

\begin{figure}
\centerline{
\epsfig{file=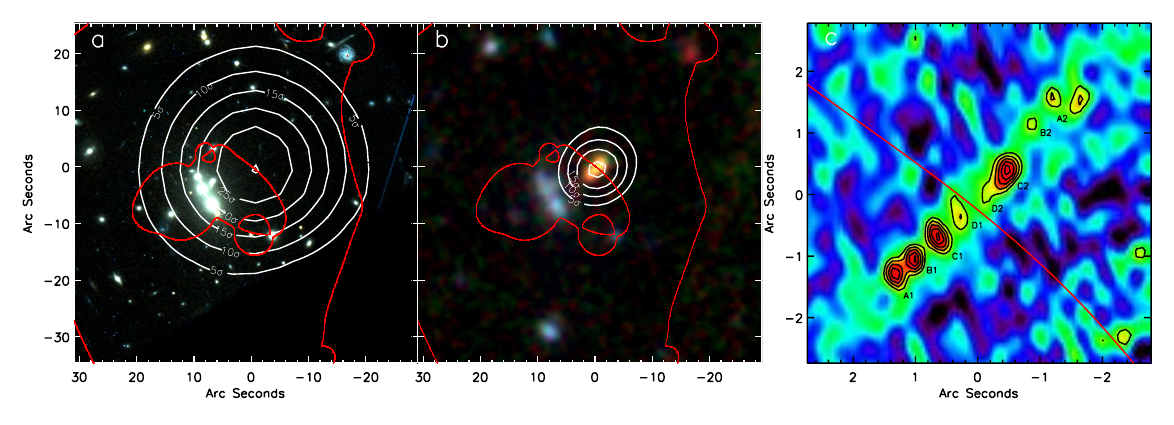,width=\textwidth}
}
\caption{(Left) Hubble Space Telescope V, I -band color image of MACSJ2135-010217 with white contours denoting the 870 micron emission of SMMJ2135-0102 with flux 
$106.0 \pm 7.0\,$mJy as observed by APEX/LABOCA.  The optical counter part is faint with $I_{AB}$=23.6$\pm$0.2.  The solid red lines denote the $z=2.326$ radial and tangential critical curves from the best-fit lens model.  (Center) True color IRAC 3.6, 4.5, 8.0 micron image of the cluster core with contours denoting the 350 micron emission from APEX/LABOCA. The mid-infrared counterpart is clearly visible as an extended red galaxy centered at the sub-mm position. (Right) SMA 870 micron image of the galaxy. The map shows eight individual components, separated by up to 4'' in projection. The red line is the same $z=2.326$ radial critical curve.  Components (A,B,C,D) represent two mirror images of the galaxy, each comprising four separate emission regions reflected about the lensing critical curve. Figure from Swinbank et al. (2010).}
\label{smmj2135}
\end{figure}

However, the most spectacular lensed sub-mm galaxy is certainly the recently discovered SMMJ2135-0102 at redshift z=2.3259 that is gravitationally magnified by a factor of 32 by the massive cluster MACSJ2135-010217 (Swinbank et al. 2010; Figure~\ref{smmj2135}).  This large magnification, when combined with high-resolution sub-mm imaging, resolves the star-forming regions at a linear scale of just $\sim$100 parsec.  The luminosity densities of these star-forming regions are comparable 
to the dense cores of giant molecular clouds in the local Universe, but they are $\sim$100$\times$ larger and 10$^{7}$ times more luminous. The star formation processes at $z\sim$2 in this vigorously star-forming galaxy appear to be similar to those seen in local galaxies even though the energetics are unlike anything found in the present-day Universe. 

In the sub-mm domain, lensing has proven to be truly useful in revealing details about high-redshift sources that would otherwise be impossible, even with the next generation of large aperture telescopes.

\subsection{Mid-Infrared Lensing Survey}

In the late 90's, the ISOCAM camera on the ESA Infrared Space Observatory (ISO) targeted a number of massive cluster lenses. The motivation for these observations was 
to probe the faint and distant Mid-Infrared galaxy population and their contribution to the cosmic mid-infrared background radiation. In particular a few well-known massive 
cluster lenses were imaged deeply by ISO at 7 and 15 micron. The deepest ISO observation of a cluster targeted Abell 2390 (Altieri et al. 1999). Cross-identification of the 
numerous mid-infrared sources with optical and near-infrared data showed that almost all 15 micron sources were identified as lensed distant galaxies. These observations 
allowed the computation of number counts in both the 7 and 15 micron bands and led to the ruling out of non-evolutionary models, and favoring very strong number count evolution. By combining the data on three massive clusters (Abell 370, Abell 2218 and Abell 2390), Metcalfe et al. (2003) detected a total of 145 mid-infrared sources, and after a very careful lensing correction derived the intrinsic counts of the background source population. It was found that roughly 70\% of the 15 micron sources are lensed background galaxies. Of sources detected only at 7 micron, 95\% are cluster galaxies in this sample. Of the 15 sub-mm sources already identified within the mapped regions of the three clusters, 7 were detected at 15 micron. Flux selected subsets of the field sources above the 80\% and 50\% completeness limits were used to derive source counts to a lensing corrected sensitivity level of 30 micro-Jy at 15 micron, and 14 micro-Jy at 7 micron. The source counts, corrected for the effects of completeness, contamination by cluster sources and lensing, confirmed and extended earlier findings of an excess by a factor of ten in the 15 micron population with respect to source models with no evolution, with a redshift distribution that spans between $z=0.4$ and $z=1.5$. 

\subsection{Lensed Extremely Red Objects}

\begin{figure}
\centerline{
\epsfig{file=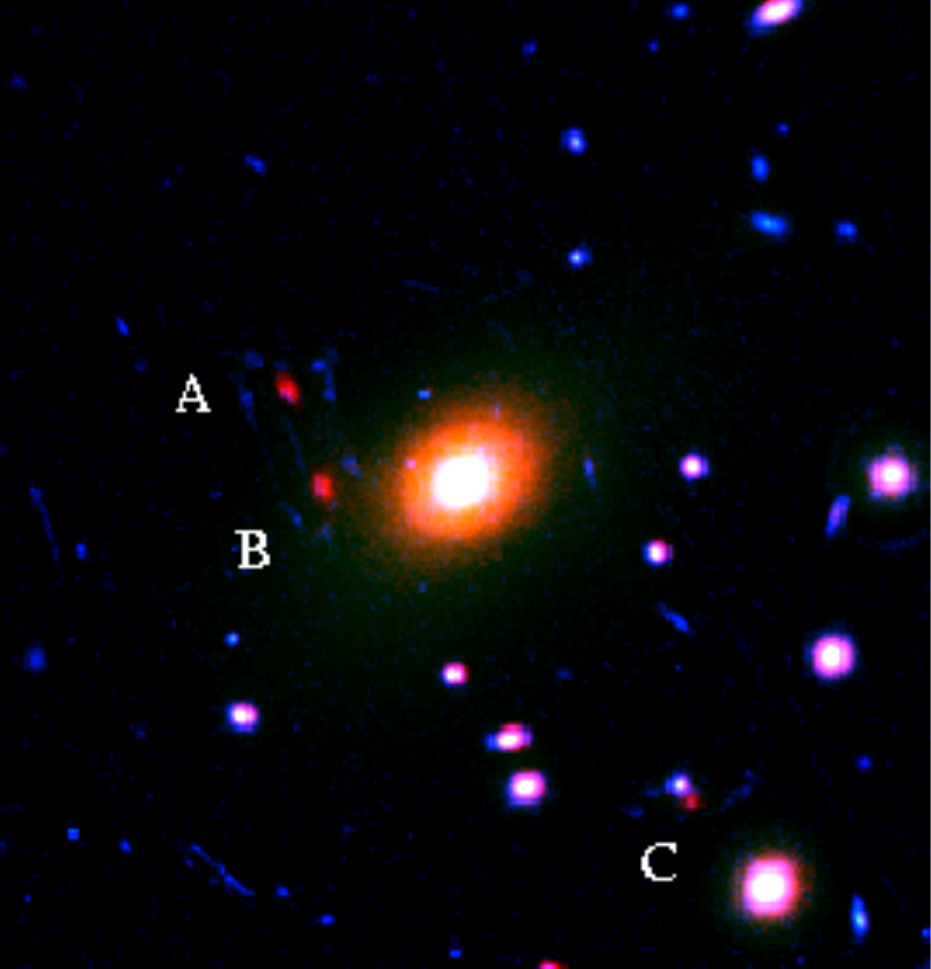,width=0.5\textwidth}
}
\caption{True color $R,K$ view  ($50''\times50''$) of the Abell 68 cluster core
(combining \emph{HST}  and  near-infrared UKIRT  data).   The
bright elliptical  galaxy in  the center of  the frame is  the central
galaxy  of the  cluster.  Three images  of  ERO\,J003707 are  clearly
visible and are marked as A, B and  C.  Each image comprises a central red
``bulge'',  surrounded by numerous  fainter blue  knots of  current or
recent star formation (from Smith et al. 2002).
\label{a68RK}
}
\end{figure}

The benefits of lensing have also been used to search for Extremely Red Objects (EROs) behind a sample
of 10 X-ray luminous galaxy clusters (Smith et al.\ 2002a, 2002b) imaged by both the WFPC2 camera (using F702W filter) 
and UKIRT in the K-band. EROs are galaxies with $R-K>5.3$ as defined by Daddi et al. (2000) as the criterion to select distant 
elliptical galaxies. The other more stringent definition with $R-K>6$  was adopted by Thompson et al. (1999).
In these clusters a total of about 60 EROs have been identified so far allowing the number counts of these
rare objects to be computed down to about 2 magnitudes fainter than previous work published at that time.
The exploitation of the lensing effect has also permitted a more accurate study of the morphology of these peculiar galaxies, 
revealing in some cases spectacular disky components already in place at fairly early times  (Figure~\ref{a68RK}). In particular, for 
the case of the multiply imaged ERO\,J003707, morphological  and photometric analyses reveal an $L^{*}$ early-type 
disk-galaxy. It has been estimated that $\sim10$\% of  EROs with  $R-K>5.3$ and  $K<21$ have similar properties.  The unique association 
of passive EROs with elliptical galaxies therefore appears to be too simplistic and has been reconsidered.  More recent work on searching for  lensed EROs 
was conducted in A1835 and AC114 (Schaerer et al. 2007) taking advantage of complete wavelength coverage including \hst, ground based 
and Spitzer data. They found in these observations that most of the EROs were, in fact, young dusty star-bursts at $z\sim 2-3$.

\subsection{Lensed Lyman-$\alpha$ Emitters}

One of the exciting  current ventures is to map the critical region of massive
clusters (Figure~\ref{figjpk19}) in order to search for Lyman-alpha emitters at very high
redshifts ($z>4$), compute their number density, derive their luminosity
function and therefore characterize this population. By pushing to very high-redshift $z>7$
one should get closer to re-ionization, and the increase in neutral gas content of the Universe should block 
Lyman-$\alpha$ photons. Thus at some point one should expect a strong evolution of the Lyman-$\alpha$ emitter 
luminosity function. Two approaches have been pursued in the search for Lyman-$\alpha$ emitters, a direct search 
through intense spectroscopy, and an indirect one that relies on conducting narrow-band imaging searches in which the 
wavelength range is tuned to probe a specific redshift window.

\subsubsection{Spectroscopic critical line mapping}

The first dedicated critical line mapping using spectroscopy was conducted at Keck, using the long slit mode of LRIS (Santos et al. 2004). Nine 
intermediate redshift, massive clusters with good lensing mass models were carefully selected, and a number of long-slit observations was conducted, 
sliding the long slit across the critical line region thus achieving magnification factors generally greater than 10. Eleven emission-line candidates were 
located in the range $2.2<z<5.6$ with Lyman-$\alpha$ as the line identification. The selection function of the survey takes into account the varying intrinsic 
Lyman-$\alpha$ line sensitivity as a function of wavelength and sky position. By virtue of the strong magnification factor, these measurements provide constraints 
on the Lyman-$\alpha$ luminosity function to unprecedented limits of 10$^{40}$ erg/s, corresponding to a star formation rate of 0.01 M$_{\odot}$/yr. Combining these 
lensing results with other surveys, limited to higher luminosities, Santos et al. (2004) argue that there exists evidence for the suppression of star formation in low-mass 
halos, as predicted in popular models of galaxy formation. The highest redshift Lyman-$\alpha$ emitter discovered in this survey is the $z=5.576$ pair in 
the cluster Abell 2218 (Ellis et al. 2001).  High-resolution spectroscopic follow-up confirmed the lensing hypothesis of the LRIS discovery by identifying the second image. The unlensed 
source appears to be a very faint source with (I$\sim$30)and is compact in nature ($<$150 h$^{-1}_{65}$ pc).  This source is a promising 
candidate for an isolated $\sim$10$^{6}$ M$_{\odot}$ system seen likely producing its first generation of stars close to the epoch of reionization.

Pushing to higher redshifts than $z>7$ requires an infra-red spectrograph. In a challenging experiment, Stark et al. (2007) blindly surveyed the 
critical line region of nine massive clusters using the Keck/NIRSPEC long slit. The magnification boost ranges from 10 to 50$\times$ for a background galaxy between 
$\sim 8<z<\sim 10$, thus pushing the sensitivity limits to unprecedented low fluxes (10$^{41}$-10$^{42}$ erg/s)
for this redshift range. This survey identified  six promising ($>$5$\,\sigma$) candidate Lyman-$\alpha$ emitters that lie between $z=8.7$ and $z=10.2$. Lower redshift line interpretations were mostly excluded through the non-detection of secondary emission in further spectroscopy undertaken with LRIS and NIRSPEC.
Nonetheless, it is considered plausible that at least two of the candidates are likely at $z\sim 9$. If true, then given the small volume surveyed, this 
suggests there is an abundance of low-luminosity star-forming sources at $z\sim 8-10$ that could provide a significant proportion of the UV photons necessary 
for cosmic reionization. A parallel study was conducted using the SINFONI 3D spectrograph at VLT, and three of the Keck/NIRSPEC candidates were re-observed - as part of the SINFONI critical line mapping program. However, no confirmation of the Keck/NIRSPEC detected lines was found, casting some doubt on the real redshifts of these particular sources. The results of this survey are presented in Clement et al. (2011). Future 8-10m class instruments such as MUSE in the visible and  MOSFIRE, KMOS and EMIR in the near infrared should, thanks to their higher multiplexing, provide new opportunities to further conduct critical line surveys and find more robustely numerous high-redshift Lyman-$\alpha$ emitters.

\subsubsection{Narrow-band searches}

\begin{figure}
\centerline{
\epsfig{file=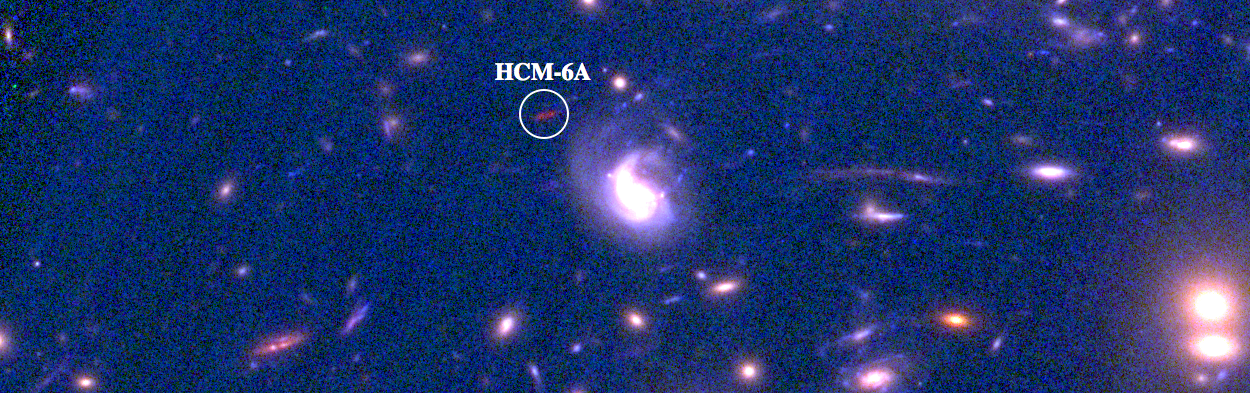,width=\textwidth}
}
\caption{\hst\  (ACS+WFC3) color $V,I,J$ view of the HCM-6A $z=6.56$ Lyman-$\alpha$ emitter located near the core of the Abell 370 cluster. Note the strong detection in the F110W (J-band) filter and its bimodal light distribution. The region shown covers 60''$\times$20''.
\label{hcm6a}
}
\end{figure}

An alternative to direct spectroscopy of Lyman-$\alpha$ emitters, is to conduct narrow-band imaging. 
Although this technique has been very popular in blank fields, only a few such observations have been
 conducted in the direction of cluster cores. Hu et al. (2002) have discovered a redshift $z=6.56$ 
 galaxy lying behind the cluster Abell 370 (Figure~\ref{hcm6a}). The object nicknamed HCM-6A was found 
 in a narrow-band imaging survey using a 118\,\AA bandpass filter centered at 9152\,\AA using LRIS 
 on the Keck telescope. At the time of discovery, HCM-6A was the first galaxy to be confirmed at 
 redshift $z>6$ (its observed equivalent width is 190\,\AA, with a flux of 2.7\,10$^{-17}$ 
 erg/cm$^{2}$/s). Using the detailed lensing model of this cluster, a lensing amplification factor 
 of 4.5 was estimated as the source is situated about 1 arcminute away from the cluster center. 
 This discovery suggested that the re-ionizing epoch of the Universe lies beyond $z\sim 6.6$. 
 Follow-up of this source with Spitzer (Chary et al. 2005) and in millimeter with MAMBO-2 
 (Boone et al. 2007 - which provided an upper limit at 1.2mm) have helped derive some physical 
 parameters with relatively high accuracy considering the distance of this source. Even more 
 ambitious was the narrow J-band filter NB119 survey (corresponding to Lyman-$\alpha$ at $z\sim 9$) 
 nicknamed the `$z$ equals nine' (ZEN) survey conducted towards three massive lensing clusters: 
 Abell clusters A1689, A1835 and AC114 (Willis et al. 2008). However, no sources consistent with a 
 narrow-band excess were found and no detection in bluer deep optical was reported.  The total 
 coverage of the ZEN survey sampled a volume at $z\sim9$ of approximately 1700 co-moving Mpc$^{3}$ 
 to a Ly-$\alpha$ emission luminosity of 10$^{43}$ erg/s. The limits from this survey still 
 offer the best constraints at this redshift.

\subsection{Lyman-break Galaxies}

\begin{figure}
\centerline{
\epsfig{file=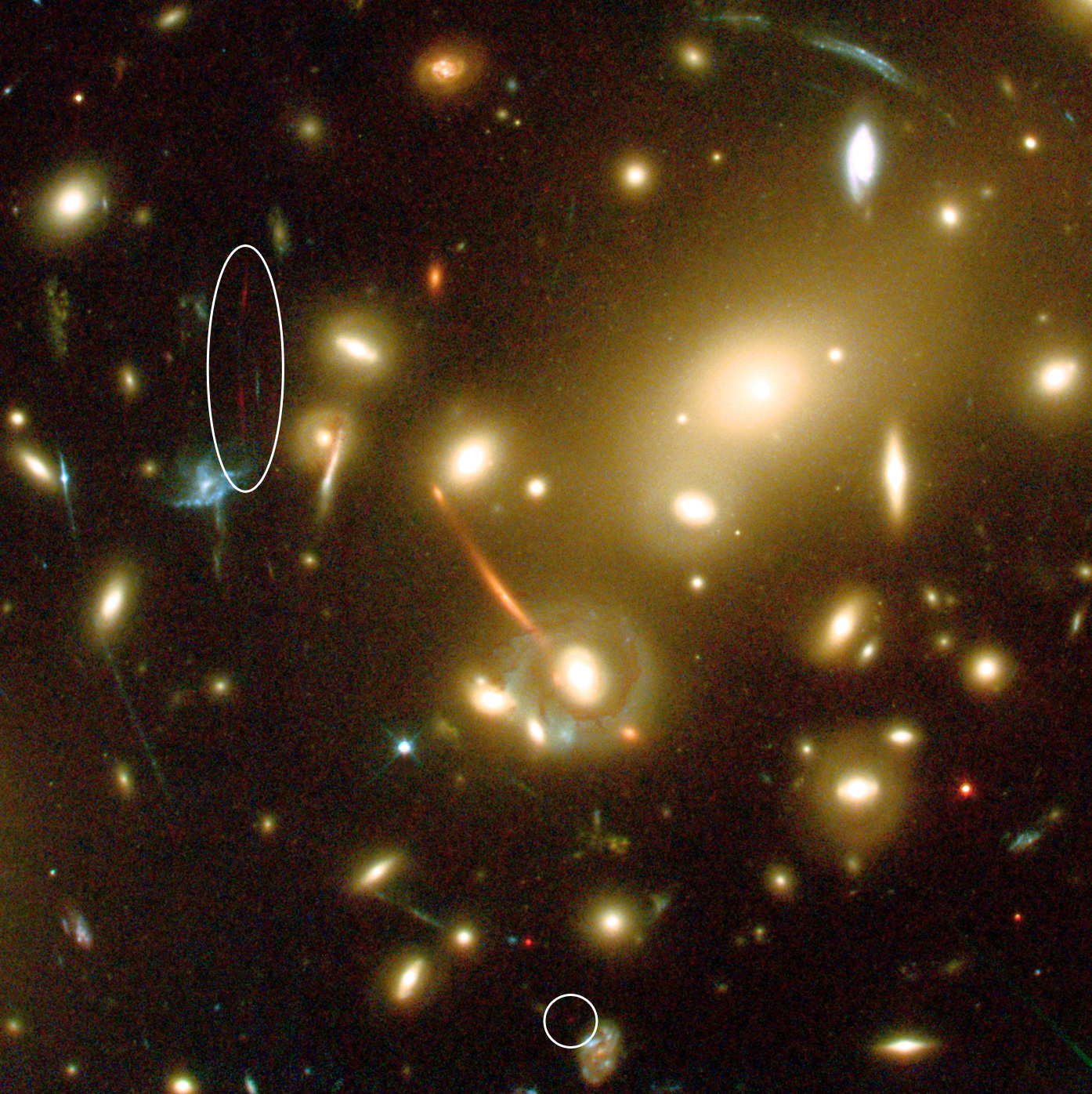,width=0.7\textwidth}
}
\caption{\hst\  ACS color $V,I,z$ view of Abell 2218  showing the triply-imaged Lyman-break galaxy at $z\sim 6.8$ (Kneib et al. 2004).
\label{kneib2004hiz}
}
\end{figure}

As the number density of Lyman-break galaxies (at  $3<z<3.5$) is typically half a galaxy per square arcminute down to $R=25$ (Steidel et al. 1996), a massive cluster 
will likely magnify one of them, and if we consider a wider range in redshift, the probability that one of them is multiply imaged is almost guaranteed. The first giant 
arc in Cl\,2244-04 at $z=2.24$ is considered to be the first Lyman-break galaxy detected in a cluster lens, although at the time of its discovery in the 
late 80's, this galaxy was not classified as such. As part of the CNOC survey, Yee et al. (1996) identified a ``proto-galaxy'' at $z=2.72$, the galaxy cB58 
in the cluster MS1512+36. Interestingly, they assumed that this object although being very close to the cluster center was unlikely to be lensed. But soon after, 
Seitz et al. (1998) demonstrated that cB58 was highly lensed, magnified by more than a factor of 50, thanks to the identification of its counter image. The estimate of such a high magnification led to a number of follow-up studies at high spectral resolution to further constrain the physical properties of this high-redshift galaxy ({\it e.g.} Pettini et al. 2002). At 
about the same time, the triple arc at $z=2.515$ in Abell 2218 (Ebbels et al. 1996) was the first recognized Lyman Break galaxy lensed by a massive cluster, however 
its magnification is only $\sim 15\times$, much less than that of cB58. Shortly afterward in the course of a spectroscopic cluster galaxy survey of MS1358+62, Franx et al. (1997) 
discovered a Lyman-break galaxy at $z=4.92$ multiply imaged by the cluster. Further study and modeling by Swinbank et al. (2010) derived a magnification factor for the 
brightest image of $12.5\pm 2$. At the time of discovery the arc in MS1358+62 was the most distant galaxy known. In the massive cluster Abell 2390, Frye \& Broadhurst (1998) 
and Pell{\' o} et al. (1999) independently found a $z=4.04$ pair, strongly lensed by the cluster. These high-redshift discoveries have demonstrated the potential of discovering 
even higher redshift galaxies lensed by massive clusters. Thanks to deep F850LP/ACS data, following-up the $z=5.56$ Lyman-$\alpha$ galaxy pair of Ellis et al. (2001), 
Kneib et al. (2004) found an $i$-band dropout detected in $z$-band (see Figure \ref{kneib2004hiz}). Detection with NICMOS confirmed a $z\sim6.8$ redshift, however a NIRSPEC/Keck spectrum failed to 
detect a Lyman-$\alpha$ line, but Spitzer IRAC 3.6 and 4.5 micron detections (Egami et al. 2005) provided strong constraints on the age of the underlying stellar population, 
making it one of the best studied objects at this redshift.

Using the ESO/VLT instruments FORS and ISAAC, a deep imaging survey of the clusters AC114 and A1835 was 
conducted by Richard et al. (2006) to search for lensed optical and near infrared dropout galaxies. In this work, they identified 26 optical dropout candidates in both 
A1835 and AC114 (with $H\sim 23.5-24.0$). Half of these candidates show an SED compatible with star-forming galaxies at $z>6$, and 6 of them are likely intermediate-redshift 
extremely red objects based on luminosity considerations. With this dataset a first attempt was made to characterize the luminosity function of these high-redshift galaxies, that are 
not well constrained by deeper \hst/NICMOS observations of the HUDF (Hubble Ultra Deep Field). 
This work lead to the study by Richard et al. (2008) of a further 6 massive clusters with \hst\  using the NICMOS camera and complemented by Spitzer observations. The survey 
yielded 10 z-band and 2 J-band dropout candidates to photometric limits of J$_{110}\sim$26.2 AB (5$\sigma$). By taking into account the magnifications afforded by the clusters, they 
probed the presence of $z>7$ sources to unlensed limits of J$_{110}\sim$30 AB, fainter than those charted in the HUDF. Taking into account the various limitations of this work, they concluded that about half of the sample of z-band dropouts are at high-redshift. An ambitious infrared spectroscopic campaign undertaken with the NIRSPEC spectrograph at the Keck Observatory for seven of the most promising candidates failed to detect any Ly-$\alpha$ emission. 

Behind Abell 1689, using \hst/NICMOS Bradley et al. (2008) found 
a bright $H=24.7$ $z\sim 7.6$ galaxy candidate: A1689-zD1. This source is 1.3 mag brighter than any known z850-dropout galaxy  (thanks to a cluster magnification factor of $\sim 9.3\times$). Nevertheless, no spectroscopic observations have yet confirmed the redshift of this candidate. In the more recent years, discoveries have been reported using either the 
ground based ESO/Hawk-I infrared imager (Laporte et al. 2011) or the new WFC3 camera installed in May 2009 onboard \hst\  (Bradley et al. 2011; Kneib et al. 2011; Paraficz et al. 2012). In the long term, the James Webb Space Telescope (JWST) and the Extremely Large Telescopes will uncover 
large numbers of these very high-redshift systems, enabling the study of their sizes, morphologies and physical parameters (e.g. Wyithe et al. 2011; Salvaterra et al. 2011).

\subsection{Far Infra-Red Lensing Surveys}

With the launch of the Herschel Space Observatory in May 2009, a new window to  the Universe has been opened. The Herschel Lensing Survey (HLS) conducted 
deep PACS and SPIRE imaging of 44 massive clusters of galaxies. These observations complement the observation of 10 massive clusters by the GTO teams.
 In particular, it is foreseen that the strong gravitational lensing power of these clusters will enable penetration through the confusion noise, which sets the ultimate limit on our ability to probe the Universe with Herschel. Although the analysis of this large dataset is still in progress, some early results were presented in the A\&A Herschel special issue in spring 2010.
In particular, Egami et al. (2010) summarized the major results from the science demonstration phase observations of the Bullet cluster ($z = 0.297$). The study of two strongly lensed and distorted galaxies at $z = 2.8$ and 3.2 and the detection of the Sunyaev-Zel'dovich (SZ) effect increment of the cluster with the SPIRE data have been reported.

\begin{figure}
\centerline{
\epsfig{file=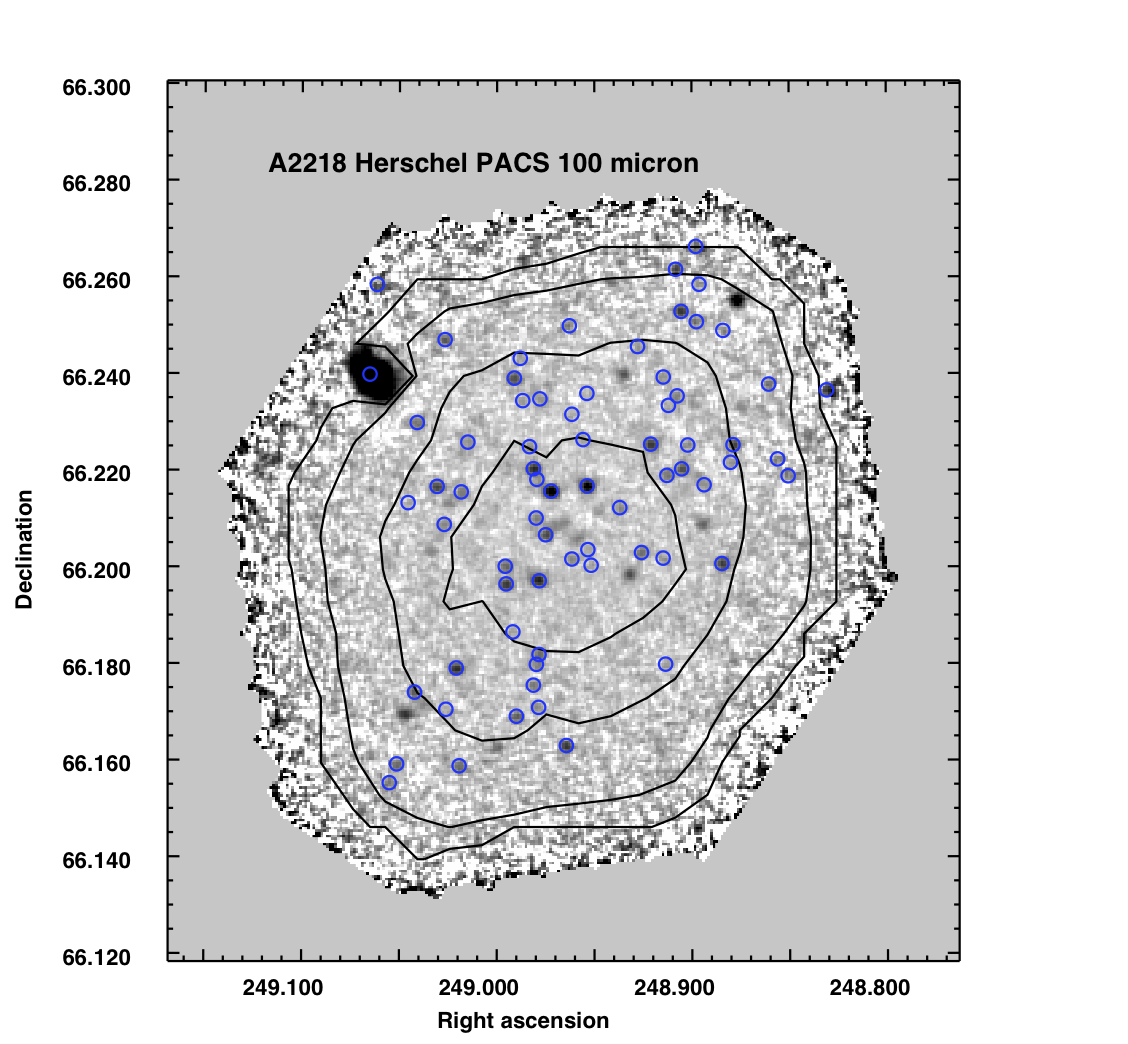,width=0.45\textwidth} \ \ \
\epsfig{file=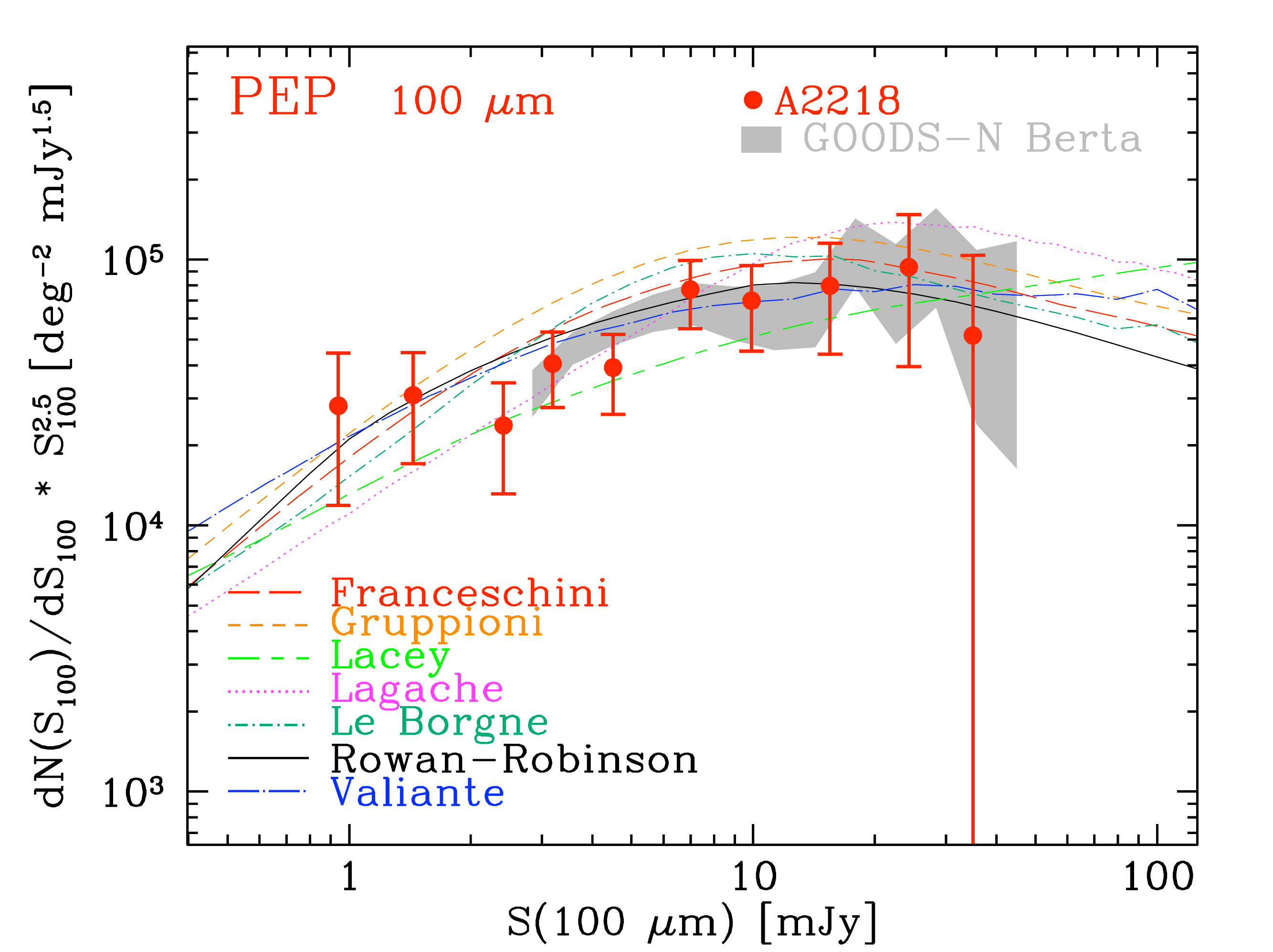,width=0.55\textwidth}
}
\caption{(LEFT) PACS 100 ?m map toward Abell 2218 with lensed and foreground sources marked with blue circles, other sources are identified cluster members. Overlaid contours in black show the rms contours at 0.7, 1.0, 2.0 and 4.0 mJy. (RIGHT)  Number counts at 100 micron with lensing correction (red filled circles), normalized to the Euclidean slope , against the prediction of various synthetic counts models. Errors refer to pure Poisson statistics at 68\% C.L. GOODS-N counts are contrasted in the shaded area (from Berta et al. (2010)). (Figures from Altieri et al 2010).
\label{a2218herschel}
}
\end{figure}

By looking at Abell 2218, Altieri et al. (2010) studied the population of intrinsically faint infrared galaxies that lie below the sensitivity and confusion limits using ultra-deep PACS 100 and 160 micron observations (Figure~\ref{a2218herschel}). They derived (unlensed) source counts down to a flux density of 1 mJy at 100 micron and 2 mJy at 160 micron. In particular, the slope of the counts below the turnover of the Euclidean-normalized differential curve could be constrained in both bands and was found to be consistent with most of the recent evolutionary models. By integrating the number counts over the flux range accessed by exploiting lensing by Abell 2218 they retrieved a cosmic infrared background surface brightness of $\sim8.0$ and $\sim9.9$ nW/m$^{2}$/sr, in the 110 and 160 micron bands respectively. By combining the Abell 2218 results with wider/shallower fields, the source fluxes correspond to $\sim 60$\% and $\sim 90$\% of the DIRBE cosmic infrared background at 100 and 160 micron. These first Herschel results from HLS and the GTO sample will certainly expand as the data are getting analyzed and we can envision numerous follow-up observations at optical and near-infrared wavelengths as well as with ALMA.

As part as the second call for observations another interesting lensing search has been implemented in the Herschel observing program to find a larger number of exceptionally bright lensed galaxies such as the one found by Swinbank et al. (2010). This project is aiming to conduct a SPIRE snapshot survey of $\sim300$ X-ray-selected massive galaxy clusters to discover the few extreme magnified objects that will then be very easy to follow-up at various wavelengths for an in-depth study. Although this survey is not yet finished,
a number of highly amplified SPIRE sources have been identified, and intensive multi-wavelength
follow-up are in progress.

\subsection{Cluster Lensed Supernovae}

In the last two decades, Supernovae (SNe) have been used for several astrophysical and cosmological applications. In particular, core collapse SNe trace the star formation history while the standard candle property of Type Ia SNe can be used for probing the expansion history of the Universe ({\it e.g.} Riess et al 1998, Perlmutter et al 1999, Amanullah et al. 2010). One of the focus on current SNe research is
to probe the distant Universe. However, one strong limitation is the light collecting power of existing telescopes.

A possible alternative to current investigation is to target these SNe in the field of view of massive clusters. Although, the idea is not new and was first discussed by Narasimha \& Chitre (1988) and then by
Kovner \& Paczynski (1988), it is only recently that SNe observations in cluster fields has become more popular (Kolatt \& Bartelmann 1998, Sullivan et al. 2000; Gal-Yam et al. 2002).
The most interesting locations are of course the strong lensing regions of clusters where the amplification is the largest, and were SNe could be multiply imaged offering the possibility to measure the time delay between the different images. 

At first SN searches were done at optical wavelengths where SNe typically emit most of their light. For example, Gal-Yam et al. (2002) using archival \hst\ imaging of 9 clusters, in which they discovered two or three likely cluster SNe and three other SNe, with one background to a cluster at redshift $z=0.985$. More recently, Sharon et al (2010) in a dedicated SNe \hst\ multi-epoch ACS $I$-band survey of 16 massive clusters (ranging from $z=0.5$ to $z=0.9$) have discovered 24 SNe, with 8 of them being background to these clusters (the highest SN redshift found is at $z=1.12$). However, none of those lensed SNe are in the regions of multiple-images.
At even higher cluster redshift $z\sim 1$, the Supernova Cosmology Project (PI: Perlmutter) has targeted 25 clusters through an \hst\ multi-epoch program
in which nine clusters and twenty other (foreground or background) SNe have been discovered (Dawson et al. 2009). However, the main focus of these cluster multi-epoch surveys was essentially geared toward the discovery and study of cluster type Ia SNe, and thus were not optimized to benefit from the cluster lens magnification.

On the contrary, Gunnarsson \& Goobar (2003) presented the feasibility of detecting high-z SNe along the line of sight of massive clusters, in particular focusing on the SNe detection in the near-infrared.
Using a dedicated VLT/ISAAC multi-epoch SN survey, Stanishev et al. (2009) and  Goobar et al. (2009)  reported the discovery of a highly amplified SN at $z\sim 0.6$ behind the well-studied  Abell 1689 cluster.
More recently, using the new VLT/Hawk-I infrared camera, Amanullah et al (2011) found one of the most distant SNe ever found at  $z = 1.703$  (measured through X-Shooter spectroscopy of the galaxy host) thanks to the large magnification ($\sim 4.3\pm0.3$) of the massive
cluster Abell 1689. This study demonstrated that further SNe follow-up may lead to important new discoveries.


\section{Cosmological constraints from cluster lensing}

In this section, we discuss 3 powerful cluster lensing based methods at various stages of development and
application that may provide competitive and important constraints on cosmological
parameters. These are: cosmography using several sets of multiple-images lensed by the 
same cluster; the abundance of arcs and the statistics of lensed image triplets. For cosmography 
and triplet statistics purely geometric constraints are obtained via the ratio of angular diameter distances, 
whereas the abundance of arcs provides potentially strong constraints on the growth of structures and primordial non-gaussianity.

\subsection{Cosmography with multiple-images}

\begin{figure}
\centerline{
\epsfig{file=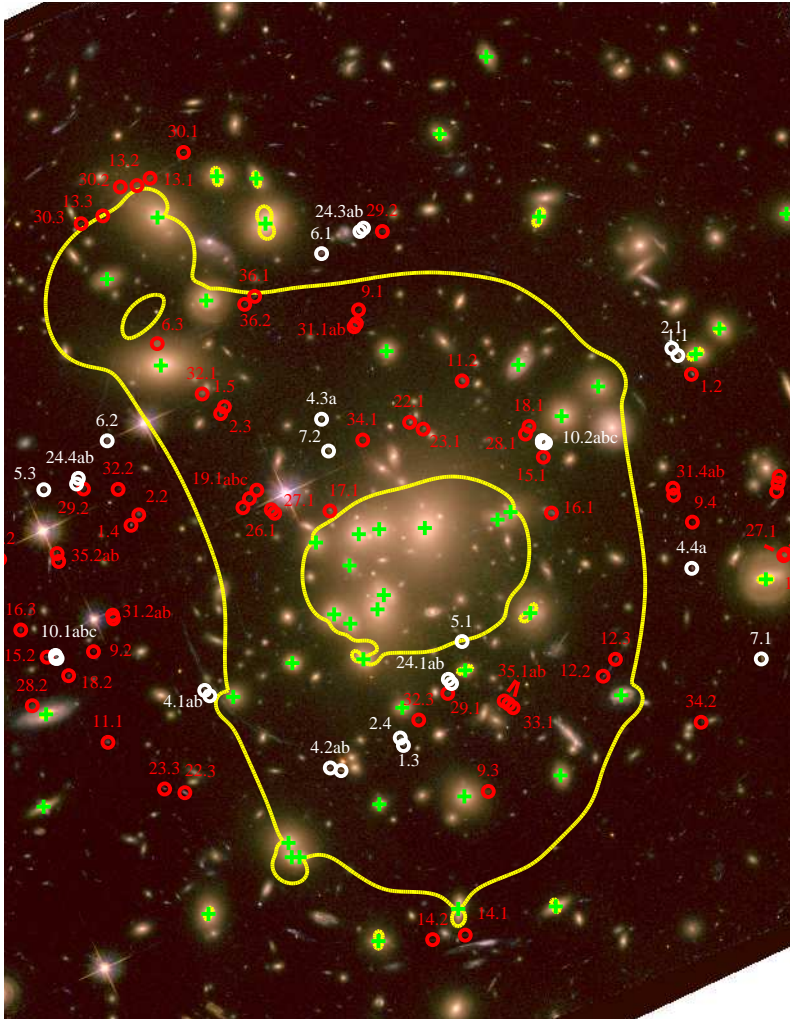,width=0.75\textwidth}
}
\caption{The critical lines for a source at $z =
3$ are over-plotted in yellow on the {\it HST ACS} image of
Abell~1689. The lensing mass model used is the one from which
we derived cosmological constraints. In addition to two large-scale clumps and 
the BCG, this model includes the contribution of 58 cluster galaxies. The positions
of cluster galaxies are marked with green crosses. Over-plotted in
white are the 28 multiple-images arising from 12 families 
used in their work; the red circles mark the positions of the
rejected images (Figure from Jullo et al. 2010).} 
\label{a1689jullo} 
\end{figure}

Measurements of the Hubble diagram for Type Ia Supernovae (SNIa)(Riess et al. 1998;
Perlmutter et al. 1999) combined with
constraints from the Wilkinson Microwave Anisotropy Probe ({\it
WMAP5}) (Spergel et al. 2003), cosmic shear 
observations (Bacon, Refregier \& Ellis 2000; Kaiser et al. 2000; 
van Waerbeke et al. 2000; Wittman et al. 2000; Semboloni et al. 2006), cluster 
baryon fraction (Allen et al. 2004), cluster abundances (Vikhlinin et al. 2009) and 
baryon acoustic oscillations (BAO) from galaxy surveys (Efstathiou et al. 2002; Seljak et al. 2005;
Eisenstein et al. 2005) suggest that $\sim72$\% of the total energy density of the Universe
is in the form of an unknown constituent with negative pressure - the
so-called dark energy, that powers the measured accelerating
expansion.  These observations probe the equation-of-state parameter
$w_{\rm X}$, defined as the ratio of pressure to energy density,
through its effect on the expansion history of the Universe and the growth of structures.

Constraining the geometry and matter content of the Universe using
multiple sets of arcs has been explored in cluster lenses using different techniques
(Paczyn\'ski \& Gorski 1981; Link \& Pierce 1998; Cooray 1999; Golse et al.
2002; Sereno 2002; Sereno \& Longo 2004; Soucail et al. 2004; Dalal et al. 2005; Meneghetti et 
al. 2005a,b; Maccio 2005; Gilmore \& Natarajan 2009; Jullo et al. 2010).

As shown in Section~1, the lensing deflection produced in the image of
a background source depends on the detailed mass distribution of the
cluster as well as on the ratio of angular diameter distances. The cosmological
dependence arises from the angular diameter distance ratios that encapsulate
the geometry of the Universe and are a function of both $\Omega_{\rm m}$ and
$\Omega_{\rm X}$. 

The most promising technique, is using multiple sets of arcs with measured redshifts. By taking 
the ratio of their respective Einstein radii and marginalizing over parameters of the mass 
distribution, one can in principle constrain the cosmological parameters
$\Omega_{\rm m}$ and $\Omega_{\rm X}$. In this method, the angular
diameter distance ratios for two images from different sources defines
the `family ratio' $\Xi$, from the cosmological dependence of which
constraints on $\Omega_{\rm m}$ and $w_{\rm X}$ are extracted:
\begin{equation}
    \Xi(z_{\rm L},z_{\rm s1},z_{\rm s2};\Omega_{\rm m},\Omega_{\rm X},w_{\rm X})=\frac{D(z_{\rm L},z_{\rm
s1})}{D(0,z_{\rm s1})}\frac{D(0,z_{\rm s2})}{D(z_{\rm L},z_{\rm
s2})},
\end{equation}
where $z_{\rm L}$ is the lens redshift, $z_{\rm s1}$ and $z_{\rm
s2}$ are the two source redshifts, and $D(z_{1},z_{2})$ is the
angular diameter distance.

\begin{figure}
\centerline{
\epsfig{file=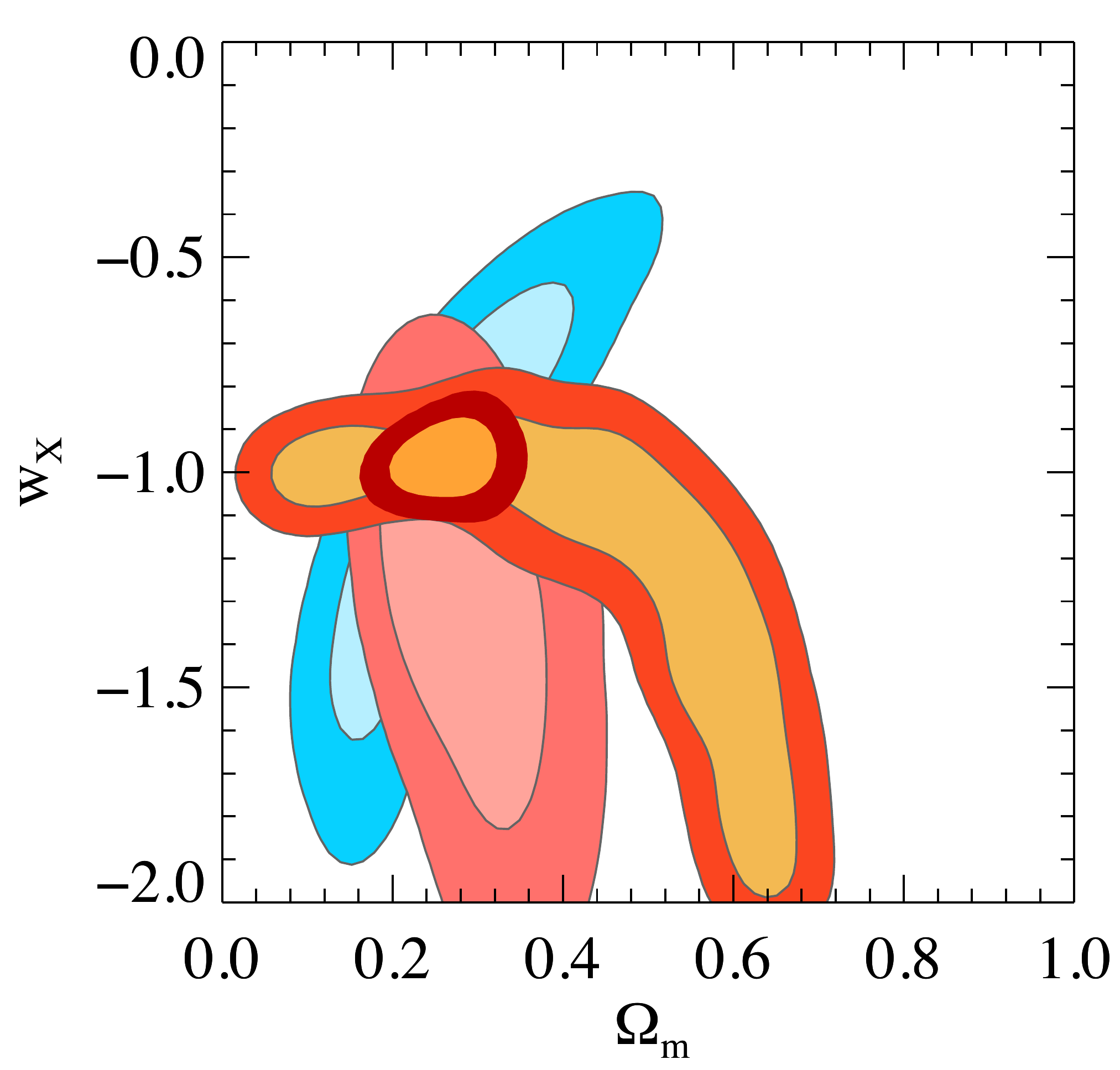,width=0.75\textwidth}
}
\caption{The current cosmological constraints in the ($\Omega_{\rm m}$, 
$\Omega_\lambda$) based on the best-fit model of Abell 1689 from Jullo et al. (2010):
the results from combining cosmological constraints from {\it WMAP5} $+$ evolution of X-ray clusters 
$+$ cluster strong lensing (cluster only methods); the 1 and 2$\sigma$ contours
are plotted, blue contours - constraints from {\it WMAP5}, pink
contours - X-ray clusters, orange contours - cluster strong
lensing.
\label{cosmoCSL}
}
\end{figure}

Link \& Pierce (1998) showed that the cosmological sensitivity of the
angular size-redshift relation could be exploited using sources at
distinct redshifts and developed a methodology to simultaneously
invert the lens and derive cosmological constraints.
Golse et al. (2002) using simulated cluster data, showed that the recovery of cosmological parameters 
was feasible with at least 3 sets of multiple-images for a single cluster. Soucail et al. (2004) 
then applied the technique to the lensing cluster
Abell 2218 using 4 systems of multiple-images at distinct redshifts,
and found ($\Omega_{\rm m}<0.37$, $w_{X}<-0.80$) assuming a flat Universe.

Jullo et al. (2010) have presented the results of the first application of this
method to the massive lensing cluster Abell 1689 at $z=0.184$ (see Figure \ref{a1689jullo}).  Based on images from the Advanced
Camera for Surveys ({\it ACS}) aboard the Hubble Space Telescope ({\it
HST}) this cluster has 114 multiple-images from 34 unique background
galaxies, 24 of which have secure spectroscopic redshifts (ranging
from $z \sim 1$ to $z \sim 5$) obtained with the Very Large Telescope
(VLT) and Keck Telescope spectrographs (Broadhurst et al. 2005; Limousin
et al. 2007). Their parametric model has a total
of 21 free parameters consisting of two large-scale potentials, a
galaxy-scale potential for the central brightest cluster galaxy (BCG),
and includes the modeling of $58$ of the brightest cluster
galaxies. The contribution of substructure in the lens plane and along the 
line of sight is explicitly included (see D'Aloisio \& Natarajan 2011 for a detailed
discussion of the systematics).  Combining the lensing derived cosmological constraints 
with those from X-ray clusters and the
{\it Wilkinson Microwave Anisotropy Probe} 5-year data gives
$\Omega_{\rm m} = 0.25 \pm 0.05$ and $w_{\rm X} = -0.97 \pm 0.07$
which are consistent with results from other methods (see Figure \ref{cosmoCSL}). Inclusion of this
work with all other techniques available brings down the current
$2\sigma$ contours on the dark energy equation of state parameter
$w_{\rm X}$ by about 30\%.

\begin{figure}
\centerline{
\epsfig{file=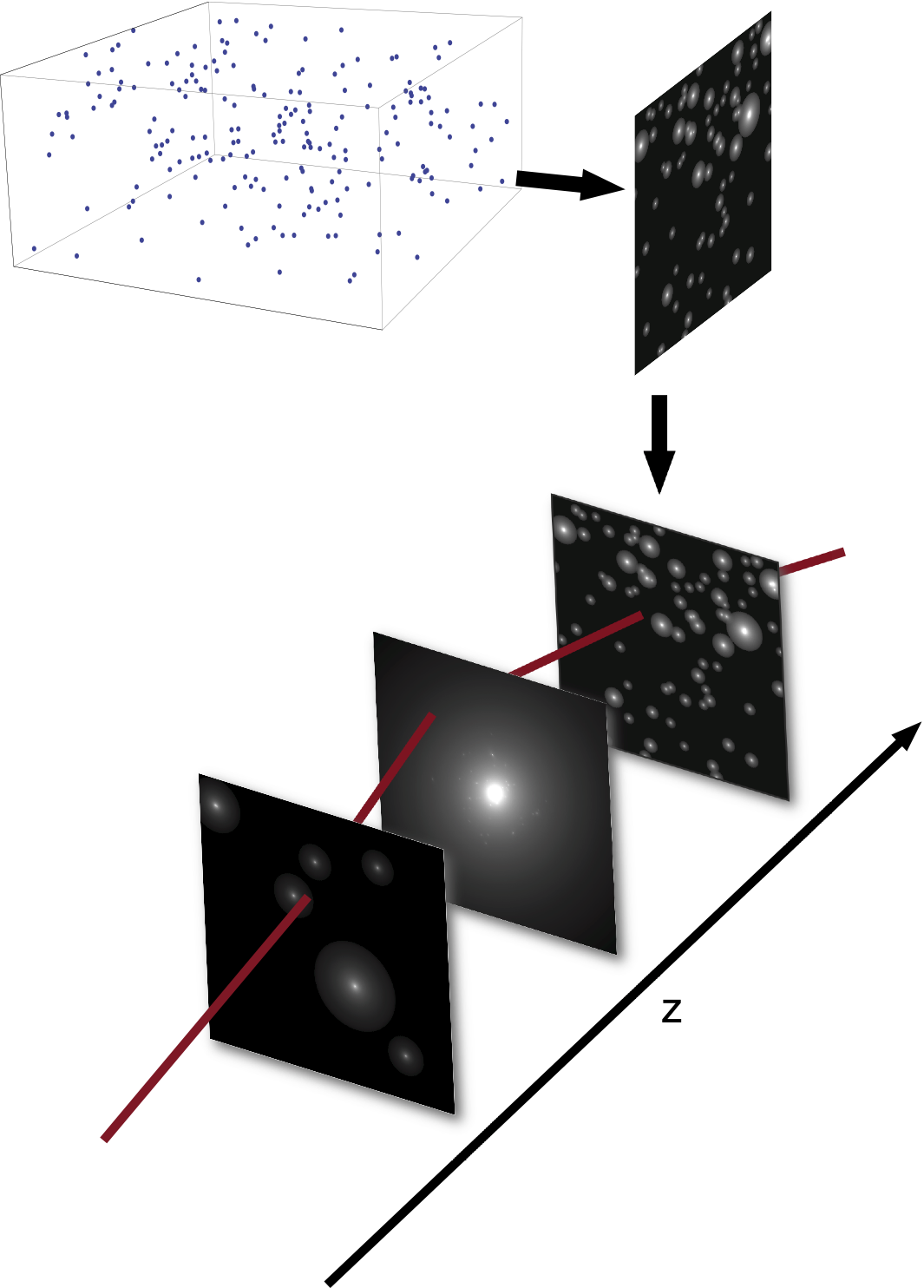,width=0.75\textwidth}
}
\caption{Schematic diagram illustrating the creation of lens planes to
quantify the lensing effects of halos along the line of sight. A rectangular slice of the
Millennium Simulation box is taken and the locations of halos are
projected along the long axis and analytic NFW potentials are placed
on those positions.  The NFW parameters are obtained through scaling
relations with mass and redshift. The lens plane is inserted at the
appropriate redshift and a multi-plane lensing algorithm is used to
trace rays. Many lens planes between $z=0$ and $z=5$ are used to
estimate the systematic errors at the positions of multiple-images
from the structure along the line of sight.}
\label{lossubstructure}
\end{figure}

As with all techniques, an accurate inventory of the key systematics and their contribution to
the error budget is also the challenge for this technique. The two significant current limitations 
arise from: (i) accounting appropriately for the lensing effect of the uncorrelated line 
of sight substructure (see schematic in Figure \ref{lossubstructure}) and (ii) the simplifying scaling relations assumed to relate galaxy total mass 
to galaxy light. However, the results from Abell 1689 are extremely encouraging and the future prospects
for this method look promising due to the power from combining several clusters at various
redshifts.

\subsection{Arc statistics and Primordial Non-gaussianity}

The production of giant arcs by lensing clusters is ubiquitously observed. The abundance of massive clusters available to do so is sensitive both to the expansion history and initial conditions of the Universe.  Given the scaling of the lensing efficiency with redshift, it is known that the frequency of giant-arc formation depends on the abundance and characteristics of galaxy-clusters roughly half-way to the sources. Cluster physics, cosmological effects and the properties of the high-redshift source population  all play a role in determining the abundance of giant arcs, however, isolating these effects is difficult. It was originally claimed by  Bartemann et al. (1998) that the $\Lambda$CDM model predicted approximately an order of magnitude fewer arcs than seen in observations.  Subsequent studies  ({\it e.g.} Zaritsky \& Gonzalez 2003; Gladders et al. 2003) substantiated this claim of a `giant-arc problem'. This mis-match between observations and the concordance cosmological model predictions suggest that either the Bartelmann et al. (1998) analysis was lacking a crucial component of properties exhibited by real cluster lenses and the source population
(Williams et al 1999) or that the concordance cosmology is in fact inconsistent with the observed abundance of giant arcs.  A significant amount of effort has been expended toward understanding the most important characteristics of arc-producing clusters, and how they may not be typical of the general cluster 
population ({\it e.g.} Hennawi et al. 2007; Meneghetti et al. 2010; Fedeli et al. 2010).  
Other studies focused on effects that were not captured in early simulations. 
The mass contribution of central galaxies appears to have a significant effect, though not enough to entirely resolve the Bartelmann et  al. (1998) disagreement alone (Meneghetti et al. 2003; Dalal et al. 2004).
The probability of giant-arc formation increases with source redshift, therefore the overall abundance is sensitive to uncertainties in the high-redshift tail 
of the source-redshift distribution (Wambsganss et al. 2004; Dalal et al. 2004; Li et al. 2005). However, none of these effects can account for the observed discrepancy. 
On the other hand, taking into account a realistic source population and observational effects, Horesh et al. (2005) claimed that the clusters predicted by $\Lambda$CDM have the same arc production efficiency as the observed clusters. The effects of baryonic physics, such as cooling and star formation, on central mass distributions have also been investigated (Meneghetti et al. 2010). 

The amplitude of the linear matter power spectrum plays a critical role in determining how severe and if there is a giant-arc problem or not.  Observations 
seem to be converging on $\sigma_8\approx0.8$ (Fu et al. 2008; Vikhlinin et al. 2009; Komatsu et al. 2011), while most numerical studies on the giant-arc abundance to 
date have assumed $\sigma_8=0.9$.  It is likely that adjusting $\sigma_8$ from $0.9$ to $0.8$ will lower the predicted giant-arc abundance significantly, increasing tension with observations (Li et al. 2006; Fedeli et al. 2008). The cosmological model may play a role here.  In arguing that the giant-arc problem may be 
unavoidable if $\sigma_8\approx 0.8$, Fedeli et al. (2008) mention in passing that early dark energy or non-Gaussian initial conditions may provide ``a way out."  
The effects of dark energy on giant-arc statistics have been investigated in Bartelmann et al. (2003), Maccio (2005), Meneghetti et al. (2005a,b,c), 
and Fedeli \& Bartelmann (2007).  
On the other hand, the possible effects of non-Gaussian initial conditions have only been explored recently by D'Aloisio \& Natarajan (2011). They argue that primordial 
non-Gaussianity (PNG) can affect the probability of giant-arc formation in \emph{at least} two ways.\footnote{The PNG model considered here is the simplest one that gives rise to a non-zero 3-point correlation function.}  First, PNG can lead to an enhanced or diminished abundance of 
galaxy clusters, depending on the particular model ({\it e.g.} Matarrese et al. 2000; LoVerde et al. 2008; Dalal et al. 2008), which would lead to a change in the number of 
supercritical lenses that are available in the appropriate redshift range.  Secondly, PNG is expected to influence the central densities of halos (Avila-Reese et al. 2003; 
Oguri \& Blandford 2009; Smith et al. 2010). Since lensing cross sections are sensitive to central densities, we expect corresponding changes in them as well. If a 
cluster-lens cannot produce arcs with length-to-width ratios above some threshold, then its cross section for giant-arc production is zero.  Roughly speaking, this 
corresponds to a minimum mass required to produce giant arcs.  Owing to the effects on central densities, we expect PNG to alter this minimum mass threshold as well. 

 \begin{figure}
\begin{center}
\epsfig{file=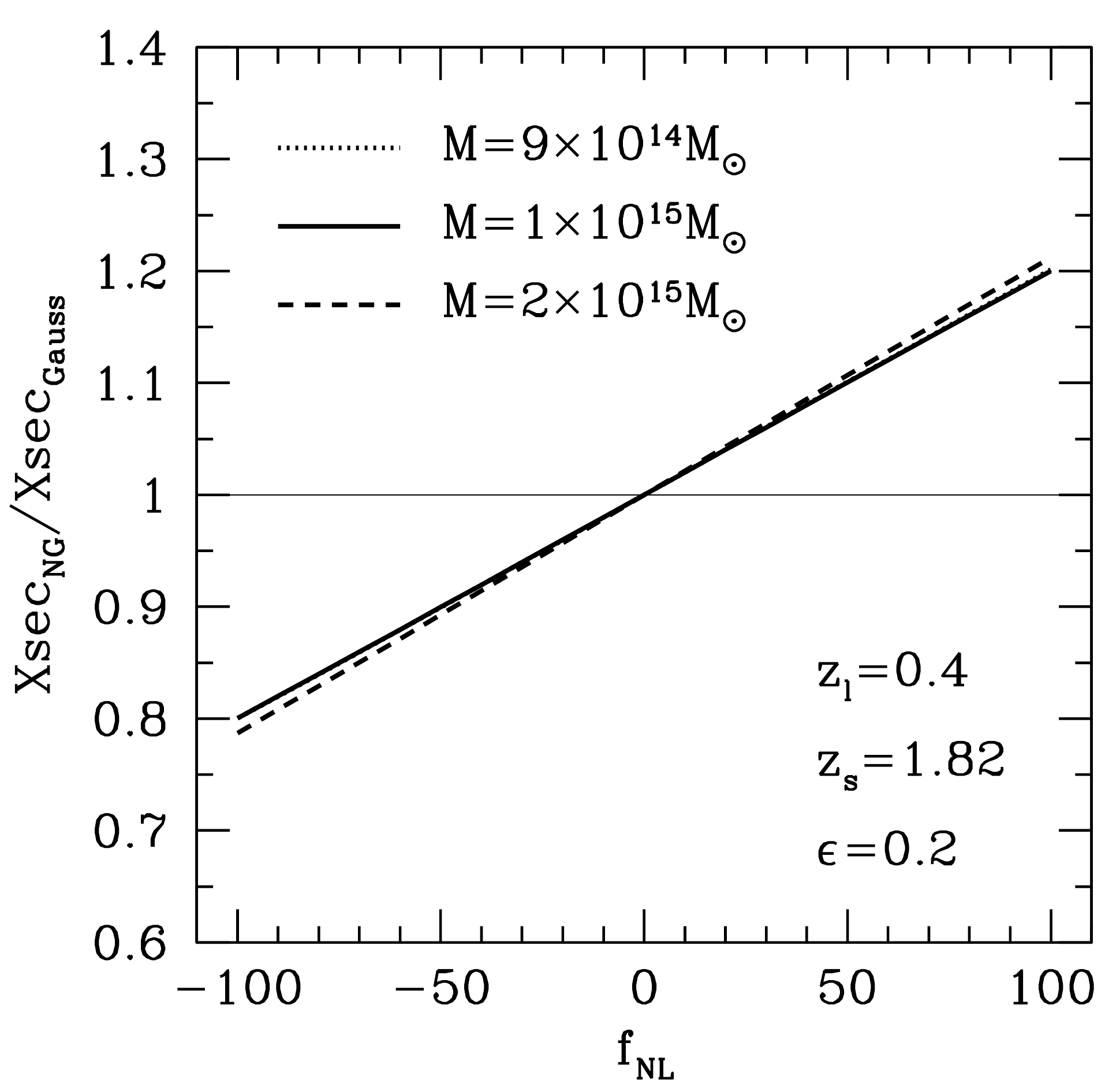,width=0.75\textwidth}
\end{center}
\caption{The ratio of giant-arc cross sections in the case of non-Gaussian and Gaussian initial conditions.    Halos have enhanced central densities in models with $f_{\rm NL} > 0$.  Their giant-arc cross sections are therefore increased relative to the Gaussian case and vice versa (Figure from D'Aloisio \& Natarajan 2011).}
\label{fnl}
 \end{figure}

D'Aloisio \& Natarajan (2011) quantify the impact of non-Gaussian initial conditions with the local bispectrum shape on the predicted frequency of giant arcs. Non-Gaussianity is generally expressed in terms of $\fNL$ that characterizes the amplitude of non-Gaussianity in the primordial curvature perturbation (Komatsu \& Spergel 2001).
  Using a path-integral formulation of the excursion set formalism, extending a semi-analytic model for calculating halo concentrations to the case of PNG, they show that massive halos tend to collapse earlier in models with positive $\fNL$, relative to the Gaussian case, leading to enhanced concentration parameters. The converse is true for $\fNL < 0$.  In addition to these effects, which change the lensing cross sections, non-Gaussianity also modifies the abundance of supercritical clusters available for lensing.  These combined effects work together to either enhance ($\fNL > 0$) or suppress ($\fNL < 0 $) the probability of giant-arc formation (see Figure \ref{fnl}).  Using the best value and $95\%$ confidence levels currently available from the Wilkinson Microwave Anisotropy Probe, they report that the giant-arc optical depth for sources at $z_s \sim 2$ is enhanced by $\sim20\%$ and $\sim45\%$ for $\fNL = 32$ and $74$ respectively. Conversely they report a suppression of $\sim5\%$ for $\fNL = -10$.  These differences translate to similar relative changes in the predicted all-sky number of giant arcs. Ideally the goal is to use giant-arc statistics to constrain small scale PNG. The prospects are extremely promising given upcoming all sky surveys planned by future deep wide-field imaging surveys such as foreseen by the Dark Energy Survey (DES), the Large Synoptic Survey Telescope (LSST) and future wide-field space mission ({\it e.g.} EUCLID).

\subsection{Triplet statistics}

Triplet statistics offer an interesting geometrical method that uses the weak gravitational lensing effects of clusters to constrain the cosmological parameters $\Omega_{\rm m}$ and $\Omega_{\Lambda}$ (Gautret, Fort \& Mellier 2000). For each background galaxy, a foreground lensing cluster induces a magnification that depends on the local convergence $\kappa$  and shear terms $\gamma_1$ and $\gamma_2$ and on the cosmological parameters through the angular diameter distance ratio $D_{LS}/D_{OS}$. To disentangle the effects of these three quantities, the ellipticities of each triplet of galaxies located at about the same apparent position in the lens plane (although at three distinct redshifts) needs to be compared. The simultaneous knowledge of ellipticities and redshifts of each triplet enable the  building of a purely geometrical estimator $G(\Omega_{\rm m}, \Omega_{\Lambda})$ that is independent of the lens potential. This estimator $G$ has the simple form of the determinant of a 3x3 matrix built with the triplet values of $D_{LS}/D_{OS}$ and observed ellipticities. 

When $G$ is averaged over many triplets of galaxies, it provides a global function which converges to zero for the true values of the cosmological parameters. However, in order to apply this method the various sources of statistical noise need to be
quantified. The linear form of $G $ with respect to the measured ellipticity of each galaxy implies that the different sources of noise contributing to $G$ decrease as $1/\sqrt{N}$, where N is the total number of observed lensed galaxies. From simulations that incorporate realistic geometries and convergences for lensing clusters and a redshift distribution for galaxies, the results are promising for a sample of 100 clusters. These 100 clusters essentially need to be imaged in multiple bands to obtain accurate photometric redshifts for the triplets. With next generation cosmological surveys the observational data needed for this sample size would not be impossible to obtain. 


\section{Comparison of observed lensing cluster properties with theoretical predictions}

With the growing success of gravitational lensing analysis of clusters, it has become
possible to compare and test theoretical predictions against observations. With the rapid
progress in high-resolution cosmological simulations of dark matter we now have a unique 
opportunity to directly compare properties of cluster dark matter halos derived from 
lensing studies. Many important physical questions with regard to the internal structures of 
halos, their dynamical evolution and the granularity of dark matter can now be tackled: the assembly 
process (role of merging sub-clusters); lensing cross sections; efficiency of lensing and super-lenses; 
selection effects; mass profiles; density profiles; ellipticity; alignments;  abundances, and the 
mass concentration. We briefly outline below the results of recent studies on this topic.

\subsection{Internal structure of cluster halos}

In cosmological simulations of structure formation it is found that the density profiles of dark matter halos are well fit over many decades in 
mass from cluster mass scales down to dwarf galaxy-scales by the Navarro-Frenk-White profile (see Appendix A.3 for details). By combining strong and weak lensing constraints, as discussed above it has become possible to probe the mass profile of the clusters on scales of 0.1Ð5 Mpc, thus providing a valuable test of the universal form proposed by NFW on large scales ({\it e.g.} Okabe 
et al. 2010, Umetsu et al. 2011). As for the inner density profile slopes there appears to be similarly a large degree of variation, some like Cl0024+1654 (Kneib et al. 2003; Tu et al. 2008; Limousin et al. 2008) adequately  fit the NFW form, others like  RXJ1347-11  are found to have slopes shallower than the NFW prediction (Newman et al. 2011, Umetsu et al. 2011), while others like MS2137-23 are found to have steeper slopes (Sand et al. 2004). One caveat with the NFW prediction is that the functional form is derived from dark matter only simulations, whereas in reality it is clear that baryons in the inner regions close to the cD/BCG play a significant role both in terms of the mass budget and modifications to density profile slopes in the very center.

\begin{figure}
\centerline{
\includegraphics[angle=0,width=10cm]{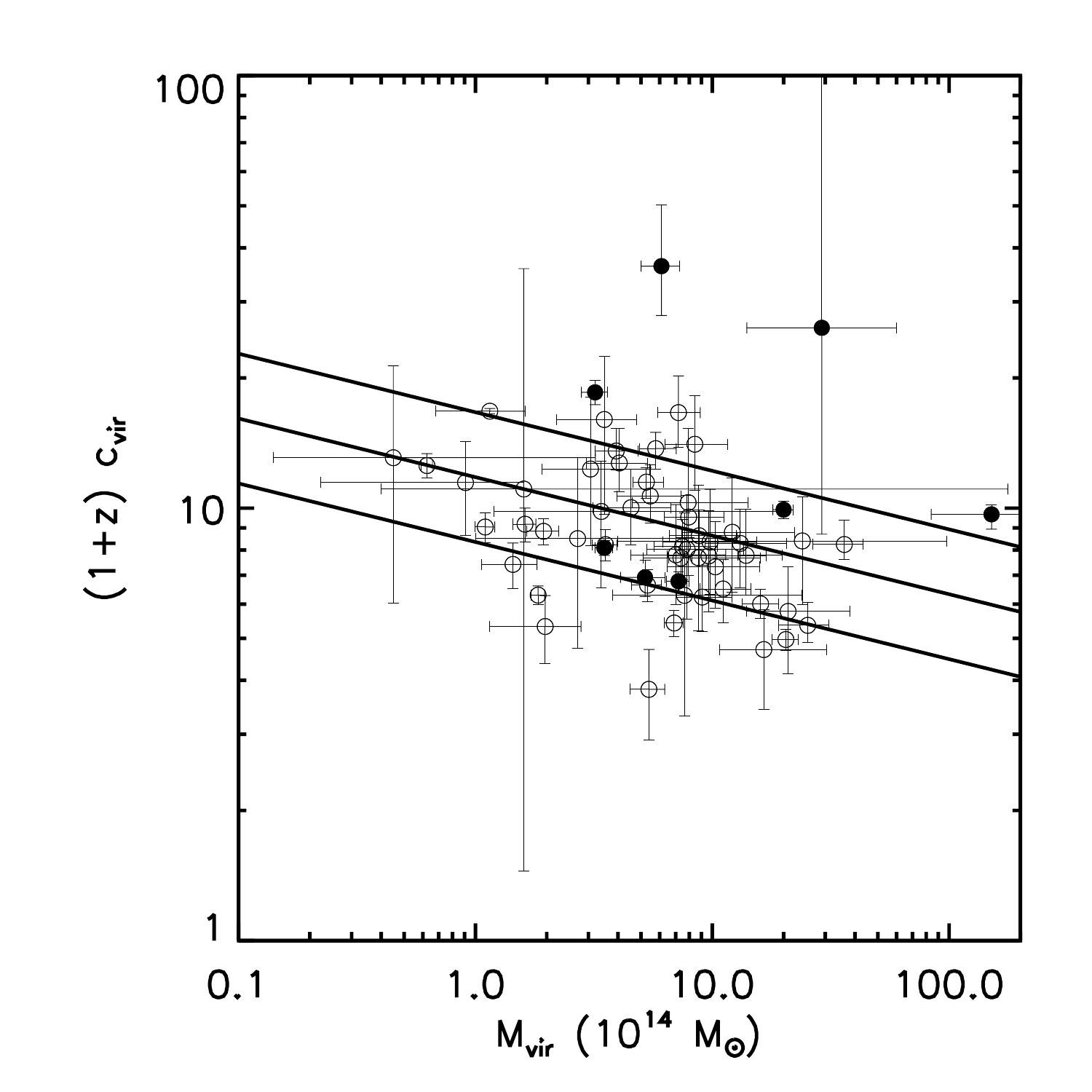}
}
\caption{Observed cluster concentrations and virial masses derived
  from lensing (filled circles) and X-ray (open circles)
  measurements.  For reference, the solid lines depict the best-fit
  power law to our complete sample and its 1-$\sigma$ scatter. The
  lensing concentrations appear systematically higher than the X-ray
  concentrations, and a Kolmogorov-Smirnov test confirms that the
  lensing results likely belong to a different parent distribution. This figure is
  from Comerford \& Natarajan 2007.
  \label{cmrelation}
  }
\end{figure}

Lensing clusters are preferentially more significantly concentrated than all clusters (see Figure \ref{cmrelation} and Comerford \& Natarajan 2007; but also: Broadhurst et al. 2008; Oguri et 
al. 2009; and Meneghetti et al. 2011) and they typically tend to be outliers on the concentration-mass relationship predicted for clusters in the $\Lambda$CDM model. The origin of this enhanced concentration parameter is likely due to: i) high incidence of projected line of sight structures for massive lensing clusters; ii) elongated shapes that enhance lensing efficiency, factors that might observationally bias lensing 
selection; iii) baryons that could play an important role in the inner regions.

\subsection{Mass function of substructure in cluster halos}

Combining observed strong and weak lensing and exploiting galaxy-galaxy lensing inside clusters, it has been possible to map 
the granularity of the dark matter distribution (Natarajan \& Kneib 1997; Natarajan \& Springel 2005; Natarajan, DeLucia \& Springel 2007) in 
clusters and compare them to predictions from the Millennium cosmological simulation (Springel et al. 2005). This is done  by attributing local anisotropies in the observed shear field to the presence of dark matter sub-halos (Natarajan et al. 2009). The mass 
function thus derived  for several clusters agrees well with  that predicted entirely independently from high-resolution cosmological simulations of 
structure formation in the standard $\Lambda$CDM paradigm over the mass range $10^{11} - 10^{13}\,M_{\odot}$. The comparison was made with clusters that 
form in the Millennium Simulation (Springel et al. 2005). This excellent agreement of the mass function derived from these 2 independent methods 
demonstrates that there is no substructure problem (which was claimed earlier) on cluster scales in $\Lambda$CDM. This is a significant result as a substructure 
crisis has been claimed on galaxy-scales. Since $\Lambda$CDM is a self-similar theory, if the substructure problem had been endemic to the model, it would have 
been replicated on cluster scales. This suggests that the substructure discrepancy on galaxy-scales arises from the galaxy formation process or from some hitherto undiscovered coupling between baryons and dark matter particles. Therefore, 
lensing clusters have provided unanticipated insights into the dark matter model. Moving on from the Millennium Simulation, state of the art at the present
time is the Mare Nostrum simulation which is promising in terms of mass resolution and larger volume probed and offers a new test-bed for comparison
with lensing data from cluster surveys like the CLASH $Hubble$ survey (Meneghetti et al. 2011).

\begin{figure}
\resizebox{16cm}{!}
{\includegraphics[angle=90]{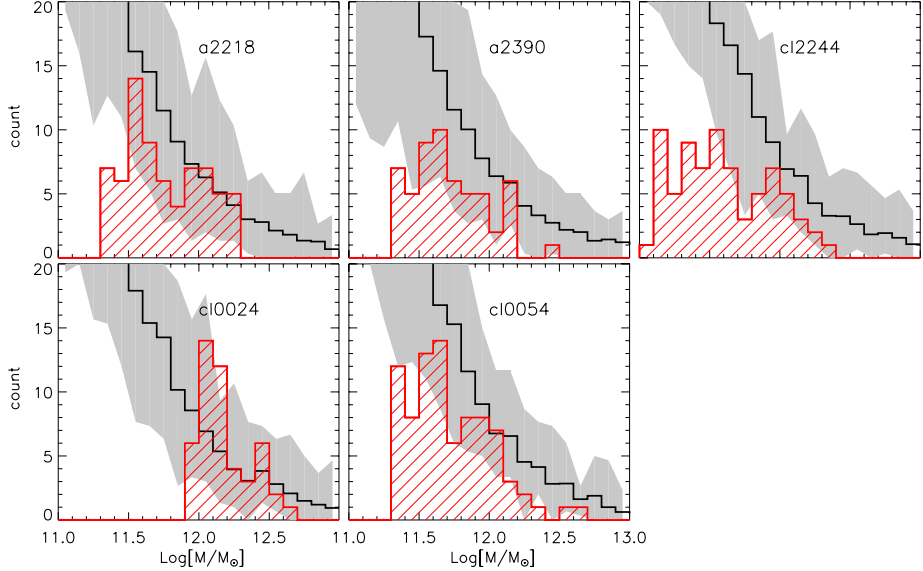}}
\caption{Comparison between substructure mass function retrieved from
  the galaxy-galaxy lensing analysis (red shaded histograms) and
  results from haloes selected from the Millennium Simulation.  The
  black solid line in each panel represents the average sub-halo mass
  function of haloes selected at the redshift of the observed lensing
  cluster (see text for details).  The grey shaded region represents,
  for each value of the sub-halo mass, the min-max number of
  substructures found in the simulated haloes (Figure from Natarajan, DeLucia \& Springel 2007).
  \label{nds2007}
  }
\end{figure}

\subsection{Dynamical evolution of cluster halos}

\begin{figure}
\epsfig{file=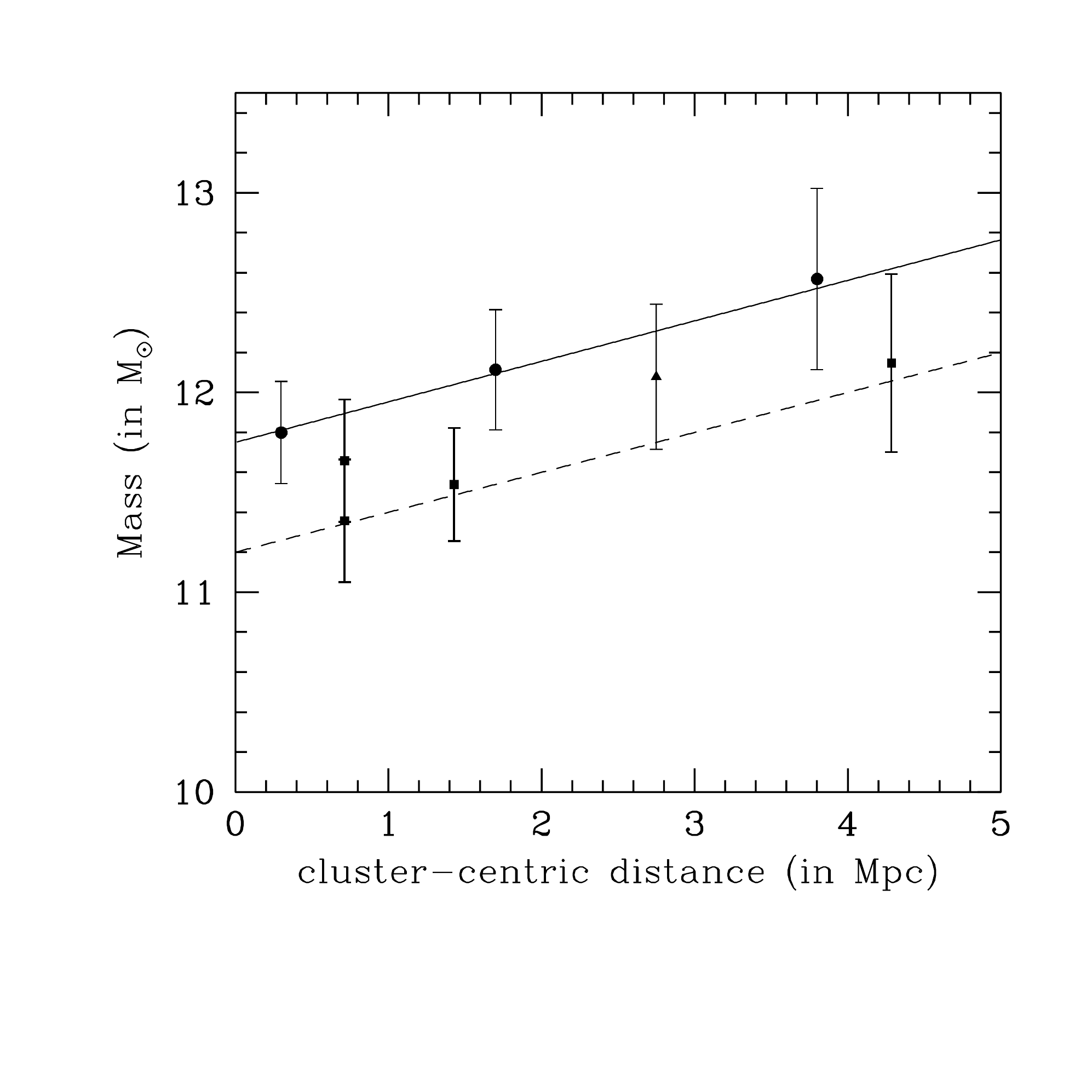,width=\textwidth}
\caption{Variation of the mass of a dark matter sub-halo that hosts
an early-type L$^*$ galaxy as a function of cluster centric
radius. The results from the likelihood analysis are used to derive
the sub-halo mass for the galaxy-galaxy lensing results and the
counterparts are derived from the Millennium simulation with an
embedded semi-analytic galaxy formation model. This enables selection
of dark matter halos that host a single L$^*$ galaxy akin to our
assumption in the lensing analysis. The solid circles are the data
points from the galaxy-galaxy lensing analysis and the solid squares
are from the Millennium simulation. The upper solid square in the core
region marks the value of the sub-halo mass with correction by a factor
of 2 as found in Natarajan, De Lucia \& Springel (2007). The solid
triangle is the galaxy-galaxy lensing data point for the sub-halo
associated with a late-type $L^*$ galaxy. The radial trend derived
from lensing is in very good agreement with simulations and demonstrate that
tidal stripping is operational with higher efficiency in the central regions as
expected.
\label{tidalstripping}
}
\end{figure}

Exploiting strong and weak gravitational lensing signals inferred from panoramic Hubble Space Telescope imaging data, high-resolution 
reconstructions of the mass distributions  are now available for clusters ranging from $z=0.2 - 0.5$. Applying galaxy-galaxy lensing techniques inside clusters
the fate of dark matter sub-halos can now be tracked as a function of projected cluster-centric radius out to 1-5 Mpc, well beyond the virial radius 
in some cases.  There is now clear detection of the statistical lensing signal of dark matter sub-halos associated with both early-type and late-type galaxies in clusters. In fact, it  appears now that late-type galaxies in clusters (which dominate the numbers in the outskirts but are rare in the inner regions of the cluster) also 
possess individual dark matter halos (Treu et al. 2002; Limousin et al. 2005, 2007;  Moran et al. 2006; Natarajan et al. 2009). In the case of 
the cluster Cl0024+1656 that has been studied to beyond the virial radius, the mass of a fiducial dark matter halo that hosts an early-type 
$L^*$ galaxy varies from $M = 6.3 \pm 2.7 \times 10^{11}$ M$_{\odot}$ 
 within $r < 0.6\, Mpc$, to $1.3 \pm 0.8 \times 10^{12}$ M$_{\odot}$ within $r < 2.9\,$Mpc, and increases further to 
 $M = 3.7 \pm 1.4 \times 10^{12}$ M$_{\odot}$ in the outskirts. 
 The mass of a typical dark matter sub-halo that hosts an $L^*$ galaxy increases with projected cluster-centric radius in line with 
 expectations from the tidal stripping hypothesis. Early-type galaxies appear to be hosted on average in more massive dark matter sub-halos compared to late-type galaxies. Early-type galaxies also trace the overall mass distribution of the cluster whereas late-type galaxies are biased tracers. The findings in this 
 cluster and others are interpreted as evidence for the active re-distribution of mass 
 via tidal stripping in galaxy clusters. Upon comparison of the masses of dark matter sub-halos as a function of projected cluster-centric with the equivalent mass function derived from clusters in the Millennium Run very good agreement is found (see Figure \ref{nds2007} and Natarajan, De Lucia \& Springel 2007). However, simulated sub-halos appear to be more efficiently stripped than lensing observations suggest (see Figure \ref{tidalstripping}). This is likely an artifact of comparison with a dark matter only simulation. Future simulations that simultaneously follow the detailed evolution of the baryonic component during cluster assembly will be needed for a more detailed comparison. Lensing has proved to be a powerful probe of how clusters assemble and grow, and it appears that our findings ratify the $\Lambda$CDM paradigm, hierarchical growth of structure and the key role played by tidal stripping during cluster assembly. 

\subsection{Constraints on the nature of dark matter}

While it is clear that clusters are vast repositories of dark matter, the nature of dark matter remains
elusive. A plethora of astronomical observations from the early Universe to the present time are consistent
with the dark matter being a cold, collision-less fluid that does not couple to baryons. However, there is potential
for dark matter self-interactions and lensing observations offer a unique window to probe this further (e.g. Miralda-EscudŽ 2002).
Limits on the dark matter interaction cross section can be placed from lensing observation of clusters, however,
these are currently not particularly constraining or illuminating. Two distinct arguments have been used to
obtain limits that strongly support the collision-less nature of dark matter. One involves the distribution of the sizes of tidally truncated sub-halos in clusters (Natarajan et  al. 2002); and the second involves estimates  from the separation between the dark matter and X-ray gas in the extreme merging system, the Bullet Cluster (Clowe et al. 2006; Brada{\v c} et al. 2006) wherein they find $\sigma/m < 4 {\rm gm}^{-1}{\rm cm}^2$ assuming that the two colliding sub-clusters experienced a head-on collision in the  plane of the sky. Similar results were also found from the so-called ``Baby Bullet'' cluster (Brada{\v c} et al. 2008). Exploring these merging clusters is certainly an avenue where lensing observations may provide constraints and insights on the nature of dark matter.

\section{Future prospects}

Since the discovery of giant arcs in the late 80's gravitational lensing by clusters of galaxies has 
now become a powerful {\it cosmological} tool. 

We list below some possible new avenues for the next exciting discoveries in the coming years using gravitational lensing in clusters, assuming improvements in the {\it data quality and  data volume}:
\begin{itemize}
\item Dedicated lensing surveys of well defined massive cluster samples
can probe in a systematic way the cluster mass distribution at high-resolution 
from galaxy-scales out to the virial radius using both weak and strong lensing. 
In particular, we can hope to study the build-up of the mass in clusters as a function of time 
and disentangle the time scales on which the segregation between the different mass components
occurs during the assembly process. One could also envision being able to systematically trace the filamentary structures linking massive clusters that are simply the nodes of the cosmic web seen in numerical simulations of the formation and evolution of structure in the Universe.
\item From wide field imaging surveys: {\it mass selected} clusters can be identified, but
likely the more interesting prospect is the ability to investigate the dark matter mass versus stellar-mass relation
and its evolution with time, thus providing useful cosmological constraints on the
growth of structure in the Universe, as well as on the underlying geometric cosmological parameters.
\item  With larger cluster samples, better constraints will be 
obtained on density profile slopes in the inner and outer regions of clusters thus permitting
to robustly test theoretical predictions of the $\Lambda$CDM model.
\item Using numerous multiple-images with measured redshifts in a number of massive clusters, 
we should be able to provide geometrical 
constraints of Dark Energy in a complementary way to other cosmological probes.
\item Measuring the time-delays of temporally variable phenomena such as Supernovae or AGN
when observed behind well-known massive lensing clusters, will lead
to measurement of the Hubble parameter $H(z)$, in a similar way as multiple quasars behind galaxies, 
but with much improved accuracy. However, in order to have a time-delay of a limited number of years, these transients event must be located close to the critical lines. While, the likelihood of detecting 
multiple-images of such transient phenomena is  extremely low, it is not insignificant 
provided there is a steady increase in the volume of the Universe probed with time.
\item Finally, massive cluster lenses will always be the {\it unique
  places} to probe the high-redshift Universe, as they offer enhanced sensitivities at
{\it all wavelengths} and enable mapping the detailed morphology and physical properties of 
the most distant galaxies in the Universe. 
\end{itemize}

In the near future lensing observations will likely be geared towards optical and near-infrared imaging exploiting the next generation of ground based experiments (DES, LSST, TMT, E-ELT), and spaced based observatories (JWST, EUCLID/WFIRST).

However, in the long run, it is not too unreasonable to think that cluster lensing observations may be well conducted in the radio domain. In such an event we foresee up to an order of magnitude improvement in lensing measurement that will result from combining information on galaxy shapes with velocity field data (Blain 2002; Morales 2006). Such preliminary developments will certainly come first with ALMA, but only centimeter radio interferometers such as SKA (Square Kilometer Array) will allow the exploration of these techniques on cosmological scales.


\section*{Acknowledgments}   

The phenomenon of gravitational lensing is one of the most beautiful predictions of Einstein's 
Theory of General Relativity. We feel very privileged to have had the opportunity to work and
contribute to this area. In particular, during and since the start of our 
scientific collaboration, our goal has been to inter-weave the observational and theoretical aspects 
of lensing by clusters to enable the exploitation of the potency of this technique. Much of the work 
reviewed here is heavily based and biased toward recent results, those obtained since JPK's PhD
thesis on cluster lenses in 1993; and PN's work on various theoretical aspects of cluster lenses 
during and since her PhD in 1999. This review consolidates the rapid progress made thus far and 
provides a  snap shot  of this exciting field as it stands today.  We would like to thank our many colleagues 
for the interesting work and fruitful discussions at various conferences, workshops and meetings over the 
past decade that have provided us a better understanding of our Universe and have transformed cluster 
lenses into a  powerful cosmological tool. JPK acknowledges support from CNRS.
PN acknowledges the John Simon Guggenheim Foundation for a
Guggenheim Fellowship and the Rockefeller Foundation for a residency at the Bellagio Rockefeller Center 
during the tenure of which this review was completed. She would also like to thank the Institute for Theory and
Computation at Harvard University for hosting her during her fellowship year.


\section*{Appendix A: Parametric mass distributions used to model clusters}

Parametric profiles have been extremely successful in modeling cluster mass 
distributions derived from observed lensing data. A key advantage of parametric
models is their flexibility, as they can be used to probe the granularity of the
mass distribution on a range of spatial scales.  For the case of clusters, this 
enables combining strong and weak lensing data that derive from different
regions of clusters in an optimal fashion.  Below we outline the lensing properties 
of three most commonly used mass distributions: the circular Singular Isothermal 
Sphere [SIS];  the truncated isothermal mass distribution with a core that can be
easily extended to the elliptical case [PIEMD] and the Navarro-Frenk-White [NFW] 
profile. While most mass distributions can be generalized to the elliptical case, there 
are not always simple and convenient analytic expressions available for lensing quantities as readily as for
the PIEMD model. The availability of analytic expressions for the surface mass 
density, shear and magnification have made the PIEMD a popular choice for 
modeling lensing clusters.

\subsubsection*{A.1 The Singular Isothermal Sphere}

The primary motivation for the circular singular isothermal sphere (SIS)
 profile derives from the good fit that it provides to the observed approximately 
 flat rotation curves of disk galaxies. Flat rotation curves can be 
reproduced with a model density profile that scales as $\rho \propto
r^{-2}$.  Such a profile with a constant velocity dispersion as a function 
of radius appears to provide a good fit to cluster scale halo lenses as well 
(see e.g.  Binney \& Tremaine 1987 for more  details).  The projected surface 
mass density of the SIS is given by:
\begin{equation}\label{sigsis}
\Sigma(R)=\frac{\sigma^2_v}{2GR},
\end{equation}
where $R$ is the distance from the center of the lens in the
projected lens plane and where $\sigma_v$ is the one-dimensional
velocity dispersion of `particles' that trace the gravitational potential
of the mass distribution. The dimensionless surface
mass density or convergence is defined in the usual way in units of
the critical surface density. For the case of the SIS we have:
\begin{equation}\label{ksis}
\kappa(\theta)=\frac{\theta_E}{2\theta};\,\,\,\,\gamma(\theta)=\frac{\theta_E}{2\theta},
\end{equation}
where $\theta=R/D_{OL}$ is the angular distance from lens center in the
sky plane and where $\theta_E$ is the Einstein deflection angle,
defined as
\begin{equation}
\theta_E=4\pi\left(\frac{\sigma_v}{c}\right)^2\frac{D_{LS}}{D_{OS}}.
\end{equation}

Lensing properties of SIS lens model in a nutshell:
\begin{itemize}
\item  the magnification and the shear are of the same magnitude; $\kappa=\gamma$ and evaluated 
at  the Einstein radius $\kappa=\gamma={1\over 2}$;
\item the tangential critical line is the Einstein ring,
and the radial critical line is reduced to the central point;
\item the central mass density is infinite, and the total mass is also infinite.
\end{itemize}

\subsubsection*{A.2 Truncated Isothermal Distribution with a Core}

Although the SIS is the simplest mass distribution, it is unphysical as it 
has an infinite central density, an infinite total mass, and therefore cannot adequately match true 
mass distributions. More complex mass distributions have therefore been 
developed to provide more realistic fits to observed clusters. The
Truncated Isothermal Distribution with a Core, which has finite mass and a finite central
density is quite popular as a lensing model, and one that we have used extensively and successfully in modeling 
cluster lenses.

The density distribution for this model is given by:
 \begin{eqnarray}
\Sigma(R)\,=\,{\Sigma_0 R_0  R_{t} \over {R_{t} - R_0}}
({1 \over \sqrt{R_0^2+R^2}}\,-\,{1 \over \sqrt{R_t^2+R^2}}),
 \end{eqnarray}
with a model core-radius $R_0$ and a truncation radius $R_t\,\gg\, R_0$.

The useful feature of this model, is the ability to reproduce a large range of mass
distributions from cluster scales to galaxy-scales by varying only the ratio $\eta$, that is defined as
$\eta=R_t/R_0$. There also exists a simple relation between the truncation radius of the mass
distribution and the effective radius $R_{\rm e}$ of the light distribution for the case of 
elliptical galaxies:
\begin{equation}
R_t\sim {4\over3} R_{\rm e}.
\end{equation}

Furthermore, this simple circular model can be easily generalized 
to the elliptical case (Kassiola \& Kovner 1993;  Kneib et al. 1996) by 
re-defining the radial coordinate $R$ as follows:
 \begin{eqnarray}
R^2\,=\,({x^2 \over  (1+\epsilon)^2}\,+\,{y^2 \over (1-\epsilon)^2})\,;
\ \ \epsilon= {a-b \over a+b},
\end{eqnarray}
Interestingly, all the lensing quantities can be expressed analytically (although using complex numbers) 
and the expressions for the 
same were first derived in Kassiola \& Kovner (1993).

The mass enclosed within radius $R$ for the model is given by:
\begin{eqnarray}
M(R)={2\pi\Sigma_0 R_0 R_{t}\over {R_{t}-R_0}}
[\,\sqrt{R_0^2+R^2}\,-\,\sqrt{R_t^2+R^2}\,+\,(R_t-R_0)\,],
\end{eqnarray}
and the total mass, which is finite, is:
\begin{eqnarray}
{M_{\infty}}\,=\,{2 \pi {\Sigma_0} {R_0} {R_t}}.
\end{eqnarray}
Calculating $\kappa$, $\gamma$ and $g$, we have,
\begin{eqnarray}
\kappa(R)\,=\,{\kappa_0}\,{{R_0} \over {(1 - {R_0/R_t})}}\,
({1 \over {\sqrt{({R_0^2}+{R^2})}}}\,-\,{1
\over {\sqrt{({R_t^2}+{R^2})}}})\,\,\,,
\end{eqnarray}
with: 
\begin{eqnarray}
2\kappa_0\,=\,\Sigma_0\,{4\pi G \over c^2}\,{D_{\rm LS}D_{\rm OL}
  \over D_{\rm OS}},
\end{eqnarray}
where $D_{\rm LS}$, $D_{\rm OS}$ and $D_{\rm OL}$ are respectively
the lens-source, observer-source and observer-lens angular diameter distances.

To obtain the reduced shear $g(R)$, given the magnification $\kappa(R)$, we solve
Laplace's equation for the projected potential $\varphi$, and evaluate the components of 
the amplification matrix following which we can proceed to solve directly for 
$\gamma(R)$, and then $g(R)$. 
\begin{eqnarray}
\varphi\,&=&\, \nonumber 2{\kappa_0}[\sqrt{R_0^2+R^2}\,-\,\sqrt{R_t^2+R^2}\,+
(R_0-R_t) \ln R\, \\ \nonumber \,\, 
&-&
\,R_0\ln\,[R_0^2+R_0\sqrt{R_0^2+R^2}]\,+
\,R_t\ln\,[R_t^2+R_t\sqrt{R_t^2+R^2}] ].\\
\end{eqnarray}
We can then derive the shear $\gamma(R)$;
\begin{eqnarray}
\gamma(R)\,
&=&\,\nonumber 
\kappa_0[\,-{1 \over \sqrt{R^2 + R_0^2}}\,
        +\,{2 \over R^2}(\sqrt{R^2 + R_0^2}-R_0)\,\\
\nonumber 
&+&\,{1 \over {\sqrt{R^2 + R_t^2}}}\,-\,
{2 \over R^2}(\sqrt{R^2 + R_t^2} - R_t)\,].\\
\end{eqnarray}

Scaling this relation by $R_t$ gives for $R_0<R<R_t$:
\begin{eqnarray}
\gamma(R/R_t)\propto {\Sigma_0 \over \eta-1} {{R_t} \over
  R}\,\sim\,{\sigma^2 \over R}.
\end{eqnarray}
where $\sigma$ is the velocity dispersion (note this is similar to the SIS case).

At larger radius, for $R_0<R_t<R$:
\begin{eqnarray}
\gamma(R/r_t)\propto {\Sigma_0\over\eta} {{R_t}^2 \over
  R^2}\,\sim\,{{M_{\rm tot}}
  \over {R^2}},
\end{eqnarray}
where ${M_{\rm tot}}$ is the total mass. In the limit that $R\,\gg\,R_t$, we have, 
\begin{eqnarray} 
\gamma(R)\,=\,{{3 \kappa_{0}}  \over {2
{R^3}}}\,[{R_{0}^2}\,-\,{R_{t}^2}]\,+\,{{2 {\kappa_0}} 
\over {R^2}} [{{R_t}\,-\,{R_0}}],
\end{eqnarray}

Lensing properties of the truncated isothermal distribution with a core in a nutshell:
\begin{itemize}
\item  $\kappa \neq \gamma$;
\item the tangential critical line once again corresponds to
the Einstein ring, and the radial critical line is a circle interior to the Einstein ring;
\item the central mass density is finite, and the total mass is also finite.
\end{itemize}

\subsubsection*{A.3 The Navarro-Frenk-White Model}

Although the truncated isothermal distribution with a core is very popular, it has never
been fitted to the results of numerical simulations, in contrast to the universal ``NFW''
density profile (Navarro, Frenk \& White 1997). In simulations of structure formation and 
evolution in the Universe, the NFW profile was found to be a good fit for a wide range of
 dark matter halo masses from $10^9 -10^{15}\,M_{\odot}$. The spherical NFW density 
 profile has the following form:
\begin{equation}
\rho(r) = \frac{\rho_s}{(r/r_s)(1+r/r_s)^2},
\end{equation}
where $\rho_s$ and $r_s$ are free parameters. It is often convenient to characterize the profile 
with the concentration parameter, $c_{\rm vir} = r_{\rm vir}/r_s$ where $r_{\rm vir}$ is the virial radius.  By integrating 
the profile out to $r_{\rm vir}$ and using $m_{\rm vir} = 200\rho_c(z)~4\pi \,r_{\rm vir}^3/3$, where $m_{\rm vir}$ is 
defined to be the virial mass and $\rho_c$ is the critical density of the universe, the concentration parameter can be related to $\rho_s$.

We now proceed to calculate the lensing properties of the NFW profile (more details can be
found in Wright \& Brainerd 2000). In the thin lens approximation, $z$ is defined as the optical axis and $\Phi(R,z)$ the 
three-dimensional Newtonian gravitational potential -- where $r=\sqrt{R^2+z^2}$. The reduced
two-dimensional lens potential in the plane of the sky is given by:
\begin{equation}
\varphi(\vec{\theta})=\frac{2}{c^2}\frac{D_\mathrm{LS}}
{D_\mathrm{OL}D_\mathrm{OS}}\int\limits_{-\infty}^{+\infty} 
\Phi(D_\mathrm{OL}\,\theta,z)\,dz,
\label{phi}
\end{equation}
\noindent where $\vec{\theta}=(\theta_1,\theta_2)$ is the angular position
in the image plane.

\noindent For convenience we introduce the
dimensionless radial coordinates $\vec{x}=(x_1,x_2)=\vec{R}/r_s=
\vec{\theta}/\theta_s$ where $\theta_s=r_s/D_\mathrm{OL}$. In the case
of an axially symmetric lens, the relations become simpler, as the
position vector can be replaced by its norm. The surface mass density
then becomes
\begin{equation}
\Sigma(x)=\int\limits_{-\infty}^{+\infty}\rho(r_s\,x,z)dz=2\rho_c r_s F(x).
\label{sigma}
\end{equation}

\noindent with
\begin{equation*}
F(x)=
\begin{cases}
\displaystyle{\frac{1}{x^2-1}\left(1-\frac{1}{\sqrt{1-x^2}}\mathrm{arcch} \frac{1}{x}\right)} & (x<1)\\
\displaystyle{\frac{1}{3}} & (x=1)\\
\displaystyle{\frac{1}{x^2-1}\left(1-\frac{1}{\sqrt{x^2-1}}
\arccos{\frac{1}{x}}\right)}
 & (x>1).
\end{cases}
\end{equation*}

\noindent and  the mean surface density inside the dimensionless radius $x$ is 
\begin{equation}
\overline{\Sigma}(x)=\displaystyle{\frac{1}{\pi x^2}\int\limits_{0}^{x}
2\pi x\Sigma(x)dx=4\rho_c r_s \frac{g(x)}{x^2}},
\label{sigma_moy}
\end{equation}

\noindent with
\begin{equation*}
g(x)=
\begin{cases}
\displaystyle{\ln{\frac{x}{2}}+\frac{1}{\sqrt{1-x^2}}\mathrm{arcch}
\frac{1}{x}} & (x<1)\\
\displaystyle{1+\ln{\frac{1}{2}}} & (x=1)\\
\displaystyle{\ln{\frac{x}{2}}+\frac{1}{\sqrt{x^2-1}}\arccos{\frac{1}{x}}}
 & (x>1).
 \end{cases}
\end{equation*}

The lensing functions $\vec{\alpha}, \kappa$ and $\gamma$ also have
simple expressions:
\begin{equation}
\left\lbrace
\begin{array}{llcl}
\vec{\alpha}(x)& = & \theta \, 
\displaystyle{\frac{\overline{\Sigma}(x)}{\Sigma_\mathrm{crit}}}
& = 4\kappa_s \, 
\displaystyle{\frac{\theta}{x^2}}g(x)\vec{e}_x \\
\kappa(x) & = & 
\displaystyle{\frac{\Sigma(x)}{\Sigma_\mathrm{crit}}}
& = 2\kappa_s \, F(x)\\
\gamma(x) & = &
\displaystyle{\frac{\overline{\Sigma}(x)-\Sigma(x)}{\Sigma_\mathrm{crit}}}
& = 2\kappa_s \left({\displaystyle{\frac{2g(x)}{x^2}-F(x)}}\right)
\end{array}
\right.
\label{sigma_nfw}
\end{equation}
with $\kappa_s= \rho_c r_s\Sigma_\mathrm{crit}^{-1}$. 
Noting $\vec{\nabla}_{\vec{x}}\,\alpha(x)=
(\partial_{x_1}\alpha,\partial_{x_2}\alpha)$ and $\phi=\arctan(x_2/x_1)$, 
we obtain some useful
relations for the following that hold for any circular mass distribution (Golse \& Kneib 2002):
\begin{equation}
\left\lbrace
\begin{array}{lll}
\kappa(x) & = & 
\displaystyle{\frac{1}{2\theta_s}\left(
\frac{\alpha(x)}{x}+\frac{\partial_{x_1}\alpha(\vec{x})}{\cos\phi}\right)}
\\[0.1cm]
\gamma(x) & = &
\displaystyle{\frac{1}{2\theta_s}\left(
\frac{\alpha(x)}{x}-\frac{\partial_{x_1}\alpha(\vec{x})}{\cos\phi}\right)}
\\[0.1cm]
\displaystyle{\frac{\partial_{x_1}\alpha(\vec{x})}{\cos\phi}} & = & 
\displaystyle{\frac{\partial_{x_2}\alpha(\vec{x})}{\sin\phi}}
\\[0.1cm]
\kappa(x)+\gamma(x) & = & \displaystyle{\frac{\alpha(x)}{\theta_s\,x}}
\end{array}
\right.
\label{useful_kg}
\end{equation}

By integrating the
deflection angle we obtain the lens potential $\varphi(x)$:
\begin{equation}
\varphi(x)=2\kappa_s\theta_s^2\,h(x),
\label{phi_h}
\end{equation}

\noindent where
\begin{equation}
h(x)=
\begin{cases}
\displaystyle{\ln^2{\frac{x}{2}}-\mathrm{arcch}^2
\frac{1}{x}} & (x<1)\\
\displaystyle{\ln^2{\frac{x}{2}}+\arccos^2{\frac{1}{x}}}
 & (x\ge1).
\end{cases}
\end{equation}

The velocity dispersion $\sigma(r)$ of this potential, computed with the Jeans
equation for an isotropic velocity distribution,
gives an unrealistic central velocity dispersion $\sigma(0)=0$.
In order to compare the pseudo-elliptical NFW potential with other
potentials, we define a scaling parameter $v_c$ (characteristic velocity)
in terms of the parameters of the NFW profile as follows:
\begin{equation}
v_c^2=\frac{8}{3}\mathrm{G}r_s^2\rho_c.
\label{sigma_c}
\end{equation}
Using the value of the critical density for closure of the Universe
$\rho_\mathrm{crit}=3H_0^2/8\pi\mathrm{G}$, we find
\begin{eqnarray}
 & & \frac{\rho_c}{\rho_\mathrm{crit}}=\frac{v_c^2}{H_0^2 r_s^2}=
1.8~10^3\,h^{-2}\times\, \left(\frac{r_s}{150~\mathrm{kpc}}\right)^{-2}
\left(\frac{v_c}{2000~\mathrm{km~s}^{-1}}\right)^2\nonumber.
\end{eqnarray}

Lensing properties of the NFW model in a nutshell:
\begin{itemize}
\item  $\kappa \neq \gamma$;
\item the tangential critical line is the Einstein ring radius,
and the radial critical line is a circle interior to the Einstein ring;
\item the central mass density is infinite, and the total mass is also infinite (however, only the
virial mass is of interest and that is calculable), and the central velocity dispersion is vanishing at the center.
\end{itemize}

\subsubsection*{A.4 Flexion for the Singular Isothermal Sphere}

For a Schwarzschild lens: by definition the first flexion is zero
everywhere except at the origin. This is of course due to the fact that
the gradient of the convergence is zero. A Schwarzchild lens does
produce "arciness" in the image. This effect is captured by construction by
the second flexion. Expressions for the first and second
flexion generated by the mass distribution of a singular
isothermal sphere have been provided by Bacon et al. (2005). We
reproduce their notation and consistent with our description in section 2.4.
The flexion produced by the SIS at angle $\theta$, measured from the centre
of the lens is given by:
\begin{equation}\label{fsis}
{\bf {\cal F}} = -\left[\frac{\theta_E}{2\theta^2}\right]e^{i\phi},
\end{equation}
where $\phi$ is the position angle with respect to the lens.
The first flexion ${\cal F}$ for the SIS profile can be written
as a vector and its direction points radially inward.

\begin{figure}
\epsfig{figure=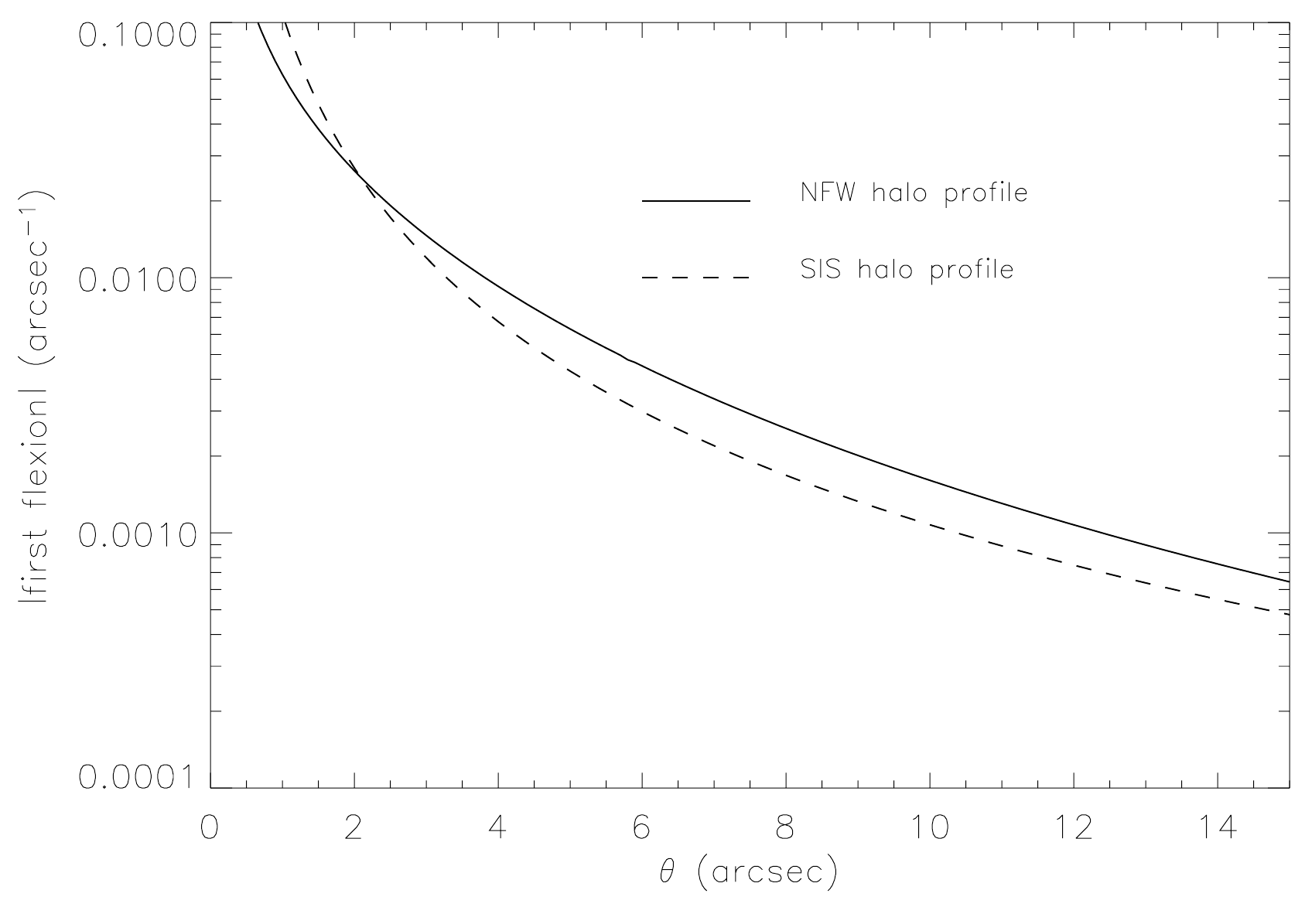,width=0.5\textwidth,angle=0}\label{lognfwfig}
\epsfig{figure=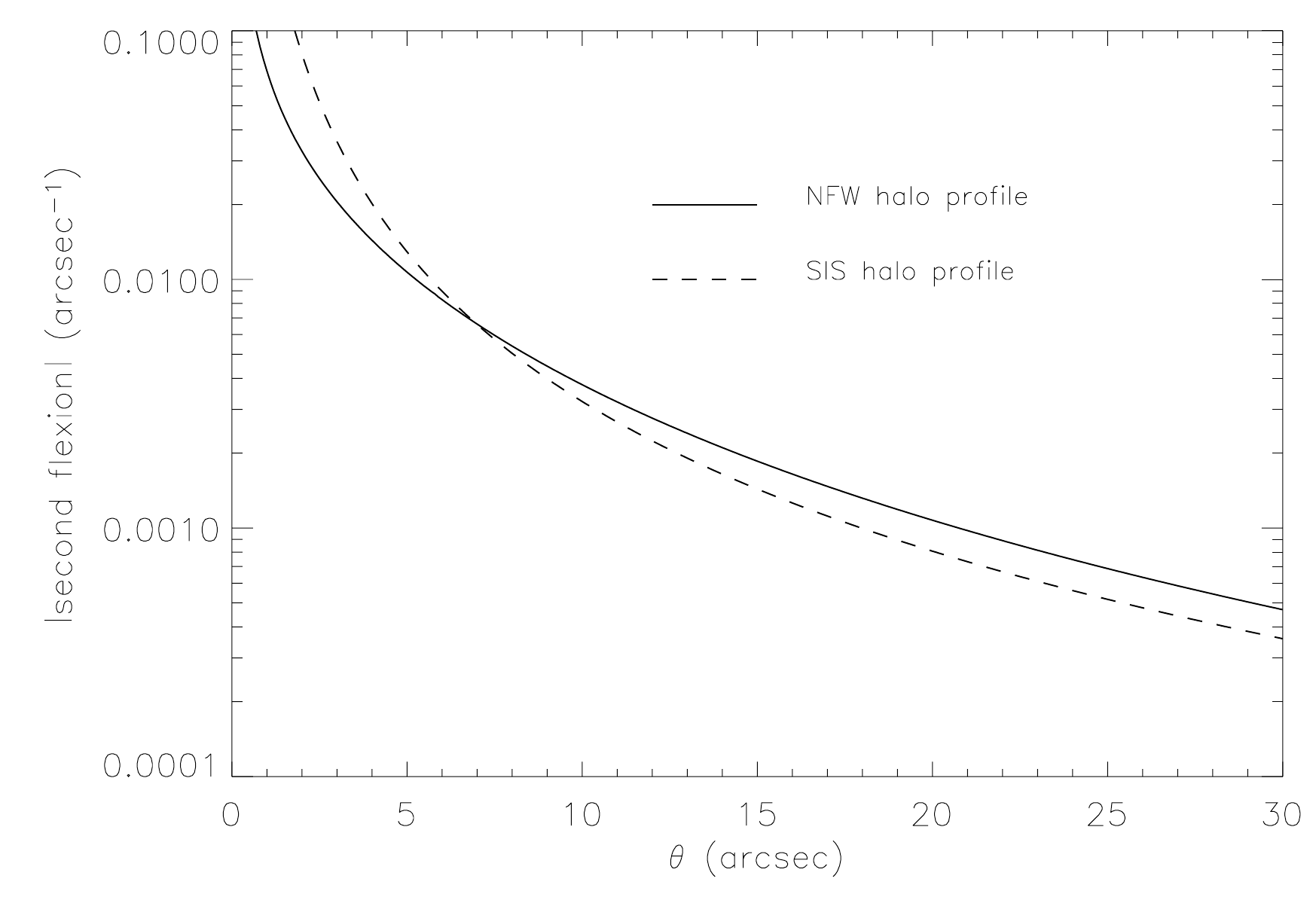,width=0.5\textwidth,angle=0}
\caption{Top: Comparison of the magnitude of first flexion due to a
NFW and a SIS halo of $M_{200}=1\times 10^{12} h^{-1}M_{\odot}$ at
redshift $z_{lens}=0.35$. Middle: A similar $\flex$ comparison but
this time the SIS halo has $M_{200}=1.8 \times 10^{12}
h^{-1}M_{\odot}$. Bottom: The magnitude of $\sflex$ for a NFW and a
SIS halo of $M_{200} = 1\times 10^{12}h^{-1}M_{\odot}$, where the
doubling in scale of the angular separation axis highlights the larger
range and amplitude of the second flexion.}
\end{figure}

The second flexion ${\cal G}$, as per the notation of Bacon et al. (2005)
is given by:
\begin{equation}
{\cal G}=\frac{3\theta_E}{2\theta^2}e^{3i\phi}\ .
\end{equation}
${\cal G}$ has a larger peak value than the first flexion for the
SIS, although both ${\cal F}$ and ${\cal G}$ fall off with the same power
law index away from the lens. For the explicit derivation of the flexion for more
complicated density profiles, namely the softened isothermal sphere
and the cosmologically motivated Navarro-Frenk-White profile, see 
Figure \ref{bacon2006} for a comparison as well as 
Bacon et al. (2006).

\end{document}